\documentclass[phd,10pt,norunningheaders,pagenumTR,noappendixpart,oneside,bold,letterpaper]{ubcthesis}

\newcommand{\chap}{\chapter}

\newcommand{\bie}{\begin{itemize}}
\newcommand{\eie}{\end{itemize}}
\newcommand{\ben}{\begin{enumerate}}
\newcommand{\een}{\end{enumerate}}
\newcommand{\btab}{\begin{tabular}}
\newcommand{\etab}{\end{tabular}}

\newcommand{\lab}{\label}

\newcommand{\beq}{\begin{equation}}
\newcommand{\eeq}{\end{equation}}
\newcommand{\bea}{\begin{eqnarray}}
\newcommand{\eea}{\end{eqnarray}}
\newcommand{\ba}{\begin{array}}
\newcommand{\ea}{\end{array}}

\newcommand{\na}{\nabla}
\newcommand{\pa}{\partial}

\newcommand{\fr}{\frac}
\newcommand{\sr}{\sqrt}
\newcommand{\ha}{\fr{1}{2}}

\renewcommand{\L}{\mathcal{L}}

\newcommand{\Screll}{{\mathcal L}}

\newcommand{\al}{\alpha}
\newcommand{\bt}{\beta}
\newcommand{\ga}{\gamma}
\newcommand{\de}{\delta}
\newcommand{\ep}{\epsilon}

\newcommand{\ze}{\zeta}
\newcommand{\et}{\eta}
\newcommand{\te}{\theta}

\newcommand{\ka}{\kappa}
\newcommand{\la}{\lambda}

\newcommand{\ph}{\phi}

\newcommand{\ps}{\psi}
\newcommand{\om}{\omega}
\newcommand{\Ga}{\Gamma}
\newcommand{\De}{\Delta}

\newcommand{\La}{\Lambda}
\newcommand{\Si}{\Sigma}

\newcommand{\Om}{\Omega}

\newcommand{\lp}{\left.}
\newcommand{\rp}{\right.}
\newcommand{\lb}{\left(}
\newcommand{\rb}{\right)}
\newcommand{\lhb}{\left[}
\newcommand{\rhb}{\right]}
\newcommand{\lbr}{\left\{}
\newcommand{\rbr}{\right\}}

\usepackage{setspace}
\usepackage{graphicx}
\usepackage{amsmath,amssymb,latexsym} 
\usepackage{ifthen} 

\usepackage{amsfonts}
\usepackage[dvips]{color}
\usepackage{color}
\usepackage{rotating,graphics,psfrag,epsfig}
\usepackage{lscape}
\usepackage{mathrsfs}
\usepackage{alltt}

\usepackage{float}
\floatstyle{boxed}
\newfloat{algorithm}{thp}{txt}
\floatname{algorithm}{Algorithm}

\usepackage{afterpage}

\newcommand{\resetcounters}{ 
	\setcounter{equation}{0} 
	\setcounter{figure}{0} 
	\setcounter{table}{0}   }       
\newcommand{\highQ}{false}

\def\lsep{\itemsep 0.05in}

\input epsf  
\epsfverbosetrue
\usepackage{psfrag}

\institution{The University Of British Columbia}
\institutionaddress{Vancouver}

\faculty{The Faculty of Graduate Studies}
\program{Physics}

\previousdegree{B.Sc., University of Brasilia, 1998}
\previousdegree{M.Sc., University of Brasilia, 2002}
  
\title{A Numerical Study of Boson Star Binaries}
\author{Bruno Coutinho Mundim}

\copyrightyear{2010}
\submitdate{\monthname\ \number\year}

\renewcommand\thepart         {\Roman{part}}
\renewcommand\thechapter      {\arabic{chapter}}
\renewcommand\thesection      {\thechapter.\arabic{section}}
\renewcommand\thesubsection   {\thesection.\arabic{subsection}}
\renewcommand\thesubsubsection{\thesubsection.\arabic{subsubsection}}
\renewcommand\theparagraph    {\thesubsubsection.\arabic{paragraph}}

\begin{document}
\frontmatter                    
 \singlespacing
 \maketitle
\begin{abstract}

This thesis describes a numerical study of binary boson stars within 
the context of an approximation to general relativity.  Boson 
stars, which are static, gravitationally bound configurations of a massive
complex scalar field, can be made gravitationally compact.  Astrophysically,
the study of gravitationally compact binaries---in which each constituent
is either a neutron star or a black hole---and especially the merger 
of the constituents that generically results from 
gravitational wave emission, continues to be of great interest. 
Such mergers are among the most energetic phenomena thought to occur in
our universe.  They typically emit copious amounts of gravitational 
radiation, and are thus excellent candidates for early detection by 
current and future gravitational wave experiments.

The approximation we adopt places certain
restrictions on the dynamical variables of general relativity 
(conformal flatness of the 3-metric), and on the time-slicing of the 
spacetime (maximal slicing), and has been previously used in the simulation 
of neutron stars mergers.  
The resulting modeling problem requires the solution of a coupled nonlinear 
system of 4 hyperbolic, and 5 elliptic partial differential equations (PDEs)
in three space dimensions and time.  We approximately solve this system as 
an initial-boundary value problem, using finite difference techniques and 
well known, computationally efficient numerical algorithms such as the 
multigrid method in the case of the elliptic equations.  Careful attention
is paid to the issue of code validation, and a key part of the thesis is 
the demonstration that, as the basic scale of finite difference discretization
is reduced, our numerical code generates results that converge to a solution
of the continuum system of PDEs as desired.  

The thesis concludes with a discussion of results from some initial 
explorations of the orbital dynamics of boson star binaries. 
In particular, we describe calculations in which
motion of such a binary is followed for more than two orbital periods, 
which is a significant advance over previous studies.
We also present results from computations in which the boson stars merge,
and where there is evidence for black hole formation.

\end{abstract}
 
\tableofcontents
\listoftables
\listoffigures
\chapter{List of Abbreviations}

\noindent
ADM: refers to the authors of the formalism, R.~Arnowitt, S.~Deser and C.~W.~Misner.\\

\noindent
AMR: Adaptive Mesh Refinement.\\

\noindent
CFA: Conformally Flat Approximation.\\

\noindent
EKG: Einstein-Klein-Gordon System of Equations.\\

\noindent
FDA: Finite Difference Approximation.\\

\noindent
FDE: Finite Difference Equation.\\

\noindent
NGS: Newton-Gauss-Seidel Iterative Technique.\\

\noindent
ODE: Ordinary Differential Equation.\\

\noindent
PDE: Partial Differential Equation.\\

\chapter{Acknowledgements}

First and foremost I would like to express my profound gratitude for my 
supervisor, Matthew W. Choptuik. Thank you very much for all your support
and wise guidance over the years this project took place. Second, I would
like to thank all the members of my graduate committee: William G. Unruh, 
Jeremy S. Heyl and Ingrid Stairs for all their input on this project 
and for asking all those hard questions that made me think deeper 
about physics in general and general relativity in particular. 
Also I would like to thank Bill Unruh for fostering interesting discussions 
taking place in the gravity group meeting. I enjoyed being part of it and
certainly will miss it a lot.

I should mention that this research was supported by NSERC and CIFAR 
as well as MPI-AEI.  I would like to thank Gerhard Huisken and 
the Albert Einstein Institute for their hospitality and
support during the time that I spent there.

It was great being a member of the numerical relativity group at UBC.
I learned a lot over the years with the innumerable interactions among
different members of the group. In particular, I would like to thank
the members of the ``old crew'': I\~{n}aki Olabarrieta and Kevin Lai 
for their patience in answering all my questions when I first started. 
My thanks goes to Roland Stevenson as well for sharing his programming 
expertise. Special thanks for the members of the ``new crew'', 
Jason Penner and Benjamin Gutierrez, for their promptness to help on the 
most diverse issues. Thank you guys for all your friendship.

I would like to thank my friends and colleagues from UBC for all their
friendship and for making my stay at UBC so vibrant. Special thanks for
my friend Sanaz Vafaei for all her support during the writing of this
manuscript. I am deeply indebted to several of friends in Brazil that have 
actively encouraged me to pursue my dreams.

I am grateful to my wife, Janine Kurtz, for all her dedication, love,
companionship and patience during my final steps to conclude this project.
Thanks for helping me to get through this difficult time.

I wish to thank my entire family for providing me a loving and nurturing
environment. Without their support and sacrifices during my
education, I would not have made it so far. I would like to dedicate 
this thesis to my parents, Elson R. Mundim and Selmara C. Mundim, for
all their love, support and encouragement during my entire life.

\mainmatter
 \resetcounters

\chapter{Introduction} \lab{intro}

One of the most revolutionary predictions of general
relativity---Einstein's spectacularly successful theory 
of the gravitational interaction---is the existence of gravitational
waves.
Binary systems composed of gravitationally compact 
objects~\footnote{An object with mass, $M$, is gravitationally compact if its radius, $R$, 
is close to its Schwarzschild radius, $R_S$, defined by $R_S=2GM/c^2$, where $G$ 
is Newton's gravitational constant, and $c$ is the speed of light.  For (spherical) 
black holes $R=R_S$, whereas for neutron stars 
$R_S/R$ is typically in the range $0.04$--$0.27$.}, 
such as two black holes, two neutron stars  or 
one neutron star and one black hole,
are among the most promising sources
of these waves.   According to Einstein's theory, such binaries 
have orbits which decay due to the emission of gravitational 
radiation.  This decay is thought to ultimately result in a violent
``plunge and merger'' of the two objects, characterized by
an extremely strong and dynamical gravitational field, and the 
emission of an especially strong burst of gravitational radiation. 
An international network of laser interferometer 
detectors (e.g. LIGO \cite{Kawabe:2008zz},
VIRGO \cite{Acernese:2008zzc}, TAMA300 \cite{Takahashi:2008zzb} and 
GEO600 \cite{Willke:2007zz}) has been
constructed to detect these bursts (as well as other sources),
where the detection 
sensitivity is best for compact objects 
having masses of the order 1 to 10 times that of 
the sun. However, even for such strong
sources, the terrestrial signal strength from a typical event 
(not likely in our own galaxy, or even in our local group of galaxies,
on the time scale of years) is expected 
to be much less than the inherent noise in the instruments.  
Faced with this situation, the most promising technique to extract the
signals from the noise is matched filtering \cite{Cutler:1994ys}, 
which involves comparison of the measured signal against a known (precomputed) 
family of waveforms.  Construction of such template waveforms is thus an 
urgent problem, and it requires {\em accurate} theoretical modelling of 
the process of compact binary inspiral.

Due to the complexity and nonlinearity of the Einstein field 
equations for the relativistic 
gravitational field---to which one must add the 
governing equations for any matter fields which are involved (such as 
a perfect fluid in the case of neutron stars)---accurate modelling 
of the late phases of binary inspiral and merger requires a numerical
approach. 
Now, as we will discuss in more detail 
shortly (Sec.~\ref{sec:overview_binaries}), the computer simulation of 
strongly interacting compact binaries has seen tremendous advances over 
the past few years.  This is especially true for the case of the inspiral and 
collision of two black holes, which although a formidable problem, 
is simplified relative to its neutron star counterpart by the fact that it does 
not involve matter fields.   For neutron stars, the need to 
solve the equations of general relativistic hydrodynamics in concert
with the Einstein equations leads to a host of other difficulties, 
including
the need to deal with shocks, turbulence and uncertainties concerning the equation
of state.  Moreover, the parameter space describing the generic
collision of two compact object has {\em many} dimensions, so that 
even with improved simulation techniques, identifying and extracting the 
key physics from the calculations will present a huge computational 
challenge for years to come.   Here we should emphasize that calculations 
of interacting binaries must be done in three spatial dimensions and time,
so that even a single calculation that is adequately resolved requires 
the use of high performance computing facilities.

Given this state of affairs there is considerable motivation to look
for simplified ``toy models'', with the hope and expectation
that they can provide insight into aspects of the 
compact binary problem, especially for the neutron star
case.  We have already noted that the plunge and merger phase of 
neutron stars in inspiral is characterized by a strong and dynamical 
gravitational field.   At least heuristically, the dynamics
is dominated by the bulk motion of the two stars, so that 
localized features in the matter---such as individual shocks, or 
small-scale turbulence---should have relatively little impact 
on the overall dynamics, or on the gravitational radiation which 
is produced.  We thus look for a matter model which can describe 
gravitationally compact objects, but where the equations 
of motion for the matter are easier to treat computationally than those for a 
relativistic fluid. 

The studies in this thesis involve precisely such a 
matter model.  Specifically, we adopt a single, massive, complex 
scalar field as a matter source.  The model
admits localized, time-independent, gravitationally-bound configurations
known as {\em boson stars}, which, through an appropriate choice 
of parameters, {\em can} be made compact.  
We can also set up initial data representing two stars, with subsequent 
evolution describing a variety of different kinds of encounters.
The equation of motion for the scalar field is simply 
the general relativistic wave equation, or Klein-Gordon equation,
whose numerical solution using finite difference techniques
is quite straightforward.

Moreover, in this thesis we take the ``toy model'' approach one step further. 
Although it is certainly possible to simulate boson stars in a fully
general relativistic setting~\cite{balakrishna:phd,Guzman:2004jw,Balakrishna:2006ru,
Palenzuela:2006wp,Palenzuela:2007dm}, we opt to study them
within the context of a relatively simple approximation to general 
relativity which has previously been used to study 
strong-gravity effects in several scenarios of astrophysical interest,
including the interaction of neutron stars.

More specifically, and as will be discussed in detail
in Chap.~\ref{formalism},  this approximation is based on the
\emph{conformally flat condition}, or CFC, for the spatial part of the metric (which 
is the fundamental dynamical variable that describes the gravitational field),
along with the \emph{maximal slicing condition}, which fixes the
time coordinatization of the spacetime.
Although the literature has traditionally used the acronym CFC for 
the resulting approximation to Einsteinian gravity,
we prefer to use
CFA, for \emph{conformally flat approximation}. This stresses the
fact that we
\emph{are} dealing with an approximation to general relativity.  Also, we
emphasize that although neither acronym makes explicit reference to the
maximal slicing condition, that choice of time coordinatization has always
been an essential ingredient of the approximation, and is so here.

The CFA is based on the heuristic assumption
that the dynamical degrees of freedom of the gravitational field,
i.e.~those associated with gravitational radiation, play a small role in at least
some phases of the strong field interaction of a merging binary,
and on the related fact that the amount of energy contained in the waves, expressed
as a fraction of the total mass-energy of the system, tends to be
small.
The CFA effectively eliminates the two dynamical
degrees of freedom present in general relativity, but still allows for investigation of the same kinds
of phenomena observed in the full general relativistic case.  These 
include the description of compact objects, the dynamics of their
interaction, and black hole formation.
It is also worth mentioning that it is still possible to study
gravitational wave generation within this approximation via a perturbative multipole
expansion of the metric components. Briefly, the incorporation
of radiation effects, although far from a trivial matter,
can be realized through the introduction of a radiation reaction potential
in the equations of motion for the matter model.
We also note that for spherically symmetric systems
the CFA is {\em not} an approximation, but can always be adopted through 
an appropriate choice of coordinates. Furthermore, use of the 
approximation for axially symmetric problems has indicated that
the results obtained mimic those of general relativity quite 
accurately~\cite{cook_etal:1996,Kley:1998uz,Dimmelmeier:2002bk,Dimmelmeier:2001rw}.
There is thus considerable motivation to perform additional studies of 
strong-field gravity using this approach.

The CFA was first studied in a theoretical/mathematical context by Isenberg in the
1970's \cite{Isenberg:2007zg}, and applied numerically (and independently) 
by Wilson \emph{et al.}~\cite{Wilson:1995uh,Wilson:1996ty,Mathews:1997vw,Mathews:1997nm,
matthews:1998,Marronetti:1998vm,marronetti:1998, marronetti_mathews_wilson:1999} 
in the 1990's to the study of coalescing
neutron-star binaries. In this prior work, Wilson and his collaborators presented evidence
that, for a realistic neutron-star equation of state, general
relativistic effects might cause the stars to individually collapse
to black holes prior to merging. Furthermore, they observed that
at least for some set of initial orbital parameters, strong-field effects
caused the last innermost stable circular orbit, or ISCO, to occur at a larger
separation distance, and thus at lower frequency, than was previously estimated
by post-Newtonian methods. This result had significance
for the possible detection of gravitational waves, since it placed
the frequency of radiation from coalescence closer to
the maximum sensitivity range of current laser-interferometric detectors.
However, the Wilson-Mathews compression effect was unexpected and 
controversial, and raised questions concerning the validity of the CFA.

Subsequently, 
Flanagan \cite{flanagan:1999} identified an inconsistency 
in the derivation of some of the equations of motion used in this study,
and suggested that
use of the correct equations would reduce
the crushing effect. A revised version of simulations 
was published shortly thereafter \cite{Mathews:1999km}: a key claim resulting 
from this work was that the
crushing effect was still present, although the magnitude of the observed 
effect {\em was} reduced relative to the
previous calculations. Further comparisons between a fully relativistic code
and its CFA counterpart in the context of head-on collisions of neutron stars showed
the presence of this effect and lent more credibility to the earlier 
calculations \cite{Wilson:2002yh}. However, the most recent simulations 
\cite{Miller:2005xq} aimed at studying the possible crushing phenomenon
actually indicate a \emph{decompressing} effect on the neutron stars.
Still, this result is not in direct contradiction to Wilson {\em et al}'s results
since the initial data used for the two sets of simulations differ.
For a more complete review of the history of this controversy, as well
as a possible explanation 
for the neutron star crushing effect, the reader should refer 
to the work by Favata \cite{Favata:2005da} and references therein.
We should also note at this point that since this thesis work was started,
new ``waveless'' formalisms have been developed \cite{Shibata:2004qz,Schaefer:2003ra},
and have been used to improve the accuracy of certain compact binary calculations~\cite{Uryu:2005vv}
in the inspiral phase, relative to fully relativistic computations.

An ultimate goal of the work started in this thesis is to 
determine to what extent the CFA {\em is} a good approximation 
for the modelling of {\em general} compact binaries.
From this point of view,  it is particularly interesting
to study the CFA within the context of a simpler matter model than that previously adopted,
and 
this provides an important motivation for our use of boson stars.
Some of the main questions we wished to address are as follows:
\begin{itemize}
\item
Would boson stars collapse individually before merger, 
or is this phenomena strongly dependent on how the matter is modelled?  
Favata \cite{Favata:2005da} posited a mechanism that would tend to
compress the
neutron stars given a particular set of conditions.
Does
that analysis apply to the binary boson star as well?
\item
How well does the CFA approximate the full general relativistic equations?  
Can we shed some light on the nature of the radiative degrees of freedom
in the full theory from a detailed study of the differences between results
obtained using the CFA and the full Einstein equations?
\item 
What is the final result of the merger? Can we compare the results to those
obtained from other techniques, including fully general relativistic
calculations? 
\item
Where is the ISCO? Do the results obtained match those
seen in collisions of fluid stars,  or can they at least be compared 
qualitatively?
\end{itemize}

In order to answer these and related questions, the Einstein equations (and ultimately the 
equations that result from adopting the CFA) have to be cast in a
form suitable for numerical computation. What follows is a brief description
of our modelling procedure; more details will be provided in subsequent chapters.

The current work makes use of the $3+1$ or ADM (Arnowitt-Deser-Misner)~\cite{ADM:1962}
decomposition of the Einstein equations. One of the main features of the $3+1$ approach, 
which underlies the majority of the work in numerical relativity~\footnote{Numerical 
relativity can be defined as the subfield of general relativity that is concerned
with the solution of the Einstein equations, as well as the equations of motion
for any matter sources, using computational methods.}, 
is that it 
provides a prescription for disentangling the
dynamical field variables from those associated with the coordinate invariance
of general relativity. 
Specifically, the spacetime metric components are grouped 
into 4 kinematical components which encode the coordinate 
freedom of the theory (the {\em lapse function} and three components of the {\em shift vector}), and
6 dynamical ones (the components of a 3-dimensional metric induced in a
constant-time hypersurface).  Now, for any specific calculation, the coordinate 
system must be completely fixed by giving prescriptions for the lapse and 
shift.  In particular, the time coordinate is fixed by a choice of the lapse
function, and, as already mentioned, in this work we follow previous studies using 
the CFA and adopt so-called maximal slicing 
(a very commonly adopted slicing condition
in numerical relativity that was originally proposed by Lichnerowicz
\cite{lichnerowicz:1944}).
A key property of maximal slicing is that its use
inhibits the focusing
of the world lines of observers that move orthogonally to the hypersurfaces:
such focusing can result in
the development of coordinate singularities, and is also associated with
the formation of physical singularities.
Another important point is that this coordinate choice generally
results in a well-posed
elliptic equation for the lapse function. 

As the name suggests, the CFA requires the spatial 3-geometry to be 
conformally flat. This 
reduces the number of independent components of the
spatial metric from 6 to 1, and the single, non-trivial function that then
defines the spatial metric is called the {\em conformal factor}.
Given the maximal slicing condition, it transpires 
that conformal flatness implies that all
non-trivial
components of the 4-metric are governed by elliptic equations.

Within the CFA, the dynamics of the gravitational field is completely
determined by the dynamics of the matter source(s), which in our case
is a complex scalar field that satisfies the Klein-Gordon equation.
Overall then, the model considered here---which we will
hereafter frequently refer to as ``our model''---is governed 
by a mixed system of 4 first-order hyperbolic partial differential 
equations (PDEs), coming from the complex Klein-Gordon equation,
and 5 elliptic PDEs.
We treat this system as an initial-boundary value problem
and the thesis focuses on the development, testing and preliminary 
use of a code for its numerical solution.  Our computational
approach is based on finite differencing~\cite{mitchell_griffiths}, 
and we use approximations that are second-order accurate in 
the grid spacing.

A numerical evolution of our model starts 
with the specification of
initial conditions for the complex scalar field.  For example, for the
relatively simple case of two identical boson stars that are initially at rest,
the solution of the coupled
Einstein/scalar field equations in spherical symmetry, and with
a time-periodic ansatz for the scalar field,  provides a static
star profile that can be duplicated and interpolated onto a
three dimensional domain. This sets initial values representing a
binary system. 
Discrete versions of the elliptic equations governing the geometric quantities
are then solved at the initial time using the multigrid method~\cite{brandt:1977}. 
The matter field values are then time advanced using a
second order discretization of the Klein-Gordon equation,
which, along with the elliptic equations for the advanced-time metric
unknowns, is solved iteratively.

Our code has been subjected to a thorough series of tests that 
assess convergence of the numerical results with respect to the basic 
discretization scale.
Additional evidence for code validity involves the use of conserved quantities, 
as well as the technique of independent
residual evaluation (which is defined and discussed in Chap.~\ref{numerical}).  
Overall, results from these tests indicate that the code is correctly solving the 
equations of motion that define the model.  Specific results calculated using 
the code include long-term evolution of an orbiting binary system, 
(more than two orbital periods), as well as high speed head-on collisions.
These results are 
promising and suggest that, especially with enhancements such as the incorporation of 
adaptive mesh refinement and capabilities for parallel execution, the code 
will be a powerful tool for investigating strong-gravity effects in the interaction
of boson stars.

We now proceed to two short overviews of subjects that are germane to 
this work.  First, in Sec.~\ref{sec:overview_binaries}, 
the current status of the numerical 
modelling of gravitationally-compact binary systems is summarized.
Second, some of the key previous studies of boson stars are reviewed
in Sec.~\ref{sec:boson_star}.

\section{Overview of Numerical Simulations of Compact Binaries} \label{sec:overview_binaries}

\subsection{Black Hole Binaries}

As mentioned previously, accurate modelling of compact star binaries
is needed to help extract the gravitational wave signal from the noise in
current and planned terrestrial gravitational wave interferometers. 
This section briefly describes the status of 
numerical relativistic modelling of these binaries by means of a quick 
survey of the pertinent literature.  We first note that there are several 
good review papers on the general subject, including the following:
\begin{itemize}
\item Rasio \& Shapiro's review of coalescing binary neutron stars~\cite{Rasio:1999ku}.
This covers work up to 1999 on coalescing binary neutron stars---using Newtonian, post-Newtonian 
and semi-relativistic approximations---and 
is highly recommended for familiarizing the interested reader with earlier research
in the field.
\item Baumgarte \& Shapiro's extensive review of the use of numerical relativity to 
model compact binaries~\cite{baumgarte:2003b}. 
\item Pretorius' more recent review on black hole collisions~\cite{Pretorius:2007nq}.
\end{itemize}

Efforts to accurately model the collision of compact objects using 
fully general relativistic calculations began 
with the work on black hole head-on collisions by Smarr and collaborators 
in the 1970's~\cite{Smarr:1976qy}.  These calculations established numerical
relativity as a sub-field of general relativity in its own right,
and indicated that one could expect 
perhaps a few percent of the total mass-energy of the system to be emitted
in gravitational waves.
In the early 1990's,
Anninos \emph{et al.}~\cite{Anninos:1993zj} revisited and extended
the  head-on calculations.  A major result from this effort was 
the demonstration that the
extracted gravitational waveform could be well-matched
to (perturbative) black hole normal modes. 
In 1993 the Binary Black Hole Grand Challenge Alliance, involving many
investigators from several different institutions in the US, was founded with the 
mission to provide stable, convergent algorithms to compute the gravitational
waveforms from black-hole collisions.  Although this research spurred 
the development of computational infrastructure that is still being 
used by the numerical relativity community, the project ended well short 
of its stated physics goals.   The most important simulations that 
resulted from the 5-year effort described the propagation of a single
black hole through a three dimensional computational domain~\cite{Cook:1997na}.
Nonetheless, this result constituted an important step 
towards the solution of the binary black hole problem, especially when 
contrasted with previous computations where the black-holes were kept fixed in the
domain by specific choices of coordinates.  
It was not until 2004 that a significant amount of orbital
motion for a black hole binary was successfully simulated, when 
Br\"{u}gmann and collaborators published results~\cite{Bruegmann:2003aw}
from computations of a full orbital period for a widely-separated black hole pair.

2005 then saw the publication of breakthrough results by Pretorius
\cite{Pretorius:2005gq,Pretorius:2006tp}, and the beginning 
of the most productive period of binary black hole research.
Pretorius presented calculations
that tracked the evolution of a binary black hole spacetime
through several orbits, the plunge and merger phase, and into the late-time
stage where the final black hole quickly rings down to a near-stationary (Kerr)
configuration.
Furthermore, Pretorius was able to extract a reasonably 
accurate gravitational wave signal from 
the whole evolution.  
Pretorius' work had an electrifying effect on research
in the field, and other groups, most notably the University of Texas at
Brownsville~(UTB)~\cite{Campanelli:2005dd,Campanelli:2006gf,
Zlochower:2005bj} and NASA/Goddard~\cite{Baker:2005vv,Baker:2006yw} teams,
 were quickly able to produce comparable results
using similar \cite{Szilagyi:2006qy} 
or different techniques (moving punctures) 
\cite{Baker:2005vv,Baker:2006yw,Campanelli:2005dd,Campanelli:2006gf,
Zlochower:2005bj,Diener:2005mg,Brugmann:2008zz}.
Moreover, it was possible to compare the various simulations and to show
that the results were generally consistent to within the level of numerical
error~\cite{Baker:2007fb}. 

Since that time, a large number of distinct simulations have been 
performed, and some interesting phenomenology has been unearthed.
Particularly noteworthy are the ``kicks'' that the final black holes 
formed from mergers can experience due to asymmetric emission of
gravitational radiation during the 
coalescence~\cite{Sopuerta:2006et,Baker:2006vn,Gonzalez:2006md,Herrmann:2007zz}.
When the merger process is itself sufficiently asymmetric---as is 
the case for the inspiralling collision of two holes with unequal 
masses---the gravitational waves produced 
carry away net linear momentum.  This imparts an equal and opposite momentum,
or kick, to the final black hole,
and the magnitude of the resulting
recoil can have astrophysical implications. For example, depending on 
the mass of the stellar cluster in which the binary is embedded,  the 
merged black hole may actually be ejected from the cluster or even the 
host galaxy.
Early calculations concerning these kicks---which used non-spinning black
holes (no intrinsic angular momentum)---have now been extended to
investigate the effects of the black hole spins on the recoil velocity of the final
black hole~\cite{Choi:2007eu,Herrmann:2007ex,Marronetti:2007wz,Baker:2007gi,Gonzalez:2007hi,Brugmann:2007zj}.
These studies have indicated that impressively large kicks ($\sim4000 {\rm km/s}$)
can result from misalignment
of the black holes' intrinsic spins and the orbital angular momentum, as well
as from large initial spins.
In addition, pioneering studies by the UTB/Rochester Institute of Technology group
\cite{Campanelli:2006uy,Campanelli:2006fg,Campanelli:2006fy,Campanelli:2007cga}
have shown that the black holes' initial spins
can have a significant impact on the orbital dynamics,
as well as on the emitted gravitational waveform at late times.

Another interesting effect that is still being studied concerns
the effect of orbital eccentricity on the gravitational radiation waveforms;
summaries of recent progress on this problem are given in
\cite{Sopuerta:2006et,Hinder:2007qu,Sperhake:2007gu,Pretorius:2007jn,Washik:2008jr}.
Moreover, comparison of the waveforms resulting from numerical simulations on
the one hand, and
post-Newtonian calculations on the other, is also a rapidly developing industry
\cite{Baker:2006kr,Buonanno:2006ui,Berti:2007fi,Hannam:2007ik,Baker:2008mj}.
Especially notable in this regard are
the very high accuracy simulations, recently reported
by the Cornell-Caltech group,
that use pseudo-spectral techniques~\cite{Scheel:2006gg,Pfeiffer:2007yz}.
Gravitational waveforms from some of these studies have now been compared
to those calculated from post-Newtonian techniques \cite{Boyle:2007ft}.
These comparisons provide conclusive evidence that 
the best numerical results can now provide  more
accurate waveforms for the late phases of inspiral than those computed using
post-Newtonian methods.

Finally, it is worth mentioning two studies
of relativistic (high speed) black hole collisions
\cite{Sperhake:2008ga,Shibata:2008rq}. This type of interaction is of
particular current interest since it ties in with the phenomenology of
possible mini-black hole
formation in new generations of particle accelerators such as the
Large Hadron Collider at CERN. 
Moreover, ultrarelativistic black-hole collisions represent largely
unexplored territory where fundamental theoretical issues might be 
addressed.  These include the possibilities of naked singularity
formation, as well as the production of an end state containing
three or more black holes, starting from a two black hole initial 
state.  Although the preliminary results
show no evidence for naked singularities, they {\em do} provide explicit 
demonstration of the copious amount of gravitational
radiation that can be emitted from such encounters.  Specifically, 
in some of the cases simulated, as much as 25\% of the initial
rest-mass energy is emitted as gravitational waves. This 
approaches the upper bound of 29\% estimated 
by Penrose~\cite{penrose_maxrad:1974,D'Eath:1992hb}.

\subsection{Neutron Star Binaries}

The first fully relativistic simulation of binary neutron star inspiral 
and merger was performed 
by Shibata and Ury\={u} in 2000~\cite{Shibata:1999wm}.
Adopting a simple equation of state, they examined the issue of 
the nature of the end state from the coalescence of two equal-mass stars.
Perhaps not surprisingly, their work indicated that the final 
configuration depended on the compactness of the initial neutron stars: 
more compact stars would lead to black hole formation, while those 
less condensed would merge to form a high-mass, differentially
rotating neutron star. 
In 2002, subsequent work by the same authors, 
using increased resolution in the 
simulations, produced gravitational waveform estimates for the mergers~\cite{Shibata:2002jb}.
Additional research 
by Shibata and his collaborators \cite{Shibata:2003ga} has focused 
on unequal mass binaries, as well as on the use of several different approaches
to mimic realistic equations of state \cite{Shibata:2005ss,Shibata:2006nm}.

Work by other groups has followed, and there have now been several
reports of stable long term simulations that extend over a few 
inspiraling orbits through to merger (the end state is typically a 
black hole)~\cite{Duez:2002bn,Miller:2003vc,Anderson:2007kz,Baiotti:2008ra}.
It is worth noting that in all of the work mentioned this far,
effects due to magnetic fields were neglected.   However, within the past 
year or so, preliminary results involving magnetized neutron star binaries have 
started to appear~\cite{Liu:2008xy,Anderson:2008zp}.  Finally,
three dimensional general relativistic simulations of mixed binaries---where one
constituent is a black hole and the other is a neutron-star---are also underway; 
see~\cite{Shibata:2006ks,Shibata:2006bs,
Loffler:2006nu,Etienne:2007jg,Yamamoto:2008js} and references therein.

\section{An Overview of Boson Stars} \label{sec:boson_star}

The concept of a boson star can be traced back to work by
Wheeler and collaborators~\cite{wheeler:1955,Power:1957zz} who considered 
self-gravitating ``lumps'' of massless fields (electromagnetic or 
gravitational), which they dubbed \emph{geons}.  
Unfortunately such configurations were soon found to be unstable and 
research on them quickly subsided.
However, the geon idea provided impetus for the 
investigation of scalar fields in the context of general relativity. 
The earliest published work of direct relevance to our study is due to 
Kaup~\cite{kaup:1968}, who pioneered the study of self-gravitating 
configurations of a massive complex scalar field, governed by 
the coupled system of Einstein and Klein-Gordon (EKG) equations.
The main result of his work, which assumed spherical 
symmetry and a harmonic ansatz for the time dependence
of the scalar field, was to determine the equilibrium 
states of the system.  These are precisely the states that
we now know as boson stars (Kaup referred to them as ``scalar geons'').
He emphasized 
the similarities with neutron (fluid) stars, and 
established the conditions under which the stars would 
be stable (or unstable) to nonadiabatic radial perturbations.
Indeed, Kaup was primarily interested in the issue of stability, 
and how resistant to gravitational collapse the massive 
complex scalar field was.

At about the same time, Ruffini and Bonazzola~\cite{ruffini:1969} 
investigated the same spherically symmetric system,
and were also able to determine the equilibrium
boson star solutions.
However, in contrast to Kaup, these authors viewed 
the model as representing a semi-classical approach in
which the Klein-Gordon field was fundamentally quantum 
mechanical. 
From considerations that parallel the calculations performed 
by Chandrasekhar for fluid stars~\cite{Chandrasekhar:1931ih}, 
one of their main results was the notion of a
critical mass for boson stars; that is a mass above which 
the pressure support from the star (supposedly arising from
the Heisenberg uncertainly principle) would be insufficient
to resist gravitational collapse.  They showed that the 
critical mass for bosonic matter scaled differently with 
the particle mass, $m$, than it did for fermionic (neutron star)
matter, and that for any reasonable values of $m$, the 
resulting sizes and mass of the boson stars would be so 
small as to be astrophysically inconsequential.

However, many years later, a seminal paper by Colpi 
and collaborators~\cite{colpi:1986} 
showed that if the scalar field was endowed with a nonlinear 
self-interaction potential, then by suitable choice of the 
coupling constants appearing in the nonlinear terms, boson stars 
with masses comparable to neutron stars could be achieved using
a plausible value for $m$.  
Specifically, for the case of a $\lambda \vert \phi \vert^4$ potential, 
and for sufficiently large values of the dimensionless coupling 
constant~\footnote{Colpi and collaborators adopt $c=\hbar=1$.},
$\Lambda=\lambda \, \hbar c/4\pi G m^2$,
the structure of the static solutions was found to change
significantly relative to the stars with $\lambda=0$ that had originally 
been studied.  
For large $\Lambda$ the stellar density profile exhibits a relatively
slow decline out to some coupling-dependent characteristic radius, and 
this leads to a significant enhancement of the star's mass.  In fact, it was 
demonstrated in~\cite{colpi:1986} that as $\Lambda\to\infty$ the 
maximum stellar mass has the {\em same} scaling dependence on the particle 
mass, $m$, as for fermionic stars.
Although this result led to revived interest 
in astrophysical implications of boson stars---including the 
possibility that they might constitute at least some fraction of the 
dark matter that is currently unaccounted for---we must emphasize 
that no fundamental scalar boson has yet been detected, nor is 
there any credible evidence for the existence of boson stars.

The work by Colpi~{\em et.~al.}~was followed by a burst of 
activity in boson star research, 
with several groups carrying out
investigations focused mainly on their stability 
properties~\cite{lee:1987,Friedberg:1987,
Gleiser:1988rq,ferrell_gleiser:1989,gleiser_watkins:1989,lee_pang:1989}. 
In 1991, Jetzer published a comprehensive review of the 
subject~\cite{Jetzer:1991jr}, and this remains a very useful general 
reference, especially for newcomers, as does the review by Lee
and Pang~\cite{Lee:1991ax}.

What one might identify as the 
modern era of boson star research began 
with work by Seidel and Suen~\cite{seidel_suen:1990}.
These authors considered the {\em dynamical} evolution of  
the full spherically-symmetric Einstein-Klein-Gordon equations,
and were thus able to directly address issues such as 
dynamical stability.  Their numerical experiments 
indicated that the stars, parametrized by the central modulus,
$\phi_0$, of the scalar field, belonged to two ``branches'':
one stable, the other unstable.  Further, and in accord 
with analogous fluid-star calculations, as well as perturbation
theory, the change in stability was found to coincide with 
the boson star that had maximal mass.
When perturbed, stars that were on the stable branch were found to 
oscillate and radiate scalar radiation with a particular frequency, 
eventually settling down to a somewhat less massive stable star.
On the other hand, stars on the unstable branch that were perturbed
would either undergo gravitational collapse to a black hole or
radiate some scalar field and migrate to a configuration on 
the stable branch with smaller mass.  
Later, in collaboration with Balakrishna, Seidel and 
Suen~\cite{Balakrishna:1998} extended these studies to include 
excited states~\footnote{The boson stars discussed thus far, and which constitute
the focus of this thesis, are so-called {\em ground state} boson star, where the 
spherically-symmetric scalar field profile is nodeless. The excited states have
one or more nodes (zero-crossings).} and boson stars
with nonlinear self-interactions.  
The dynamics of spherically symmetric boson stars have also been studied 
by other authors, including Choptuik and 
collaborators~\cite{hawley_choptuik:2000,Lai:2007tj}, who have investigated 
the nature of critical gravitational collapse~\cite{gundlach:2003} in this 
context.

In addition to the spherically symmetric work, the past decade 
or so has seen the accumulation of a considerable literature 
on axially symmetric boson stars.  Since most of this work is not 
closely related to ours, we direct the interested reader
to Schunck and Mielke~\cite{schunck:2003}, which remains the most recent
comprehensive review of boson stars.

Finally, there have been a few previous efforts in which 
the dynamical evolution of boson stars has been studied in three 
spatial dimensions.  This includes work by 
Balakrishna~{\em et.~al.}~\cite{Guzman:2004jw} and 
Palenzuela~{\em et.~al.} which, in both cases, involved the 
solution of the full Einstein equations.

\section{Summary of Thesis}

The remainder of this thesis is organized as follows:

Chap.~\ref{formalism} discusses the principal formalisms and 
(continuum) approximations used in this thesis. 
A quick overview of general relativity is given, primarily to fix notation,
and to introduce key geometrical quantities (metric, curvature, etc.).  
This is followed by a review of the $3+1$ decomposition of spacetime,
in which the 4-geometry of spacetime is viewed as the time
history of the 3-geometry of a spacelike hypersurface. 
We then proceed to a description of our matter model (a massive complex 
scalar field) which includes a derivation of the associated equations of motion
for the scalar field.
This is followed by a section devoted to a general conformal 
decomposition of the $3+1$ equations. 
The conformally flat approximation is then introduced,
and the final form of the equations 
of motion for our model is derived.  Related issues such as the boundary conditions 
and the ADM mass are also discussed.

Chap.~\ref{initialdata} is concerned with the generation of initial data 
for our model. 
We first give a description of the construction of spherically symmetric, ground 
state boson stars. 
We then detail the procedures 
we used to interpolate one of two stars onto
the three dimensional computational domain, and to provide the stars 
with initial velocities through the application of approximate Lorentz boosts.

Chap.~\ref{numerical} begins with a review of some basic concepts related 
to the finite difference approximation (FDA) 
of partial differential equations (PDEs).  This is followed by a description
of a simple example that illustrates the specific scheme we used to 
discretize our hyperbolic PDEs.
We then discuss the strategies and techniques we use to 
establish the validity of our code and to assess the 
quality of the numerical results.   The chapter concludes with 
a brief overview of the multigrid method in general, as well
as a description of the specific multigrid algorithm we 
implemented to solve our discretized elliptic PDEs.

The majority of the original research contributions in this thesis are 
contained in Chap.~\ref{results}. 
The chapter starts with a discussion of the construction of the numerical 
code, including a description of the overall algorithmic flow.  This 
is followed by the results from a thorough series of code tests that 
employed several different types of initial data.
The dynamics of head-on 
collisions of boson stars is then considered. The last part of the chapter 
discusses results related to orbital dynamics.  We perform 
a modest parameter space survey and identify three distinct endstates 
for the calculations:  1) long term orbital motion, 2) merger to 
a conjectured rotating and pulsating boson star, and 3) merger to
a conjectured black hole.

App.~\ref{ap:bsidpa} documents the use of a publicly-available 
FORTRAN subroutine that generates initial data for a spherically-symmetric
boson star, and which implements a general polynomial self-interaction 
term for the scalar field. We feel that this routine should be of significant 
utility to others who wish to study the dynamics of general relativistic
boson stars.

App.~\ref{ap:stencil} lists all of the finite difference formulae we 
have used in the thesis and includes a note on the regularization 
of differential operators in curvilinear coordinate systems prior 
to finite-differencing.

App.~\ref{ap:pdefda} contains brief documentation of a set of Maple
procedures which facilitate the construction and coding of 
finite-difference discretizations of generic systems of PDEs. Again, some 
effort has been expended in designing the routine so that it is potentially 
useful in other contexts, and for other researchers.

Finally, 
App.~\ref{ap:ngs} contains a brief review of the relaxation techniques
that we have used in this work.

\section{Conventions, Notation and Units}\lab{notation}

We adopt the abstract index notation for tensors as defined and discussed 
in Wald~\cite{wald}. 
In particular, letters from the beginning of the Latin alphabet, 
$\{a,b,c,...\}$, denote abstract indices.  We then use two sets 
of indices for tensor {\em components}: Greek indices $\{\mu, \nu, ...\}$ 
range over the spacetime values $0, 1, 2, 3$ (where $0$ is the time index),
while the subset of Latin indices $\{i,j,k,l,m,n\}$ range over the spatial
values $1,2,3$.  The Einstein summation convention applies to both of these 
component index types.  We also adopt Wald's sign conventions, so, in 
particular,
the metric signature is $(-,+,+,+)$. 

The totally symmetric and
totally antisymmetric parts of a tensor of type $(0,2)$ are defined 
by
\beq
T_{(ab)}= \ha (T_{ab}+T_{ba}) \, ,
\eeq
and
\beq
T_{[ab]}= \ha (T_{ab}-T_{ba}) \, ,
\eeq
respectively. 

Additionally, we use a common terminology from computational science in which 
the sets of time-dependent PDEs with dependence on 1, 2 and 3 spatial 
dimensions (independent variables) are referred to as 1D, 2D and 3D,
respectively.  In particular, PDEs describing spherically symmetric 
systems in which the field variables depend on time and a radial 
coordinate, $r$, are referred to as 1D. 

A variety of differentiation operators are used below. 
$\na_a$ and $D_a$ denote the covariant derivative operators compatible 
with the spacetime metric, $g_{ab}$, and induced hypersurface metric,
$\ga_{ab}$, respectively. 
The Lie derivative along a vector field $v^a$ is denoted by $\Screll_v$.
Ordinary derivatives are represented by different notations according to the context.
For example the ordinary derivative of the function $f=f(x,y,z)$ 
with respect to 
the coordinate $x$ will be denoted in one of the following ways: 
${\pa f}/{\pa x} \equiv
\pa_x f \equiv f_{,x}$. For time dependent functions, $g=g(t,x,y,z)$, we 
also sometimes denote temporal differentiation using an overdot, so
$ \dot{g} \equiv{\pa g}/{\pa t} \equiv \pa_t g$. Similarly, 
 for functions of one spatial variable, $g=g(t,r)$ or $g=g(r)$, we 
sometimes use the prime notation for the first (spatial) derivative: 
$g'\equiv \pa g/\pa r$ or $g'\equiv dg/dr$.
Finally, a variant of geometric units is used throughout this thesis.  
Specifically, Newton's gravitational
 constant, $G$, the speed
of light in vacuum, $c$, and the scalar field mass parameter, $m$, are all set to unity: $G=c=m=1$.

 \resetcounters
\def\calH{{\mathcal H}}
\def\calT{{\mathcal T}}
\def\calR{{\mathcal R}}
\def\calG{{\mathcal G}}
\def\H#1#2{{\mathcal H}^{#1}{}_{#2}}
\def\CFA{{\rm CFA}}

\chap{Formalism and Equations of Motion} \lab{formalism}

As described in the introductory chapter, this thesis focuses on the 
numerical solution of a model that describes the gravitational
interaction of boson stars within an approximation to Einstein's 
general theory of relativity.  The bulk of this chapter is
devoted to a discussion of the formalism 
underlying the model, and the derivation of the set of partial
differential equations (PDEs) which govern it.
We start with a brief review of the general relativistic field 
equations in Sec.~\ref{sec:GR_field_equations}. We then proceed 
to a discussion of the so called $3+1$ (or ADM) formalism
which underlies most of the computational work~\footnote{Here we
note that an alternate formulation of Einstein's equations, known as 
the generalized harmonic approach, has also had a major impact on numerical
relativity in the past few years.} in which systems of general 
relativistic partial differential equations are solved as initial
value problems (or initial-boundary values problems). 
In Sec.~\ref{sec:matter_model} we discuss 
the matter content of our model, which is a massive complex scalar field. 
The equation of motion for the scalar field is derived, as are
related quantities, including its stress-energy tensor.
Sec.~\ref{sec:CFC_Maximal_slicing} then provides a definition
of the approximation we make to the Einstein equations, as 
well as a detailed derivation of the PDEs that, using this approximation,
then govern our model.  The key element in the approximation is the 
demand that the 3-metric (which is the fundamental dynamical variable
in the $3+1$ approach to general relativity) be conformal to a {\em flat}
metric.  In addition to this requirement, a specific choice of time 
coordinatization---known as maximal slicing---is 
made~(Sec.~\ref{subsec:max_slice}).  This combination
of an assumption of conformal flatness and a choice of maximal slicing
leads to what we call the conformally flat approximation (or CFA)
of general relativity.

Principally for the sake of completeness, Sec.~\ref{subsec:conf_dec}
contains a review of a specific approach to the conformal decomposition
of the $3+1$ form of the Einstein equations.  Once this is in hand, 
it is a relatively straightforward matter to impose conformal flatness,
adopt maximal slicing and then derive the equations of motion for our 
model (Sec.~\ref{subsec:conf_flat}).
Ultimately this results in a system of nonlinear PDEs of mixed 
elliptic-hyperbolic type that are written in 
Cartesian coordinates (Sec.~\ref{subsec:cartesian}).
In Sec.~\ref{subsec:bdy_cond} we discuss three different strategies that 
we have investigated for the implementation of 
boundary conditions for this system.  This is followed by a definition
of the ADM mass and a short description of our use of that quantity
as a diagnostic in our computations (Sec.~\ref{subsec:mass}).
Sec.~\ref{sec:EOM_overview} then concludes the chapter with a recap 
of the equations of motion, as well as a synopsis of how they were 
solved numerically.  Thus, the reader interested
only in the final form of the equations of motion may wish to proceed 
directly to that section.

\section{The Field Equations of General Relativity} \label{sec:GR_field_equations}
Space and time in general relativity are modelled as a four dimensional Lorentzian 
manifold, $\mathcal{M}$.  Each point in the spacetime manifold corresponds 
to a physical event. The notion of distance between two points or, in the spacetime
case, the interval between two events, is encoded in a symmetric, non-degenerate,
tensor field of type $(0,2)$, the \emph{metric} $g_{ab}$.  The metric is thus the 
fundamental entity used to quantify the geometry of spacetime, and gravitational
effects arise due to the fact that this geometry is, in general, curved.

Let $\{x^\mu\}$ be a coordinate system that we assume covers that part 
of the manifold which is of interest. The vector and dual vector 
bases associated with this
coordinate system are called coordinate bases and are usually written as
$\{\lb \pa/ \pa x^\mu \rb^a \}$ and $\{\lb dx^{\mu} \rb_a\}$, respectively. 
Therefore the
metric components 
in this coordinate basis can be defined by:
\beq \label{eq:metric}
g_{ab} \equiv ds^2 = g_{\mu \nu} \lb dx^{\mu}\rb_a \lb dx^{\nu}\rb_b .
\eeq

The covariant derivative map, $\na_a$, is written in terms of the ordinary (partial)
derivative map, $\pa_a$, via
\beq \label{eq:cov}
\na_a v^b = \pa_a v^b + \Ga^b{}_{ac}v^c ,
\eeq
where $ \Ga^b{}_{ac}$ is the \emph{Christoffel symbol} and $v^b$ is an arbitrary vector. 
Throughout this thesis we use a covariant derivative that is compatible with the 
metric, i.e.~that satisfies
$\na_a g_{bc}=0$. 
The
Christoffel symbol can then be calculated in terms of the ordinary derivative 
operator:
\beq \label{eq:chris}
\Ga^c{}_{ab}=\ha g^{cd} \lb  \pa_a g_{bd} + \pa_b g_{ad} - \pa_d g_{ab}  \rb .
\eeq

The intrinsic notion of curvature of the spacetime can be made mathematically 
precise by considering the parallel transport of vectors along curves in 
the spacetime.~\footnote{See Chap.~3 of 
\cite{wald} for a complete and detailed discussion of this material.}
For example, a spacetime is curved in a region if an arbitrary vector that is parallel
transported along a closed path that bounds that region experiences an 
overall rotation.
Another consequence of
curvature is that the result of parallel transporting a vector is, in general,
path dependent.
In fact, the failure of successive applications of the covariant derivative to commute
captures the path-dependence of parallel transport (along with the notion of 
curvature) and is quantified by the so called intrinsic curvature tensor, 
the \emph{Riemann curvature tensor}, defined by:
\beq \label{eq:riemann}
2 \na_{[a}\na_{b]} w_c =  R_{abc}{}^d \, w_d .
\eeq
Contractions of the Riemann tensor give rise to the \emph{Ricci tensor}:
\beq \label{eq:riccit}
R_{ac}=R_{adc}{}^d \, ,
\eeq
and to the \emph{Ricci scalar} or \emph{scalar curvature}:
\beq \label{eq:riccis}
R=g^{ab}R_{ab} .
\eeq

A key feature of the Riemann curvature tensor is that it satisfies the contracted 
\emph{Bianchi identity} which can be expressed as the fact that the 
\emph{Einstein tensor}, defined in terms of 
the Ricci tensor and scalar curvature as 
\beq \label{eq:einsteint}
G_{ab} = R_{ab} - \ha  g_{ab} R \, ,
\eeq
has vanishing divergence:
\beq \label{eq:bianchi}
\na^a G_{ab} = 0 .
\eeq

Given a metric, $g_{ab}$, the Riemann curvature tensor can be written in terms of 
the Christoffel symbols through the following tensorial equation:
\beq \label{eq:riemann_chris}
R_{abc}{}^d = -2 \lb   \pa_{[a} \Ga^d{}_{b]c} - \Ga^e{}_{c[a}\Ga^d{}_{b]e} \rb .
\eeq

Once a coordinate basis is chosen, the Riemann tensor components can
be computed from the metric components and its various derivatives in that 
basis. 
In particular, since from~(\ref{eq:chris})
the Christoffel symbol 
components involve first derivatives of the metric components, the 
components of the curvature tensor generally involve second derivatives 
of the metric. 

The curvature of spacetime arises in response to the distribution of matter-energy 
in the spacetime, which includes contributions from the gravitational field itself,
as well as those from any matter fields present in the spacetime. 
For any given matter field, 
the coupling to gravity is described by the stress-energy-momentum tensor 
(stress-energy for short): $T_{ab}$. In many cases, including that 
considered in this thesis, the equation of motion for the matter field 
can be derived from the vanishing of the divergence of the stress tensor,
\beq \label{eq:matter_eom}
\na^a T_{ab} = 0 \, .
\eeq

The \emph{Einstein field equations} expresses the intimate relationship between the physical 
phenomena taking place in the spacetime and the curvature of the spacetime
geometry itself. In their most compact and elegant form the field equations are:
\begin{equation}\label{eq:einstein}
G_{ab} \equiv R_{ab} - \fr{1} {2} g_{ab} R = \ka T_{ab} ,
\end{equation}
where the constant $\ka$ depends on the system of units, but is also chosen
so that in the weak-field limit, the Newtonian description of gravity is recovered.
In the geometric units used in this thesis we have $\ka=8 \pi$.

Once a coordinate basis is chosen, the Einstein equations can be cast as 
a system of $10$ nonlinear second order partial differential equations 
for the metric components, $g_{\mu\nu}$, in the chosen coordinate system.
Since the metric
signature is Lorentzian, i.e.~$(-,+,+,+)$, this system of partial differential
equations has a hyperbolic or wave-like character. 

Although there have been many ``exact'' (closed form) solutions of 
the Einstein equations discovered over the years, very few of these are
considered to be of significant physical interest, and even fewer
are thought to be relevant to astrophysics.  Typically, exact solutions
are obtained by demanding symmetries: the spherically symmetric 
Schwarzschild solution for a single black hole, as well as the 
Kerr solution, which is axially symmetric and generalizes Schwarzschild
to the case of rotating black holes, are among the most important 
examples.  However, for situations in which 1) the gravitational field 
is strong and highly dynamical, 2) there are no symmetries, and
3) one does not have substantial {\em a priori} information about 
the expected nature of the solution, 
numerical analysis provides the only currently viable general approach
for solving the Einstein equations.  This, of course, is precisely
the situation that confronts us in our current study of interacting 
boson stars.

We end this brief overview of general relativity, with the observation
that the complex, tensorial nature of the Einstein field equations has 
fostered the development of {\em many} different approaches to their 
expression and analysis.
We will refer to any such approach as a {\em formalism} for 
general relativity, and in this thesis will focus exclusively on formalisms 
that are appropriate
for solving the field equations as an initial-value problem.  That is, 
we are interested in approaches where the state of the gravitational
field---defined in terms of some set of dynamical 
variables for the field---can be prescribed at some initial time.
The initial data will then be evolved to the future (or 
past) using partial differential equations of motion for the 
dynamical variables.
A given formalism must thus include prescriptions for 1) the decomposition
of spacetime into space and time, 2) the choice of dynamical and auxiliary 
variables, and 3) the choice of time and space coordinates.  Crucially,
it must also provide a specific form of the equations of motion that 
will be suitable for use in numerical computations.  As mentioned 
previously, the $3+1$ (or ADM) formalism has formed the 
basis for most work to date in numerical relativity. It is also
the fundamental approach that we adopt here, and is the 
topic of the next section.~\footnote{Here we should emphasize that 
developments over the past decade or so---both theoretical and 
computational---have
made it clear that the traditional form of the $3+1$ equations is {\em not}
generally suitable for fully 3D numerical calculations.  This is due 
to the fact that this form of the field equations is not strongly 
hyperbolic, and thus does not lead, in general cases, to a well posed 
initial value problem.  However, this fact is not relevant to our work 
since as a result of the approximation we adopt, the quantities describing 
the gravitational field are all governed by {\em elliptic} rather than 
hyperbolic PDEs.}

\newpage

\section{The $3+1$ (ADM) Formalism} \label{sec:adm}

As we have just indicated, 
in the $3+1$ (or ADM) formalism, spacetime is decomposed into 
space and time.~\footnote{The development in this section closely parallels
that of York in~\cite{york:1978}.}
Fundamental
to this approach is the choice of a timelike unit vector field, $t^a$,
in the spacetime and a foliation of spacelike hypersurfaces, $\Si_t$,
parametrized by
a time function. 
The timelike vector field is chosen such that its 
integral curves represent the time coordinate (or time function), $t$, 
throughout the spacetime, 
i.e. such that $t^a\na_a t=1$. In essence the vector field and the time function are 
chosen to enable the definition of the notion of dynamical evolution 
for quantities defined on the hypersurfaces.
In order to make the last statement precise, 
it is necessary to decompose spacetime vector fields (as well as tensor
fields of higher rank)
into (vector) pieces that are either defined 1) exclusively on any given
hypersurface, or 2) in a 
direction normal to the hypersurface.  For example, for the vector 
field $t^a$ itself we write
\beq \label{eq:t}
t^a = \al n^a + \bt^a \, ,
\eeq
where $n^a$ is the future-directed, timelike unit vector
field normal to $\Sigma_t$ 
(thus satisfying $g_{ab} n^a n^b = -1$ with our convention for the 
metric signature), $\alpha$ is the function that gives 
the component of $t^a$ in the normal direction, and 
$\bt^a$ is a vector field that resides on the hypersurface
(thus satisfying $g_{ab} n^a \bt^b = 0$).
In $3+1$ parlance, $\alpha$ and $\bt^a$ are known as the 
\emph{lapse function} and {\em shift vector}, respectively. 

From $g_{ab} n^a n^b = -1$ and $g_{ab} n^a \bt^b = 0$ it follows 
from~(\ref{eq:t}) that  
\beq \label{eq:lapse}
\al = - g_{ab} t^a n^b \, .
\eeq
Introducing the projection tensor, 
$\ga^a{}_{b}$, defined by
\beq\label{eq:projection}
\ga^a{}_{b} = \de^a{}_{b} + n^a n_b ,
\eeq
and also sometimes denoted $\perp^a{}_b$,
it follows from~(\ref{eq:t}) and (\ref{eq:projection}) that the shift vector 
is given by
\beq \label{eq:shift}
\bt^a = \ga^a{}_{b} t^b \, .
\eeq

We also note that in order for $t^a \na_a t = 1$ to be 
satisfied, (\ref{eq:lapse}) implies that we must have 
\beq\label{eq:normal}
n_a \equiv g_{ab}n^b = -\al \na_a t .
\eeq
 
The decomposition of vectors into ``temporal'' and ``spatial'' parts can be 
readily generalized to tensors of arbitrary rank. 
Following York~\cite{york:1978}, for any vector field, $W^a$, we define
\beq
\label{eq:vupproj}
	W^{\hat n} = - W^a n_a \, ,
\eeq
and, in general, any upstairs ${\hat n}$ index denotes that the original
tensor index has been contracted with $-n_a$.  On the 
other hand, for a dual vector field, $W_a$, York defines
\beq
\label{eq:vdownproj}
	W_{\hat n} =  + W_a n^a \, ,
\eeq
and then any downstairs ${\hat n}$ index denotes contraction with $+n^a$.

Of special interest are tensors which have been completely projected
onto the hypersurface.  For a  general tensor,
$T^{ab \ldots}{}_{cd \ldots}$ we write
\beq \label{eq:proj_tensor}
\perp T^{ab \ldots}{}_{cd \ldots} \equiv 
                    \ga^a{}_{e} \, \ga^b{}_{f} \ldots  \ga^g{}_{c} \, \ga^h{}_{d} \ldots
      T^{ef \ldots}{}_{gh \ldots}
\eeq
and $\perp T^{ab \ldots}{}_{cd \ldots}$ is called a {\em spatial 
tensor} since 
\beq
n_a\perp T^{ab \ldots}{}_{cd \ldots} = 
n_b\perp T^{ab \ldots}{}_{cd \ldots} \ldots =
n^c\perp T^{ab \ldots}{}_{cd \ldots} = 
n^d\perp T^{ab \ldots}{}_{cd \ldots} \ldots = 0 \, .
\eeq
One particularly important spatial tensor
is the induced three dimensional metric, $\ga_{ab}$, of the 
hypersurface
\beq\label{eq:gamma}
\ga_{ab} = g_{ab} + n_a n_b \, ,
\eeq
which, the reader should note, is also given by lowering the upstairs 
index of the projection tensor, $\perp^a{}_b \equiv \ga^a{}_b$.

In order to describe parallel transport of spatial tensors and  
curvature within the hypersurface, a covariant derivative operator 
must be defined. 
A natural choice is to project the four dimensional covariant derivative 
onto the spacelike hypersurface, leading to the 
definition of a three dimensional covariant derivative operator, $D_a$:
\beq \label{eq:cov_D}
D_a \equiv \perp \na_a .
\eeq
The Riemann curvature tensor on the hypersurface is defined analogously to its 
four dimensional counterpart:
\beq \label{eq:riemann3d}
2 D_{[a} D_{b]} w_c =  \mathcal{R}_{abc}{}^d w_d ,
\eeq
while the Ricci tensor and Ricci scalar are obtained by the usual contractions
of the Riemann curvature tensor:
\beq \label{eq:ricci3d}
\mathcal{R}_{ab} =  \mathcal{R}_{acb}{}^c \qquad \textrm{and} \qquad
\mathcal{R}=\mathcal{R}^a{}_{a} \, .
\eeq

As mentioned previously, the Riemann curvature tensor describes the 
curvature {\em intrinsic} to a manifold.  In the current case, which 
involves the embedding of three-dimensional hypersurfaces in a four-dimensional
spacetime, there is a second type of curvature, known as the 
\emph{extrinsic} curvature, that quantifies the embedding.  
Since the orientation
of the hypersurface within the spacetime is related to the unit normal vector, 
$n^a$, the covariant derivative of $n^a$ thus characterizes nearby changes 
in the orientation. 
The extrinsic curvature tensor, $K_{ab}$, can therefore be defined as the 
projection of the covariant derivative of the dual vector field associated to
the normal vector field:
\beq \label{eq:extrinsic}
K_{ab} = - \perp \na_a n_b = - \ha \perp \Screll_n g_{ab} = -\ha \Screll_n \ga_{ab} .
\eeq

Since both sides of the Einstein equations must be decomposed in the 3+1 approach,
we must also consider various projections of the stress-energy
tensor along the normal $n^a$ and onto the hypersurface, $\Si_t$. First note 
that the stress-energy tensor $T_{ab}$ is a type $(0,2)$ symmetric tensor. 
A generic tensor of this type can be decomposed in the following way in the 
3+1 formalism:
\beq \label{eq:stress_dec}
T_{ab}=\perp T_{ab}-2n_{(a}\perp T_{b) \hat n} + n_a n_b T_{\hat n \hat n} .
\eeq
We rewrite the above as
\beq
T_{ab}=S_{ab}-2 J_{(a}n_{b)} + \rho n_a n_b ,
\eeq
where the quantities, $\rho$, $J_a$ and $S_{ab}$ are defined by
\bea 
\rho& \equiv& T_{\hat n \hat n} = T_{ab} n^a n^b = T^{ab} n_a n_b \label{eq:rho}, \\
J_a & \equiv&  \perp  T_{a \hat n} = \perp ( T_{ab}n^b) \label{eq:Ja},\\ 
J^a & \equiv&  \perp  T^{a \hat n} = - \perp ( T^{ab}n_b) \label{eq:JA},\\ 
S_{ab}&\equiv& \perp T_{ab} \label{eq:Sab},
\eea
Physically, 
$\rho$ is interpreted as the local energy density, $J^a$ as the momentum 
density,
and $S_{ab}$ as the spatial stress tensor, all measured by observers moving orthogonally
to the slices.

The several possible combinations of projections of Einstein equations along 
the ``temporal'' and ``spatial'' directions give rise to the equations of motion
in the $3+1$ form. Projecting both indices with $n^a$ we find
\beq \label{eq:ham}
\mathcal{R} + K^2 - K_{ab} K^{ab} = 16\pi\rho ,
\eeq
where $K \equiv K^a{}_{a}$ is the trace of the extrinsic curvature. 
Eq.~(\ref{eq:ham}) is also known as \emph{Hamiltonian constraint}.
On the other hand, if only one index is contracted along $n^a$, while the other
is projected onto the hypersurface, we derive a three-vector equation
known as the \emph{momentum constraint}:
\beq \label{eq:mom}
D_b K^{ab} - D^a K = 8 \pi J^a .
\eeq

We note that care must be exercised in using the covariant form of 
this equation, since 
due to the relative sign in the definitions of $J_a$ and $J^a$ 
in equations~(\ref{eq:Ja}) and (\ref{eq:JA}) (see definitions
(\ref{eq:vupproj}) and (\ref{eq:vdownproj})), we have
\bea
\perp G_{a \hat n} = - D^b K_{ab} + D_a K &=& 8\pi J_a ,\\ 
\perp G^{a \hat n} = D_b K^{ab} - D^a K &=& 8\pi J^a .
\eea
Key features of the constraint equations are the presence of \emph{only}
spatial tensors and the absence of explicit time derivative of these tensors.
They must be satisfied by $\{\ga_{ab},K_{ab}\}$ on all slices, including the initial 
slice.

The 3+1 equations that {\em do} involve time derivatives of the spatial tensors
$\{\ga_{ab},K_{ab}\}$, are thus called evolution equations.
For the spatial metric, an evolution equation follows from the 
definition of the extrinsic curvature~(\ref{eq:extrinsic}):
\beq \label{eq:g_evo}
\Screll_t \ga_{ab} = - 2 \al K_{ab} + \Screll_{\bt}\ga_{ab} .
\eeq
where $\Screll_t$ is the Lie derivative with respect to the vector field $t^a$.
We note here that the Lie derivative of a general tensor,
$T^{a_1\ldots a_k}{}_{b_1\ldots b_l}$, with respect 
to a vector field, $v^a$, is defined
by:~\footnote{We should emphasize that this expression is valid for 
{\em any} derivative operator $D_a$. The reader should refer to 
App.~C of Wald~\cite{wald} for a comprehensive discussion.}
\bea
\Screll_v T^{a_1\ldots a_k}{}_{b_1\ldots b_l} &=&
v^c D_c T^{a_1\ldots a_k}{}_{b_1\ldots b_l} - 
\sum_{i=1}^{k} T^{a_1\ldots c \ldots a_k}{}_{b_1\ldots b_l} D_c v^{a_i}
\nonumber
\\
& & + \sum_{j=1}^{l} T^{a_1\ldots a_k}{}_{b_1\ldots c \ldots b_l} D_{b_j} v^{c} \, .
\label{eq:lie_def}
\eea
 
The evolution equation for the extrinsic curvature
can be derived by
considering the projection of both indices of Einstein equations, which 
involves computation of 
$\perp R_{a \hat n a \hat n}$. After some manipulation, we find
\beq \label{eq:k_evo}
\Screll_t K^a{}_{b} = \Screll_{\bt} K^a{}_{b} - D^a D_b \al + \al      \lhb
                   \mathcal{R}^a{}_{b} + K K^a{}_{b} + 8 \pi \lb
                   \ha \ga^a{}_{b} (S - \rho) - S^a{}_{b}    \rb       \rhb .
\eeq

All of the definitions and decompositions  
discussed thus far are independent of any choice of coordinate
system. Operationally however, we must introduce coordinates,
$x^\mu \equiv (t,x^i)$, in order to cast~(\ref{eq:ham}), 
(\ref{eq:mom}), (\ref{eq:g_evo}) and 
(\ref{eq:k_evo}) as a system of partial differential equations that
can then be solved---using a numerical approach in general---for given
initial data and boundary conditions
(we remind the reader that Greek indices such
as $\mu$ range over the spacetime values, 0,1,2,3, while Latin 
indices such as $i$ are restricted 
to spatial values, 1,2 and 3).  We thus adopt such a coordinate system,
with the time coordinate, $t$, being identified with the time function.
In terms of tensor components taken with respect to the coordinate basis,
the spacetime displacement can 
be written as 
\begin{eqnarray} \label{eq:adm_metric}
ds^2 &=& g_{\mu\nu} dx^\mu dx^\nu \nonumber \\
     &=& -\al^2 dt^2 + \ga_{ij} \left(dx^i + \bt^i dt \right)
     \left(dx^j + \bt^j dt \right) \, .
\end{eqnarray}
We emphasize that the lapse function, $\alpha$, the three shift
vector components, $\bt^i$, and the six components of the symmetric 
3-metric, $\ga_{ij}$, that appear in the above expression are all functions 
of the coordinates $x^\mu\equiv(t,x^i)$.
We also note that
the lapse function can 
be interpreted as the ratio of proper time to coordinate time
for an observer travelling normally 
to the hypersurface, while the shift vector encodes the translation of
spatial coordinates from one slice to the other, again relative to 
propagation in
the normal direction.  In addition, component indices of spatial tensors 
are lowered and raised  with
the 3-metric $\ga_{ij}$ and its inverse $\ga^{ij}$, respectively, where 
$\ga^{ij}$ is defined by $\ga^{ik} \ga_{kj} = \de^i{}_{j}$. 

As discussed above,
the decomposition of the stress-energy tensor gives rises to a variety of 
energy-momentum quantities
defined by equations~(\ref{eq:rho}-\ref{eq:Sab}).  Using the relations
$n^{\mu}=(1/\al;-\bt^i/\al)$ and $n_{\mu}=(-\al;0)$, 
these quantities become
\bea 
\rho &=&  \fr{T_{00}}{\al^2} -2  \fr{\bt^i T_{0i}}{\al^2} + 
\fr{\bt^i \bt^j T_{ij}}{\al^2}=\al^2 T^{00}      \label{eq:rho_adm}, \\
J_i&=&\fr{T_{i0}}{\al} - \fr{T_{ij}\bt^j}{\al}   \label{eq:Ja_adm},\\ 
J^i&=&\al(T^{i0} + T^{00}\bt^i)                  \label{eq:JA_adm},\\ 
S_{ij}&=&T_{ij}                                  \label{eq:Sab_adm}.
\eea

The component form of the evolution equations can then be written as
\beq \label{eq:g_evo_adm}
\pa_t \ga_{ij} = - 2 \al \ga_{ik} K^k{}_{j} + 2 D_{(i} \bt_{j)} \, ,
\eeq
and
\bea
\pa_t K^i{}_{j} &=& \bt^k \pa_k K^i{}_{j} - \pa_k \bt^i K^k{}_{j} + \pa_j \bt^k K^i{}_{k}
-D^i D_j \al 
\nonumber\\
             &+& \al \lhb 
          \mathcal{R}^i{}_{j} + K K^i{}_{j} + 8 \pi 
                        \lb
                           \ha \de^i{}_{j} ( S - \rho ) - S^i{}_{j}
                        \rb
                   \rhb \label{eq:k_evo_adm} \, .
\eea
We will use these versions of the evolution equations for the components of the 
3-metric and extrinsic curvature when discussing spherically symmetric spacetimes
in Ch.~\ref{initialdata}.  However, for the purpose of performing a conformal
decomposition of the 3+1 evolution equations---which is done later in the current
chapter---we will start from slightly different forms which may be easily derived from
previous formulae using straightforward tensor calculus.  
Specifically, for the 3-metric components we can also write 
\beq
	\label{eq:g_evo_adm_cov}
	\Screll_{(t-\bt)}\ga_{ij} = -2\al K_{ij}
\eeq
which follows immediately from~(\ref{eq:g_evo}) as well as 
\beq
	\label{eq:g_evo_adm_contra}
	\Screll_{(t-\bt)}\ga^{ij} = 2\al K^{ij} \, .
\eeq
This last result is easily established from~(\ref{eq:g_evo_adm_cov}) and the fact that 
\begin{eqnarray*}
 0 &=& \Screll_{(t-\bt)} \lb \delta^i{}_k\rb = \Screll_{(t-\bt)} \lb \ga^{ij} \ga_{jk} \rb
	= \ga_{jk} \Screll_{(t-\bt)} \ga^{ij} + \ga^{ij} \Screll_{(t-\bt)} \ga_{jk} \\
   &\implies&  \ga_{jk} \Screll_{(t-\bt)} \ga^{ij} = - \ga^{ij} \Screll_{(t-\bt)} \ga_{jk} \, .
\end{eqnarray*}
For the extrinsic 
curvature components, $K_{ij}$, we have 
\beq
\Screll_{(t-\bt)} K_{ij} = 
-D_i D_j \al + \al \lhb 
          \mathcal{R}_{ij} + K K_{ij} - 2 K_{ik}K^k{}_j + 8 \pi 
                        \lb
                           \ha \ga_{ij} ( S - \rho ) - S_{ij}
                        \rb
                   \rhb \label{eq:k_evo_adm_cov} \, .
\eeq

Additionally, the constraint equations in component form are:
\bea
\mathcal{R} + K^2 - K_{ij} K^{ij} &=& 16 \pi \rho \label{eq:ham_adm} ,\\
D_j K^{ij} - D^i K &=& 8 \pi J^i                  \label{eq:mom_adm} .
\eea

Even within a specific formulation of the Einstein equations, such as the 
3+1 approach described above, the coordinate invariance of general relativity ensures 
that there are generally many distinct possibilities to solve the 
specific set of PDEs that results once the coordinate system has been fully fixed 
(full specification of the lapse and shift).  Here we are referring to the fact that 
we have more equations (4 second-order ``elliptic'' constraints + 12 first-order-in-time evolution
for a total of 16 equations) than fundamental dynamical unknowns 
(6 $g_{ij}$ + 6 $K_{ij}$ = 12 unknowns).
The interested reader is referred
to the classic paper by Piran~\cite{piran:1980} 
in which nomenclature, such as free evolution,
constrained evolution and partially-constrained evolution is defined and discussed. 
Here, the key thing to note is that the approximation (CFA)
that is adopted in this thesis 
has the advantage of providing a single, well defined set of 5 elliptic PDEs for 
5 well defined functions that completely fix the spacetime geometry.  
In this sense, and in an abuse 
of Piran's original classification, we implement a fully constrained evolution for the 
geometrical field and, further, in contrast to the full general-relativistic situation
there are {\em no} purely gravitational degrees of
freedom.  That means that in the model considered here, as is the case for any 
model that adopts the 
CFA with maximal slicing condition, all dynamics is linked to the dynamics of the matter.
This has a host of ramifications, physically, mathematically and computationally, but 
particularly given the efforts that have been expended on taming instabilities in 
free evolution approaches for the full Einstein equations, is one of the most 
attractive features of Isenberg's proposal.

\newpage
\section{The Complex Scalar Field} \label{sec:matter_model}

The matter model adopted in this thesis is a complex \emph{Klein-Gordon field}, 
which satisfies a Klein-Gordon equation as discussed in detail below.
This field represents a simple type of matter that when coupled to Einstein
gravity, or in the context of the approximation adopted in this 
thesis, admits star-like solutions.
Studies focusing on such solutions---known as boson stars---using a variety of 
techniques including numerical
analysis, have a rich history and we refer the reader to the paper by Schunck and Mielke,
\cite{schunck:2003} (and references therein), for an excellent and
thorough review of the subject as of about five years ago. 

General relativists have studied 
Klein-Gordon fields for many purposes 
over the years. As either 1) a classical field or 2) a quantum-theory of spin 0 
particles, scalar fields have been widely exploited for exploratory 
theoretical studies.
A key point is that the simplicity of scalar matter (in terms, e.g. of physical 
interpretation as well as complexity of the equations of motion),
often allows one to investigate and understand basic theoretical issues in 
Einstein gravity relatively free of 
the complications a more realistic matter model could bring in. 
This is a chief motivation for the use of a scalar field in the current work.

In the discussion below, we will refer to the system of a single Klein-Gordon
field minimally coupled~\footnote{The notion of minimal coupling 
is defined in the next section.} to Einstein gravity as the
Einstein-Klein-Gordon (EKG) system.
We note that we adopt a complex scalar field, rather than a real one, 
since it has been long known that there are no regular, static solutions 
(i.e. star-like solutions) for a real scalar field in general relativity.  
Interestingly, for us this turns out to be something of a technical point, 
since for a real field 
coupled to Einstein gravity there are quasi-static solutions known as 
``oscillons'' which have decay times that can be {\em much} longer than 
the intrinsic dynamical time~\cite{Seidel:1991zh}.  Thus in principle one could use 
a real scalar field to study some of the effects we wish to examine in 
this thesis and follow-up work.  However, for a variety of reasons,
not least including the ease with which one can generate star-like solutions, 
we prefer to work with the complex field.

Additionally, the complex field 
must interact in an non-trivial potential, which we define to include a 
mass term.  The possibilities for potential choice are endless, and have 
formed the basis for much previous work.  Again, we choose the simplest 
approach and, at least initially, adopt only a mass term. 
The boson stars modelled by scalar fields with this self-interaction potential 
are also known in the literature as \emph{mini-boson stars}.

Since mathematically we are ultimately interested in solving an approximate 
EKG system as an initial value problem, we note in passing that
a scalar field is known to admit a well-posed initial value formulation in the
following sense \cite{wald}: 

\bie
\item For an initial data in a spacelike Cauchy surface $\Si$ in a globally
hyperbolic spacetime \footnote{When the domain of dependence of a Cauchy 
surface is the whole spacetime (region of interest) then this spacetime (region)
is said to be globally hyperbolic} $(\mathcal{M},g_{ab})$, there is an open 
neighbourhood $O$ of $\Si$ such that the Klein-Gordon equation has a solution 
in $O$ and $(O,g_{ab})$ is globally hyperbolic.
\item The solution in $O$ is unique and propagates causally.
\item The solution depends continuously on the initial data.
\eie

In addition to possessing star-like solutions, a key advantage of scalar matter
relative to the more-astrophysically relevant perfect fluid case, is that 
the solutions do not tend to develop shocks or rarefaction regions. Rather,
as is expected from the structure of the equations, and has been born out by many 
\cite{seidel_suen:1990,Seidel:1991zh,choptuik:1993,Balakrishna:1998,hawley_choptuik:2000,Hawley:2002zn,rousseau:master,graxi:2003b,cwlai:phd,Guzman:2004jw,Balakrishna:2006ru,Palenzuela:2006wp,Olabarrieta:2007di,Lai:2007tj,Palenzuela:2007dm,Choptuik:2009ww}
previous numerical studies, 
solutions tend to remain as smooth as the initial data, except at 
actual physical singularities (produced, for example, by gravitational 
collapse).

\subsection{Einstein-Klein-Gordon System}

One route to study matter models in general relativity 
is to postulate equations of motion for the matter, derive a suitable 
stress-energy tensor $T_{ab}$ compatible with those equations and 
then use the Einstein equations to relate the matter 
distribution to the spacetime curvature through this stress-energy tensor.
However, and as already mentioned, in many cases 
local conservation of the stress-tensor, implies the matter equation
of motion.
In such cases it is essentially sufficient to postulate
the stress-energy tensor $T_{ab}$ for the matter model in order to study
the coupled system of matter distribution and spacetime geometry.

Additionally, and for a variety of reasons, it is often useful to adopt 
a Lagrangian (or variational) approach to Einstein equations and we 
will do so here.  Here, a basic observation is that
the vacuum Einstein equations can be obtained from the functional derivative 
of the so called Hilbert action functional:
\beq \label{eq:hilbert_action}
S_G[g^{ab}]= \int_{\mathcal{M}} \mathcal{L}_G = \int_{\mathcal{M}} \sqrt{-g} R ,
\eeq
where $\mathcal{L}_G = \sqrt{-g} R$ is the Einstein Lagrangian density and $R$ the
Ricci scalar. It is a standard exercise to show that 
the functional derivative of the action with respect to the inverse metric $g^{ab}$ is
\beq
\fr{\de S_G}{\de g^{ab}} = \sqrt{-g} G_{ab},
\eeq
which then clearly yields the vacuum Einstein equations $G_{ab}=0$ when 
the field configuration satisfies the action extremization condition:
\beq
\fr{\de S_G}{\de g^{ab}} = 0 .
\eeq

In order to obtain a coupled matter-gravity system, one then simply adds to
the matter Lagrangian density to the Hilbert term (this is the so-called 
minimal coupling prescription).  We thus have
\beq
\mathcal{L} = \mathcal{L}_G + \al_M \mathcal{L}_M ,
\eeq
where $\al_M$ is a coupling constant that can typically be rescaled through 
a redefinition of the matter fields. 
In the case of the Einstein-Klein-Gordon
system one conventional choice that we adopt here is $\al_{KG} = 16\pi$.  

The stress-energy tensor can now be 
calculated as the variation of the matter action with respect to the 
inverse metric field $g^{ab}$. Specifically, one has 
\beq \label{eq:stress_tensor_def}
T_{ab}= - \fr{\al_M}{8\pi} \fr{1}{\sqrt{-g}} \fr{\de S_M}{\de g^{ab}} ,
\eeq
where $S_M$ is the action functional for the matter field $M$ (understood here
as a generic collection of matter fields and their higher order covariant 
derivatives):  
\beq \label{eq:matter_action}
S_M[g^{ab},M]= \int_{\mathcal{M}} \mathcal{L}_M  .
\eeq
Finally, variations of the action $S_M$ with respect to the matter fields themselves 
generate the equations of motion for the matter.

For the reasons discussed above, we now restrict attention to matter consisting 
of a single complex scalar field, $\Phi$.
We write the field as 
\bea \label{eq:complex_field}
\Phi=\phi_{1}+i\phi_{2}=\phi_{0} \exp(i \theta),
\eea
where $\phi_1$, $\phi_2$, $\phi_0$ and $\theta$ are real-valued functions of 
the spacetime coordinates $x^{\mu}$.  The Lagrangian density associated with this 
field is 
\bea \label{eq:lagrangian_complex}
\mathcal{L}_{\Phi}&=&-\frac{1}{2}\sqrt{-g} \lb g^{ab}\nabla_{a}\Phi\nabla_{b}\Phi^{*}
                                               + U(|\Phi|^2) \rb ,
\eea
where $U(|\Phi|^2)$ is the scalar field self-interaction potential.  As also
discussed above, we will eventually specialize to the case where $U$ contains 
only a mass term
\bea
U(|\Phi|^2)=m^{2} \Phi\Phi^{*}=m^{2} \phi_0^2 = m^{2} ( \phi_1^2 + \phi_2^2 ) ,
\eea
but for the time being we will continue the discussion 
in terms of general potentials.

We now rewrite the 
Lagrangian (\ref{eq:lagrangian_complex}) 
in terms of the real-valued quantities defined in~(\ref{eq:complex_field}):
\bea
\mathcal{L}_{\Phi}&=&-\frac{1}{2}\sqrt{-g}\lb g^{ab}\nabla_{a}\phi_1\nabla_{b}\phi_1 +
                           g^{ab}\nabla_{a}\phi_2\nabla_{b}\phi_2 +
                           U(\phi_0^2) \rb.
\eea
Klein-Gordon equations for each real valued component 
($\phi_A \in \{\phi_1,\phi_2\}$)
can then be obtained by the usual variational procedure, yielding
\beq \label{eq:klein-gordon}
       \square \phi_{A} - \fr{dU(\phi_0^2)}{d\phi_0^2} \phi_{A} = 0 \qquad 
       \textrm{or} \qquad 
       g^{ab}\na_a \na_b \phi_{A} - \fr{dU(\phi_0^2)}{d\phi_0^2} \phi_{A} 
       = 0, \qquad  A=1,2,
\eeq
where $\square \equiv g^{ab}\na_a \na_b $ is the general relativistic D'Alambertian 
operator.

Once a coordinate system is chosen, each of the above scalar Klein-Gordon equations
is a second-order-in-time PDE.  In keeping with the 3+1 spirit, it is often
conventional to recast these equations in first-order-in-time form, and we do so 
here.  One specific way of doing this is to pass to the Hamiltonian description 
of the system in the standard fashion.
Namely, we consider the Lagrangian as a function
of the field and its spatial and time derivatives; we define a conjugate momentum 
associated with the field; we write down the Hamiltonian functional from the Lagrangian
by performing a Legendre transformation for the conjugate momentum; and we 
then evaluate the 
Hamilton evolution equations from the Lagrangian. Full details of this procedure can 
be found in standard texts such as  Wald \cite{wald}, and here we simply summarize 
the results for the scalar field.

Since the scalar-field Lagrangian~(\ref{eq:lagrangian_complex})
does not contain time derivatives higher than first order, 
the conjugate momentum associated with each component of the scalar field 
can be defined as:
\beq
\Pi_{A} \equiv \frac{\delta(\sqrt{-g}L_{\phi_{A}})}{\delta\dot{\phi_{A}}},
\eeq
or, more explicitly
\beq
\Pi_A = \fr{\sr{-g}}{\al^2} \lhb \dot \phi_A - \bt^i \pa_i \phi_A \rhb ,
\eeq
where the overdot denotes differentiation
with respect to the time coordinate.

The dynamical equations of motion~(\ref{eq:klein-gordon}) can be rewritten 
in terms of these conjugate fields, leading to four first-order-in-time 
partial differential equations for the two conjugate pairs of variables
$\{\phi_A,\Pi_A\}$ (where $A = 1, 2$): 
\bea
\partial_{t}\phi_{A}&=&\frac{\alpha^{2}}{\sqrt{-g}}\Pi_{A}+\beta^{i}\partial_{i}\phi_{A}
\label{eq:phidot},\\
\partial_{t}\Pi_{A}&=&\partial_{i}(\beta^{i}\Pi_{A})+\partial_{i}(\sqrt{-g}
\gamma^{ij}\partial_{j}\phi_{A})-\sqrt{-g}\fr{dU(\phi_0^2)}{d\phi_0^2} \phi_{A}
\label{eq:pidot} .
\eea
These last equations 
can be further manipulated using the following relationship between the 
determinants of the spacetime and spatial metrics:
\footnote{This 
relation is derived from the definition of inverse metric: 
$g^{00} =- \al^{-2} = (-1)^{0+0} det(\ga_{ij})/det(g_{\mu\nu})= \ga/g$.}
\beq
\sr{-g} \equiv \al \sr{\ga} ,
\eeq
yielding
\bea
\pa_t\phi_A&=&\fr{\al}{\sr{\ga}}\Pi_A +\bt^i\pa_i\phi_A \label{eq:phidot_adm},\\
\pa_t\Pi_A&=&\pa_i(\bt^i\Pi_A)+\pa_i(\al\sr{\ga}\ga^{ij}\pa_j\phi_A)
-\al\sr{\ga} \fr{dU(\phi_0^2)}{d\phi_0^2} \phi_A \label{eq:pidot_adm}.
\eea

Having obtained equations of motion for the scalar field, we now consider 
computation of the stress-energy tensor and the 3+1 quantities derived from it.
Using the variational prescription sketched above we find
\beq
\label{eq:Tab}
T_{ab}= \frac{1}{2} \lhb \nabla_{a}\Phi\nabla_{b}\Phi^{*}+
\nabla_{b}\Phi\nabla_{a}\Phi^{*}- g_{ab}(g^{cd}\nabla_{c}\Phi\nabla_{d}\Phi^{*}
+U(\Phi\Phi^{*})) \rhb ,
\eeq
which, adopting a coordinate basis, and working with the real-valued 
field components becomes
\beq
T_{\al\bt}= \sum_{A=1}^2 \frac{1}{2}\lhb \pa_{\al}\phi_A\pa_{\bt}\phi_A+
\pa_{\bt}\phi_A\pa_{\al}\phi_A- g_{\al\bt}g^{\mu\nu}\pa_{\mu}\phi_A
\pa_{\nu}\phi_A \rhb -\frac{1}{2} g_{\al\bt} U(\phi_0^2) .
\eeq
From the above, and using~(\ref{eq:rho_adm})-(\ref{eq:Sab_adm}),
we compute the 3+1 stress-energy quantities and find:
\bea
\rho &=& \fr{1}{2}\sum_{A=1}^2 \lhb \fr{\Pi_A^2}{\ga}+\ga^{ij}\pa_i\phi_A
\pa_j\phi_A \rhb + \fr{1}{2} U(\phi_0^2) 
,\label{eq:rho_adm1}\\
J_i&=&\sum_{A=1}^2 \lhb   \fr{\Pi_A}{\sr{\ga}}\pa_i\phi_A \rhb 
,\label{eq:Ja_adm1}\\
J^i&=&\sum_{A=1}^2 \lhb -  \fr{\Pi_A}{\sr{\ga}}\ga^{ij}\pa_j\phi_A \rhb 
,\label{eq:JA_adm1}\\
S_{ij}&=&\fr{1}{2}\sum_{A=1}^2 \lbr 2\pa_i\phi_A\pa_j\phi_A + \ga_{ij}
\lhb \fr{\Pi_A^2}{\ga} - \ga^{pq} \pa_p\phi_A\pa_q\phi_A 
\rhb \rbr- \fr{1}{2} \ga_{ij} U(\phi_0^2) 
.\label{eq:Sab_adm1}
\eea
Additionally, we need to compute the trace of the spatial stress tensor,
$S\equiv S^i{}_i$ as well as the combination $\rho + S$.  These are given by
\bea
S^i{}_i&=&\fr{1}{2}\sum_{A=1}^2 \lhb 3\fr{\Pi_A^2}{\ga}-\ga^{ij}\pa_i\phi_{A}
\pa_j\phi_A  \rhb - \fr{3}{2} U(\phi_0^2) 
,\label{eq:S_adm1}\\
\rho+S&=&\sum_{A=1}^2 \lhb 2\fr{\Pi_A^2}{\ga} \rhb 
- U(\phi_0^2)
.\label{eq:Sphro_adm1}
\eea

\subsection{Noether Charge}

The invariance of the Klein-Gordon Lagrangian density, 
Eq.~(\ref{eq:lagrangian_complex}), under a global $U(1)$ symmetry transformation
$\Phi \rightarrow \Phi e^{i\ep}$ gives rise to a conserved current density according
to Noether's theorem. Roughly, this result can be obtained as follows:

First, consider the Klein-Gordon action as a functional of the inverse metric,
the scalar field and its first covariant derivative, instead of the inverse metric and
the scalar field alone, as in the last subsection:
\beq
S_{KG}[g^{ab},\Phi,\na_a \Phi]= \int_{\mathcal{M}} \mathcal{L}_{KG}(g^{ab},\Phi,\na_a\Phi).
\eeq
Also note that for a scalar field we have
$\na_a \Phi = \pa_a \Phi$.
Defining the variation of a functional or function with respect to a parameter $\ep$
as:
\beq
\de S[\Phi_{\ep}] \equiv \fr{dS}{d\ep}\bigg|_{\ep=0} \qquad  \textrm{and} 
\qquad  \de \Phi \equiv \fr{d\Phi}{d\ep}\bigg|_{\ep=0} ,
\eeq
the variation of the Klein-Gordon action functional can then be expanded as:
\beq \label{eq:KG_action_variation}
\de S_{KG} = \int_{\mathcal{M}} \fr{\de S_{KG}}{\de g^{ab}} \de g^{ab}
                            +  \fr{\de S_{KG}}{\de \Phi} \de \Phi 
                            +  \fr{\de S_{KG}}{\de (\pa_a \Phi)} \de (\pa_a \Phi).
\eeq
Our interest here is in variations that keep the action functional constant; that is,
variations such that $\de S_{KG} = 0$. The inverse metric $g^{ab}$ is invariant 
under the action of a $U(1)$ transformation: 
$\de g^{ab} = \fr{dg^{ab}}{d\ep}|_{\ep=0} =0$, and the first term in the 
Eq.~(\ref{eq:KG_action_variation}) drops out. Further simplification of
Eq.~(\ref{eq:KG_action_variation}) results from noting that the variation with respect to 
the field derivative can be rewritten as:
\beq
\de(\pa_a \Phi) = \fr{d(\pa_a\Phi)}{d\ep}\bigg|_{\ep=0} = \pa_a \lb 
\fr{d\Phi}{d\ep} \rb \bigg|_{\ep=0} = \pa_a (\de\Phi) ,
\eeq
since ordinary derivatives commute.
Inserting the above relationship in Eq.~(\ref{eq:KG_action_variation}), we 
have 
after some simple algebraic manipulation:
\beq
\de S_{KG} = \int_{\mathcal{M}} \lbr \lhb \fr{\de S_{KG}}{\de \Phi} 
- \pa_a \lb \fr{\de S_{KG}}{\de (\pa_a \Phi)} \rb \rhb \de \Phi
+ \pa_a \lb \fr{\de S_{KG}}{\de (\pa_a \Phi)} \de \Phi \rb \rbr .
\eeq
The first term of the equation above is simply the Klein-Gordon equation of motion
which vanishes identically. The second term is a total divergence that can 
be converted to a surface term using Stokes theorem, and which also has to 
vanish if
the action is supposed to be invariant under the field variation $\de\Phi$.
This then implies that the current density, $j^a$, associated with the $U(1)$ symmetry
and defined by 
\beq \label{eq:current_density}
j^a \equiv \fr{\de S_{KG}}{\de (\pa_a \Phi)} \de \Phi ,
\eeq
is conserved
\beq \label{eq:current_density1}
\pa_a j^a = 0 .
\eeq

A conserved (Noether) charge, $Q_N$, is associated with the ``time'' component 
of the current density:
\beq
Q_N = \int_{\Si_t} j^t ,
\eeq
where $\Si_t$ is a spacelike hypersurface as previously, 
and a fixed volume element on $\Si_t$, $\mathbf{e}$,
is understood in the integration.

To compute the explicit form of the Noether current 
we apply Eq.~(\ref{eq:current_density}) to the Klein-Gordon 
Lagrangian~(\ref{eq:lagrangian_complex}), obtaining
\beq
\de \mathcal{L}_{KG} \equiv \pa_a \lb 
\fr{\de S_{KG}}{\de (\pa_a \Phi)} \de \Phi  +
\fr{\de S_{KG}}{\de (\pa_a \Phi^{*})} \de \Phi^{*} 
                                  \rb  =
\pa_a \lb -\ha \sqrt{-g} g^{ab} (\pa_b \Phi^{*} \de \Phi + 
                                 \pa_b \Phi \de \Phi^{*}) \rb ,
\eeq
so the current density is
\beq
j^a =  -\ha \sqrt{-g} g^{ab} (\pa_b \Phi^{*} \de \Phi +
                                 \pa_b \Phi \de \Phi^{*}) .
\eeq
For an infinitesimal $U(1)$ transformation, 
$\Phi \rightarrow \Phi + i \ep \Phi$, we have
\beq
\de \Phi = i \ep \Phi  \qquad \textrm{and} \qquad
\de \Phi^{*} = -i \ep \Phi^{*} , 
\eeq
and we have
\beq
j^a =  -i\ha \sqrt{-g} g^{ab} (\Phi \pa_b \Phi^{*}  -
                               \Phi^{*} \pa_b \Phi  ),
\eeq
where the constant $\ep$ has been factored out. Using the component form
of the field, $\Phi = \phi_1 + i\phi_2$, this can also be
expressed as:
\beq
j^a =  \sqrt{-g} g^{ab} (\phi_2 \pa_b \phi_1  -
                               \phi_1 \pa_b \phi_2 ).
\eeq
Finally the time component of the current density in a 3+1 coordinate 
basis assumes the following form:
\bea
j^t &=& \sqrt{-g} \lhb g^{tt} (\phi_2 \pa_t \phi_1 - \phi_1 \pa_t \phi_2) + 
                       g^{ti} (\phi_2 \pa_i \phi_1 - \phi_1 \pa_i \phi_2)\rhb \nonumber \\
    &=& \sqrt{-g} \lhb -\fr{1}{\al^2} (\phi_2 \pa_t \phi_1 - \phi_1 \pa_t \phi_2) + 
         \fr{\bt^i}{\al^2} (\phi_2 \pa_i \phi_1 - \phi_1 \pa_i \phi_2)\rhb \nonumber \\
    &=& \phi_1 \Pi_2 - \phi_2 \Pi_1 ,
\eea
where Eq.~(\ref{eq:phidot}) was used to simplify the second line of the 
above, and to express $\pa_t \phi_A$ in terms of their respective 
conjugate momenta $\Pi_A$. 
Choosing the fixed volume element $\mathbf{e}$ on $\Si_t$ to be the 
coordinate volume element $d^3x$, the Noether charge can be written as:
\beq
Q_N = \int_{\Si_t} (\phi_1 \Pi_2 - \phi_2 \Pi_1) \, d^3x
\eeq
and can be expected to be conserved: i.e.~to have the same value on 
each slice $\Si_t$ of the 
spacetime foliation. This expression is used in Chap.~\ref{results}  
as one diagnostic to ensure that the numerical code used 
to solve our model system is producing sensible results.

 \newpage  
\section{The Conformally Flat Approximation (CFA)} \label{sec:CFC_Maximal_slicing}

This section provides a detailed description of the conformally 
flat approximation (CFA) of general relativity.  Our definition of 
the CFA includes a particular choice of time coordinate, known 
as maximal slicing, which is briefly discussed
in~Sec.~\ref{subsec:max_slice}. 
We continue in Sec.~\ref{subsec:conf_dec} with a detailed review 
of a specific conformal decomposition of the $3+1$ Einstein equations.
Once the conformal Einstein equations have been derived, we introduce 
the assumption of conformal flatness in~\ref{subsec:conf_flat} and
derive the simplified set of field equations that result.  These 
equations are covariant with respect to a choice of spatial coordinates,
and in Sec.~\ref{subsec:cartesian} we fix those coordinates to be 
Cartesian. This then yields the actual PDEs that we solve numerically.
Sec.~\ref{subsec:bdy_cond} discusses three different 
approaches we investigated for imposing boundary conditions on the PDEs.
Finally, Sec.~\ref{subsec:mass} describes the definition and calculation of 
the ADM mass, which we use as a diagnostic quantity in our computations.

\subsection{Maximal Slicing} \label{subsec:max_slice}

In the initial value, or Cauchy, formulation formulation of Einstein's
equations, the choice of time coordinate is related to the choice of the 
spacelike hypersurfaces that foliate spacetime since the hypersurfaces 
are level surfaces of the time coordinate. 
As briefly discussed in Sec.~\ref{sec:adm}, the embedding 
of these three dimensional hypersurfaces in the four dimensional spacetime 
is described by the extrinsic curvature $K_{ab}$. It is therefore natural
to choose the time coordinate by imposing a condition on the extrinsic 
curvature. One particular choice widely used in numerical relativity 
demands that trace of the extrinsic curvature tensor vanish:
\beq \label{eq:slicing}
K \equiv K^a{}{}_{a} = 0 \qquad  \textrm{and}  \qquad  \pa_t K = 0.
\eeq

This choice is called \emph{maximal slicing} and it is a particularly 
useful slicing for numerical computation since it tends to avoid 
spacetime singularities---such as those that arise from gravitational
collapse of matter to a black hole---by ``freezing'' the 
evolution in regions close to locations where such singularities
are developing~\cite{estabrook_etal:1973}. 
Considering a congruence of worldlines for a family of observers 
travelling normally to the hypersurfaces, the maximal slicing condition
implies that the expansion of the congruence vanishes, inhibiting 
focusing of the worldlines as well as the formation of caustics. 
Moreover, as the name suggests, and due to the non-Euclidean signature
of spacetime, $K = 0$ slices have {\em maximal} volume with respect 
to small, but arbitrary, deformations of the 
hypersurfaces.  The interested reader is directed to reference~\cite{MTW}
for a more detailed discussion of this choice of time slicing and its 
properties.

\subsection{Conformal Decomposition of the Einstein Equations} \label{subsec:conf_dec}

Conformal decompositions of the Einstein equations in the context of the 
$3+1$ formalism were first introduced by Lichnerowicz in 1944 
\cite{lichnerowicz:1944}, who proposed a specific conformal decomposition 
of the Hamiltonian constraint. The goal of the decomposition was to 
write the constraint as a manifestly elliptic partial differential
equation for a specific part of the 3-metric---namely an overall scale,
or conformal factor---and then to establish existence and uniqueness 
of solutions of the PDE.
In the 1970's and early 1980's, York and his 
collaborators (most notably \'O Murchadha)~\cite{York:1971hw,York:1972sj,York:1972pr,York:1973ia,O'Murchadha:1974nc,O'Murchadha:1974nd,Bowen:1980yu,york:1983},
made significant and highly influential advances of Lichnerowicz's work 
through a general program aimed at understanding which 
of the basic dynamical variables in the 3+1 approach should be 
freely specified at the initial time, and which should be fixed by 
the constraints.
An analogy was made to electromagnetism, where the 
field can be decomposed into longitudinal and 
transverse parts, with the former representing the gauge degree of freedom,
and the latter encoding the dynamical (radiative) content of the 
theory.  Building on previous work by Deser and 
others~\cite{deser:1967,berger_ebin:1969}, York~\cite{York:1973ia} 
thus considered similar decompositions for the case  of the
symmetric rank-2 tensor fields that appear in the $3+1$ approach:
namely, the spatial metric and the extrinsic curvature.
The main idea was that a covariant~\footnote{Here ``covariant'' means 
covariant with respect to coordinate transformations in the 
hypersurfaces, i.e.~with respect to spatial diffeomorphisms.} decomposition of 
these tensors could yield at least a formal (or perhaps approximate)
solution to the constraint equations, as well as providing a route 
to establishing existence and uniqueness of the solutions.
In addition, it was hoped that the process would lead to an identification
of the ``true'' dynamical degrees of freedom of the gravitational 
field within the $3+1$ framework.  Ideally, these degrees 
of freedom were to be covariant and freely specifiable, and were to 
encapsulate the radiative content of general relativity.  
York's work resulted in a specific conformal decomposition in which 
the transverse-traceless part of the dynamical 
variables, $\{\ga_{ab},K_{ab}\}$, were to represent the radiative 
degrees of freedom.  

Now, there is a strong argument to be made that the goals of York's effort 
were not entirely successful, especially in terms of identifying 
radiative degrees of freedom on a single slice in the strong-field
regime. However there can be no doubt that his program was an absolutely 
crucial development for numerical relativists since it allowed the 
constraint equations to be written as a set of coupled, nonlinear 
elliptic PDEs, for which existence and uniqueness {\em could} be 
established. Moreover the equations could be tackled and solved 
using standard numerical approaches for elliptic systems.  
Most importantly for this thesis, the conformal transverse-traceless 
decomposition lies at the heart of the approximation to the Einstein 
equations that we adopt for our model.  

Thus, reemphasizing that the 
physical interpretation of York's approach is still a matter of
debate, we now proceed to work through the details of the decomposition.
\footnote{The reader who is interested in more details 
concerning transverse-traceless decompositions and their relationship
to gravitational radiation in general is directed to the 
review paper by Thorne~\cite{thorne:1980}.}
We start with the general case in which no assumptions are made about 
the 3-metric, $\gamma_{ab}$, and then specialize in~Sec.~\ref{subsec:conf_flat} 
to the case of a conformally flat  $\gamma_{ab}$.  Our discussion
parallels lecture notes on the $3+1$ approach due to 
Gourgoulhon~\cite{Gourgoulhon:2007ue} and, as mentioned previously,
we include the development here largely for the sake of completeness.

The first step towards the conformal decomposition of the Einstein 
equations in $3+1$ form is an investigation of how each of the fields
that appear in the equations changes under a conformal transformation. 
Specifically, we must consider the action of conformal scalings on 
the components of 1) the 3-metric, $\ga_{ij}$,
2) the connection, $C^i{}_{jk}$, associated with the spatial covariant 
derivative $D_i$, 3) the 3-Ricci tensor, ${\cal R}_{ij}$, and, finally,
4) the extrinsic curvature $K_{ij}$.  Each of these is discussed in
turn below, after which results are assembled and used in the 
decomposition of the field equations themselves.

For the case of the induced spatial metric, $\ga_{ij}$, 
conformal decomposition means that we introduce a base metric,
$\tilde{\ga}_{ij}$, and a strictly positive function, 
$\psi\equiv\psi(x^\mu)$, known as the {\em conformal factor}, and then
write
\beq \label{eq:conf_metric}
\ga_{ij} \equiv \psi^4 \tilde{\ga}_{ij} \, .
\eeq 
We note that $\tilde{\ga}_{ij}$ is frequently known as the {\em conformal
metric} and emphasize that it has no direct physical interpretation. 
Defining the inverse of the conformal metric, $\tilde{\ga}^{jk}$, in 
the usual way, so that $\tilde{\ga}_{ij}\tilde{\ga}^{jk} = \de^k{}_i$,
and noting that  $\ga_{ij}\ga^{jk} = \de^k{}_i$ we have
\beq \label{eq:inv_conf_metric}
\ga^{ij} = \psi^{-4} \tilde{\ga}^{ij}.
\eeq 

Proceeding to the connection, we first recall that
any covariant derivative, $D_i$, can be defined in terms of the ordinary 
derivative operator, $\pa_i$ via
\beq
D_i v^j = \pa_i v^j + \Ga^j{}_{ik} v^k,
\eeq
where $v^j$ are the components of a spatial vector 
and $\Ga^j{}_{ik}$ are the Christoffel 
symbols associated with $D_i$. 
If the covariant derivative is chosen to be compatible with the 
spatial metric, as will be done here, then we have $D_i\ga_{jk}=0$ and the 
Christoffel symbols can be calculated using the usual formula:
\beq
\Ga^j{}_{ik} = \ha \ga^{jl} \lb \pa_i \ga_{kl} + \pa_k \ga_{il} - \pa_l \ga_{ik} \rb \, .
\eeq
Similarly, choosing a covariant derivative $\tilde{D}_i$ compatible
with the conformal metric, so that $\tilde{D}_i\tilde{\ga}_{jk}=0$, we 
have:
\beq
{\tilde \Ga}^j{}_{ik} = \ha {\tilde \ga}^{jl} \lb \pa_i {\tilde \ga}_{kl} + \pa_k {\tilde \ga}_{il} - \pa_l {\tilde \ga}_{ik} \rb \, .
\eeq
The two covariant 
derivatives $D_i$ and ${\tilde D}_i$ can be related through the 
connection tensor, $C^j{}_{ik}$:
\beq \label{eq:cov_D_connection}
D_i v^j = \tilde{D}_i v^j + C^j{}_{ik} v^k ,
\eeq
where $C^j{}_{ik}$ is calculated in terms of conformal covariant 
derivatives as follows:
\beq \label{eq:connection}
C^j{}_{ik} = \ha \ga^{jl} \lb \tilde{D}_i \ga_{kl} + \tilde{D}_k \ga_{il} - 
                               \tilde{D}_l \ga_{ik} \rb \, .
\eeq
This last expression can be rewritten in terms of the conformal metric
using equations~(\ref{eq:conf_metric})~and~(\ref{eq:inv_conf_metric})
along with the fact that $\tilde{D}_i\tilde{\ga}_{jk}=0$.  After some 
manipulation, we find:
\beq\label{eq:connection_conf}
C^j{}_{ik} = 2 \tilde{\ga}^{jl} \lhb \tilde{D}_i \lb \ln{\psi} \rb \tilde{\ga}_{kl} + 
                                      \tilde{D}_k \lb \ln{\psi} \rb \tilde{\ga}_{il} -
                                      \tilde{D}_l \lb \ln{\psi} \rb \tilde{\ga}_{ik} \rhb .
\eeq
The contraction of the above equation on its first and second indices,
or, due to the symmetry $C^j{}_{ik} = C^j{}_{ki}$, on its first 
and third indices, can be used to provide a useful expression for the 
divergence of a vector in terms of a conformal divergence of an
appropriate conformal scaling of the vector:
\beq \label{eq:div_vector}
D_i v^i = \tilde{D}_i v^i + C^i{}_{ik} v^k = \tilde{D}_i v^i + 6 \tilde{D}_i \lb 
\ln{\psi} \rb v^i =  \fr{1}{\psi^6} \tilde{D}_i \lb \psi^6 v^i \rb.
\eeq

We next reexpress the spatial Ricci tensor, $\mathcal{R}_{ik}$, and 
Ricci scalar, $\mathcal{R}$, in terms of their 
conformal counterparts, $\tilde{\mathcal{R}}_{ik}$, and $\tilde{\mathcal{R}}$,
as well as additional terms involving conformal derivatives of $\psi$. 
To that end we start from the definition of the spatial Riemann tensor 
and then rewrite it in terms of the connection tensor
$C^j{}_{ik}$. From the definition~(\ref{eq:riemann3d}) we have
\beq
\label{eq:rijkl}
\mathcal{R}_{ijk}{}^{l} w_l = 2 D_{[i} D_{j]} w_k  \, .
\eeq
In addition, the covariant derivative of a tensor of 
type $(p,q)$ is given by
\beq \label{eq:cov_tensor}
D_k T^{i_1 \ldots i_p}{}_{j_1 \ldots j_q} = 
\tilde{D}_k T^{i_1 \ldots i_p}{}_{j_1 \ldots j_q} 
+ \sum_{l=1}^{p} C^{i_l}{}_{ks} T^{i_1 \ldots s \ldots i_p}{}_{j_1 \ldots j_q}
- \sum_{l=1}^{q} C^{s}{}_{kj_l} T^{i_1 \ldots i_p}{}_{j_1 \ldots s \ldots j_q}
\, .
\eeq 
Using~(\ref{eq:cov_tensor}) in (\ref{eq:rijkl}), and performing some 
algebraic simplifications we find:
\beq \label{eq:riemann_conf}
\mathcal{R}_{ijk}{}^{l} = \tilde{\mathcal{R}}_{ijk}{}^{l}
-2 \tilde{D}_{[i} C^l{}_{j]k} + 2 C^m{}_{k[i} C^l{}_{j]m} \, ,
\eeq
The Ricci tensor is now given by contracting this last equation on its 
second and fourth indices, yielding 
\beq \label{eq:riccit_conf}
\mathcal{R}_{ik} \equiv \mathcal{R}_{ilk}{}^{l}= \tilde{\mathcal{R}}_{ik} 
-2 \tilde{D}_{[i} C^l{}_{l]k} + 2 C^m{}_{k[i} C^l{}_{l]m} \, .
\eeq
Using~(\ref{eq:connection_conf}) in (\ref{eq:riccit_conf}, we have the 
desired relationship between the physical and conformal spatial Ricci
tensors.
\bea 
\mathcal{R}_{ik} &=& \tilde{\mathcal{R}}_{ik} 
-2 \tilde{D}_i \tilde{D}_k \lb \ln{\psi} \rb
-2 \tilde{\ga}_{ik} \tilde{\ga}^{lm}  \tilde{D}_l \tilde{D}_m \lb \ln{\psi} \rb \nonumber\\
&+&4 \tilde{D}_i  \lb \ln{\psi} \rb \tilde{D}_k \lb \ln{\psi} \rb
-4 \tilde{\ga}_{ik} \tilde{\ga}^{lm}  
\tilde{D}_l  \lb \ln{\psi} \rb \tilde{D}_m \lb \ln{\psi} \rb \, .     \label{eq:riccit_conf1}
\eea
Taking the trace of this equation then provides an expression relating 
the physical and conformal Ricci scalars:
\bea \label{eq:ricci_conf}
\mathcal{R} \equiv \ga^{ik} \mathcal{R}_{ik} &=& \psi^{-4} \lhb 
\tilde{\mathcal{R}}
-8 \tilde{\ga}^{ik}  \tilde{D}_i \tilde{D}_k \lb \ln{\psi} \rb
-8 \tilde{\ga}^{ik} \tilde{D}_i  \lb \ln{\psi} \rb \tilde{D}_k \lb \ln{\psi} \rb 
                                                         \rhb  \nonumber \\
&=& \psi^{-4}\tilde{\mathcal{R}}-8\psi^{-5}\tilde{\ga}^{ik} \tilde{D}_i \tilde{D}_k \psi \, ,
\eea
where $\tilde{\mathcal{R}} \equiv \tilde{\ga}^{ik} \tilde{\mathcal{R}}_{ik}$.
One sees immediately from this last equation one of the crucial results of 
Lichnerowicz's approach: namely that the conformal transformation ``pulls 
out'' from the extremely complicated expression for $\mathcal{R}$ (when 
written out in full for the case of a generic 3-metric) a term
that is proportional to the covariant Laplacian of the conformal factor,
$\tilde{\ga}^{ik} \tilde{D}_i \tilde{D}_k \psi$

Before moving on to a conformal treatment of the Einstein equations 
themselves, we need to perform a conformal decomposition of the 
extrinsic curvature tensor, $K_{ij}$.  In anticipation of our use 
of the decomposition in conjunction with maximal slicing, 
$K\equiv K^i{}_i=0$,
it is convenient to first write the tensor as a sum of traceless 
and traceful pieces:
\begin{eqnarray}
\label{eq:extr_curv_trace_dec}
K_{ij} &=& A_{ij} + \fr{1}{3} K \ga_{ij},\\
\label{eq:extr_curv_trace_dec_contra}
K^{ij} &=& A^{ij} + \fr{1}{3} K \ga^{ij},
\end{eqnarray}
where $A_{ij}$ and $A^{ij}$ are, by definition, the traceless parts of 
$K_{ij}$ and $K^{ij}$, respectively, so that
$\ga^{ij} A_{ij}=\ga_{ij}A^{ij}=0$.  
In the following it will eventually be $A^{ij}$--- i.e.
the traceless part of $K^{ij}$---that we will conformally transform 
according to 
\beq
	A^{ij} \equiv \psi^s \tilde{A}^{ij} \, .
\eeq
The non-zero conformal power, $s$, appearing in this definition can be chosen 
so that convenient mathematical relationships result. In this 
regard it turns out to be natural to start from the evolution 
equations~(\ref{eq:g_evo_adm_cov})
for the spatial metric components. We recall that these are given by
\beq \label{eq:g_evo_adm_lie}
\Screll_m \ga_{ij} \equiv \Screll_{(t-\bt)} \ga_{ij} = -2 \al K_{ij} \, ,
\eeq
where $m^a = t^a -\bt^a = \al n^a$.  Using~(\ref{eq:extr_curv_trace_dec})
and~(\ref{eq:conf_metric}) in the above, we find
\beq \label{eq:conf_evo}
\Screll_m \tilde{\ga}_{ij} = -2 \al \psi^{-4} A_{ij} - \fr{2}{3} \al K \tilde{\ga}_{ij}
-4\tilde{\ga}_{ij} \Screll_m \lb \ln{\psi} \rb .
\eeq
Note that this equation involves time derivatives (through the Lie 
derivatives along $m^a$) for both the conformal metric and the 
conformal factor itself, and in this sense is a ``coupled'' evolution 
equation for $\tilde{\ga}_{ij}$ and $\psi$.  Since we wish to treat 
these quantities as dynamically independent, we must perform some 
manipulation to get decoupled evolution equations.
We start by taking the trace of~(\ref{eq:conf_evo}) by 
contracting both sides with $\tilde{\ga}_{ij}$, and making use of 
$\tilde{\ga}^{ij}A_{ij}=\psi^{-4}\ga^{ij}A_{ij}=0$:
\beq \label{eq:conf_evo_trace}
\tilde{\ga}^{ij} \Screll_m \tilde{\ga}_{ij} = -2 \al K -12 \Screll_m \lb \ln{\psi} \rb .
\eeq
We now use the well known formula 
\beq
\label{eq:ddet}
\de(\ln{\det{M}}) \equiv \textrm{tr } (M^{-1} \cdot \de M)
\eeq
where $M$ is an invertible matrix, $\delta$ represents an arbitrary 
derivative operator, tr denotes the trace operation and $\cdot$ is
matrix multiplication, to rewrite the left hand side 
of~(\ref{eq:conf_evo_trace}) as
\beq
\label{eq:conf_evo_a}
\tilde{\ga}^{ij} \Screll_m \tilde{\ga}_{ij} = \Screll_m \lb \ln{\det{\tilde{\ga}_{ij}}} 
 \rb = \Screll_m \lb \ln{\tilde{\ga}} \rb \, ,
\eeq
where ${\tilde \ga}\equiv{\rm det}(\tilde{\ga}_{ij})$.
Now, in the approach to the conformal decomposition of the Einstein equations
that we are following, a key demand is that the determinant of the 
conformal spatial metric be Lie dragged from hypersurface to hypersurface 
along the vector field $t^a$:
\beq
\label{eq:conf_evo_b}
	\Screll_t \tilde{\ga}=0 \, .
\eeq
Using~(\ref{eq:conf_evo_b}) in~(\ref{eq:conf_evo_a}) we have
\beq
\label{eq:conf_evo_c}
\tilde{\ga}^{ij} \Screll_m \tilde{\ga}_{ij} = - \Screll_{\bt} \lb \ln{\tilde{\ga}} \rb.
\eeq
Moreover, again using~(\ref{eq:ddet}) we can write 
\beq
\label{eq:conf_evo_d}
	- \Screll_{\bt} \lb \ln{\tilde{\ga }} \rb =
	- \tilde{\ga}^{ij} \Screll_{\bt} \tilde{\ga}_{ij} \, ,
\eeq
so that~(\ref{eq:conf_evo_c}) becomes
\beq
\label{eq:conf_evo_e}
\tilde{\ga}^{ij} \Screll_m \tilde{\ga}_{ij} = 
- \tilde{\ga}^{ij} \Screll_{\bt} \tilde{\ga}_{ij} = - 2 \tilde{D}_i \bt^i ,
\eeq
where we have used the definition~(\ref{eq:lie_def}) in the last 
equality above.
We can then use this last result to eliminate the 
term $\tilde{\ga}^{ij} \Screll_m \tilde{\ga}_{ij}$ that appears in 
Eq.~(\ref{eq:conf_evo_trace}).  Performing the substitution and
some trivial manipulation and rearrangement, we have (recalling 
that  $m^a = t^a - \beta^a$)
\beq \label{eq:conf_factor_evo}
\Screll_m \ln{\psi} = \lb \pa_t - \Screll_{\bt} \rb \ln{\psi} = \fr{1}{6} \lb \tilde{D}_i \bt^i - \al K\rb \, .
\eeq
This then provides the desired (decoupled) evolution equation for 
the conformal factor, $\psi$.  

Additionally, (\ref{eq:conf_factor_evo}) can now be 
used in~(\ref{eq:conf_evo}) to 
yield a set of evolution equations for the conformal metric components:
\begin{eqnarray}
	\nonumber
	\lb \pa_t - \Screll_{\bt} \rb  \tilde{\ga}_{ij} 
   &=& -2 \al \psi^{-4} A_{ij} - \fr{2}{3} \al K \tilde{\ga}_{ij}
	-4\tilde{\ga}_{ij} \Screll_m \lb \ln{\psi} \rb \\
	\nonumber
	&=& -2 \al \psi^{-4} A_{ij} - \fr{2}{3} \al K \tilde{\ga}_{ij}
	-\frac{2}{3}\tilde{\ga}_{ij} \lb {\tilde D}_k \beta^k - \al K \rb \\
	\label{eq:conf_metric_evo}
	&=& -2 \al \tilde{A}_{ij}
	- \fr{2}{3} \tilde{\ga}_{ij} \tilde{D}_k \bt^k ,
\end{eqnarray}
where we have (suggestively) defined $\tilde{A}_{ij} \equiv \psi^{-4} A_{ij}$. 
Note that $\tilde{A}_{ij}$ 
is traceless
\beq
\tilde{\ga}^{ij}\tilde{A}_{ij}=\psi^4\ga^{ij}\psi^{-4}A_{ij}=0 \, ,
\eeq
and that the contravariant 
components, ${\tilde A}^{ij}$, are given by
\beq \label{eq:Aij_def}
{\tilde A}^{ij} \equiv {\tilde \ga}^{ik} {\tilde \ga}^{jl}{\tilde A}_{kl}
A^{ij} = \psi^{-4} A^{ij},
\eeq
Thus, the conformal decomposition of the evolution equation for the spatial 
metric suggests that we choose $s=-4$ for the conformal exponent in 
\beq
	A^{ij} \equiv \psi^s \tilde{A}^{ij} \, .
\eeq
It is worth emphasizing, however, that the choice $s=-4$ is not unique in terms
of leading to simplifications (or ``naturalness'') in the conformal equations. 
Another common scaling that was first adopted by Lichnerowicz 
\cite{lichnerowicz:1944} is $s=-10$. This choice also arises naturally from a decomposition of one of the Einstein 
equations, but this time when the decomposition is applied to the momentum 
constraint equations. Since we do not use this specific decomposition in the current
work, the reader who is interested in further details concerning it is directed to
Gourgoulhon's lectures \cite{Gourgoulhon:2007ue}.

We now turn to the task of deriving conformal evolution equations for 
${\tilde A}_{ij}$ and $K$.
This will be done in two stages. We will 
first use the $3+1$ evolution equations for $K_{ij}$, as given by (\ref{eq:k_evo_adm_cov}),
to derive evolution equations for the trace-free components, $A_{ij}$, and the trace, $K$.  
These equations will be expressed in terms of physical (i.e.~non conformally scaled) quantities, but will then be used in the second stage to derive update equations for
${\tilde A}_{ij}$ and $K$ that are written in terms of the conformally scaled variables. 

We thus start from~(\ref{eq:k_evo_adm_cov}), which we recall reads
\beq \label{eq:k_evo_adm_decomp}
\Screll_m K_{ij} = -D_i D_j \al + \al \lhb \mathcal{R}_{ij} + K K_{ij} -2K_{ik}K^k{}_j 
     + 8\pi \lb \ha \ga_{ij} (S-\rho) - S_{ij} \rb \rhb .
\eeq
Now from definition~(\ref{eq:extr_curv_trace_dec}) of the decomposition of $K_{ij}$, we have $K_{ij} = A_{ij} + \frac13 K \ga_{ij}$,
so
\beq \label{eq:lie_kij}
\Screll_m K_{ij} = \Screll_m A_{ij} + \fr{1}{3} \ga_{ij} \Screll_m K +  \fr{1}{3} K
\Screll_m \ga_{ij} = \Screll_m A_{ij} + \fr{1}{3} \ga_{ij} \Screll_m K - \fr{2}{3}\al K
K_{ij} ,
\eeq
where we have used~(\ref{eq:g_evo_adm_lie}) in the last step. Furthermore, we have
\beq \label{eq:lie_k}
\Screll_m K = \Screll_m \lb \ga^{ij} K_{ij} \rb = \ga^{ij} \Screll_m  K_{ij} +
K_{ij} \Screll_m \ga^{ij} = \ga^{ij} \Screll_m  K_{ij} + 2 \al K_{ij}  K^{ij},
\eeq
where we have used~(\ref{eq:g_evo_adm_contra})
\beq \label{eq:g_evo_adm_lie_contra}
\Screll_m \ga^{ij} \equiv \Screll_{(t-\bt)} \ga^{ij} = 2 \al K^{kj}  \, .
\eeq
In addition, taking the trace of~(\ref{eq:k_evo_adm_decomp}) we find
\beq \label{eq:k_evo_adm_decomp_trace}
\ga^{ij} \Screll_m  K_{ij} = -\ga^{ij}  D_i D_j \al + \al \lhb \mathcal{R} + K^2 
-2K_{ij}K^{ij} + 4\pi \lb S - 3\rho \rb \rhb .
\eeq
Substitution of this last result in~(\ref{eq:lie_k}) gives
\beq \label{eq:lie_k1}
\Screll_m K = -\ga^{ij}  D_i D_j \al + \al \lhb \mathcal{R} + K^2
+ 4\pi \lb S - 3\rho \rb \rhb . 
\eeq
Now the 3-Ricci scalar $\mathcal{R}$ is an extremely complicated function of $\ga_{ij}$ 
and its first and second derivatives.  It is thus computationally advantageous to use the 
Hamiltonian constraint~(\ref{eq:ham_adm}) in the form
\beq
	\mathcal{R} = -K^2 + K_{ij}K^{ij} + 16\pi \rho
\eeq
to eliminate $\mathcal{R}$ from~(\ref{eq:lie_k1}).  The yields the evolution equation
for $K$ in its final form: 
\beq \label{eq:lie_k2}
\Screll_m K \equiv \lb \pa_t - \Screll_{\bt} \rb K 
= -\ga^{ij}  D_i D_j \al + \al \lhb K_{ij}K^{ij}
+ 4\pi \lb S + \rho \rb \rhb  .
\eeq

We now continue by deriving the evolution equation for the the trace free components,
$A_{ij}$, of the physical extrinsic curvature tensor.  We first solve~(\ref{eq:lie_kij}) 
for $\Screll_m A_{ij}$:
\beq
	\Screll_m A_{ij} = \Screll_m K_{ij} - \frac13 \ga_{ij} \Screll_m K + \frac23 \al K K_{ij}
	\, .
\eeq
Using~(\ref{eq:k_evo_adm_decomp}) and (\ref{eq:lie_k2}) to replace the terms $\Screll_m K_{ij}$ and $\Screll_m K$, respectively, we find
\bea
\Screll_m A_{ij} = -  D_i D_j \al &+& 
\al \lhb \mathcal{R}_{ij} + \fr{5}{3} K K_{ij} -2K_{ik}K^k{}_j
                   - 8\pi \lb  S_{ij} - \fr{1}{3} \ga_{ij} S  \rb \rhb \nonumber\\
&+&\fr{1}{3} \ga_{ij} \lhb  \ga^{lk} D_l D_k \al -\al \lb \mathcal{R} + K^2 \rb \rhb
\label{eq:Aij_evo} \, .
\eea
Using the decompositions
\bea
	K_{ij} &=& A_{ij} + \frac13 K \ga_{ij} \, ,\\
   K^i{}_j &=& A^i{}_j + \frac 13 K \delta^i{}_j \, ,
\eea
in the above, we find, after some manipulation
\bea
\Screll_m A_{ij} = -  D_i D_j \al &+& 
\al \lhb \mathcal{R}_{ij} + \fr{1}{3} K A_{ij} -2A_{ik}A^k{}_j
                   - 8\pi \lb  S_{ij} - \fr{1}{3} \ga_{ij} S  \rb \rhb \nonumber\\
&+&\fr{1}{3} \ga_{ij} \lb  \ga^{lk} D_l D_k \al -\al  \mathcal{R} \rb
\label{eq:Aij_evo1} .
\eea
This is the final form of the evolution equation for $A_{ij}$, which together with 
the evolution equation~(\ref{eq:lie_k2}) provides an equivalent system to the original
$3+1$ equation~(\ref{eq:k_evo_adm_decomp}) for $K_{ij}$.  

We now proceed to the second stage of the current calculation which involves 
reexpressing~(\ref{eq:Aij_evo1}) and (\ref{eq:lie_k2}) in terms of the conformally
rescaled variables.  We begin by considering the left hand side of~(\ref{eq:Aij_evo1}),
which can be rewritten as
\beq
\Screll_m A_{ij} = \Screll_m \lb \psi^4  \tilde{A}_{ij} \rb =  
\psi^4 \Screll_m \tilde{A}_{ij} +4\psi^3 \Screll_m \psi \tilde{A}_{ij} .
\eeq
Using~(\ref{eq:conf_factor_evo}) to eliminate the $\Screll_m \psi$ term in this
last expression we have
\beq
\Screll_m A_{ij} = \psi^4 \lhb \Screll_m \tilde{A}_{ij} + \fr{2}{3} \tilde{A}_{ij}
\lb \tilde{D}_l \bt^l - \al K  \rb\rhb .
\eeq
Replacing the left hand side of~(\ref{eq:Aij_evo1}) with the right hand side of the 
above equation, and solving for $\Screll_m {\tilde A}_{ij}$, we find
\bea
\Screll_m \tilde{A}_{ij} &=& \psi^{-4} \lbr -  D_i D_j \al  + 
\al \lhb \mathcal{R}_{ij} + \fr{1}{3} K A_{ij} -2A_{ik}A^k_{\; j}
                   - 8\pi \lb  S_{ij} - \fr{1}{3} \ga_{ij} S  \rb \rhb  \rp
\nonumber\\
&+& \lp \fr{1}{3} \ga_{ij} \lb  \ga^{lk} D_l D_k \al -\al  \mathcal{R} \rb \rbr
- \fr{2}{3} \tilde{A}_{ij} \lb \tilde{D}_l \bt^l - \al K  \rb  .
\label{eq:Aij_tilde_evo}
\eea

Proceeding, we must also rewrite the terms involving the physical covariant derivatives 
of the lapse function, $\al$, using the derivative operator, ${\tilde D}_i$, 
which is compatible with the conformal metric $\tilde{\ga}_{ij}$.  Specifically, we have
\bea
D_i D_j \al &=& \tilde{D}_i \lb D_j \al \rb - C^k_{\; ij} \lb D_k \al \rb
\nonumber\\
&=& \tilde{D}_i \tilde{D}_j \al - C^k_{\; ij}  \tilde{D}_k \al 
\nonumber\\
&=& \tilde{D}_i \tilde{D}_j \al - 2 \tilde{D}_k \al \lhb 
\tilde{D}_i \lb \ln{\psi} \rb \de^k{}_j +
\tilde{D}_j \lb \ln{\psi} \rb \de^k{}_i -
\tilde{\ga}_{ij} \tilde{\ga}^{kl} \tilde{D}_l \lb \ln{\psi} \rb 
                                                    \rhb
\nonumber\\
&=& \tilde{D}_i \tilde{D}_j \al 
- 2  \tilde{D}_i \lb \ln{\psi} \rb \tilde{D}_j \al
- 2  \tilde{D}_j \lb \ln{\psi} \rb \tilde{D}_i \al
+ 2\tilde{\ga}_{ij} \tilde{\ga}^{kl} \tilde{D}_l \lb \ln{\psi} \rb \tilde{D}_k \al
\label{eq:double_lps} \, ,
\eea
where the definitions~(\ref{eq:cov_tensor}) and~(\ref{eq:connection_conf}) 
were used in the first and third 
lines, respectively. 
We also need the trace of this equation, which is given by
\beq
\label{eq:double_lps_trace}
\ga^{ij} D_i D_j \al = \psi^{-4} \tilde{\ga}^{ij} D_i D_j \al = \psi^{-4} \lhb
\tilde{\ga}^{ij} \tilde{D}_i \tilde{D}_j \al 
+2 \tilde{\ga}^{ij} \tilde{D}_i \lb \ln{\psi} \rb \tilde{D}_j \al \rhb .
\eeq

We have now expressed all of the terms that appear in the evolution equation~(\ref{eq:Aij_tilde_evo}) for ${\tilde A}_{ij}$ in terms of conformally
scaled quantities. Assembling results, we find after a considerable amount of algebraic 
simplification that
\bea
\lb \pa_t - \Screll_{\bt} \rb \tilde{A}_{ij} &=& \psi^{-4} \lbr
-\tilde{D}_i \tilde{D}_j \al
+ 4  \tilde{D}_{(i} \lb \ln{\psi} \rb \tilde{D}_{j)} \al
+ \fr{1}{3} \tilde{\ga}_{ij} \tilde{\ga}^{kl} \lb 
\tilde{D}_k \tilde{D}_l \al
-4\tilde{D}_k \lb \ln{\psi} \rb \tilde{D}_l \al \rb \rp
\nonumber \\
&+& \al \lhb \tilde{\mathcal{R}}_{ij}-\fr{1}{3}\tilde{\ga}_{ij} \tilde{\mathcal{R}}
-2 \tilde{D}_i \tilde{D}_j \lb \ln{\psi} \rb 
+4\tilde{D}_i \lb \ln{\psi} \rb \tilde{D}_j \lb \ln{\psi} \rb \rp
\nonumber \\
&+&\lp \lp \fr{2}{3} \tilde{\ga}_{ij} \tilde{\ga}^{kl} \lb
\tilde{D}_k \tilde{D}_l \lb \ln{\psi} \rb 
-2 \tilde{D}_k \lb \ln{\psi} \rb \tilde{D}_l \lb \ln{\psi} \rb  \rb
\rhb \rbr
\nonumber \\
&+&\al \lhb K \tilde{A}_{ij} -2 \tilde{\ga}^{kl}\tilde{A}_{ik}\tilde{A}_{lj} 
-8\pi \lb \psi^{-4}S_{ij}-\fr{1}{3}\tilde{\ga}_{ij} S \rb \rhb
-\fr{2}{3} \tilde{A}_{ij} \tilde{D}_l \bt^l  \, .
\label{eq:Aij_tilde_evo_conf}
\eea
Here we have used equations~(\ref{eq:riccit_conf1}), (\ref{eq:ricci_conf}), (\ref{eq:double_lps}) and (\ref{eq:double_lps_trace}) as well as 
$A_{ij} = \psi^4 {\tilde A}_{ij}$.

We can now also write down the evolution equation for $K$ in terms of the conformally
scaled variables.  To do so we first note that from the 
decompositions~(\ref{eq:extr_curv_trace_dec}) and {\ref{eq:extr_curv_trace_dec_contra}),
and using the fact that $A_{ij}$ is traceless, 
we have 
\beq \label{eq:kijkij_conf}
K_{ij} K^{ij} = \lb A_{ij} + \fr{1}{3} K \ga_{ij} \rb 
                \lb A^{ij} + \fr{1}{3} K \ga^{ij}  \rb
              = \tilde{A}_{ij} \tilde{A}^{ij}+\fr{1}{3} K^2 ,
\eeq
Using this last result as well as~(\ref{eq:double_lps_trace}) in~(\ref{eq:lie_k2}),
we find
\beq \label{eq:lie_k2_conf}
 \lb \pa_t - \Screll_{\bt} \rb K 
= - \psi^{-4} \tilde{\ga}^{kl} \lhb  \tilde{D}_k \tilde{D}_l \al 
+2  \tilde{D}_k \lb \ln{\psi} \rb \tilde{D}_l \al \rhb
+ \al \lhb  \tilde{A}_{kl} \tilde{A}^{kl}+\fr{1}{3} K^2
+ 4\pi \lb S + \rho \rb \rhb  .
\eeq
Eqs.~(\ref{eq:Aij_tilde_evo_conf}) and (\ref{eq:lie_k2_conf}) are our desired evolution
equations for the extrinsic curvature quantities within the conformal framework. 

Having dealt with the evolution equations for the spatial metric and extrinsic curvature,
we are left with the conformal treatment of the constraint equations. 

First, using~(\ref{eq:ricci_conf}) and (\ref{eq:kijkij_conf}) in the
Hamiltonian constraint~(\ref{eq:ham_adm}), we readily find
\beq \label{eq:ham_adm_conf} 
\tilde{\ga}^{kl} \tilde{D}_k \tilde{D}_l \psi -\fr{1}{8}\psi\tilde{\mathcal{R}} 
+ \lb \fr{1}{8} \tilde{A}_{kl} \tilde{A}^{kl} 
-\fr{1}{12} K^2 +2\pi \rho \rb \psi^5 = 0 \, .
\eeq

Second, in order to rewrite the momentum constraint, we first calculate
\bea
D_j K^{ij} &=& D_j A^{ij} + \fr{1}{3} \ga^{ij} D_j K
\nonumber\\
&=& \tilde{D}_j A^{ij}+C^i_{\; jk}A^{jk}+C^j_{\; jk}A^{ik}+\fr{1}{3}\psi^{-4}
\tilde{\ga}^{ij} \tilde{D}_j K
\nonumber\\
&=& \tilde{D}_j A^{ij}+4 \tilde{D}_j \lb \ln{\psi} \rb A^{ij} 
+ 6\tilde{D}_j\lb \ln{\psi} \rb A^{ij}+\fr{1}{3}\psi^{-4}\tilde{D}^i K
\nonumber\\
&=& \tilde{D}_j A^{ij}+\psi^{-10} \tilde{D}_j \lb \psi^{10} A^{ij}\rb  
+\fr{1}{3}\psi^{-4}\tilde{D}^i K \, .
\label{eq:kij_div}
\eea
In the above derivation we have used the following: (\ref{eq:extr_curv_trace_dec}) in writing 
down the first line; (\ref{eq:cov_tensor}), as well as the fact that the 
actions of $D_i$ and ${\tilde D}_i$ on scalar functions such as $K$
are identical, in going from the first to the second line; expression~(\ref{eq:connection_conf}) for the connection, and the fact 
that $A_{ij}$ is traceless in going from the second to the third line.
Using~(\ref{eq:kij_div}) in 
the momentum constraint~(\ref{eq:mom_adm}) then yields
\beq
\tilde{D}_j A^{ij}+\psi^{-10} \tilde{D}_j \lb \psi^{10} A^{ij}\rb 
-\fr{2}{3}\psi^{-4} \tilde{D}^i K = 8 \pi J^i ,    \label{eq:mom_adm_conf} 
\eeq
and using $A^{ij} = \psi^{-4} {\tilde A}^{ij}$
we have
\beq
\tilde{D}_j \tilde{A}^{ij}+6 \tilde{D}_j \lb \ln{\psi} \rb \tilde{A}^{ij}
-\fr{2}{3} \tilde{D}^i K = 8 \pi \psi^4 J^i  .    \label{eq:mom_adm_conf1} 
\eeq

We have now completed the task of conformally decomposing the Einstein field equations
within the $3+1$ formalism, and it is thus appropriate to collect the results.

We have the following set of evolution equations for the conformal
factor, $\psi$, the conformal metric $\ga_{ij}$, the trace of the 
extrinsic curvature, $K$, and the conformally scaled, trace-free part of the 
extrinsic curvature, ${\tilde A}_{ij}$:
\beq
\label{eq:conf_factor_evo_gather}
\lb \pa_t - \Screll_{\bt} \rb \psi = \fr{1}{6} \psi \lb \tilde{D}_i \bt^i - \al K\rb , 
\eeq
\beq
\label{eq:conf_metric_evo_gather}
\lb \pa_t - \Screll_{\bt} \rb \tilde{\ga}_{ij} = -2 \al \tilde{A}_{ij}
- \fr{2}{3} \tilde{\ga}_{ij} \tilde{D}_k \bt^k , 
\eeq
\beq
\label{eq:lie_k2_conf_gather}
 \lb \pa_t - \Screll_{\bt} \rb K 
 = - \psi^{-4} \tilde{\ga}^{kl} \lhb  \tilde{D}_k \tilde{D}_l \al 
+2  \tilde{D}_k \lb \ln{\psi} \rb \tilde{D}_l \al \rhb
+ \al \lhb  \tilde{A}_{kl} \tilde{A}^{kl}+\fr{1}{3} K^2
+ 4\pi \lb S + \rho \rb \rhb , 
\eeq
\bea
\lb \pa_t - \Screll_{\bt} \rb \tilde{A}_{ij} &=& \psi^{-4} \lbr
-\tilde{D}_i \tilde{D}_j \al
+ 4  \tilde{D}_{(i} \lb \ln{\psi} \rb \tilde{D}_{j)} \al
+ \fr{1}{3} \tilde{\ga}_{ij} \tilde{\ga}^{kl} \lb 
\tilde{D}_k \tilde{D}_l \al
-4\tilde{D}_k \lb \ln{\psi} \rb \tilde{D}_l \al \rb \rp
\nonumber \\
&+& \al \lhb \tilde{\mathcal{R}}_{ij}-\fr{1}{3}\tilde{\ga}_{ij} \tilde{\mathcal{R}}
-2 \tilde{D}_i \tilde{D}_j \lb \ln{\psi} \rb 
+4\tilde{D}_i \lb \ln{\psi} \rb \tilde{D}_j \lb \ln{\psi} \rb \rp
\nonumber \\
&+&\lp \lp \fr{2}{3} \tilde{\ga}_{ij} \tilde{\ga}^{kl} \lb
\tilde{D}_k \tilde{D}_l \lb \ln{\psi} \rb 
-2 \tilde{D}_k \lb \ln{\psi} \rb \tilde{D}_l \lb \ln{\psi} \rb  \rb
\rhb \rbr
\nonumber \\
&+&\al \lhb K \tilde{A}_{ij} -2 \tilde{\ga}^{kl}\tilde{A}_{ik}\tilde{A}_{lj} 
-8\pi \lb \psi^{-4}S_{ij}-\fr{1}{3}\tilde{\ga}_{ij} S \rb \rhb
-\fr{2}{3} \tilde{A}_{ij} \tilde{D}_l \bt^l .
\label{eq:Aij_tilde_evo_conf_gather}
\eea
We also have the Hamiltonian and momentum constraints:
\beq \label{eq:ham_adm_conf_gather} 
\tilde{\ga}^{kl} \tilde{D}_k \tilde{D}_l \psi -\fr{1}{8}\psi\tilde{\mathcal{R}} 
+ \lb \fr{1}{8} \tilde{A}_{kl} \tilde{A}^{kl} 
-\fr{1}{12} K^2 +2\pi \rho \rb \psi^5 = 0 ,
\eeq
\beq
\tilde{D}_j \tilde{A}^{ij}+6 \tilde{D}_j \lb \ln{\psi} \rb \tilde{A}^{ij}
-\fr{2}{3} \tilde{D}^i K = 8 \pi \psi^4 J^i .     
	\label{eq:mom_adm_conf1_gather} 
\eeq

The above system of $3+1$ conformal Einstein equations is to be solved
for the conformal unknowns
$\psi$, $\tilde{\ga}_{ij}$, $K$ and $\tilde{A}_{ij}$.
Once the conformal variables have been computed, the physical 
quantities---namely the spatial 
metric $\ga_{ij}$ and the extrinsic curvature $K_{ij}$---can be recovered 
from~(\ref{eq:conf_metric}), (\ref{eq:extr_curv_trace_dec}) and 
$A_{ij} = \psi^4 {\tilde A}_{ij}$:
\bea 
\ga_{ij}&=&\psi^4 \tilde{\ga}_{ij},\\ 
K_{ij}&=&\psi^4 \lb \tilde{A}_{ij} + \fr{1}{3} K \tilde{\ga}_{ij} \rb .
\eea 

\subsection{The CFA Equations} \label{subsec:conf_flat}

We are now in a position to derive the equations that define 
the approximation of general relativity that we adopt in this thesis.
As discussed previously (Sec.~\ref{subsec:conf_dec}), a key element of the 
approximation is the demand that the 3-metric, $\ga_{ij}$, be conformally 
flat, so that ${\tilde \ga}_{ij}$ is the constant, flat 3-metric,
which we denote ${\hat \ga}_{ij}$.  This motivates 
the nomenclature ``Conformally Flat Approximation'' (CFA) that we have 
adopted.  However, we emphasize that we combine the demand of 
conformal flatness with the particular time coordinatization
given by the maximal slicing condition, which requires that 
$K=0$ on each hypersurface.  Together, conformal flatness and maximal 
slicing result in a significant simplification of the system
of conformal Einstein equations that was displayed at the end 
of the previous section.
First, denoting by ${\hat D}_i$ the covariant derivative compatible with
the flat 3-metric ${\hat \ga}_{ij}$, and using $K = 0$, 
the evolution equation~(\ref{eq:conf_factor_evo_gather}) for the conformal
factor $\psi$ reduces to
\beq \label{eq:conf_factor_evo_flat}
\lb \pa_t - \Screll_{\bt} \rb \psi = \fr{1}{6} \psi \hat{D}_i \bt^i  \, .
\eeq

Next, the evolution equation~(\ref{eq:conf_metric_evo_gather}}) for 
$\hat{\ga}_{ij}$, becomes 
\beq \label{eq:conf_metric_evo_flat}
\lb \pa_t - \Screll_{\bt} \rb \hat{\ga}_{ij} = -2 \al \tilde{A}_{ij}
- \fr{2}{3} \hat{\ga}_{ij} \hat{D}_k \bt^k \, .
\eeq

Continuing, since $K$ must identically vanish on all slices, we have 
$(\partial_t - \Screll_\bt) K = 0$. Thus 
equation~(\ref{eq:lie_k2_conf_gather}) becomes 
\beq \label{eq:lie_k2_conf_flat}
0 
= - \psi^{-4} \hat{\ga}^{kl} \lhb  \hat{D}_k \hat{D}_l \al 
+2  \hat{D}_k \lb \ln{\psi} \rb \hat{D}_l \al \rhb
+ \al \lhb  \tilde{A}_{kl} \tilde{A}^{kl}
+ 4\pi \lb S + \rho \rb \rhb   ,
\eeq
which is now interpreted as an elliptic equation that the lapse function,
$\al$, must satisfy on each slice in order that the hypersurfaces remain
maximal.

Proceeding to equation~(\ref{eq:Aij_tilde_evo_conf_gather}), we note 
that since the conformal metric is flat, the associated Ricci
tensor, ${\tilde {\mathcal R}}_{ij} = {\hat {\mathcal R}}_{ij}$
and Ricci scalar, ${\tilde {\mathcal R}} = {\hat {\mathcal R}}$, both
vanish.  Using this fact, along with $K=0$, 
the evolution equation for 
$\tilde{A}_{ij}$ becomes
\bea
\lb \pa_t - \Screll_{\bt} \rb \tilde{A}_{ij} &=& \psi^{-4} \lbr
-\hat{D}_i \hat{D}_j \al
+ 4  \hat{D}_{(i} \lb \ln{\psi} \rb \hat{D}_{j)} \al
+ \fr{1}{3} \hat{\ga}_{ij} \hat{\ga}^{kl} \lb 
\hat{D}_k \hat{D}_l \al
-4\hat{D}_k \lb \ln{\psi} \rb \hat{D}_l \al \rb \rp
\nonumber \\
&+& \al \lhb 
-2 \hat{D}_i \hat{D}_j \lb \ln{\psi} \rb 
+4\hat{D}_i \lb \ln{\psi} \rb \hat{D}_j \lb \ln{\psi} \rb \rp
\nonumber \\
&+&\lp \lp \fr{2}{3} \hat{\ga}_{ij} \hat{\ga}^{kl} \lb
\hat{D}_k \hat{D}_l \lb \ln{\psi} \rb 
-2 \hat{D}_k \lb \ln{\psi} \rb \hat{D}_l \lb \ln{\psi} \rb  \rb
\rhb \rbr
\nonumber \\
&+&\al \lhb  -2 \hat{\ga}^{kl}\tilde{A}_{ik}\tilde{A}_{lj} 
-8\pi \lb \psi^{-4}S_{ij}-\fr{1}{3}\hat{\ga}_{ij} S \rb \rhb
-\fr{2}{3} \tilde{A}_{ij} \hat{D}_l \bt^l \, .
\label{eq:Aij_tilde_evo_conf_flat}
\eea

Turning to the Hamiltonian constraint~(\ref{eq:ham_adm_conf_gather}),
and again using ${\tilde {\mathcal R}} = 0$ and $K=0$, we find
\beq \label{eq:ham_adm_conf_flat} 
\hat{\ga}^{kl} \hat{D}_k \hat{D}_l \psi 
+ \lb \fr{1}{8} \tilde{A}_{kl} \tilde{A}^{kl} 
+2\pi \rho \rb \psi^5 = 0  , .
\eeq
 
Finally the momentum constraint~(\ref{eq:mom_adm_conf1_gather}) becomes
\beq
\hat{D}_j \tilde{A}^{ij}+6 \hat{D}_j \lb \ln{\psi} \rb \tilde{A}^{ij}
 = 8 \pi \psi^4 J^i      \label{eq:mom_adm_conf1_flat} \, ,
\eeq
where once again $K=0$ has been used.  

Eq.~(\ref{eq:conf_metric_evo_flat}) merits further attention. Since 
the conformal 3-metric is constant from slice to slice we have
$\pa_t \hat{\ga}_{ij} =0$.  
Additionally, Eq.~(\ref{eq:lie_def}) tells us that we can write
\beq
	\Screll_\bt {\hat \ga}_{ij} = \hat{D}_k {\hat \ga}_{ij} +
   \hat{\ga}_{kj} \hat{D}_i \bt^k + \hat{\ga}_{ik} \hat{D}_j \bt ^k =
   \hat{\ga}_{kj} \hat{D}_i \bt^k + \hat{\ga}_{ik} \hat{D}_j \bt ^k \, ,
\eeq
where we have used $\hat{D}_k {\hat \ga}_{ij}=0$ in the 
last step.  Using this last equation along with $\pa_t \hat{\ga}_{ij} =0$
in~(\ref{eq:conf_metric_evo_flat}) we find 
\beq \label{eq:Aij_down_flat}
 \tilde{A}_{ij} =
\fr{1}{2\al} \lb   \hat{\ga}_{kj} \hat{D}_i \bt^k + \hat{\ga}_{ik} \hat{D}_j \bt^k  
- \fr{2}{3} \hat{\ga}_{ij} \hat{D}_k \bt^k \rb \, .
\eeq
We can also raise 
both indices of~(\ref{eq:Aij_down_flat})---using the inverse flat metric, 
${\hat \ga}^{ij}$---to get
\beq \label{eq:Aij_up_flat}
 \tilde{A}^{ij} =
\fr{1}{2\al} \lb   \hat{\ga}^{ik} \hat{D}_k \bt^j + \hat{\ga}^{jk} \hat{D}_k \bt^i  
- \fr{2}{3} \hat{\ga}^{ij} \hat{D}_k \bt^k \rb .
\eeq
We thus observe that having made the requirement of conformal flatness,
we can view the components of $\tilde{A}^{ij}$ as being 
{\em derived} from the shift vector components, $\beta^k$.
Now our general inventory of geometric variables within the context of the 
conformal
$3+1$ approach included the 4 kinematical quantities $\al$ and $\bt^k$,
and the 12 dynamical fields $\psi$, ${\tilde \ga}_{ij}$, $K$ and 
${\tilde A}_{ij}$.  We now see that of these, once conformal flatness 
has been imposed and we have chosen $K=0$, only $\al$, $\bt^k$ and 
$\psi$ remain as independent variables.  We have already noted 
that~(\ref{eq:lie_k2_conf_flat}) provides an elliptic equation that 
fixes $\al$ on each slice, and we will display the final form
of that equation momentarily.  We now proceed to show that the constraints
provide similar equations for $\psi$ and $\beta^k$.

To do so, we first consider the divergence of $\tilde{A}^{ij}$.  
Using~(\ref{eq:Aij_up_flat}) we calculate
\bea
\hat{D}_j \tilde{A}^{ij} &=& \fr{1}{2\al} \hat{D}_j \lb   
\hat{\ga}^{ik} \hat{D}_k \bt^j + \hat{\ga}^{jk} \hat{D}_k \bt^i
- \fr{2}{3} \hat{\ga}^{ij} \hat{D}_k \bt^k \rb 
- \fr{1}{\al} \hat{D}_j \al \tilde{A}^{ij} 
\nonumber\\
&=& \fr{1}{2\al} \lb \hat{\ga}^{jk} \hat{D}_k \hat{D}_j \bt^i 
+ \fr{1}{3} \hat{\ga}^{ik} \hat{D}_k \hat{D}_j \bt^j 
- 2\tilde{A}^{ij} \hat{D}_j \al \rb \label{eq:Aij_div_flat} .
\eea

Using~(\ref{eq:Aij_div_flat}) and (\ref{eq:Aij_up_flat}) 
in the momentum constraint~(\ref{eq:mom_adm_conf1_flat}) and rearranging,
we find
\bea
\hat{\ga}^{jk} \hat{D}_k \hat{D}_j \bt^i &=& 
- \fr{1}{3} \hat{\ga}^{ik} \hat{D}_k \hat{D}_j \bt^j 
+2\tilde{A}^{ij} \lhb \hat{D}_j \al -6\al \hat{D}_j\lb \ln{\psi} \rb \rhb
+16 \pi \al \psi^4 J^i
\nonumber\\
&=& - \fr{1}{3} \hat{\ga}^{ik} \hat{D}_k \hat{D}_j \bt^j
+\lhb \hat{\ga}^{ik} \hat{D}_k \bt^j + \hat{\ga}^{jk} \hat{D}_k \bt^i
- \fr{2}{3} \hat{\ga}^{ij} \hat{D}_k \bt^k  \rhb
\hat{D}_j \lhb \ln{\lb\fr{\al}{\psi^6}\rb}\rhb
\nonumber\\
&+&16 \pi \al \psi^4 J^i \label{eq:ellipt_shift} \, ,
\eea
which are 3 elliptic equations for the 3 shift vector components, $\bt^k$.
Next, we use (\ref{eq:Aij_down_flat}) and (\ref{eq:Aij_up_flat}) 
to rewrite the product $\tilde{A}_{ij} \tilde{A}^{ij}$ in terms of 
covariant derivatives of the shift vector:
\beq \label{eq:AijAij_flat}
\tilde{A}_{ij} \tilde{A}^{ij} = \fr{1}{2\al} \lb 
\hat{\ga}_{kj}  \hat{\ga}^{il} \hat{D}_i \bt^k \hat{D}_l \bt^j +
\hat{D}_k \bt^i \hat{D}_i \bt^k 
- \fr{2}{3} \hat{D}_i \bt^i \hat{D}_j \bt^j \rb . \nonumber
\eeq
Substituting the right hand side of this equation into the 
Hamiltonian constraint~(\ref{eq:ham_adm_conf_flat}) and rearranging, we have
\beq \label{eq:ellipt_psi} 
\hat{\ga}^{kl} \hat{D}_k \hat{D}_l \psi 
= - \lhb \fr{1}{16\al}\lb 
\hat{\ga}_{kj}  \hat{\ga}^{il} \hat{D}_i \bt^k \hat{D}_l \bt^j +
\hat{D}_k \bt^i \hat{D}_i \bt^k
- \fr{2}{3} \hat{D}_i \bt^i \hat{D}_j \bt^j \rb 
-2\pi \rho \rhb \psi^5 .
\eeq
which is an elliptic equation for the conformal factor, $\psi$.

Finally, (\ref{eq:lie_k2_conf_flat}), which we recall arises from the demand 
that $K\equiv0$  can be written as
\bea 
\hat{\ga}^{kl} \hat{D}_k \hat{D}_l \al  &=&
-2 \hat{\ga}^{kl} \hat{D}_k \lb \ln{\psi} \rb \hat{D}_l \al 
+ \psi^4 \al \lhb  \tilde{A}_{kl} \tilde{A}^{kl}
+ 4\pi \lb S + \rho \rb \rhb  
\nonumber\\
&=& -2 \hat{\ga}^{kl} \hat{D}_k \lb \ln{\psi} \rb \hat{D}_l \al
+ \ha \psi^4 \lhb \hat{\ga}_{kj}  \hat{\ga}^{il} \hat{D}_i \bt^k \hat{D}_l \bt^j +
\hat{D}_k \bt^i \hat{D}_i \bt^k
- \fr{2}{3} \hat{D}_i \bt^i \hat{D}_j \bt^j \rhb 
\nonumber\\
&+& 4\pi \psi^4 \al \lb S + \rho \rb \label{eq:ellipt_lapse} \, ,
\eea
and is, as stated previously, an elliptic equation for the lapse function,
$\al$.

We emphasize that a sequence of solutions of 
equations~(\ref{eq:ellipt_shift})--(\ref{eq:ellipt_lapse})---computed on 
each hypersurface of the 
spacetime---completely determines the 4-dimensional 
metric, $g_{\mu\nu}$. Specifically, we have
\begin{eqnarray} \label{eq:adm_metric_conf_flat}
ds^2 &=& g_{\mu\nu} dx^\mu dx^\nu 
\nonumber \\
     &=& -\al^2 dt^2 + \ga_{ij} \left(dx^i + \bt^i dt \right)
     \left(dx^j + \bt^j dt \right)
\nonumber \\
     &=& -\al^2 dt^2 + \psi^4 \hat{\ga}_{ij} \left(dx^i + \bt^i dt \right)
     \left(dx^j + \bt^j dt \right) .
\end{eqnarray}
Table~\ref{table:KLG_EOM} summarizes the steps taken in this chapter
to produce the CFA equations of motion.  It includes enumerations of 
the full Einstein-Klein-Gordon equations within both the standard and 
conformal $3+1$ approaches, as well as the subset of the conformal 
equations that were used to derive the CFA system.

In concluding this section, we note that the system 
(\ref{eq:ellipt_shift})--(\ref{eq:ellipt_lapse}) is still written 
in a generally 3-covariant form---that is, the equations are valid for 
any set of spatial coordinates that we might choose to adopt in
conjunction with a flat 3-metric.  In order to recast the system as 
a specific set of PDEs that can be solved numerically, we must first
fix the spatial coordinates, and this is the topic of the next section.
\begin{landscape}
\headsep = 1.0in
\hoffset = -0.25in
\begin{table}[hpt]
\begin{tabular}{|c|c|c|} \hline
EKG eqns. in $3+1$ form. & 
Conformally decomposed EKG eqns. & 
CFA equations of motion. \\ \hline
\small
$ \begin{aligned}
\mathcal{R} + K^2 - K_{ij} K^{ij} = & \  16 \pi \rho \ (\ref{eq:ham_adm}) \\ & \\
D_j K^{ij} - D^i K = & \  8 \pi J^i                  \ (\ref{eq:mom_adm}) 
\end{aligned} $
&  
\small
$ \begin{aligned}
\tilde{\ga}^{kl} \tilde{D}_k \tilde{D}_l \psi -\fr{1}{8}\psi\tilde{\mathcal{R}} 
+ \lb \fr{1}{8} \tilde{A}_{kl} \tilde{A}^{kl} 
-\fr{1}{12} K^2 
 \rb \psi^5 =& \  -2\pi \rho \psi^5 \ (\ref{eq:ham_adm_conf_gather}) \\ 

\tilde{D}_j \tilde{A}^{ij}+6 \tilde{D}_j \lb \ln{\psi} \rb \tilde{A}^{ij}
-\fr{2}{3} \tilde{D}^i K =& \  8 \pi \psi^4 J^i  \  (\ref{eq:mom_adm_conf1_gather}) 
\end{aligned} $
& 
\small
$ \begin{aligned}
\hat{\ga}^{kl} \hat{D}_k \hat{D}_l \psi =& \  
- \lb \fr{1}{8} \tilde{A}_{kl} \tilde{A}^{kl} 
+2\pi \rho \rb \psi^5 \quad (\ref{eq:ham_adm_conf_flat})  \\

\hat{\ga}^{jk} \hat{D}_k \hat{D}_j \bt^i =&  \  
- \fr{1}{3} \hat{\ga}^{ik} \hat{D}_k \hat{D}_j \bt^j  
+2\tilde{A}^{ij} \lhb \hat{D}_j \al \right. \\
-& \left. 6\al \hat{D}_j\lb \ln{\psi} \rb \rhb
+16 \pi \al \psi^4 J^i \quad  (\ref{eq:ellipt_shift}) 
\end{aligned} $
\\ \hline
\small
$ \begin{aligned}
\Screll_{(t-\bt)}  \ga_{ij} =& \  - 2 \al \ga_{ik} K^k{}_{j} 
\ (\ref{eq:g_evo_adm}) \\ & \\ & \\

\Screll_{(t-\bt)} K^i{}_{j} =& \  
-D^i D_j \al   \\
+& \  \al \lhb  
\mathcal{R}^i{}_{j} 
+  K K^i{}_{j} \rp \\
+& \  8 \pi 
\lb
\ha \de^i{}_{j} ( S - \rho ) \rp \\ 
-& \lp \lp S^i{}_{j}
\rb
\rhb \ \ (\ref{eq:k_evo_adm}) \\ & \\ & \\ & \\ & \\ & \\ & \\ & 
\end{aligned} $
& 
\small
$ \begin{aligned}
\lb \pa_t - \Screll_{\bt} \rb \psi =& \  \fr{1}{6} \psi \lb \tilde{D}_i \bt^i - \al K\rb 
\quad (\ref{eq:conf_factor_evo_gather}) \\

\lb \pa_t - \Screll_{\bt} \rb \tilde{\ga}_{ij} =& \  -2 \al \tilde{A}_{ij}
- \fr{2}{3} \tilde{\ga}_{ij} \tilde{D}_k \bt^k 
\quad (\ref{eq:conf_metric_evo_gather}) \\

\lb \pa_t - \Screll_{\bt} \rb K =& 
- \psi^{-4} \tilde{\ga}^{kl} \lhb  \tilde{D}_k \tilde{D}_l \al 
+2  \tilde{D}_k \lb \ln{\psi} \rb \tilde{D}_l \al \rhb \\
+& \  \al \lhb  \tilde{A}_{kl} \tilde{A}^{kl}+\fr{1}{3} K^2
+ 4\pi \lb S + \rho \rb \rhb  
\quad (\ref{eq:lie_k2_conf_gather}) \\

\lb \pa_t - \Screll_{\bt} \rb \tilde{A}_{ij} =& \  \psi^{-4} \lbr
-\tilde{D}_i \tilde{D}_j \al
+ 4  \tilde{D}_{(i} \lb \ln{\psi} \rb \tilde{D}_{j)} \al \rp
\\
+& \  \fr{1}{3} \tilde{\ga}_{ij} \tilde{\ga}^{kl} \lb 
\tilde{D}_k \tilde{D}_l \al
-4\tilde{D}_k \lb \ln{\psi} \rb \tilde{D}_l \al \rb 
\\
+& \al \lhb \tilde{\mathcal{R}}_{ij}-\fr{1}{3}\tilde{\ga}_{ij} \tilde{\mathcal{R}}
-2 \tilde{D}_i \tilde{D}_j \lb \ln{\psi} \rb \rp 
\\
+& \ 4\tilde{D}_i \lb \ln{\psi} \rb \tilde{D}_j \lb \ln{\psi} \rb 
\\
+&\lp \lp \fr{2}{3} \tilde{\ga}_{ij} \tilde{\ga}^{kl} \lb
\tilde{D}_k \tilde{D}_l \lb \ln{\psi} \rb 
-2 \tilde{D}_k \lb \ln{\psi} \rb \tilde{D}_l \lb \ln{\psi} \rb  \rb
\rhb \rbr
\\
+&\al \lhb K \tilde{A}_{ij} -2 \tilde{\ga}^{kl}\tilde{A}_{ik}\tilde{A}_{lj} 
-8\pi \lb \psi^{-4}S_{ij}-\fr{1}{3}\tilde{\ga}_{ij} S \rb \rhb
\\
-& \ \fr{2}{3} \tilde{A}_{ij} \tilde{D}_l \bt^l 
\quad (\ref{eq:Aij_tilde_evo_conf_gather})
\end{aligned} $
& 
\small
$ \begin{aligned}
\color{red}
\Screll_{(t-\bt)} \psi =& \  \color{red} \fr{1}{6} \psi \hat{D}_i \bt^i  
\quad (\ref{eq:conf_factor_evo_flat}) \\

\tilde{A}_{ij} =& \ 
\fr{1}{2\al} \lb   \hat{\ga}_{kj} \hat{D}_i \bt^k + \hat{\ga}_{ik} \hat{D}_j \bt^k  
\rp \\
 -& \lp \fr{2}{3} \hat{\ga}_{ij} \hat{D}_k \bt^k \rb
\quad (\ref{eq:Aij_down_flat}) \\

\hat{\ga}^{kl} \hat{D}_k \hat{D}_l \al  =& \ 
-2 \hat{\ga}^{kl} \hat{D}_k \lb \ln{\psi} \rb \hat{D}_l \al \\ 
+& \  \psi^4 \al \lhb  \tilde{A}_{kl} \tilde{A}^{kl}
+ 4\pi \lb S + \rho \rb \rhb  
\quad (\ref{eq:ellipt_lapse}) \\

\color{red}
\Screll_{(t-\bt)} \tilde{A}_{ij} =& \ 
\color{red}
\psi^{-4} \lbr -\hat{D}_i \hat{D}_j \al
+ 4  \hat{D}_{(i} \lb \ln{\psi} \rb \hat{D}_{j)} \al \rp 
\\
\color{red}
+& 
\color{red}
\fr{1}{3} \hat{\ga}_{ij} \hat{\ga}^{kl} \lb 
\hat{D}_k \hat{D}_l \al
-4\hat{D}_k \lb \ln{\psi} \rb \hat{D}_l \al \rb 
\\
\color{red}
+& \  
\color{red}
\al \lhb -2 \hat{D}_i \hat{D}_j \lb \ln{\psi} \rb 
+4\hat{D}_i \lb \ln{\psi} \rb \hat{D}_j \lb \ln{\psi} \rb \rp
\\
\color{red}
+&
\color{red}
\lp \lp \fr{2}{3} \hat{\ga}_{ij} \hat{\ga}^{kl} \lb
\hat{D}_k \hat{D}_l \lb \ln{\psi} \rb 
-2 \hat{D}_k \lb \ln{\psi} \rb \hat{D}_l \lb \ln{\psi} \rb  \rb
\rhb \rbr
\\
\color{red}
+& \ 
\color{red}
\al \lhb  -2 \hat{\ga}^{kl}\tilde{A}_{ik}\tilde{A}_{lj} 
-8\pi \lb \psi^{-4}S_{ij}-\fr{1}{3}\hat{\ga}_{ij} S \rb \rhb 
\\
\color{red}
-& \
\color{red}
\fr{2}{3} \tilde{A}_{ij} \hat{D}_l \bt^l 
\quad (\ref{eq:Aij_tilde_evo_conf_flat})

\end{aligned} $
\\ \hline
\small
$ \begin{aligned}
\pa_t\phi_A =& \ \fr{\al}{\sr{\ga}}\Pi_A +\bt^i\pa_i\phi_A \ (\ref{eq:phidot_adm})\\
\pa_t\Pi_A =& \ \pa_i(\bt^i\Pi_A) \\
+& \ \pa_i(\al\sr{\ga}\ga^{ij}\pa_j\phi_A) \\
-& \ \al\sr{\ga} \fr{dU(\phi_0^2)}{d\phi_0^2} \phi_A \  (\ref{eq:pidot_adm})
\end{aligned} $
& 
\small
$ \begin{aligned}
\pa_t\phi_A =& \ \fr{\al}{\psi^6\sr{\tilde{\ga}}}\Pi_A +\bt^i\pa_i\phi_A \\
\pa_t\Pi_A =& \ \pa_i(\bt^i\Pi_A) \\
+& \ \pa_i(\al\psi^2\sr{\tilde{\ga}}\tilde{\ga}^{ij}\pa_j\phi_A) 
- \al\psi^6\sr{\tilde{\ga}} \fr{dU(\phi_0^2)}{d\phi_0^2} \phi_A 
\end{aligned} $
& 
\small
$ \begin{aligned}
\pa_t\phi_A =& \ \fr{\al}{\psi^6\sr{\hat{\ga}}}\Pi_A +\bt^i\pa_i\phi_A \\
\pa_t\Pi_A =& \ \pa_i(\bt^i\Pi_A) \\
+& \ \pa_i(\al\psi^2\sr{\hat{\ga}}\hat{\ga}^{ij}\pa_j\phi_A) 
- \al\psi^6\sr{\hat{\ga}} \fr{dU(\phi_0^2)}{d\phi_0^2} \phi_A 
\end{aligned} $
\\ \hline
\end{tabular}
\caption[Equations of Motion.]{Equations of motion. This table summarizes
the steps taken in the derivation of the CFA equations of motion. The 
fully general relativistic $3+1$ Einstein-Klein-Gordon (EKG) equations,
and their conformal decompositions, are given in
the first two columns. The first row lists the constraint equations, while
the evolution equations for the geometric and matter variables are 
are listed in the second
and third rows, respectively. The maximal slicing condition and 
assumption of conformal flatness yield the CFA equations; these 
are given in 
the third column. Note that within the CFA the components of 
the traceless conformal extrinsic curvature components, $\tilde{A}_{ij}$,
are {\em defined} in terms of the lapse function and shift vector, and that the 
evolution equations for the conformal factor, $\psi$, and 
the $\tilde{A}_{ij}$, shown in red, are not used.}
\label{table:KLG_EOM}
\end{table}

\end{landscape}

\subsection{Specialization to Cartesian Coordinates} \label{subsec:cartesian}

The choice of spatial coordinates is frequently a crucial issue in 
numerical relativity calculations.  In cases where the spacetime 
is constrained to have an exact symmetry, or where the physical scenario has 
an approximate symmetry, coordinates adapted to the symmetry 
are often used.  Adopting such coordinates---which will generally
be curvilinear---can be advantageous not only for the significant 
simplification in the equations of motion that may result, but also 
for providing a ``natural'' representation of the solution, which in turn
can minimize computational cost.  For example, consider the problem 
of simulating a single star in 3 spatial dimensions plus time, but where 
the star remains approximately spherically symmetric. Then, irrespective
of the method we choose to discretize the continuum equations 
(i.e.~finite difference, finite element, spectral etc.), we can 
expect that for specified accuracy it will cheaper to compute a 
solution using 
spherical polar coordinates than Cartesian coordinates.

However, especially when dealing with tensor equations, the use 
of curvilinear coordinates for numerical calculations can also 
be problematic.  Such coordinate systems generically have coordinate 
singularities, such as $r=0$ and the ``$z$--axis'' in the spherical 
polar case, where special attention and care is needed when
designing and implementing the numerical approach to ensure 
that the computed solution remains smooth.   In any case,
the ultimate goal of the work described here is to simulate boson star 
binaries,  where the expected solutions will not have any specific simple 
symmetry that would motivate us to adopt some special curvilinear 
coordinate system.  We instead use a Cartesian coordinate system, 
$(x,y,z)$, which has the advantages of 
1) covering the spatial hypersurfaces 
in a smooth fashion (i.e. without any coordinate singularities or other 
pathologies) provided that there are no physical singularities on the 
slices,  and 2) inducing particularly simple forms for the differential 
operators appearing in the governing PDEs.

In addition to having Cartesian ``topology'', and in contrast to some previous 
related work (most notably that of Wilson, Mathews, Marronetti~\cite{Wilson:1995uh,Wilson:1996ty,Mathews:1997vw,Mathews:1997nm,matthews:1998, Marronetti:1998vm, marronetti_mathews_wilson:1999,Mathews:1999km,Wilson:2002yh})
we require that our coordinate system be ``asymptotically inertial'', i.e. 
we demand that at large distances from the matter sources the metric 
components approach those of flat spacetime in an inertial frame, 
$g_{\mu\nu} \to \eta_{\mu\nu}= {\rm diag} (-1,1,1,1)$.  In the work of 
Wilson and collaborators the coordinate system was typically chosen to be 
in corotation with the binary system being studied.  Finally, we emphasize 
that the topology of the hypersurfaces is taken to be $\mathbb{R}^3$, so 
that the slices are infinite in extent in all three directions.  
Naturally, this has significant implications for the numerical treatment 
of boundary conditions in our model as will be discussed in following 
sections.

Thus, from this point on, the set of field variables defining our model 
system, namely $\al$, $\psi$, $\bt^i$, $i =1,2,3$, and $\phi_A$, $\Pi_A$, 
$A=1,2$, are all to be understood to be functions of $(t,x,y,z)$. 

With the choice of Cartesian spatial coordinates, 
the conformally flat $3$-dimensional line element is simply
\beq
^{(3)}ds^2=\psi^4(dx^2+dy^2+dz^2) .
\eeq

We can now display the equations of motion for our model in 
essentially the form used for our numerical computations. 
We start with the evolution 
equations for the complex scalar field,~(\ref{eq:phidot_adm}) and 
(\ref{eq:pidot_adm}), which become
\bea
\partial_{t}\phi_{A}&=&\frac{\alpha}{\psi^6}\Pi_{A}+\beta^{i}\partial_{i}\phi_{A}
\label{eq:phidot_adm_cart},\\
\partial_{t}\Pi_{A}&=&\pa_x \lb \bt^x\Pi_A+\al\psi^2\pa_x\phi_A \rb
                     +\pa_y \lb \bt^y\Pi_A+\al\psi^2\pa_y\phi_A \rb \nonumber \\
                   &+&\pa_z \lb \bt^z\Pi_A+\al\psi^2\pa_z\phi_A \rb 
                     -\al\psi^6\fr{dU(\phi_0^2)}{d\phi_0^2}\phi_A
\label{eq:pidot_adm_cart} .
\eea

Continuing, the energy-momentum quantities defined 
by~(\ref{eq:rho_adm1})-(\ref{eq:Sphro_adm1}) are
\bea
\rho &=& \fr{1}{2}\sum_{A=1}^2 \lhb \fr{\Pi_A^2}{\psi^{12}}+ \fr{1}{\psi^4}
\lhb \lb \pa_x\phi_A \rb^2 +\lb \pa_y\phi_A \rb^2 + \lb \pa_z\phi_A \rb^2
\rhb  \rhb +\fr{1}{2} U(\phi_0^2) ,
\label{eq:rho_adm1_cart}\\
J_i&=&\sum_{A=1}^2 \lhb   \fr{\Pi_A}{\psi^6}\pa_i\phi_A \rhb, 
\label{eq:Ja_adm1_cart}\\
J^i&=&\sum_{A=1}^2 \lhb -  \fr{\Pi_A}{\psi^{10}}\pa_i\phi_A \rhb, 
\label{eq:JA_adm1_cart}\\
S_{ij}&=&\fr{1}{2}\sum_{A=1}^2 \lbr 2\pa_i\phi_A\pa_j\phi_A  
 + \psi^4\de_{ij}
\lhb \fr{\Pi_A^2}{\psi^{12}}\right. \right.  \notag \\
&&\left. \left. - \fr{1}{\psi^4} 
\lhb \lb \pa_x\phi_A \rb^2 +\lb \pa_y\phi_A \rb^2 + \lb \pa_z\phi_A \rb^2
\rhb 
\rhb \rbr-\fr{1}{2}\psi^4\de_{ij}U(\phi_0^2), 
\label{eq:Sab_adm1_cart} \\
S^i_i&=&\fr{1}{2}\sum_{A=1}^2 \lhb 3\fr{\Pi_A^2}{\psi^{12}}- \fr{1}{\psi^4}
\lhb \lb \pa_x\phi_A \rb^2 +\lb \pa_y\phi_A \rb^2 + \lb \pa_z\phi_A \rb^2
\rhb
 \rhb- \fr{3}{2}U(\phi_0^2)
\label{eq:S_adm1_cart},\\
\rho+S&=&\sum_{A=1}^2 \lhb 2\fr{\Pi_A^2}{\psi^{12}} 
 \rhb- U(\phi_0^2)
\label{eq:Sphro_adm1_cart}.
\eea

Next, we have the equations that constrain the geometric variables.
The maximal slicing condition~(\ref{eq:ellipt_lapse}) for the lapse function becomes
\beq\label{eq:lapse_cartesian}
\fr{\pa^2 \alpha}{\pa x^2} + \fr{\pa^2 \alpha}{\pa y^2} +\fr{\pa^2 \alpha}{\pa z^2}  = 
-\frac{2}{\psi} \lhb \fr{\pa \psi}{\pa x} \fr{\pa \al}{\pa x} + 
                     \fr{\pa \psi}{\pa y} \fr{\pa \al}{\pa y} +
                     \fr{\pa \psi}{\pa z} \fr{\pa \al}{\pa z}     \rhb
+ \alpha \psi^4 \left( \tilde{A}_{ij} \tilde{A}^{ij} 
+ 4 \pi \left(\rho + S \right) \right) ,
\eeq
while the Hamiltonian constraint~(\ref{eq:ellipt_psi}) for $\psi$ is
\begin{equation}\label{eq:psi_cartesian}
\fr{\pa^2 \psi}{\pa x^2} + \fr{\pa^2 \psi}{\pa y^2} +\fr{\pa^2 \psi}{\pa z^2} = 
-\frac{\psi^5}{8} \left( \tilde{A}_{ij} \tilde{A}^{ij} + 16 \pi \rho \right) .
\end{equation}
For both~(\ref{eq:lapse_cartesian}) and (\ref{eq:psi_cartesian}) the term
$\tilde{A}_{ij} \tilde{A}^{ij}$ is given by
\begin{align*} \label{eq:AijAij_cartesian}
\notag
\tilde{A}_{ij} \tilde{A}^{ij}=& \fr{1}{2\al^2} \bigg[  \lb \fr{\pa \bt^x }{\pa x} \rb ^2
                                   + \lb \fr{\pa \bt^x }{\pa y} \rb ^2
                                   + \lb \fr{\pa \bt^x }{\pa z} \rb ^2 
                                   + \lb \fr{\pa \bt^y }{\pa x} \rb ^2
                                   + \lb \fr{\pa \bt^y }{\pa y} \rb ^2
                                   + \lb \fr{\pa \bt^y }{\pa z} \rb ^2 \\
\notag        &                    + \lb \fr{\pa \bt^z }{\pa x} \rb ^2
                                   + \lb \fr{\pa \bt^z }{\pa y} \rb ^2
                                   + \lb \fr{\pa \bt^z }{\pa z} \rb ^2
  + \lb \fr{\pa \bt^x}{\pa x}  \fr{\pa\bt^x}{\pa x} 
      + \fr{\pa \bt^y}{\pa x}  \fr{\pa\bt^x}{\pa y}
      + \fr{\pa \bt^z}{\pa x}  \fr{\pa\bt^x}{\pa z} \rb  \\
\notag
& + \lb \fr{\pa \bt^x}{\pa y}  \fr{\pa\bt^y }{\pa x} 
      + \fr{\pa \bt^y}{\pa y}  \fr{\pa\bt^y }{\pa y} 
      + \fr{\pa \bt^z}{\pa y}  \fr{\pa\bt^y }{\pa z} \rb 
  + \lb \fr{\pa \bt^x}{\pa z}  \fr{\pa\bt^z}{\pa x} 
      + \fr{\pa \bt^y}{\pa z}  \fr{\pa\bt^z}{\pa y} 
      + \fr{\pa \bt^z}{\pa z}  \fr{\pa\bt^z}{\pa z} \rb  \\
& - \fr{2}{3} \lb \fr{\pa \bt^x }{\pa x} 
                + \fr{\pa \bt^y }{\pa y} 
                + \fr{\pa \bt^z }{\pa z} \rb^2 
                               \bigg] .
\end{align*}
Finally, from the momentum constraints~(\ref{eq:ellipt_shift}), 
we have the 3 equations that fix the components of the shift vector:
\bea\label{eq:shift_cartesian_x}
\fr{\pa^2 \bt^x}{\pa x^2} + \fr{\pa^2 \bt^x}{\pa y^2} +\fr{\pa^2 \bt^x}{\pa z^2} &= &
- \fr{1}{3} \fr{\pa }{\pa x} \lb \fr{\pa \bt^x }{\pa x} + \fr{\pa \bt^y }{\pa y}
+\fr{\pa \bt^z }{\pa z} \rb + \al \psi^4 \, 16 \pi J^x \\ \notag 
& &
- \fr{\pa }{\pa x} \lhb \ln \lb \fr{\psi^6}{\al} \rb \rhb \lhb \fr{4}{3} 
\fr{\pa \bt^x }{\pa x} - \fr{2}{3} \lb \fr{\pa \bt^y }{\pa y} 
+ \fr{\pa \bt^z }{\pa z} \rb \rhb \\ \notag 
& &
- \fr{\pa }{\pa y} \lhb \ln \lb \fr{\psi^6}{\al} \rb \rhb \lhb
\fr{\pa \bt^x }{\pa y} + \fr{\pa \bt^y }{\pa x} \rhb
- \fr{\pa }{\pa z} \lhb \ln \lb \fr{\psi^6}{\al} \rb \rhb \lhb
\fr{\pa \bt^x }{\pa z} + \fr{\pa \bt^z }{\pa x} \rhb ,
\eea
\bea\label{eq:shift_cartesian_y}
\fr{\pa^2 \bt^y}{\pa x^2} + \fr{\pa^2 \bt^y}{\pa y^2} +\fr{\pa^2 \bt^y}{\pa z^2} &= &
- \fr{1}{3} \fr{\pa }{\pa y} \lb \fr{\pa \bt^x }{\pa x} + \fr{\pa \bt^y }{\pa y}
+\fr{\pa \bt^z }{\pa z} \rb + \al \psi^4 \, 16 \pi J^y \\ \notag 
& &
- \fr{\pa }{\pa y} \lhb \ln \lb \fr{\psi^6}{\al} \rb \rhb \lhb \fr{4}{3} 
\fr{\pa \bt^y }{\pa y} - \fr{2}{3} \lb \fr{\pa \bt^x }{\pa x} 
+ \fr{\pa \bt^z }{\pa z} \rb \rhb \\ \notag 
& &
- \fr{\pa }{\pa x} \lhb \ln \lb \fr{\psi^6}{\al} \rb \rhb \lhb
\fr{\pa \bt^x }{\pa y} + \fr{\pa \bt^y }{\pa x} \rhb
- \fr{\pa }{\pa z} \lhb \ln \lb \fr{\psi^6}{\al} \rb \rhb \lhb
\fr{\pa \bt^y }{\pa z} + \fr{\pa \bt^z }{\pa y} \rhb ,
\eea
\bea\label{eq:shift_cartesian_z}
\fr{\pa^2 \bt^z}{\pa x^2} + \fr{\pa^2 \bt^z}{\pa y^2} +\fr{\pa^2 \bt^z}{\pa z^2} &= &
- \fr{1}{3} \fr{\pa }{\pa z} \lb \fr{\pa \bt^x }{\pa x} + \fr{\pa \bt^y }{\pa y}
+\fr{\pa \bt^z }{\pa z} \rb + \al \psi^4 \, 16 \pi J^z \\ \notag
& &
- \fr{\pa }{\pa z} \lhb \ln \lb \fr{\psi^6}{\al} \rb \rhb \lhb \fr{4}{3} 
\fr{\pa \bt^z }{\pa z} - \fr{2}{3} \lb \fr{\pa \bt^x }{\pa x} 
+ \fr{\pa \bt^y }{\pa y} \rb \rhb \\ \notag 
& &
- \fr{\pa }{\pa y} \lhb \ln \lb \fr{\psi^6}{\al} \rb \rhb \lhb
\fr{\pa \bt^z }{\pa y} + \fr{\pa \bt^y }{\pa z} \rhb
- \fr{\pa }{\pa x} \lhb \ln \lb \fr{\psi^6}{\al} \rb \rhb \lhb
\fr{\pa \bt^x }{\pa z} + \fr{\pa \bt^z }{\pa x} \rhb .
\eea

The 4 hyperbolic scalar field evolution equations, 
(\ref{eq:phidot_adm_cart}) and (\ref{eq:pidot_adm_cart}), along 
with the 5 elliptic equations, 
(\ref{eq:lapse_cartesian})-(\ref{eq:shift_cartesian_z}), 
constitute the basic set of equations for our model.
This set of PDEs must of course be supplemented by boundary and 
initial conditions in order to complete the mathematical prescription of 
the model, and these will be discussed in the sections that follow.

Before moving on to that discussion though, it is worth mentioning that 
the derivation of equations of motion such as the above set is a 
non-trivial and error prone process.  It is therefore very useful to use 
symbolic manipulation software to check calculations, and we have done so. 
Specifically, after having been derived by hand, the equations presented 
in this thesis were checked using Maple~\cite{maple}, including a tensor 
manipulation package due to Choptuik~\cite{choptuik:ftp_tensorV6}.

\subsection{Boundary Conditions} \label{subsec:bdy_cond}

In this section we discuss issues related to the boundary 
conditions that are to be applied in conjunction with the equations 
of motion summarized above.   Mathematically, our model is 
to be solved as a Cauchy problem where the $t = {\rm const.}$ surfaces 
extend to spatial infinity.~\footnote{We note that our use of the 
term ``boundary conditions'' here is thus a slight abuse of nomenclature 
in the sense that the $t={\rm const}$ surfaces are edgeless, and thus have 
no boundaries.}
We will restrict attention to cases where 
the matter source (the complex scalar field) has compact support on
the initial time-slice: the hyperbolicity of the scalar wave equation
then guarantees that the matter can never reach spatial infinity, which,
of course, is appropriate from a physical point of view.  This 
restriction, combined with the demands that 1) the spacetimes we 
construct be 
asymptotically flat, 2)~that our $(x,y,z)$ coordinate system be 
``asymptotically inertial'', and 3)~that coordinate time is identical to 
proper time at infinity, provides the following set of 
boundary conditions:
\begin{eqnarray}
	\label{eq:bcfirst}
	\lim_{r\to\infty}& &\phi_A(t,x,y,z) = 0  \, , \\
	\lim_{r\to\infty}& &\Pi_A(t,x,y,z) = 0  \, , \\
	\lim_{r\to\infty}& &\psi(t,x,y,z) = 1 \, , \\
	\lim_{r\to\infty}& &\alpha(t,x,y,z) = 1 \, , \\
	\label{eq:bclast}
	\lim_{r\to\infty}& &\bt^k(t,x,y,z) = 0 \, . 
\end{eqnarray}
Here and below $r$ is defined by $r\equiv\sqrt{x^2+y^2+z^2}$.

The computational problems that arise from the fact that our
boundary conditions are naturally specified at infinity are familiar 
ones, not only in numerical relativity, but in many other areas of
numerical analysis and computational science that involve the 
solution of hyperbolic PDEs on unbounded domains.  A key, if rather
obvious observation, is that while the spatial domain is 
infinite, any specific numerical computation based on a discretization 
technique (such as finite differencing, as used in this thesis) 
must be restricted to a finite number of discrete unknowns.
Given this, there are essentially two basic strategies for dealing
with the infinite solution domain.  The first involves artificially
introducing boundaries at $x=x_{\rm min}$, $x=x_{\rm max}$,
etc. and then imposing approximate boundary conditions on the 
solution unknowns.  The second involves ``compactification'' of the 
infinite domain, using a coordinate transformation which maps
infinity to a finite value of the transformed coordinate.  The 
exact boundary conditions can then be applied on the boundaries of 
the compactified domain.  In the 
current work, we have experimented with two variations of the first approach, 
and one of the second.  
We proceed to discuss each of the approaches that we studied, and
in the the original order that they were investigated. We highlight 
specific challenges that we encountered, and 
indicate possible future directions for improvement. We also note that parts 
of the discussion rely on computational concepts and techniques that are 
discussed in detail in Chap.~\ref{numerical}.  The reader may thus wish to 
postpone a detailed study of this part of the thesis until that chapter 
has been perused.

\subsubsection{Spatial Compactification} \label{sec:space_compact}

Spatial compactification was the first strategy we considered for 
a computational treatment of the 
conditions~(\ref{eq:bcfirst})--(\ref{eq:bclast}).  As already 
mentioned, the basic idea in this case is quite simple: a smooth coordinate 
transformation is applied to map each infinite coordinate range
to a finite interval, which, without loss of generality we
can take to be $[-1,1]$. Since elements of the Jacobian matrix of 
the coordinate transformation will appear in the PDEs when written
in the new coordinates, it is reasonable to require that the
transformation be given in some simple closed form.  With this in mind 
we chose the hyperbolic tangent function to define compactified coordinates
$(\xi,\eta,\zeta)$ by
\beq \label{eq:comp_coord}
\xi(x) = \tanh{x}, \qquad \qquad \et(y) = \tanh{y}, \qquad \qquad \ze(z) = \tanh{z}.
\eeq
Clearly, this transformation maps 
$-\infty < x,y,z < \infty$ to $-1 \le \xi,\eta,\zeta \le 1$,
as desired, and the exact (Dirichlet) boundary conditions can then be set at 
$\xi,\eta,\zeta = \pm 1$.
Transforming the PDEs that govern our model requires nothing 
more than the chain rule for differentiation.  For our specific 
choice of compactification we have
\beq \label{eq:comp_diff}
\fr{\partial}{\partial x} = 
\fr{d \xi}{dx} \fr{d}{d\xi} = (1-\tanh^2{x}) \fr{d}{d\xi} 
= (1-\xi^2) \fr{d}{d\xi},
\eeq
and
\beq \label{eq:comp_diff2}
\fr{d^2}{dx^2} = (1-\xi^2) \fr{d}{d\xi} \lhb (1-\xi^2) \fr{d}{d\xi} \rhb
 = (1-\xi^2)^2 \fr{d^2}{d\xi^2} -2\xi (1-\xi^2) \fr{d}{d\xi} \, ,
\eeq
as well as analogous formulae for the $y$ and $z$ derivatives.
Using these formulae, we derive the transformed equations 
of motion simply by rewriting all of the derivatives appearing in 
(\ref{eq:phidot_adm_cart})--(\ref{eq:pidot_adm_cart}) and
(\ref{eq:lapse_cartesian})--(\ref{eq:shift_cartesian_z}),
in terms of derivatives taken with respect to the compactified variables.

We note at this point that spatial compactification of the sort we 
have just described generically leads to well-known computational
problems for hyperbolic systems.  In particular, for the case 
of finite differencing,
and assuming that the discretization is uniform in the compactified 
coordinates (i.e.~so that the spacing between grid points in each
of the coordinate directions is constant), the physical separation
between grid points becomes increasingly large as the boundaries of the 
compactified domain are approached.  This means that waves propagating
outwards invariably become very poorly resolved.  In turn, this can lead 
to various difficulties including numerical instabilities and spurious 
reflections of waves back into the interior of the domain. 

Nonetheless, spatial compactification of this type {\em has} 
proven successful in some previous time dependent calculations in numerical 
relativity.  This includes a study of the Gregory-Laflamme instability
of black strings~\cite{Choptuik:2003qd}, as well as Pretorius' 
ground-breaking work on binary black hole inspiral and 
merger~\cite{Pretorius:2005gq,Pretorius:2004jg}.  In both of these 
instances, computational difficulties of the sort mentioned above were 
kept under control through the addition of numerical dissipation.

Given the previous discussion, as well as the 
fact that compactification {\em is} routinely used with great success
in the numerical analysis of elliptic PDEs, it is somewhat ironic
that it was problems associated
with the solution of the elliptic equations of our model that ultimately
forced us to abandon compactification.  
Noting that the specific difficulties we encountered are well documented
in the numerical analysis literature, we nonetheless feel it important
to discuss them here in some detail.  We are especially motivated
by the fact that there appear to be very few instances where 
elliptic systems have been solved in numerical 
relativity using compactification in conjunction with finite difference
techniques.

In order to understand the nature and source of the 
problem we ran into, it suffices to consider a model elliptic
PDE, namely the Poisson equation:
\beq \label{eq:poisson}
\na^2 u(x,y,z) = \rho(x,y,z) \qquad \Longleftrightarrow \qquad  
\fr{\pa^2 u}{\pa x^2} + \fr{\pa^2 u}{\pa y^2} + \fr{\pa^2 u}{\pa z^2} = \rho \, .
\eeq
Here $u$ is the unknown function while $\rho$ is a 
specified source function, which we will require to have compact support.  
In analogy with the elliptic equations of 
our model system we will consider the solution of~(\ref{eq:poisson})
on~$\mathbb{R}^3$ and require that 
\begin{equation}
	\lim_{r\to\infty} u(x,y,z) = 0  \, , \label{eq:poisson_bdy}
\end{equation} 
where, again, $r \equiv \sqrt{x^2+y^2+z^2}$.
We now transform~(\ref{eq:poisson}) to the compactified coordinate system
defined by~(\ref{eq:comp_coord}) using (\ref{eq:comp_diff2}) 
and the corresponding 
formulae for the second $y$ and $z$ derivatives.
We get 
\begin{multline} \label{eq:poisson_compact}
(1-\xi^2)^2 \fr{\pa^2 \bar{u}}{\pa \xi^2} +
(1-\et^2)^2 \fr{\pa^2 \bar{u}}{\pa \et^2} +
(1-\ze^2)^2 \fr{\pa^2 \bar{u}}{\pa \ze^2} -
\\
2\xi (1-\xi^2)\fr{\pa \bar{u}}{\pa \xi} -
2\et (1-\et^2)\fr{\pa \bar{u}}{\pa \et} -
2\ze (1-\ze^2)\fr{\pa \bar{u}}{\pa \ze} = \rho(\xi,\et,\ze) ,
\end{multline}
where $\bar{u} = \bar{u}(\xi,\et,\ze) = u(x,y,z)$.  
The boundary conditions~(\ref{eq:poisson_bdy}) now become 
\def\ub{{\bar u}}
\begin{equation}
	\ub(1,\et,\ze)  =
	\ub(-1,\et,\ze)  =
	\ub(\xi,1,\ze)  =
	\ub(\xi,-1,\ze)  =
	\ub(\xi,\et,1)  =
	\ub(\xi,\et,-1) = 0
\end{equation}
that is, they are simply Dirichlet conditions imposed on the boundaries 
of the domain $-1 \le \xi, \et, \ze \le 1$, precisely as we have for 
the elliptic equations that govern the metric variables in
our model when expressed in the $(\xi,\et,\ze)$ system.  Now, clearly,
the transformation to compactified coordinates increases
the algebraic complexity of any elliptic PDE to which it is applied,
and introduces additional terms involving first derivatives of 
the unknown.  However, dealing with these complications within 
the context of a finite difference approach
is not difficult in principle, and, at least naively, it seems
reasonable to expect that the extra 
computational cost involved will be more than offset by the increase 
in accuracy and improved convergence properties that the use of 
exact Dirichlet conditions will provide.

Unfortunately, these expectations are {\em not} met when one factors
in the combination of 1) the specific technique that we have adopted to 
solve the discretized elliptic equations, and 2) the need to eventually
have a code that runs in parallel on multi-processor architectures.
We can summarize the situation as follows.  As described in detail in 
Chap.~\ref{numerical}, we have chosen the multigrid method to solve our 
finite-differenced elliptic PDEs due to its efficiency: in particular,
it is the only available technique that can solve discretized forms
of general nonlinear elliptic systems using $O(N)$ calculations, where 
$N$ is the total number of discrete unknowns.   However, as we
will now discuss, the use of 
compactified coordinates induces a major stumbling block to the 
parallelization of the multigrid algorithm.

Consider the following schematic form for a 
Poisson-like equation, where the coordinate system $(x,y,z)$ may 
be some general curvilinear system such as $(\xi,\et,\ze)$
\beq \label{eq:poisson_anisotropic}
A(x,y,z) \fr{\pa^2 u}{\pa x^2} + B(x,y,z) \fr{\pa^2 u}{\pa y^2} + 
C(x,y,z) \fr{\pa^2 u}{\pa z^2} = \rho(x,y,z).
\eeq
From the coefficient functions $A(x,y,z)$, $B(x,y,z)$ and $C(x,y,z)$ we
can construct the ratios $|A/B|$, $|A/C|$ and $|B/C|$.  If any of these ratios 
exhibit large variations in magnitude on the solution domain, then the
PDE is said to be \emph{anisotropic}. 
It is clear, then, that the compactified Poisson 
equation~(\ref{eq:poisson_compact}), is highly anisotropic in this sense,
since each of the ratios $|A/B|$, $|A/C|$ and $|B/C|$ ranges from 
$0$ to $\infty$ for $-1\le \xi,\et,\ze\le 1$

Now---and again as discussed in more detail in Chap.~\ref{numerical}---to 
construct 
an efficient multigrid algorithm, we must have a way to efficiently 
{\em smooth} the errors in the discrete solution independently of 
the size of the mesh spacing that is used in any given calculation. 
For equations that are {\em not} anisotropic, simple relaxation methods 
such as point-wise Newton-Gauss-Seidel (see App.~\ref{ap:ngs}) tend 
to be very effective smoothers.  Importantly, point-wise techniques 
are readily parallelized because the operation of updating any given 
unknown during a relaxation sweep requires only ``local information'':
that is the update operation only involves values of other unknowns 
which are directly coupled through the finite difference equations 
themselves.  For the type of finite differencing used in this thesis 
(see App.~\ref{ap:stencil}) this typically amounts to nearest-neighbours 
in each of the coordinate directions.

For the case of anisotropic elliptic operators, however, point-wise
relaxation will generically fail to be an effective 
smoother~\cite{brandt:1977}.  For example, consider the situation
where the ratios $|A/B|$ and $|A/C|$ for the schematic 
equation~(\ref{eq:poisson_anisotropic}) satisfy 
$|A/B| \approx |A/C| \gg 1$. Then the equation is anisotropic and 
is said to be {\em strongly coupled} in the $x$ direction.  As shown
in~\cite{brandt:1977}, point-wise relaxation tends to smooth 
only along directions of strong coupling, so in this instance, the 
point-wise approach will not provide effective smoothing along
the $y$ and $z$ directions.  Now, there are well known strategies 
for recovering a good smoother when facing anisotropy.  Chief among 
these is the use of {\em line} relaxation, which for the current 
example would involve the simultaneous update of all of the unknowns
in the $x$ direction for each discrete combination of $(y,z)$.    
Unfortunately, line relaxation destroys the locality of the smoothing 
process, and thus inhibits parallelization.  This is the main reason that we 
decided not to implement compactification in the version of our code 
that was used to generate the results described in this thesis.

Before concluding this section, however, we point out that there is 
another approach to deal with anisotropy in multigrid that seems 
very promising for our application.  This is the technique
of semi-coarsening~\cite{trottenberg,MGM_brandt_1982}, whereby the coarsening operation
inherent to all multigrid algorithms is first performed only along 
directions of strong coupling.  The coarsening process tends to 
weaken the degree of coupling in those directions, so that 
point-relaxation can again be used as a smoother.  
For compactifying transformations 
of the form~(\ref{eq:comp_coord}) the effective strong-coupling directions
will be location dependent, so that we would need to implement an algorithm
that cycles through the three coordinates $\xi$, $\et$ and $\ze$,
semi-coarsening along each direction in turn.  
However, all of the operations involved 
in this method, including the semi-coarsening itself would remain local, and 
thus the method would be amenable to parallelization.

\subsubsection{Asymptotic and Sommerfeld Boundary Conditions} \label{sec:sommerfeld}

The next strategy we considered for treating the 
boundary conditions~(\ref{eq:bcfirst})--(\ref{eq:bclast}) 
is the one that is probably the most widely used in current 3D 
numerical relativity codes.  Here,
the underlying idea is to work in (non-compactified) $(x,y,z)$
coordinates and replace the spatially infinite domain with a finite 
region defined by
\begin{equation}
	x_{\rm min} \le x \le x_{\rm max} \, , \quad\quad
	y_{\rm min} \le y \le y_{\rm max} \, , \quad\quad
	z_{\rm min} \le z \le z_{\rm max} \, ,
\end{equation}
where 
$x_{\rm min}$, $x_{\rm max}$,
$y_{\rm min}$, $y_{\rm max}$,
$z_{\rm min}$, $z_{\rm max}$,
become adjustable parameters of the computation.  We then use the 
known and/or expected behaviour of the solution unknowns as 
$r\to\infty$~\cite{wald}~\footnote{Again recall that $r\equiv\sqrt{x^2+y^2+z^2}$.}
to derive approximate conditions which can be imposed on 
the boundaries of the domain. 

For example, considering the conformal factor, $\psi$, 
asymptotic flatness of the spacetime requires that 
\beq
   \label{eq:asympt_falloff}
	\lim_{r\to\infty} \psi = 1 + \frac {k}{r} + O(r^{-2}) \, ,
\eeq
where $k$ is a constant.  Using a trick that is well known to 
numerical relativists~\cite{york-piran:1982}, we can 
convert~(\ref{eq:asympt_falloff}) to a boundary condition of mixed 
type as follows.  We first differentiate (\ref{eq:asympt_falloff}) 
with respect to $r$ to get 
\beq
	\lim_{r\to\infty} \frac{\partial \psi}{\partial r} = 
		-\frac{k}{r^2} + O(r^{-3}) \, .
\eeq
This implies that
\beq
	k = \lim_{r\to\infty} \lhb -r^2\frac{\partial \psi}{\partial r} \rhb 
	+ O(r^{-1}) \, .
\eeq
Using this result to eliminate $k$ in~(\ref{eq:asympt_falloff}) we have
\beq
	\lim_{r\to\infty} \psi = 1 - r \frac{\partial \psi}{\partial r} + 
	O(r^{-2}) \, ,
\eeq
or
\beq
	\lim_{r\to\infty} \lhb \frac{\partial \psi}{\partial r}  + 
		\frac{\psi - 1}{r} \rhb = O(r^{-3}) \, .
\eeq
Neglecting the $O(r^{-3})$ term in the above, 
the condition that we impose on $\psi$ on the boundaries of the 
finite domain is then
\beq
	\label{eq:psi_bc_mixed}
	 \frac{\partial \psi}{\partial r}  +
      \frac{\psi - 1}{r} = 0 \, .
\eeq
In terms of Cartesian coordinates, $(x,y,z)$, we have
\beq
	\nonumber
	\frac{\pa \psi}{\pa r} =
		\frac{\pa \psi}{\pa x}\frac{\pa x}{\pa r}  +
		\frac{\pa \psi}{\pa y}\frac{\pa y}{\pa r}  +
		\frac{\pa \psi}{\pa z}\frac{\pa z}{\pa r}   
	=
		\frac{x}{r} \frac{\pa \psi}{\pa x} +
		\frac{y}{r} \frac{\pa \psi}{\pa y} +
		\frac{z}{r} \frac{\pa \psi}{\pa z}  \, ,
\eeq
where the relations $\partial x/\partial r = x/r$ etc.~follow from the 
standard transformation from rectangular to spherical polar coordinates.
Using the above result in~(\ref{eq:psi_bc_mixed}) we 
have~\footnote{We note that a mixed boundary condition such
as~(\ref{eq:psi_bc_mixed_cart}), which involves the value of the function
{\em and} its normal derivative, is sometimes called a {\em Robin} 
condition.}
\beq
	\label{eq:psi_bc_mixed_cart}
	x\frac{\pa \psi}{\pa x} + 
	y\frac{\pa \psi}{\pa y} +  
	z\frac{\pa \psi}{\pa z} +  \psi - 1 = 0 \, .
\eeq

Similar boundary conditions of mixed type can be derived for the 
lapse function, $\alpha$, and the shift vector components, $\bt^k$,
from the asymptotic behaviours:
\beq
	\lim_{r\to\infty} \alpha = 1 + O(r^{-1}) \, ,
\eeq
\beq
	\lim_{r\to\infty} \beta^k = O(r^{-1}) \, .
\eeq

For the case of the scalar field variables, $\phi_A$ and $\Pi_A$, we 
also have fall-off conditions, but additionally must take into account
the radiative nature of the fields.  The situation is further 
complicated by the fact that since the field is massive, it has a 
non-trivial dispersion relationship.  The simplest---and fairly 
crude---approach is to treat the scalar field as if it were massless.
Letting $v(t,r)$ denote any of the $\phi_A$, $\Pi_A$ in the asymptotic
region $r\to\infty$, we then expect
\beq
	\label{eq:radiative_falloff}
	\lim_{r\to\infty} v(t,r) = \frac{h(t-r)}{r}
\eeq
where $h$ is a function that describes a purely {\em outgoing} disturbance
propagating at the speed of light (i.e.~with speed $c=1$ in our units).
If we impose~(\ref{eq:radiative_falloff}) as a boundary condition, then
we are requiring that there be no {\em incoming} radiation (from infinity)
at any time during the evolution: such a demand is often called a 
{\em Sommerfeld} boundary condition.

The Sommerfeld condition in the form~(\ref{eq:radiative_falloff}) cannot 
be implemented directly in a numerical calculation since $h$ is not 
a known function.  However, observing that~(\ref{eq:radiative_falloff})
implies
\beq
	\lim_{r\to\infty} \lb r v \rb = h(t-r) \, ,
\eeq
we immediately have
\beq
	\lim_{r\to\infty} 
		\lhb \frac{\pa}{\pa t} \lb r v \rb +
			\frac{\pa}{\pa r} \lb r v \rb  \rhb = 0 \, , 
\eeq
or
\beq
	\lim_{r\to\infty} \lhb
		r \frac{\pa v}{\pa t} + r \frac{\pa v}{\pa r} + v \rhb = 0 \, .
\eeq
In Cartesian coordinates, and dropping the $\lim_{r\to\infty}$ this becomes
\beq
	\label{eq:radiative_falloff_dx}
	\sqrt{x^2+y^2+z^2} \frac{\pa v}{\pa t} + 
		x \frac{\pa v}{\pa x} + y \frac{\pa v}{\pa y} + z \frac{\pa v}{\pa z} 
		= 0 \, .
\eeq
 
The remainder of this section concerns technical details and issues related
to our numerical implementation of boundary conditions exemplified 
by~(\ref{eq:psi_bc_mixed_cart}) for the metric variables, 
and (\ref{eq:radiative_falloff_dx})
for the scalar field quantities.   As such, we again note that the reader 
may first wish to consult Chap.~(\ref{numerical}) as well as 
App.~(\ref{ap:stencil}) for background information on the specific finite 
differencing techniques that we used to discretize our model.

The key piece of~(\ref{eq:radiative_falloff_dx}) on which we need to focus is 
\beq
	\label{eq:key_piece}
	x \frac{\pa v}{\pa x} + y \frac{\pa v}{\pa y} + z \frac{\pa v}{\pa z}
\eeq
Observe that an expression of  identical form appears in the boundary 
condition~(\ref{eq:psi_bc_mixed_cart}) for the conformal factor, $\ps$, and
also appears in the corresponding equations for $\al$ and $\bt^k$ that 
we have not displayed here.
Denoting the coordinates of the finite difference grid by 
\beq
(x_i,y_j,z_k), \quad i = 1 \ldots n_x, \quad j = 1 \ldots n_y, 
	\quad k = 1 \ldots n_z
\eeq
(see Sec.~\ref{sec:FDA}), we consider the boundary $x=x_{\rm min}$ which
is $x=x_1$ on the grid.  Noting that our grid uses the same (constant)
mesh spacing, $h$, in each of the coordinate directions, our 
finite difference version of~(\ref{eq:key_piece}) is
\bea
\nonumber
x_{3/2} \frac{v_{2,j,k} - v_{1,j,k}}{h} &+&
	\frac{1}{2} y_j \lb \frac{v_{1,j+1,k} - v_{1,j-1,k}}{2h} + 
	                    \frac{v_{2,j+1,k} - v_{2,j-1,k}}{2h} \rb \\
\label{key_piece_fd}
	&+&
\nonumber
	\frac{1}{2} z_k \lb \frac{v_{1,j,k+1} - v_{1,j,k-1}}{2h} + 
	                    \frac{v_{2,j,k+1} - v_{2,j,k-1}}{2h} \rb \, \\
	&&j = 2 \ldots n_y-1, \quad k = 2 \ldots n_z-1
\eea
Note that this formula is centred at the ``fictitious'' grid point 
$(x_{3/2},y_j,z_k) \equiv ((x_1 + x_2)/2,y_j,z_k)$, and that due to 
this centring the approximation of the $x$ derivative 
is $O(h^2)$ while only involving grid function values at $x_1$ and 
$x_2$.  Similar formulae are readily derived for the other boundaries
$x=x_{\rm max}$, $y=y_{\rm min}$, $y=y_{\rm max}$, $z=z_{\rm min}$ 
and $z=z_{\rm max}$.  Additionally, we stress that expressions such
as~(\ref{key_piece_fd}) cannot be used along the edges or at the corners 
of the grid.  There we use modified versions of~(\ref{key_piece_fd}) that 
maintain the basic strategy of centring the formula at a fictitious grid
point (defined as the centre of 4 points along the edges, and 8 points
at the corners).   We note that although we expect that this 
differencing technique should work adequately for the scalar field 
variables (maintaining stable evolution in particular) the current 
version of our code has some convergence problems when stringent
tolerances are set for the overall time-step 
iteration~(see Sec.~\ref{sec:num_code}).  We believe that this is due to one 
or more code bugs, but have not invested much time 
on the matter given the issue with the elliptic equations that we now discuss.

Indeed, use of the same differencing scheme
to discretize the boundary conditions---such 
as~(\ref{eq:psi_bc_mixed_cart})---imposed on the 
elliptic PDEs led to severe numerical difficulties that we 
have yet to resolve.  Once more the issue is related to our 
use of  multigrid to solve the elliptic equations.  Even though 
our basic relaxation method (used in multigrid as a smoother, as 
discussed briefly in the previous section) converged to a 
smooth solution when applied to the full set of difference equations for 
the elliptic unknowns, our multigrid algorithm would {\em not} converge.
We strongly suspect that the multigrid failure can be traced to
our use of the particular differencing strategy for the boundary 
conditions that we have outlined above.  Heuristically, the use 
of centring at the fictitious points means that the boundary 
conditions are actually being imposed in the interior of the domain, 
rather than on the boundaries {\em per se}.  
During the execution of a multigrid cycle, as coarser and coarser grids 
come into play, the locations where the boundary conditions are set 
thus penetrate further and further into the solution domain, and 
this seems to destroy the convergence of the method.

One possible remedy for this problem would be to augment the grid with
so-called ghost points, and then to impose discretized versions 
of the interior PDEs as well as the boundary conditions at the boundary 
points.  Typically, when this approach is adopted one then eliminates 
the ghost values that are referenced by the interior equations using 
the discrete boundary conditions.  However, although this strategy
works well in 1D and 2D, it is not so straightforward to implement 
in 3D, especially when the PDEs involve mixed derivatives as is 
the case for our system.  Additional complications arise due to the 
need to maintain smoothness near the boundaries during the multigrid 
process, as well as in formulating appropriate transfers within
the grid hierarchy that are compatible with the boundary conditions.

In brief, although we devoted considerable effort to the task, we 
were not able to construct a convergent multigrid solver for our 
elliptic equations in the case that discrete versions of 
asymptotic boundary conditions such as~(\ref{eq:psi_bc_mixed_cart})
were used.  However, we remain hopeful that a resolution to this problem
{\em does} exist, and plan to continue to work towards it.

\subsubsection{Dirichlet Boundary Conditions}\label{sec:bc-dirichlet}

Our third approach to the numerical treatment 
of the boundary conditions for our model, and the one that was 
incorporated in the code used to generate all of the results presented 
in Chap.~\ref{results}, is extremely simple.  We simply set the 
conditions~(\ref{eq:bcfirst})--(\ref{eq:bclast}) on the boundaries
of the finite domain defined in the previous section:
\begin{equation}
   x_{\rm min} \le x \le x_{\rm max} \, , \quad\quad
   y_{\rm min} \le y \le y_{\rm max} \, , \quad\quad
   z_{\rm min} \le z \le z_{\rm max} \, ,
\end{equation}
Specifically, we have
\bea
\nonumber
	U(t,x_{\rm min},y,z) = U(t,x_{\rm max},y,z) &=& 1 \\
\nonumber
	U(t,x,y_{\rm min},z) = U(t,x,y_{\rm max},z) &=& 1 \\
	U(t,x,y,z_{\rm min}) = U(t,x,y,z_{\rm max}) &=& 1
\eea
where $U$ represents $\alpha$ or $\psi$, and
\bea
\nonumber
	V(t,x_{\rm min},y,z) = V(t,x_{\rm max},y,z) &=& 0 \\
\nonumber
	V(t,x,y_{\rm min},z) = V(t,x,y_{\rm max},z) &=& 0 \\
	V(t,x,y,z_{\rm min}) = V(t,x,y,z_{\rm max}) &=& 0
\eea
where $V$ is any of the $\bt^k$, $\phi_A$ or $\Pi_A$.
With this choice, we were able to construct a convergent multigrid 
solver for the elliptic equations, and, as described in 
Chap.~\ref{results} our numerical evolutions
of the coupled elliptic-hyperbolic system of PDEs governing our 
model were stable and convergent.  

Although this choice certainly constitutes quite a crude approximation,
especially for the domain sizes that we have used in our calculations 
to date, we note that we ultimately intend to incorporate
adaptive mesh refinement techniques~\cite{BO:1984} into our code. 
Particularly for scenarios involving the interaction of two boson
stars, we anticipate that fine grids will be needed only in the 
central region where the stars interact. The hope is then that a ``telescoping''
sequence of ever coarser grids will allow us to increase the 
physical size of the numerical solution domain to the point where the 
use of Dirichlet conditions contributes a relatively minor amount 
to the overall error in the solutions.  Again, however, this is  
a matter for future investigation.

\subsection{ADM/York Mass} \label{subsec:mass}

In the Hamiltonian formulation of general relativity, the total spacetime 
energy---also called the ADM mass---is associated with the 
numerical value of the Hamiltonian, which at any time, $t$, is
to be computed on a surface, $\pa V$, at spatial 
infinity.
The surface encloses the entire volume $V$ of 
the spacelike hypersurface $\Si_t$.  
For an asymptotically flat spacetime, 
the total energy on $\Si_t$ 
can be expressed as a surface integral involving derivatives of 
the components of the 3-metric at spatial infinity.  In terms of the $3+1$ variables
and Cartesian coordinates, this integral is 
\beq \label{eq:M_ADM}
M_{{\rm ADM}} \equiv E \equiv \fr{H_{\infty}}{16\pi} = \fr{1}{16\pi} 
\lim_{r\rightarrow \infty}{\oint_{\pa V} \lb \fr{\pa \ga_{ij}}{\pa x^i} 
                                           - \fr{\pa \ga_{ii}}{\pa x^j} \rb N^j dS} \, .
\eeq
Here $r=\sqrt{x^2+y^2+z^2}$, $H_{\infty}$ is the numerical
value of the Hamiltonian at spatial infinity,
$N^j$ is a outwards-directed unit vector normal to the surface $\pa V$, 
and $dS$ is the area element on the surface. That is, $dS = \sqrt{q} \, 
d^2y$, where $q$ is the determinant of the induced metric on
$\pa V$ and $d^2y$ are differentials associated with the coordinates on
the surface.  For example, if $\pa V$ is a spherical 
surface at $i^0$, then adopting the usual spherical polar coordinates,
$(r,\theta,\phi)$ the measure is simply $dS = \sqrt{q} d^2y 
= r^2 \sin{\theta} \, d\theta d\phi$ and the unit normal, $N^i$,
has components $(N^r,N^\theta,N^\phi)=(1,0,0)$.

In situations such as ours, where the outer boundaries of the computational
domain are {\em not} in the vicinity of spatial infinity, it is problematic
to use~(\ref{eq:M_ADM}) to compute a good estimate for the ADM mass. 
Fortunately there is an alternate expression for $M_{\rm ADM}$ 
due to \'O~Murchadha and York \cite{OMurchadha:1974pq} that provides 
the basis for a more accurate calculation of the mass in our simulations.
\'O~Murchadha and York investigated the relationship between the ADM masses
of two spatial metrics that were conformally related, 
i.e. for $\ga_{ij} = \psi^4 \tilde{\ga}_{ij}$. 
Starting from (\ref{eq:M_ADM}),
they showed that the difference in energy associated 
with the two metrics is given by the 
following volume integral over the spacelike hypersurface:
\beq
16 \pi \lb M_{\rm ADM} - \tilde{M}_{\rm ADM} \rb= -8 \int_V \sqrt{\tilde{\ga}} 
\lb \tilde{\ga}^{ij} \pa_i \pa_j \psi \rb \; d^3 x .
\eeq
This expression takes a particularly convenient form when the conformal 
metric is flat, since in that case the conformal ADM mass, 
$\tilde{M}_{\rm ADM}$, vanishes 
and we have~\footnote{\'O~Murchadha and York thus noted that this result 
provided an alternate definition of $M_{\rm ADM}$ that was valid for 
for any asymptotically conformally flat hypersurface.}
\beq
16 \pi  M_{\rm ADM} = -8 \int_V \sqrt{\hat{\ga}} 
\lb \hat{\ga}^{ij} \pa_i \pa_j \psi \rb \; d^3 x .
\eeq
Written in Cartesian coordinates this becomes
\beq \label{eq:york_mass}
M_{\rm ADM} = - \fr{1}{2\pi} \int_V 
\lb  \psi_{,xx} + \psi_{,yy} + \psi_{,zz} \rb \; d^3 x \, .
\eeq
In the simulations of our model problem we evaluate the above volume
integral numerically, and use the value of $M_{\rm ADM}$ computed 
in this way as a diagnostic quantity.  In this regard we mention that 
the question of whether $M_{\rm ADM}$ is exactly conserved in our 
model remains, to our knowledge, an open one.  It is thus noteworthy
that our simulations to date suggest that it {\em is} a conserved
quantity. However in the absence of a theorem to this effect, we 
emphasize that our use of the 
conservation of $M_{\rm ADM}$ during an evolution as a measure of 
the accuracy of the calculation should be viewed with caution.

\newpage
\section{Overview of the Equations of Motion} \label{sec:EOM_overview}

In previous sections we have discussed the derivation of the 
PDEs that govern our model in considerable detail. Here we provide
a recap of that development and redisplay the equations in a schematic
form that emphasizes the principal parts of the various differential
operators involved.
In addition, we provide a brief commentary concerning 
some of the properties of the PDEs as well as few details about how 
they were solved numerically.  We also note that we use a standard 
tensor calculus notation for partial differentiation below: namely
that $f_{,x} = \pa f/\pa x,\, f_{,xx} = \pa^2 f/\pa x^2$ etc.

In summary, our equations of motion constitute a mixed elliptic-hyperbolic
system of 9 nonlinear PDEs.  There are $5$ quasi-linear, elliptic 
equations for the geometric variables, which include
the lapse function, $\al$, the conformal 
factor, $\psi$, and the three components of the shift vector, $\bt^i$.
In addition, there are $4$ quasi-linear, hyperbolic equations 
which govern the components of the complex scalar field, 
$\phi_1$ and $\phi_2$, and their conjugate momenta, $\Pi_1$ and $\Pi_2$.

The elliptic equations were derived as follows. The maximal slicing 
condition for $\alpha$ resulted from taking the trace of the 
evolution equation for the extrinsic curvature $K^i{}_j$, and demanding
that the right hand side vanish.  Following some manipulation (including 
the use of the Hamiltonian constraint to eliminate the 3-Ricci scalar)
this led to equation~(\ref{eq:lapse_cartesian}) which can be 
written schematically as:
\beq
\al_{,xx} + \al_{,yy} + \al_{,zz} = N_{\al}\lb \al,\al_{,j},\psi,\psi_{,j},
                                       \bt^i_{\; ,j}, \phi_A,\Pi_A  \rb \, .
\eeq
Here $N_{\al}$ is a function which is nonlinear in many of its arguments,
including the first-order derivatives of $\alpha$, $\psi$ and $\bt^k$.

The elliptic equation~(\ref{eq:psi_cartesian}) that governs 
the conformal factor, $\psi$, came from the Hamiltonian 
constraint and has the form:
\beq
\psi_{,xx} + \psi_{,yy} + \psi_{,zz} = N_{\psi}\lb \al,\psi,
                                    \bt^i_{\; ,j}, \phi_A,  \phi_{A,j}, \Pi_A  \rb ,
\eeq
where $N_{\psi}$ is again nonlinear, but in this case, the key derivative
nonlinearities involve only the shift vector components.

Additionally, the momentum constraints were
used to derive~(\ref{eq:shift_cartesian_x})-(\ref{eq:shift_cartesian_z}), 
which are elliptic equations for 
the components of the shift vector $\bt^k = (\bt^x,\bt^y,\bt^z)$.  These 
take the form
\beq
\fr{4}{3} \bt^x_{\; ,xx} + \bt^x_{\; ,yy} + \bt^x_{\; ,zz} 
+\fr{1}{3} \lb \bt^y_{\; ,yx} + \bt^z_{\; ,zx} \rb
= N_{\bt^x}\lb \al,\al_{,j}
,\psi,\psi_{,j},\bt^x_{\; ,j},
\bt^y_{\; ,y},\bt^y_{\; ,x},
\bt^z_{\; ,z},\bt^z_{\; ,x},
\phi_{A,x}, \Pi_A  \rb ,
\eeq
\beq
\bt^y_{\; ,xx} + \fr{4}{3} \bt^y_{\; ,yy} + \bt^y_{\; ,zz} 
+\fr{1}{3} \lb \bt^x_{\; ,xy} + \bt^z_{\; ,zy} \rb
= N_{\bt^y}\lb \al,\al_{,j}
,\psi,\psi_{,j},\bt^y_{\; ,j},
\bt^z_{\; ,z},\bt^z_{\; ,y},
\bt^x_{\; ,x},\bt^x_{\; ,y},
\phi_{A,y}, \Pi_A  \rb ,
\eeq
\beq
\bt^z_{\; ,xx} + \bt^z_{\; ,yy} + \fr{4}{3} \bt^z_{\; ,zz} 
+\fr{1}{3} \lb \bt^x_{\; ,xz} + \bt^y_{\; ,yz} \rb
= N_{\bt^z}\lb \al,\al_{,j}
,\psi,\psi_{,j},\bt^z_{\; ,j},
\bt^x_{\; ,x},\bt^x_{\; ,z},
\bt^y_{\; ,y},\bt^y_{\; ,z},
\phi_{A,z}, \Pi_A  \rb  \, .
\eeq
Here, the right-hand-side functions $N_{\bt^x}$, $N_{\bt^y}$ and 
$N_{\bt^z}$, although generally nonlinear, are linear in the 
first order derivatives of the $\bt^k$.

These five elliptic PDEs have to be supplemented by boundary 
conditions. As discussed in Sec.~\ref{subsec:bdy_cond}, for all of the results 
reported in this thesis, we used Dirichlet boundary conditions 
as follows
\beq
\al|_B=1, \qquad \psi|_B=1, \qquad \bt^x|_B=0, \qquad \bt^y|_B=0, 
	\qquad \bt^z|_B=0.
\eeq
where the notation $f|_B$ means ``$f$ evaluated on the boundary of the 
(finite) solution domain''. 

Numerically, and as discussed in more detail in Chap.~\ref{numerical} and
App.~\ref{ap:stencil}, the elliptic equations were discretized using second
order finite difference methods, on a uniform mesh (constant spacing
$h$ in all three of the coordinate directions).  The resulting 
discrete equations were then solved using a multigrid algorithm which 
is described in~Sec.~\ref{sec:MG}.

The hyperbolic part of our model consists of the four first-order-in-time
PDEs~(\ref{eq:phidot_adm_cart})--(\ref{eq:pidot_adm_cart}), all of 
which originate from the general relativistic
Klein-Gordon equation for a complex scalar field, $\phi = \phi_1 + i\phi_2$,
where $\phi_1$ and $\phi_2$ are real-valued.  These equations are of the form
\beq
\dot{\phi}_A = N_{\phi_A}\lb \al,\psi,\bt^i,\phi_{A,j}, \Pi_A \rb
\eeq
and 
\beq
\dot{\Pi}_A = N_{\Pi_A}\lb \al,\al_{,j},\psi,\psi_{,j},\bt^i,\bt^i_{\; ,i},
\phi_A, \phi_{A,j},\phi_{A,jj}, \Pi_A, \Pi_{A,j} \rb ,
\eeq
for $A=1,2$, and where $\Pi_A$ are the momenta conjugate to $\phi_A$.
In this case we observe that the source functions $N_{\phi_A}$ and $N_{\Pi_A}$  
are {\em linear} in the scalar field quantities themselves.~\footnote{This 
assumes that the scalar field potential has only a mass term, which is 
the case for the calculations reported here.}
As with the elliptic equations, we impose Dirichlet boundary conditions
on the hyperbolic variables:
\beq
(\phi_A)|_B = 0 \qquad \textrm{and} \qquad (\Pi_A)|_B = 0.
\eeq

The hyperbolic equations were also discretized via second 
order finite difference techniques, and on the same grid used
for the elliptic PDEs.  More specifically, we used an implicit Crank-Nicholson 
scheme~(Sec.~\ref{sec:CN} for details), and the resulting
algebraic equations were solved with a point-wise Newton-Gauss-Seidel 
iteration~(App.~\ref{ap:ngs} for 
details). 
Again, Chap.~\ref{numerical} and App.~\ref{ap:stencil} provide many more details 
concerning all of the major numerical techniques and algorithms we used in
the current work.

The reader will note that we have not yet discussed the issue of initial
data for our model which, of course, is an absolutely crucial
part of the complete specification of any initial-boundary value problem.
Modulo potential difficulties due to the nonlinearity of the equations,
we can, in principle, specify {\em arbitrary} values for the scalar 
fields, $\phi_A$, and the conjugate momenta, $\Pi_A$, as long as those values 
are sufficiently differentiable,~\footnote{Mathematically, it would suffice to 
have initial data that is twice differentiable, but especially given
that we are using centred finite difference techniques, it will
be more convenient to require the data to be smooth---i.e. infinitely
differentiable.} and compatible with the boundary conditions. 
The elliptic equations can then be solved to determine the initial 
values for the metric functions.  The future (or past) development 
of the initial data will then be given by the solution of the coupled
elliptic-hyperbolic system, subject to the boundary conditions.

However, our principal aim is to study the interaction of boson stars,
which, we recall, are localized, gravitationally bound configurations 
of the scalar field that, in isolation, produce time-independent
gravitational fields.  The process of determining initial data for 
even a single such star is a non-trivial matter: generating initial
states for binaries adds a few more complications.
We thus devote the entire next chapter to the issue of computing 
this specific class of initial data.

 \chap{Initial Data} \label{initialdata}

In this chapter we describe the computation of initial configurations for 
our model that contain boson stars.  Although we are most interested in 
the case of initial data describing boson star binaries, the problem of 
determining data for a single star is interesting in its own right, and, 
of course, forms the basis for the task of constructing an initial state 
for two stars.  To generate data for one star, we adopt a particular 
ansatz, detailed in Sec.~\ref{sec:id-ansatz}, in which both the spacetime 
and the scalar field, $\phi$, are spherically symmetric, and where due to 
an assumed harmonic time-dependence for $\phi$, the spacetime is 
time independent.

Now, it transpires that for spherically symmetric spacetimes, the 
requirement of conformal 3-flatness does {\em not} imply that we are 
considering an approximation of the full Einstein equations, as it does in 
the general 3D case.  Rather, as we will discuss in Sec.~\ref{sec:id-mi}, 
in spherical symmetry conformal flatness amounts to a specific choice of 
radial coordinate, $r$, that we can always make, and which we can further 
combine with maximal slicing to completely fix the spherical coordinate
system.
For this reason the first five sections below treat the full 
Einstein equations coupled to a complex scalar field (with some general 
self-interaction potential), and we thus begin in Sec.~\ref{sec:id-ss} 
with a review of the $3+1$ equations for spherically symmetric spacetimes.  
Additionally, since it is easiest to determine the boson star solutions 
using a so-called areal radial coordinate, $R$, and then transform the 
solutions to the coordinate $r$ in which the 3-metric is conformally flat, 
we discuss the relevant Einstein equations in both coordinate systems 
in Secs.~\ref{sec:id-pa} and \ref{sec:id-mi}.  Sec.~\ref{sec:id-ansatz} 
then defines the ansatz used to generate a single boson star. This reduces 
the field equations to a system of ordinary differential equations ODEs for 
the scalar field and metric variables, which is further an eigenvalue 
problem.  We briefly discuss how the system is solved, and then in 
Sec.~\ref{sec:id-family} comment on some of the properties of the solutions.

The last two sections of the chapter then return focus to the 3D case, and 
the use of the results from the spherically symmetric general relativistic 
calculations to provide initial conditions for our model.  
Sec.~\ref{sec:id-1d3d} describes the straightforward process that we use to 
set up initial data in Cartesian coordinates for one or two stars, and 
Sec.~\ref{sec:id-boost} details the further transformations of the field 
quantities that are used to give stars non-vanishing velocities at the 
initial time.

We emphasize that apart from the material in Secs.~\ref{sec:id-1d3d} and 
\ref{sec:id-boost}, none of what is discussed here is original to this work.  
Indeed, general relativistic spherically symmetric boson stars have been 
studied by a large number of authors, so there is a considerable 
literature on the subject~(see \cite{schunck:2003} for a review).  In particular, 
much of the presentation in~Secs.~\ref{sec:id-ss}--\ref{sec:id-family} follows 
Chap.~$4$ of Lai's PhD thesis \cite{cwlai:phd} (which the interested reader 
can consult for additional details) as well as unpublished lecture notes 
due to Choptuik \cite{choptuik:nr_lecture}.

\section{Spherically Symmetric Spacetime}\label{sec:id-ss}

In general relativity, spherical symmetry can be precisely defined in 
terms of the symmetry group, $SO(3)$, whose group orbits are 2-spheres,
and which, physically, is associated 
with rotations in three dimensional space~\cite{wald}.
In particular, spherical
symmetry dictates that the spacetime metric, $g_{ab}$, be invariant under 
the action of $SO(3)$.  Choosing a set of coordinates $(t,r,\te,\phi)$
which is adapted to the symmetry, the metric induced on any 
2 sphere (i.e.~any collection of events defined by $t = {\rm const.}$,
$r={\rm const.}$), is some multiple of the metric on the unit
2-sphere, which itself is given by
\beq
d\Om^2 = d\te^2 + \sin^2{\te} d\phi^2 .
\eeq
It is straightforward to argue that 
a completely general form for the line element on a $t={\rm const}$ 
hypersurface in spherical symmetry is then 
\beq
\label{eq:id-3ds2}
^{(3)}ds^2= a^2(t,r)\, dr^2 + r^2 b^2(t,r) \,d\Om^2 \, ,
\eeq
where we stress that the functions $a$ and $b$ depend only on $t$ and 
$r$.  Similarly, it can be readily established that the most 
general line element for spherically spacetime can be written in the 
3+1 form
\begin{equation}
\label{eq:id-4ds2}
  ds^2 = ( -\alpha(t,r)^2 + a^2\bt(t,r)^2)\,dt^2
       + 2a^2\bt \, dtdr + a^2 dr^2 + r^2 b^2 d\Omega^2 .
\end{equation}
Here, $\alpha(t,r)$ is the lapse function as usual, and $\bt(t,r)$ is 
the single non-vanishing component of the shift vector 
$\bt^k=(\bt^r,\bt^\theta,\bt^\phi)\equiv(\bt,0,0)$.

Additionally, it is not hard to prove from~(\ref{eq:g_evo_adm}) that 
the extrinsic curvature corresponding to the spatial metric, $\ga_{ij}$, has 
only two independent components in spherical symmetry, $K^r{}_r$ 
and $K^\te{}_\te$:
\begin{equation}
  K^i{}_{ j} = \mbox{diag}(K^r{}_{ r}(t,r), K^{\theta}{}_{ \theta}(t,r), 
K^{\theta}{}_{ \theta}(t,r)) .
\end{equation}

Before writing down the $3+1$ Einstein equations for spherically 
symmetric spacetimes, it is convenient to introduce 
auxiliary fields, $\Phi(t,r)$ and $\Pi(t,r)$, which are defined in 
terms of the complex scalar field, $\phi$ as follows:
\begin{equation}
  \Phi(t,r) \equiv \phi'(t,r) \equiv \pa_r \phi(t,r) 
  \qquad \textrm{and} \qquad
\label{eq:id-pi}
  \Pi(t,r) \equiv \frac{a}{\alpha}(\dot{\phi} - \beta\phi') .
\end{equation}
(Here and throughout this chapter we use the overdot and prime notations 
for partial differentiation with respect to the time and radial 
coordinates respectively.)
\footnote{As a technical note, we point out that 
$\Pi$ as defined in (\ref{eq:id-pi}) is {\em not} the conjugate momentum of 
the field $\phi$; but that $ r^2 b^2 \sin(\theta) \Pi$ is. The 
definition~(\ref{eq:id-pi}) is motivated by the form of the dynamical
equations for $\Phi$ and $\Pi$ that result 
(vis. (\ref{eq:PhiDot})-(\ref{eq:PiDot})).}
In terms of these 
auxiliary variables, and using~(\ref{eq:rho_adm})-(\ref{eq:Sab_adm}), 
the non-vanishing components of the stress-energy tensor are:
\begin{align*}
  &\rho = \frac{|\Phi|^2 + |\Pi|^2}{2a^2} + \frac{U(|\phi|^2)}{2} ,
  &j_r = - \frac{\Pi^* \Phi + \Pi\Phi^*}{2a} = a^2j^r,\\
  &S^r{}_{\; r} = \rho - U(|\phi|^2),
  &S^\theta{}_{\theta} = \frac{|\Pi|^2 - |\Phi|^2}{2a^2} - \frac{U(|\phi|^2)}{2},\\
  &S^{\phi}{}_{ \phi} = S^\theta{}_{ \theta},
  &S = \frac{3|\Pi|^2 - |\Phi|^2}{2a^2} - \frac{3}{2} U(|\phi|^2).
\end{align*}
where $U(|\phi|^2)$ is the scalar field's interaction potential.
The spherically symmetric Klein Gordon equation gives us the 
following first-order-in-time evolution equations for the 
scalar field variables, $\phi$, $\Phi$ and $\Pi$:
\bea \label{eq:phiDot}
  \dot{\phi} &=& \frac{\alpha}{a} \Pi + \beta\Phi ,
  \\ \label{eq:PhiDot}
  \dot{\Phi} &=& \left( \frac{\alpha}{a}\Pi + \beta\Phi \right)' ,
  \\ \label{eq:PiDot}
  \dot{\Pi} &=& \frac1{(rb)^2} \left[ (rb)^2\left( \beta\Pi + \frac{\alpha}a \Phi \right) \right]'
   + 2\left[ \alpha K^{\theta}{}_{ \theta} - \beta\frac{(rb)'}{rb} \right]\Pi
   - \alpha a \frac{dU(|\phi|^2)}{d|\phi|^2} \phi .
\eea
Continuing, from~(\ref{eq:g_evo_adm}) we have evolution equations 
for the metric functions $a$ and $b$:
\begin{align}
  \label{eq:adot}
  \dot{a} = - \alpha a K^r{}_{ r} + (a\beta)' ,\\
  \label{eq:bdot}
  \dot{b} = - \alpha b K^{\theta}{}_{ \theta} + \frac{\beta}{r} (rb)' ,
\end{align}
and from~(\ref{eq:k_evo_adm}), evolution equations for 
the extrinsic curvature components, $K^r{}_r$ and $K^\te_\te$:
\begin{gather}
\label{eq:krdot}
  \dot{K}^r{}_{ r} = \beta {K^r{}_{ r}}' - \frac{1}{a} \left( \frac{\alpha'}{a} \right)' + \alpha\left\{ -\frac{2}{arb} \left[ \frac{(rb)'}{a} \right]' + KK^r{}_{ r} - 4\pi\left[ \frac{2|\Phi|^2}{a^2} + U(|\phi|^2) \right] \right\} , \\
\label{eq:ktdot}
  \dot{K}^{\theta}{}_{\theta} = \beta {K^{\theta}{}_{ \theta}}' + \frac{\alpha}{(rb)^2} - \frac{1}{a(rb)^2}\left[ \frac{\alpha r b}{a}(rb)' \right]' + \alpha K K^{\theta}{}_{ \theta} - 4\pi\alpha U(|\phi|^2) .
\end{gather}

Turning to the constraint equations, 
the Hamiltonian constraint~(\ref{eq:ham_adm})  reduces to
\beq
  \mathcal{R} + 4K^r{}_{ r}K^{\theta}{}_{ \theta} + 2{K^{\theta}{}_{ \theta}}^2 
= 16\pi\rho  \, ,
\eeq
which, using the explicit form of the 3-Ricci scalar, ${\mathcal R}$, becomes
\beq
\label{eq:ham_const_ss}
 - \frac{2}{arb} \left\{ \left[ \frac{(rb)'}{a} \right]' 
+ \frac{1}{rb}\left[ \left( \frac{rb}{a}(rb)' \right)' 
-a \right] \right\} + 4K^r{}_{ r}K^{\theta}{}_{ \theta} 
+ 2{K^{\theta}{}_{ \theta}}^2 = 
8\pi\left[ \frac{|\Phi|^2 + |\Pi|^2}{a^2} + U(|\phi|^2) \right] \, .
\eeq
Finally the momentum constraint~(\ref{eq:mom_adm}) is
\begin{equation}\label{eq:mom_const_ss}
  {K^{\theta}{}_{ \theta}}' + \frac{(rb)'}{rb}\left( K^{\theta}{}_{ \theta} - K^r{}_{ r} \right) = 2\pi \frac{\Pi^*\Phi + \Pi\Phi^*}{a} .
\end{equation}

We have now displayed a complete set of equations for the 
Einstein-Klein-Gordon system in spherical symmetry, which is valid for 
any system of coordinates $(t,r,\te,\phi)$ that are adapted to the 
symmetry.  In the next two sections we specialize these equations 
to the cases of two specific coordinate systems that bear on our
current work.  

\section{Maximal-Isotropic Coordinates}\label{sec:id-mi}

As we discussed in Sec.~\ref{sec:CFC_Maximal_slicing} the 
choice of maximal slicing fixes the time coordinate by demanding 
that the trace of the extrinsic curvature, $K\equiv K^i{}_i$,
vanish at all times.  To implement this condition 
in spherical symmetry then, we require $K(0,r)=0$ and ${\dot K}(t,r)$
for all $t$ and $r$.  We note that since we have 
$K^i{}_j = (K^r{}_r,K^\te{}_\te,K^\te{}_\te)$, and thus $K=K^r{}_r + 
2K^\te{}_\te$, the choice $K=0$ allows us to eliminate one of 
$K^r{}_r$ or $K^\te{}_\te$ from the overall set of
equations.  In particular, we will take
\beq
\label{eq:id-ktete}
   K^{\te}{}_{ \te} = -\ha K^r{}_{r} \, ,
\eeq
in the following.
The specification of the coordinate system is completed by fixing 
the radial coordinate.  We do this by requiring that the 3-metric
be conformally flat, so that again introducing the conformal factor,
$\psi(t,r)$, as in Sec.~\ref{subsec:conf_dec}, we have
\beq
\label{eq:id-cflat}
^{(3)}ds^2=  \psi^4 \lb dr^2 + r^2  d\Om^2 \rb .
\eeq
We reemphasize the point made earlier in this chapter
that in spherical symmetry conformal flatness amounts 
to a coordinate choice: we are always free to write the 3-metric
in the form~(\ref{eq:id-cflat}), and it does {\em not} imply that 
we are approximating Einstein's equations in some way, as it does
when applied to the generic 3D case.

In terms of the general form~(\ref{eq:id-3ds2}) for the spherically 
symmetric line element, (\ref{eq:id-cflat}) implies
\beq
\label{eq:id-cflat1}
a = b\equiv \psi^2(t,r) \, .
\eeq

Traditionally, the radial coordinate implied by~(\ref{eq:id-cflat}),
or equivalently~(\ref{eq:id-cflat1}), has been termed {\em isotropic},
and, although we  will also use that terminology, the reader should 
keep in mind that ``isotropic'' is synonymous with ``conformally flat''
in this context.

Operationally, to implement the isotropic condition~(\ref{eq:id-cflat1}),
we require that it holds on the initial hypersurface, so that 
$a(0,r)=b(0,r)$, and that ${\dot a}(t,r) = {\dot b}(t,r)$ for all $t$ and $r$.
Equating the right hand sides of (\ref{eq:adot}) and~(\ref{eq:bdot})
and using~$a(t,r)=b(t,r)$, we easily derive the following ODE for 
the shift vector component, $\bt$:
\beq
r \lb \fr{\bt}{r} \rb' = \al ( K^r{}_{r} - K^{\theta}{}_{\theta} ) \, ,
\eeq
or, using~(\ref{eq:id-ktete}),
\beq \label{eq:bt_MI}
r \lb \fr{\bt}{r} \rb' = \fr{3}{2} \al K^r{}_{r} \, .
\eeq

Returning to the maximal slicing condition, we have $K=0$ and ${\dot K}=0$.
We thus take the trace of (\ref{eq:k_evo_adm}) and equate the right hand
side of the resulting equation to 0. We then use the Hamiltonian constraint
to eliminate the 3-Ricci scalar, ${\mathcal R}$, appearing in 
the equation in favour of terms involving the extrinsic curvature 
and stress-energy components, and perform additional 
simplifications using $K=0$ and 
${\dot K}^{\te}{}_{ \te} = -{\dot K}^r{}_{r}/2$.
After some manipulation we find the following equation for the lapse:
\beq
\nonumber
\al'' + \fr{a}{\lb r b\rb^2} \lhb \fr{(rb)^2}{a} \rhb' \al'
+\lhb 4\pi a^2  U(|\phi|^2) -8\pi |\Pi|^2 -\fr{3}{2}a^2 {K^r{}_{r}}^2\rhb \al = 0 \, ,
\eeq
which using $a=b=\ps^2$, becomes
\beq \label{eq:lps_MI}
\al'' + \fr{2}{ r \psi^2} \fr{\pa}{\pa (r^2)} \lb r^2\psi^2 \rb  \al'
+\lhb 4\pi \psi^4  U(|\phi|^2) -8\pi |\Pi|^2 -\fr{3}{2}\psi^4 {K^r{}_{r}}^2\rhb \al = 0 \, .
\eeq
Here we note that the operator $\pa/\pa r^2$ takes a derivative
with respect to $r^2$.  Its use here is motivated by the need
to maintain regularity in numerical solutions of~(\ref{eq:lps_MI}), 
and is an example of a general technique first introduced to the 
numerical relativity community by Evans~\cite{evans:phd}.  The basic
observation is that as $r\to0$, we have $\psi(t,r) \to \psi_0(t) +
r^2\psi_2(r) + O(r^4)$, and thus $r^2\psi(t,r) \to r^2 \psi_0(t) + O(r^4)$.
Since the leading order term goes like $r^2$, finite difference operators
based on $\pa/\pa (r^2)$ rather than $\pa/\pa r$ tend to produce
numerical solutions that are smoother
near $r=0$.  This issue is discussed in more detail in App.~\ref{ap:stencil},
Sec.~\ref{sec:ddrp}.

Eqs.~(\ref{eq:lps_MI}) and (\ref{eq:bt_MI}) determine the kinematical
geometrical variables, $\alpha$ and $\beta$, respectively.  We now 
need equations for the dynamical geometrical variables, namely the 
conformal factor, $\psi$, and the extrinsic curvature component, $K^r{}_{r}$.
Here, due to the overdetermined nature of the Einstein equations,
which, we recall, results from the coordinate invariance of the theory, we have 
two choices for each function.  We can either use evolution equations 
derived from (\ref{eq:adot})--(\ref{eq:krdot}), or we can use the constraint
equations~(\ref{eq:ham_const_ss}) and (\ref{eq:mom_const_ss}).
Here we choose the latter option, which produces 
what is known as a \emph{fully constrained} scheme~\cite{piran:1980}.
Specifically the Hamiltonian constraint~(\ref{eq:ham_const_ss})
becomes an ODE for $\psi$:
\beq
\label{eq:ham_const_MI}
\fr{3}{\psi^5} \fr{d}{dr^3} \lb r^2 \fr{d\psi}{dr} \rb
+ \fr{3}{16}{K^{r}{}_{ r}}^2 = -\pi \lb \fr{|\Phi|^2 + |\Pi|^2}{\psi^4} 
+ U(|\phi|^2) \rb  \, ,
\eeq
while the momentum constraint~(\ref{eq:mom_const_ss}) provides an 
ODE for $K^{r}{}_{r}$:
\beq
\label{eq:mom_const_MI}
{K^{r}{}_{ r}}' + 3 \fr{(r\psi^2)'}{r\psi^2} K^{r}{}_{ r} = -\fr{4\pi}{\psi^2}
\lb \Pi^*\Phi + \Pi\Phi^* \rb  \, .
\eeq
We note that our ability to completely determine the geometric 
variables without the explicit use of any evolution equations (and,
in particular, without using any equations of the form ${\dot K}^i{}_j = 
\ldots$) is a reflection of the fact that, in spherical symmetry, the 
general relativistic gravitational field has no independent dynamics: 
genuine time dependence
of the metric~\footnote{As opposed to apparent time dependence arising
from a choice of coordinates, which can in principle be identified by 
finding a solution of the Killing equations $\Screll_\xi g_{ab}=0$, 
where $\xi$ is the sought-for timelike Killing vector (see 
Sec.~\ref{sec:id-ansatz}).} must result from time
dependence of a matter field that is coupled to gravity.

Finally, 
the specialization of (\ref{eq:phiDot})--(\ref{eq:PiDot}) to 
maximal-isotropic coordinates yields the following set of evolution 
equations for the scalar field variables:
\bea \label{eq:phiDot_MI}
  \dot{\phi} &=& \frac{\alpha}{\psi^2} \Pi + \beta\Phi ,
  \\ \label{eq:PhiDot_MI}
  \dot{\Phi} &=& \left( \frac{\alpha}{\psi^2}\Pi + \beta\Phi \right)' ,
  \\ \label{eq:PiDot_MI}
  \dot{\Pi} &=& \frac{3}{ \psi^4} \fr{d}{d r^3}\left[ r^2\psi^4\left( \beta\Pi 
+ \frac{\alpha}{\psi^2} \Phi \right) \right]
   - 2\left[\ha \alpha K^r{}_{ r} + \beta\frac{(r\psi^2)'}{r\psi^2} \right]\Pi
   - \alpha \psi^2 \frac{dU(|\phi|^2)}{d|\phi|^2} \phi .
\eea

The set of equations~(\ref{eq:bt_MI})--(\ref{eq:PiDot_MI})
fully determine the 
spherically symmetric Einstein-Klein-Gordon system in maximal-isotropic 
coordinates. Note, however, that these equations must be supplemented
with initial values and boundary conditions in order to 
generate a unique solution.  However, we have not used this 
system to generate any of the results discussed below, and 
have included the derivation of the equations of motion primarily 
for the sake of completeness.  We will thus 
conclude our discussion of this particular coordinate choice at this 
point and refer the interested reader to App.~B of Lai's 
thesis~\cite{cwlai:phd} for details of a numerical implementation of 
(\ref{eq:bt_MI})--(\ref{eq:PiDot_MI}). When we return to the 
maximal-isotropic coordinate system in~Sec.~\ref{sec:id-ansatz} it will be 
in the context of the initial data problem.

\section{Polar-Areal Coordinates}\label{sec:id-pa}

The second coordinate system for spherically symmetric spacetimes that 
we consider also uses a condition on the extrinsic curvature to 
fix the time coordinate.  In this case we demand that
\beq
\label{eq:id-polar}
K\equiv K^i{}_{ i} = K^R{}_{ R} \, ,
\eeq
where we note that we use $T$ and $R$ in this section to denote 
the time and radial coordinates, respectively, with 
${\dot u}\equiv\pa u/\pa T$ and $u'\equiv\pa u/\pa R$.
Eq.~(\ref{eq:id-polar}) is known as the 
polar slicing condition~\cite{bardeen_piran:1983}, and, as for the 
choice of maximal slicing, has the immediate consequence of reducing the 
number of independent components of the extrinsic curvature.  From 
the definition of $K$ and~(\ref{eq:id-polar}) we have
\beq
K\equiv K^{R}{}_{ R} + 2 K^{\theta}{}_{ \theta} = K^{R}{}_{ R}  \, ,
\eeq
which  implies
\beq
K^{\theta}{}_{ \theta} = 0 \, .
\eeq
This last result is then used to derive an equation that must
be satisfied by the lapse function, $\alpha(T,R)$, at any time $T$.
Specifically, we require that the initial data satisfy 
$K^\te{}_\te(0,R) = 0$, and then impose ${\dot K}^\te{}_\te(T,R)=0$ for 
all $T$ and $R$:
\beq
K^{\theta}{}_{ \theta}(T,R)=\dot{K}^{\theta}{}_{ \theta}(T,R) = 0 \, .
\eeq

As the name ``polar-areal'' suggests, the spatial coordinate is 
fixed by demanding that $R$ directly measure the proper surface
area of any $T={\rm const.}$, $R={\rm const.}$ 2-sphere.  Now, 
in terms of our general form~(\ref{eq:id-3ds2}) for the 3-metric, this area,
$A(T,R)$, is given by 
\beq
	A(T,R) = 4\pi R^2 b(T,R)^2 \, ,
\eeq
so if $R$ is to be areal, we must have $b(T,R)\equiv1$.   Thus, 
the choice of radial coordinate eliminates another of the geometric
dynamical variables from the system.  Moreover, the general evolution
equation~(\ref{eq:bdot}) reads 
\beq
	\label{eq:bdot_areal}
	{\dot b} = -\alpha b K^\te{}_\te + \frac{\bt}{R}\lb R b\rb'
\eeq
in the current case.  Parallelling the implementation of the other 
coordinate conditions discussed thus far, to enforce 
$b(T,R)\equiv1$ we require $b(0,R)=1$ at $T=0$, and ${\dot b(T,R)}=0$
for all $T$ and $R$. Using these relations, as well as $K^\te{}_\te(T,R)=0$ 
in~(\ref{eq:bdot_areal}), we find that the shift vector component, $\beta$,
must identically vanish.  Thus the spacetime line-element in 
polar-areal coordinates is simply
\beq \label{eq:polar_areal_gij}
ds^2 =  -\alpha^2(T,R) dT^2 + a^2(T,R) dR^2 + R^2 d\Omega^2  \, ,
\eeq
which can be viewed as a natural generalization of the familiar Schwarzschild
line element 
\beq
	ds^2 = -\lb 1 - \frac{2M}{R} \rb dT^2 + 
			\lb 1 - \frac{2M}{R} \rb^{-1} dR^2 + R^2 d\Omega^2 \, ,
\eeq
to the case of time dependent spherical geometries.  We further note 
that the elimination of {\em three} of the geometric quantities---namely
$b$, $K^\te{}_\te$ and $\beta$---using {\em two} coordinate conditions
is a special feature of the polar-areal system.  

Not surprisingly, given the simple (and diagonal) form of the line
element~(\ref{eq:polar_areal_gij}), the resulting $3+1$ Einstein 
equations also assume a very simple form.  Once more, since only 
two dynamical geometrical variables remain---the metric function,
$a$, and the extrinsic curvature component, $K^R{}_R$---we can 
use the constraint equations in lieu of evolution equations to 
determine them.

Using $b=1$ and $K^\te{}_\te=0$
in the general form of the Hamiltonian constraint~(\ref{eq:ham_const_ss}),
we find that $a$ must satisfy the ODE
\beq \label{eq:aprime_ham}
a'=\ha \lbr \fr{a}{R} \lb 1-a^2\rb+4\pi R a \lhb |\Phi|^2+|\Pi|^2 
+a^2 U(|\phi|^2) \rhb \rbr \, ,
\eeq
while the momentum constraint~(\ref{eq:mom_const_ss}) provides 
an {\em algebraic} equation for $K^{R}{}_R$:
\beq \label{eq:Krr_mom}
K^{R}{}_{ R}=-\fr{2\pi R}{a} \lb \Pi^*\Phi + \Pi\Phi^*  \rb \, .
\eeq

As discussed above, the polar slicing condition requires that 
$K^\te{}_\te(T,R)= 0$ and ${\dot K}^\te{}_\te(T,R)=0$.  Setting 
the right hand side of~(\ref{eq:ktdot}) to 0, and using $K=0$, $\bt=0$ 
and $b=1$, we derive the following ODE for the lapse:
\beq \label{eq:lpsprime_slicing}
\al'=\fr{\al}{2} \lbr \fr{a^2-1}{R}+4\pi R \lhb |\Phi|^2+|\Pi|^2
-a^2 U(|\phi|^2) \rhb\rbr \, .
\eeq
Finally, for the scalar field variables, using $b=1$, $K^\te{}_\te=0$ and 
$\beta=0$ in~(\ref{eq:phiDot})--(\ref{eq:PiDot}) we find:
\bea \label{eq:phiDot_PA}
  \dot{\phi} &=& \frac{\alpha}{a} \Pi  ,
  \\ \label{eq:PhiDot_PA}
  \dot{\Phi} &=& \left( \frac{\alpha}{a}\Pi  \right)' ,
  \\ \label{eq:PiDot_PA}
  \dot{\Pi} &=& 3 \fr{\pa}{\pa (R^3)}\left[ R^2 \frac{\alpha}{a} \Phi \right]
   - \alpha a \frac{dU(|\phi|^2)}{d|\phi|^2} \phi \, .
\eea

The set~(\ref{eq:aprime_ham})--(\ref{eq:PiDot_PA}) constitutes
a sufficient set of equations for the spherically 
symmetric Einstein-Klein-Gordon system in polar-areal 
coordinates (again, modulo initial values and boundary 
conditions).
In the next section we use a further simplified subset of these 
equations to compute solutions representing single boson stars.

\section{Constructing Boson Stars: The Static Ansatz} \label{sec:id-ansatz}

A spherically symmetric, localized, time independent 
configuration of matter captures the simplest notion of a star. 
Provided that the matter is regular everywhere, we should expect
such a configuration to produce a gravitational field that is also
spherically symmetric, time independent and globally regular.  Moreover,
if we are to be able to study the time evolution of these objects to
any significant degree, then they should also be dynamically stable.
For the case of a complex field it is not possible to construct
such states on the basis of time independence of $\phi$ itself.  Indeed,
Friedberg, Lee and Pang \cite{Friedberg:1987} demonstrated that 
in order for a boson star to be in a (minimal energy) ground state---a 
necessary 
condition for stability---$\phi$ must have harmonic
time dependence.  Thus, we adopt the following ansatz for the complex
scalar field
\begin{equation} \label{eq:SS_ansatz}
  \phi(t,r) = \phi_0(r)\,e^{-i\omega t} \, ,
\end{equation}
where $\omega$ is assumed to be a non-negative real constant, and
where, for the time being, $(t,r)$ are a general set of coordinates 
for spherically symmetric, time dependent spacetimes as discussed in 
Sec.~\ref{sec:id-ss}.
As can be quickly verified by examination of the form of~(\ref{eq:Tab}), 
all components of the stress-tensor, $T_{\mu\nu}$, become time independent
under this assumption.  Consequently, the spacetime (i.e. the gravitational
field) that is produced {\em can} also be expected to be time independent. 

Now, we note that the most generic sense of ``time independence'' in 
general relativity means that the spacetime has a timelike Killing 
vector field, in which case the spacetime is said to be {\em stationary}.  
Additionally, there is a more restrictive definition which requires 
that we be able to foliate the spacetime with hypersurfaces, $\Sigma_t$,
such that the Killing vector is everywhere orthogonal to these slices.
In this case the spacetime is said to be {\em static} and it is 
straightforward to show that in coordinates $(t,x^i)$ adapted to the 
timelike symmetry, the metric components $g_{\mu\nu}$ must be invariant
under the ``time reflection'' symmetry, $t\to-t$~\cite{wald}.  An example
of a spacetime which is stationary but not static is provided 
by an axially symmetric star composed of self-gravitating fluid 
that is in rotation about some symmetry axis, and which thus has
some net angular momentum.  Although one {\em can} construct boson star 
models that have angular momentum, we will not consider them here, 
so the demand that spacetime be static is the appropriate choice.
We observe that, assuming that $t$ is adapted to the timelike 
symmetry---which we will hereafter require---(\ref{eq:SS_ansatz}) is 
compatible with this demand.  We will therefore refer to 
it as the ``static ansatz''.

We now consider what the ansatz implies for the spherically symmetric 
$3+1$ equations displayed in the two previous sections.  First, the static
requirement immediately implies that all of the metric components,
$g_{\mu\nu}$, must be time independent functions, and that 
in terms of the general $3+1$ form of the line 
element~(\ref{eq:id-4ds2}), the shift vector component
$\beta$, must vanish:
\beq \label{eq:SS_shift}
\bt(t,r) = 0 \, .
\eeq
We observe that this last condition is always satisfied in polar-areal 
coordinates, but that it needs to be imposed explicitly when
working in maximal-isotropic coordinates.

Since the specific metric functions $a(t,r)$ and $b(t,r)$ must 
have vanishing time derivatives, and we also have $\beta=0$, 
Eqs.~(\ref{eq:adot}) and~(\ref{eq:bdot}) then imply that the 
extrinsic curvature components also vanish:
\beq
K^r{}_{ r} = K^{\theta}{}_{ \theta} = 0 \, .
\eeq
As a consequence, for the case of a static solution,
polar slicing is also maximal, and vice versa, so that to perform 
a coordinate transformation of a static spacetime from 
polar-areal coordinates to the maximal-areal system, we only 
have to transform the radial coordinate.

Considering the scalar field variables, we note that~(\ref{eq:SS_ansatz}) 
results in the following expressions for the time and radial derivatives
of $\phi$:
\bea
\dot{\phi}(t,r)&=&-i\om\phi_0(r) e^{-i\om t}, \\
\phi'(t,r)&=&\phi'_0(r) e^{-i\om t} \, .
\eea
From these relations we then have
\bea
\Pi(t,r)&=&-i\om \fr{a}{\al}\phi_0(r) e^{-i\om t}\equiv \Pi_0(r) e^{-i\om t}, \\
\label{eq:id-Phitr}
\Phi(t,r)&=&\phi'_0(r) e^{-i\om t}\equiv \Phi_0(r) e^{-i\om t} \, .
\eea
We also have
\beq
\Pi^*\Phi+\Pi\Phi^*=0\,,
\eeq
which means that the momentum constraint~(\ref{eq:mom_const_ss})
is satisfied identically.

We now adopt polar-areal coordinates $(t,R)$  (where we use $t$ rather
than $T$ in view of the above observation that polar and maximal slicings
are identical for static spacetimes), and use the above results in 
the set of equations for the spherically symmetric EKG system
that was given in the previous section.
After some manipulation, the 
Hamiltonian constraint (\ref{eq:aprime_ham}), the slicing condition
(\ref{eq:lpsprime_slicing}), and the evolution equation for the 
scalar field~(\ref{eq:PiDot_PA}), 
become the following system of coupled ODEs:
\bea 
a'(R)&=&\ha \lbr \fr{a}{R} \lb 1-a^2\rb
+4\pi R a \lhb \Phi_0^2+\om^2\fr{a^2}{\al^2}\phi_0^2
+a^2 U(\phi_0^2) \rhb \rbr \label{eq:aprime_ham_IVP},\\ 
\al'(R)&=&\fr{\al}{2} \lbr \fr{a^2-1}{R}+4\pi R \lhb \Phi_0^2
+\om^2\fr{a^2}{\al^2}\phi_0^2
-a^2 U(\phi_0^2) \rhb\rbr \label{eq:lpsprime_slicing_IVP},\\
\phi_0'(R)&=&\Phi_0 \label{eq:Phi_IVP},\\
\Phi_0'(R)&=& -\lb 1+a^2-4\pi R^2 a^2 U(\phi_0^2) \rb \fr{\Phi_0}{R}
+ a^2 \lhb \frac{dU(\phi_0^2)}{d\phi_0^2} - \fr{\om^2}{\al^2}\rhb \phi_0 .
\label{eq:Phiprime_IVP}
\eea
Note that a prime now denotes {\em ordinary} differentiation and 
that the equation $\phi_0' = \Phi_0$ follows from the 
{\em definition} of $\Phi_0$ 
in~(\ref{eq:id-Phitr}).
We will subsequently refer to this set of equations as the polar-areal ODE 
system.

We observe that this system contains terms such as $a(1-a^2)/R$ and 
$\Phi_0/R$ which, naively at least, appear to be singular at the 
origin, $R=0$. 
In order to ensure regularity at the origin,
certain conditions must be imposed on $a(R)$, $\alpha(R)$ and $\phi_0(R)$.
A thorough mathematical treatment of this 
subject defines a function (or, more generally, a component of 
a tensor) to be regular at the origin
if, using Cartesian coordinates, $(x,y,z)$, it has a convergent
Taylor series in a neighbourhood of $(x,y,z)=(0,0,0)$;
i.e. at~$R=0$~\cite{bardeen_piran:1983}.
For the case at hand (see~\cite{bardeen_piran:1983,matt:phd,oliver:phd}
for additional details) this requirement turns out to imply the following
limiting forms for $a(R)$, $\al(R)$ and $\phi_0(R)$:
\bea
\label{eq:id-a-reg}
	\lim_{R\to0} a(R) &=& a_0 + a_2 R^2 + O(R^4) = 1 + a_2 R^2 + O(R^4) \, ,\\
\label{eq:id-al-reg}
	\lim_{R\to0} \al(R) &=&  \al_0 + \al_2 R^2 + O(R^4) \, , \\
\label{eq:id-phi-reg}
	\lim_{R\to0} \phi_0(R) &=& \phi_{00} + \phi_{02} R^2 + O(R^4) \, .
\eea
Here, $a_0$, $a_2$, $\al_0$, $\al_2$, $\phi_{00}$ and $\phi_{02}$ are constants,
and the fact that $a_0 = 1$ follows from the demand that spacetime 
be locally flat at $R=0$.  From~(\ref{eq:id-phi-reg}) we also have 
\beq
\label{eq:id-Phi-reg}
   \lim_{R\to0} \Phi_0(R) = 2 \phi_{02} R + O(R^3) \, .
\eeq

Using~(\ref{eq:id-a-reg})--(\ref{eq:id-Phi-reg}) 
in (\ref{eq:aprime_ham_IVP})--(\ref{eq:Phiprime_IVP}), it is straightforward 
to show that the system of ODEs {\em is} regular at $R=0$.

Since (\ref{eq:aprime_ham_IVP})--(\ref{eq:Phiprime_IVP}) is a system
of 4 ODEs, it would seem that we need 4 boundary conditions to 
generate a unique solution.  However, there is a twist here which 
arises from the fact that the ODE system is actually to be treated 
as an eigenvalue problem, where $\omega$ is the eigenvalue.  Specifically,
for any choice of interaction potential
$U(|\phi|^2)$, the solutions of 
(\ref{eq:aprime_ham_IVP})--(\ref{eq:Phiprime_IVP}) form a one-parameter
family, where the central modulus of the scalar field, $\phi_0(0)$,
is a convenient
choice for the family parameter.  Since we want the solutions of 
the ODE system to represent stars, and thus to describe (essentially)
localized distributions of matter, we must have 
\beq \label{eq:bdy_phi}
\lim_{R\to \infty} \phi_0(R) = 0 \, .
\eeq 
For any given choice of $\phi_0(0)$, and in conjunction with the other 
boundary conditions that we will enumerate below, solutions 
satisfying~(\ref{eq:bdy_phi}) will only exist for a discrete set of values 
of $\omega\equiv\omega(\phi_0(0))$ and, further, we will only be 
interested in the smallest such $\omega$, which will correspond to 
the lowest energy boson star for a given $\phi_0(0)$.  In particular the 
function $\phi_0(R)$ for that choice of $\omega$ will have no 
zero-crossings (i.e.~it will be ``nodeless''): higher energy eigenstates
will have $1, 2, \ldots$ zero-crossings but, again, will be of no concern
to us here.

Bearing this in mind, and using~(\ref{eq:id-a-reg})--(\ref{eq:id-Phi-reg}),
the following constitute a sufficient set of boundary conditions 
for the polar-areal ODE system:
\bea
\label{eq:id-a-bc}
	a(0) &=& 1 \, , \\
\label{eq:id-al-bc}
	\alpha(0) &=& \alpha_0 \, , \\
\label{eq:id-phi-bc}
	\phi_0(0) &=& \phi_{00} \, , \\
\label{eq:id-Phi-bc}
	\Phi_0(0) &=& 0 \, .
\eea
where $\phi_{00}$ is freely specifiable, as is, at least momentarily,
$\alpha_0$, but where it is to be understood 
that $\omega=\omega(\phi_0)$ must be 
determined so that the asymptotic condition~(\ref{eq:bdy_phi}) is 
also satisfied.  

We now consider the boundary condition~(\ref{eq:id-al-bc}) 
more closely.  We note that the eigenvalue $\omega$ enters in the 
system~(\ref{eq:aprime_ham_IVP})--(\ref{eq:Phiprime_IVP}) strictly
in the combination $\omega^2/\alpha^2$.  Combined with the fact that,
apart from that term, the slicing equation~(\ref{eq:lpsprime_slicing_IVP}) 
is linear and homogeneous in~$\alpha$, this means that if we generate 
a solution of~(\ref{eq:aprime_ham_IVP})--(\ref{eq:Phiprime_IVP}) subject 
to the boundary conditions~(\ref{eq:id-a-bc})--(\ref{eq:id-phi-bc}), 
then we can take 
\bea
\label{eq:id-om-rescale}
	\omega &\to& k \, \omega \, ,\\
\label{eq:id-al-rescale}
	\alpha(R) &\to& k \, \alpha(R) \, ,
\eea
where $k$ is an arbitrary positive constant, and still have a solution.
Moreover, since the profiles $a(R)$, $\phi_0(R)$ and $\Phi_0(R)$ will
be unchanged under the 
rescaling~(\ref{eq:id-om-rescale})--(\ref{eq:id-al-rescale}), 
any solutions obtained via
such transformations will correspond to the same boson star.  
This reflects that fact that when we adopt a time-slicing condition
such as polar slicing or maximal slicing, we can always freely
rescale $\alpha(t,r)$ via $\alpha(t,r) \to f(t) \, \alpha(t,r)$ where 
$f(t)$ is some arbitrary function of the time 
coordinate, and still have a solution
of Einstein's equations.  The physical interpretation
of this rescaling is that we have complete freedom as to how we 
are to assign, or reassign, specific labels to the family of 
hypersurfaces, $\Sigma_t$, that foliate the spacetime.

In practice it is a common and sensible choice to perform this 
labelling so that coordinate time and proper time coincide for 
observers located at $R=\infty$, and who are at rest in the slices.  
This means that we want 
\beq
	\lim_{R\to\infty} \alpha(R) = 1 \, .
\eeq

As was the case in Sec.~\ref{subsec:bdy_cond} we are now confronted 
with the issue of dealing with a boundary condition naturally expressed 
at infinity, which is problematic if our numerical computations can
only extend to finite values of $R$.  Fortunately, in this case there 
is a simple resolution of this issue which is based on the uniqueness
of the solution of Einstein's equations in spherically symmetry 
when no matter is present.  This is known as Birkhoff's theorem,
and the interested reader can consult
Hawking and Ellis~\cite{hawking_ellis:LSSST}, for example, for a proof.  
Here 
we assert that the solutions of 
(\ref{eq:aprime_ham_IVP})--(\ref{eq:Phiprime_IVP})---i.e. the boson
star solutions---are characterized by scalar field profiles, $\phi_0(R)$,
which fall off exponentially after some characteristic radius.  Thus,
to well within the numerical 
accuracy to which we work in this thesis, the spacetime can be 
considered to be vacuum for $R > R_{\rm max}$, where $R_{\rm max}$ 
is the limit of integration of the 
system~(\ref{eq:aprime_ham_IVP})--(\ref{eq:Phiprime_IVP}).

Therefore, in terms of the static form of the $3+1$ element we 
adopt for polar-areal coordinates:
\beq \label{eq:polar_areal_gij1}
ds^2 =  -\alpha(R)^2 dt^2 +  a(R)^2 dR^2 + R^2 d\Omega^2 ,
\eeq
Birkhoff's theorem tells us that for $R>R_{\rm max}$ we must 
be able to identify~(\ref{eq:polar_areal_gij1}) with the familiar 
Schwarzschild form:
\beq\label{eq:sch}
ds^2 =  -\lb 1-\fr{2M}{R} \rb dt^2 + \lb 1-\fr{2M}{R} \rb^{-1} dR^2 
	+ R^2 d\Omega^2 \, ,
\eeq
where the constant $M$ is the total mass of the spacetime 
(i.e.~$M=M_{\rm ADM}$).  Note that since 
\beq
	\lim_{R\to\infty} \left[  1 - \frac{2M}{R} \right] = 1 \, ,
\eeq
the constant time surfaces defined by~(\ref{eq:sch}) {\em do} satisfy the 
condition that proper and coordinate time coincide at infinity.
Comparing~(\ref{eq:polar_areal_gij1}) and (\ref{eq:sch}) we thus have 
the following limits for the metric functions $\alpha(R)$ and $a(R)$:
\bea
\lim_{R\rightarrow \infty} \al^2(R)&=&\lb 1-\fr{2M}{R} 
	\label{eq:lps_limit}\rb \, , \\
\lim_{R\rightarrow \infty} a^2(R)&=&\lb 1-\fr{2M}{R} \rb^{-1} \, ,
	\label{eq:a_limit} 
\eea
which implies that we should choose the rescaling defined 
by~(\ref{eq:id-om-rescale}) and~(\ref{eq:id-al-rescale}) so 
that
\beq \label{eq:bdy_lps}
 \al(R_{\rm max}) = \fr{1}{a(R_{\rm max})} \, .
\eeq

The correspondence $a(R) \to \lb 1 - 2M/R \rb^{-1}$, suggested by
(\ref{eq:a_limit}), also motivates the definition of a useful 
diagnostic function, the so-called \emph{mass aspect function}:
\beq \label{eq:mass_aspect}
M(R)\equiv \fr{R}{2} \lb 1 - \fr{1}{a(R)^2} \rb \, .
\eeq
It can be shown that $M(R)$ defined in this way {\em does} measure 
the gravitating mass contained within an $R={\rm const.}$ sphere, so 
for $R\to \infty$ it must limit to the total ADM mass:
\beq
\lim_{R\rightarrow \infty}M(R) = M \equiv M_{\rm ADM}.
\eeq

To summarize, to determine a specific boson star solution we 
perform the following steps:
\begin{enumerate}
	\item We choose a particular value, $\phi_0(0)$, for the central modulus
		of the scalar field.  This constitutes the specification of 
		the boundary condition~(\ref{eq:id-phi-bc}).
	\item Using this choice and the additional boundary conditions 
		\bea
		\label{eq:id-a-bc-2}
			a(0) &=& 1 \, , \\
		\label{eq:id-al-bc-2}
			\alpha(0) &=& 1 \, , \\
		\label{eq:id-Phi-bc-2}
			\Phi_0(0) &=& 0 \, ,
		\eea
		we solve the polar-areal ODE 
		system~(\ref{eq:aprime_ham_IVP})--(\ref{eq:Phiprime_IVP}) on 
		the interval $0\le R\le R_{\rm max}$
		by determining the eigenvalue~$\omega=\omega(\phi_0(0))$ such that 
		$\phi(R_{\rm max}) \to 0$.  Since $\alpha(R)$ and $\omega$
      will be rescaled, we choose $\alpha(0)=1$ arbitrarily
      and for convenience.
	\item Once the solution has been determined, we rescale $\alpha(R)$
		and $\omega$ using
		\bea
		\label{eq:id-om-rescale-2}
			\omega &\to& k \, \omega \, , \\
		\label{eq:id-al-rescale-2}
			\alpha(R) &\to& k \, \alpha(R) \, ,
		\eea
		 where $k$ is chosen so that~(\ref{eq:bdy_lps}) is satisfied.
\end{enumerate}

The only remaining technical issue to be discussed is how we 
determine the eigenvalue $\omega$ in step 2 of above procedure.
We do this with a straightforward shooting method~\cite{burden}.
This means that for any specified value of $\phi_0(0)$, we must 
first determine an initial bracket $[\om_{-},\om_{+}]$ satisfying
$\om_{-} < \omega < \om_{+}$.  Solutions computed using $\om_{-}$
and $\om_{+}$ will display distinct behaviours as $R\to R_{\rm max}$.
Specifically, we find integration with $\om=\om_{-}$ results in
$\phi_0(R)\to\infty$ as $R\to R_{\rm max}$, whereas for $\om=\om_{+}$
the integration leads to $\phi_0(R)\to -\infty$.  Once the 
initial bracket has been found, we use a bisection method~\cite{burden}
to compute increasingly accurate estimates of $\omega$, where after 
each step in the bisection, the appropriate end point of the 
interval, $[\om_{-},\om_{+}]$, is replaced with the current estimate
$(\om_{-}+\om_{+})/2$. 

The solution method that we have just described was coded 
in a FORTRAN subroutine called {\tt bsidpa} that takes care of the 
integration of the polar areal ODE system, the shooting process {\em per se} 
and the rescaling of the lapse function. 
The routine is documented in App.~\ref{ap:bsidpa} and 
we note that we have made the code available to others who might find 
use for it.

Once we have computed a boson star solution in polar-areal coordinates,
we can transform the solution to maximal-isotropic coordinates, which 
are compatible with the coordinates we use in our 3D evolution code.
As mentioned above, for static solutions the time coordinates 
in the two systems are identical, so the transformation only involves
the radial coordinates.  As shown in full detail in App.~D of Lai's
PhD thesis~\cite{cwlai:phd}, this transformation can be made by
solving the ODE:
\beq \label{eq:areal2isotropic}
\fr{dr}{dR}=a\fr{r}{R} \, ,
\eeq
subject to the boundary condition:
\beq
\label{eq:id-a2i-bc}
r|_{R=R_{\rm max}}= \lhb \lb \fr{1+\sqrt{a}}{2} \rb^2 \fr{R}{a} \rhb_{R=R_{\rm max}}  \, .
\eeq
Here $r$ and $R$ are the isotropic and 
areal radial coordinates, respectively, and $a\equiv a(R)$ is the metric 
function that appears in the static, areal form of the 3-dimensional
spherically symmetric line element
\beq
	{}^{(3)}ds^2 = a^2 dR^2 + R^2 d\Omega^2  \, .
\eeq
Once $R(r)$ has been determined, the conformal factor $\psi(r)$ is 
easily computed from
\beq \label{eq:psi1d}
\psi(r) = \sqrt{\fr{R}{r(R)}}  \,  ,
\eeq
with
\beq
\label{eq:psi1d-0}
\psi(0)=\sqrt{\lb \left.\fr{dr}{dR}\right\vert_{R=0} \rb^{-1}} \, ,
\eeq
as follows from an application of l'Hopital's rule.

\section{Boson Stars: Some Basic Properties}
\label{sec:id-family}

\begin{figure}
\begin{center}
\epsfxsize=16.0cm
\epsffile{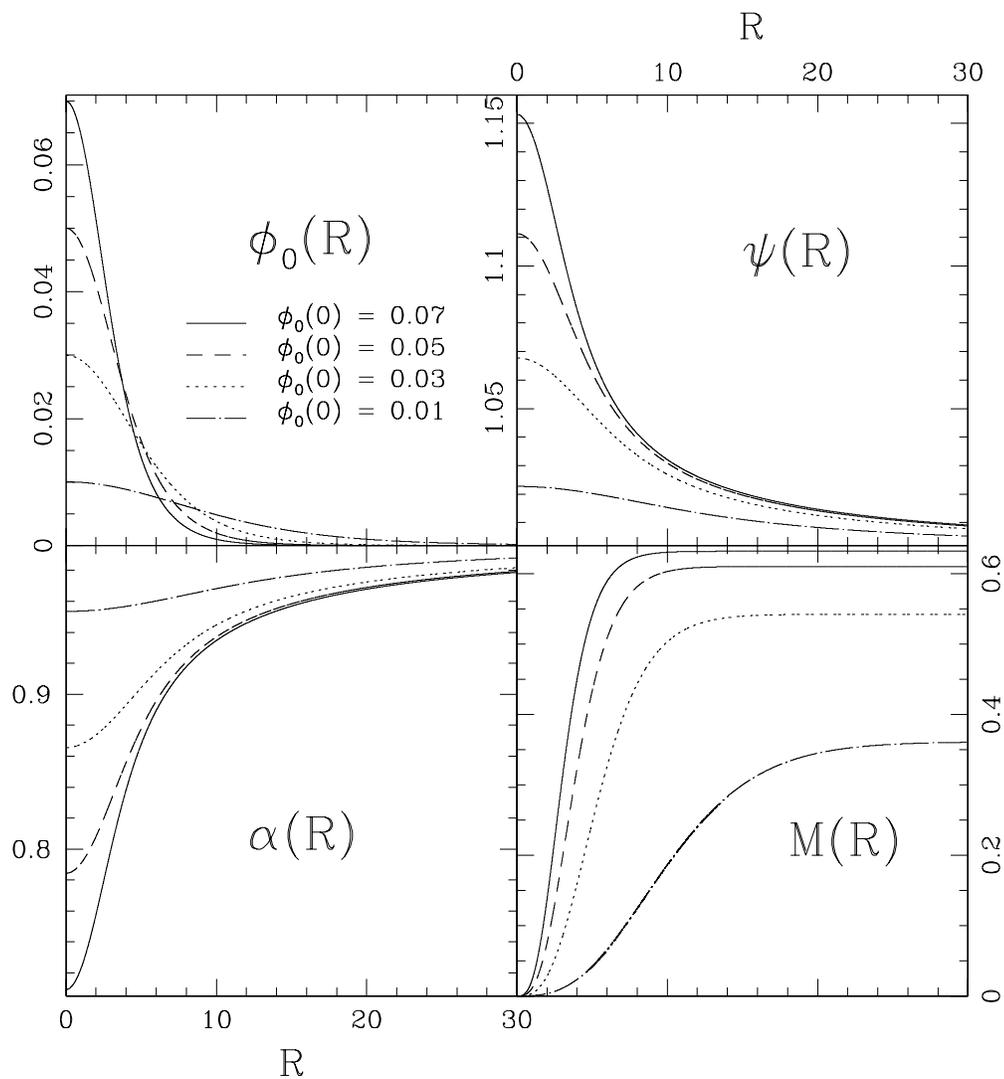}
\end{center}
\caption
[Typical Boson Star Solutions.]
{Typical Boson Star Solutions.
These plots shows the values of $\phi_0(R), \psi(R), \al(R)$ and $M(R)$ 
as a function of the areal coordinate, $R$, for boson stars defined 
by $\phi_0(0) = 0.01$, $0.03$, $0.05$ and $0.07$.  As discussed in more 
detail in the text, all of these configurations are dynamically stable
against radial perturbations.  Note that as $\phi_0(0)$ increases
the stars become smaller (more compact).
} \label{bs1d} \end{figure}

Fig.~\ref{bs1d} shows four separate solutions of the polar-areal ODE 
system, computed for central scalar field values 
$\phi_0(0)=0.01$, $0.03$, $0.05$ and $0.07$, 
with a potential $U(|\phi|^2)$ having only a mass term
\beq
	U(|\phi|^2) =  m^2 \phi^2 \, .
\eeq
Here $m$ is the mass parameter for the field, which, by our choice 
of units, satisfies $m=1$.
Each plotted solution 
represents a distinct boson star.  

We note that as $\phi_0(0)$ increases,
the stars become increasingly compact---meaning that the star's
size (defined,
for example, as the radius, $R$, at which $M(R)$ given 
by~(\ref{eq:mass_aspect}) is 99\% of 
$M_{\rm ADM}$)---{\em decreases} as $\phi_0(0)$ increases.  We also observe 
that the central values of the lapse, $\alpha(R)$, decrease as $\phi_0$
increases, while the $R=0$ values of the conformal factor, $\psi(R)$, 
increase.  All of these trends are indicative of the fact that 
the gravitational self-interaction of the scalar field strengthens 
as $\phi_0$ assumes larger values.  Further, although it is not 
immediately apparent from the plots, we assert that all of these solutions 
satisfy the appropriate asymptotic boundary conditions, namely that 
$\lim_{R\to\infty}\psi(R)=\lim_{R\to\infty}\alpha(R)=1$.
The figure also shows graphs of the mass aspect function, $M(R)$, for 
the stars, from which one can see a convergence of 
$M(R)\to M_{\rm ADM}$, with the convergence being more rapid for the 
stars defined by larger $\phi_0(0)$.   Finally we note that all four 
of these stars are dynamically stable against radial perturbations,
although, as we will discuss shortly, this is {\em not} the case 
for all solutions in the one-parameter family.

\begin{figure}
\begin{center}
\epsfxsize=16.0cm
\epsffile{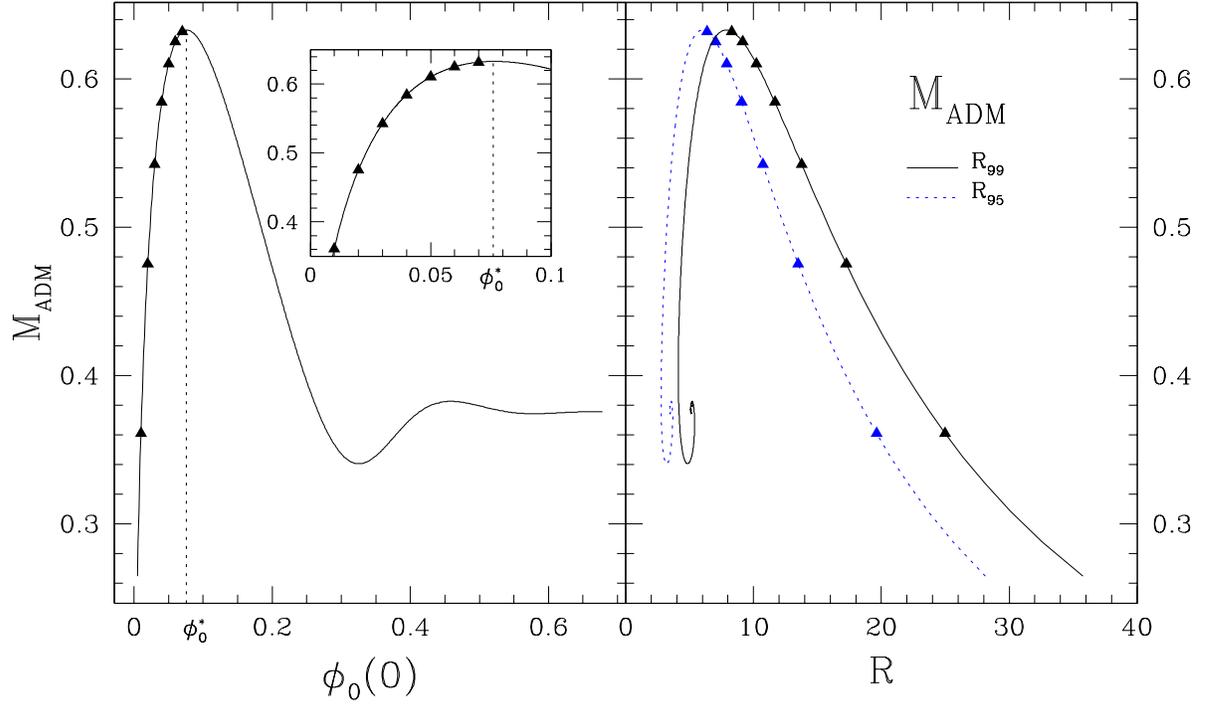}
\end{center}
\caption
[ADM Mass $M_{\rm ADM}$ as a Function of Central Scalar Field Value 
$\phi_0(0)$ and the Star Radius $R$.]
{This figure shows the ADM mass, $M_{\rm ADM}$, as a function 
of the central scalar field value, $\phi_0(0)$ (left panel),
as well as a function of two estimates, $R_{99}$, and $R_{95}$, 
of the stellar radius (right panel).  
$R_{99}$ and $R_{95}$ are defined to be the radii for
which $M(R)=0.99\,M_{\rm ADM}$ and $M(R)=0.95 \, M_{\rm ADM}$, respectively.
The $M_{\rm ADM}(\phi_0(0))$ curve has an absolute 
maximum at $\phi_0(0)=\phi_0^\star \approx0.08$ 
with $M_{\rm max}=M\lb\phi_0^\star\rb\approx0.633$.   As discussed 
in the text, the value $\phi_0^\star$ (shown as a dashed line in the 
plot) signals a change in 
dynamical stability of the solutions:  for $\phi_0(0)<\phi_0^\star$
the stars are stable to radial perturbations, while those with
$\phi_0(0) \ge \phi_0^\star$ are unstable.
The solid triangles label specific models computed with 
$\phi_0(0)=0.01$, $0.02$, $0.03$, $0.04$, $0.05$, $0.06$, and $0.07$---these 
are all on the stable branch.  The inset in the left plot 
highlights the form of $M_{\rm ADM}(\phi_0(0))$ for 
$0.01 \le \phi_0(0) \le 0.10$.  Note that for the curves in the right plot,
the triangles corresponding to $\phi_0(0)=0.01, 0.02, \ldots 0.07$ appear
right-to-left along the plots---i.e.~as noted in the caption of 
the previous figure, as well as in the text, the stellar radius {\em 
decreases} with increasing $\phi_0(0)$.
} 
\label{Madm1d} 
\end{figure}

Fig.~\ref{Madm1d} displays plots of the ADM mass, $M_{\rm ADM}$, as a 
function of the 
central scalar field value $\phi_0(0)$ (left panel), as well as a function
of two estimates of the stellar radius, $R_{99}$ and $R_{95}$ 
(right panel).  $R_{99}$ and $R_{95}$ are defined to be the radii for 
which $M(R)=0.99\,M_{\rm ADM}$ and $M(R)=0.95 \, M_{\rm ADM}$, respectively:
both of these definitions are commonly used in the boson star literature.
We note that $M(\phi_0(0))$ has an absolute maximum at 
$\phi_0(0)=\phi_0^\star \approx0.08$ 
where it attains a value $M_{\rm max}=M\lb\phi_0^\star\rb\approx0.633$
(again, recall that we work in units in which $c=G=m=1$).
In both plots the solid triangles label stars computed 
with $\phi_0(0) = 0.01$, $0.02$, $0.03$, $0.04$, $0.05$, $0.06$ and $0.07$
(left to right along the $M(\phi_0(0))$ curve, and right to left along 
the plots of $M_{\rm ADM}(R_{99})$ and $M_{\rm ADM}({R_{95}}))$.

It is possible to show from perturbation analyses, as well as from evolution of 
the full equations of motion~\cite{gleiser_watkins:1989,hawley_choptuik:2000, 
shawley:phd,cwlai:phd} that stars satisfying 
$\phi_0(0)< \phi_0^\star$ are dynamically stable against radial 
perturbations, while those defined by $\phi_0(0) \ge \phi_0^\star$ 
are dynamically unstable.  The sequences of configurations defined 
by $\phi_0(0)< \phi_0^\star$ and $\phi_0(0) \ge \phi_0^\star$ are thus 
known as the {\em stable} and {\em unstable} branches, respectively.
Moreover, with reference to the plot of $M(\phi_0(0))$ shown
in the left panel of Fig.~\ref{Madm1d}, it can be shown that 
every extremum in the plot corresponds to an additional perturbative
mode becoming unstable.  We emphasize that all of the results 
concerning boson star evolutions that are reported in this thesis 
(Chap.~\ref{results}) used stars from the stable branch.

The existence of a maximum mass for our one-parameter family of boson
stars, and the fact that there is a change of stability at 
$\phi_0(0)=\phi_0^\star$, is completely 
analogous to the Chandrasekhar limits for spherically 
symmetric white dwarfs 
and neutron stars~\cite{shapiro_teukolsky:BHWD}.
In both of those instances, above some mass limit
there is no stable static configuration and the star 
is prone to gravitational collapse (to a neutron star in the 
case of a white dwarf, and, it is widely believed, to a black hole 
in the case of a neutron star).
We note, however, that for boson stars there is no degeneracy 
pressure, as there is for the fermionic white dwarfs and neutron
stars.
In the bosonic case, the effective pressure support that counteracts 
gravity can be viewed heuristically as coming from the uncertainty 
principle if the system is studied semiclassically, or from the
dispersive nature of the Klein-Gordon wave equation when the system 
is studied classically, as it is in our current work.

\newpage
\section{Boson Stars: Representation in Cartesian Coordinates} 
\label{sec:id-1d3d}

In Sec.~\ref{sec:id-ansatz} we summarized the procedure we use to compute 
boson star data in areal coordinates, as well as the subsequent 
transformation of the data to isotropic coordinates.  As we noted there, the 
coordinate transformation is needed to provide initial conditions
in a form that is compatible with our 3D code, which assumes conformal flatness
of the 3-metric.  In addition, since the code {\em is} 3D, and based in
Cartesian coordinates, $(x,y,z)$, we must also perform some interpolation
of the field values.

Let us first consider the case where we wish to evolve a single star with
some given central field value, $\phi_0(0)$.  We start by solving the 
static polar-areal ODE system, which yields the functions $\al(R_J)$,
$a(R_J)$ and $\phi_0(R_J)$, as well as the eigenvalue for the 
solution, $\omega=\omega(\phi_0(0))$.  Here $R_J$, $J=1, 2,\ldots, N_J$
are the $N_J$ discrete values of the areal radius at which we choose 
to compute (store) the numerical solution of the ODEs. 
We then determine the coordinate transformation
to isotropic coordinates, $r_J=r(R_J)$, by solving~(\ref{eq:areal2isotropic}) 
with the boundary conditions~(\ref{eq:id-a2i-bc}).
Once this is done, 
values for the conformal factor, $\psi(r_J)$, are determined 
from~(\ref{eq:psi1d}) and~\ref{eq:psi1d-0}).  
Next, since the time coordinates are identical
in the polar-areal and maximal-isotropic systems, the lapse function 
transforms as a scalar, and we have~$\alpha(r_J)=\alpha(r(R_J))$.  Finally,
the matter field, $\phi_0$, {\em is} a scalar, so we also have 
$\phi_0(r_J)=\phi_0(r(R_J))$. 

Having computed $\psi(r)$, $a(r)$ and $\alpha(r)$, it is a simple
matter to interpolate these functions onto the discrete domain 
which our 3D code uses to solve the PDEs
derived in Chap.~\ref{formalism}.
Without 
going into the full details (these are given in the next chapter),
the finite difference grid points have coordinates $(x_i,y_j,z_k)$
where $i = 1,\ldots,n_x$, $j = 1,\ldots,n_y$, 
and $k = 1,\ldots,n_k$, so that the grid has dimensions
$n_x\times n_y\times n_z$.  We choose a point having coordinates 
$(x_0,y_0,z_0)$ that lies within the computational domain and at 
which we will ``centre'' the star.  Then for all $i$, $j$ and $k$
we define 
\beq \label{eq:cartesian_sph_transf_law}
r_{ijk}=\sqrt{(x_i-x_0)^2+(y_j-y_0)^2+(z_k-z_0)^2}
\eeq
and then set 
\bea
\al(x_i,y_j,z_k) &=& \al(r_{ijk}) \, ,\\
\ps(x_i,y_j,z_k) &=& \ps(r_{ijk}) \, ,\\
\ph_0(x_i,y_j,z_k) &=& \ph_0(r_{ijk}) \, .
\eea
Here, the values 
$\al(r_{ijk})$,
$\ps(r_{ijk})$ and
$\phi_0(r_{ijk})$ are computed using quadratic Lagrange polynomial 
interpolation~\cite{burden} in the 
values $\al(r_J)$, $\ps(r_J)$ and $\phi_0(r_J)$, respectively.  

Next, using the values $\ph_0(x_i,y_j,z_k)$, and the scalar 
ansatz~(\ref{eq:SS_ansatz}) evaluated at $t=0$, we initialize
the real and imaginary components of the scalar field by
\bea
\ph_1(x_i,y_j,z_k) &=& \ph_0(x_i,y_j,z_k) \, ,\\
\ph_2(x_i,y_j,z_k) &=& 0 \, .
\eea

In addition, the field conjugate momenta can be computed
from~(\ref{eq:phidot_adm_cart}), again, by using the static 
ansatz~(\ref{eq:SS_ansatz}) for the complex scalar field, and the fact that 
the shift vector components vanish in the static case. In terms of 
values defined above, and calculated using interpolation of the ODE
solution, we have
\bea
\Pi_1(x_i,y_j,z_k) &=& \om \fr{\psi(x_i,y_j,z_k)^6}{\al(x_i,y_j,z_k)} 
                        \ph_0(x_i,y_j,z_k) \sin(\om t) \, , \\
\Pi_2(x_i,y_j,z_k) &=& -\om \fr{\psi(x_i,y_j,z_k)^6}{\al(x_i,y_j,z_k)} 
                         \ph_0(x_i,y_j,z_k) \cos(\om t) \, .
\eea 
Since $t=0$, the above expressions
further reduce to:
\bea
\Pi_1(x_i,y_j,z_k) &=& 0 \, , \\
\Pi_2(x_i,y_j,z_k) &=& -\om \fr{\psi(x_i,y_j,z_k)^6}{\al(x_i,y_j,z_k)} 
                         \ph_0(x_i,y_j,z_k)  \, .
\eea

For the case of a binary system, the above process is performed 
for both stars, and we then superimpose the solutions as follows:
\bea
\label{eq:id-al-sup}
\al(x_i,y_j,z_k)&=&\al^{(1)}(x_i,y_j,z_k)+\al^{(2)}(x_i,y_j,z_k)-1 \, ,\\
\label{eq:id-psi-sup}
\psi(x_i,y,z_k)&=&\psi^{(1)}(x_i,y_j,z_k)+\psi^{(2)}(x_i,y_j,z_k)-1 \, ,\\
\label{eq:id-phi1-sup}
\phi_1(x_i,y_j,z_k)&=&\phi_1^{(1)}(x_i,y_j,z_k)+\phi_1^{(2)}(x_i,y_j,z_k) \, ,\\
\label{eq:id-phi2-sup}
\phi_2(x_i,y_j,z_k)&=&\phi_2^{(1)}(x_i,y_j,z_k)+\phi_2^{(2)}(x_i,y_j,z_k) \, ,\\
\label{eq:id-pi1-sup}
\Pi_1(x_i,y_j,z_k)&=&\Pi_1^{(1)}(x_i,y_j,z_k)+\Pi_1^{(2)}(x_i,y_j,z_k) \, ,\\
\label{eq:id-pi2-sup}
\Pi_2(x_i,y_j,z_k)&=&\Pi_2^{(1)}(x_i,y_j,z_k)+\Pi_2^{(2)}(x_i,y_j,z_k) \, .
\eea
Here the superscripts $(1)$ and $(2)$ refer to the interpolated solutions
for the first and second star, respectively.
In practice, when we set up data for a binary we try to ensure that
the two stars do not overlap significantly.  In this case, we can 
expect that superposition of the individual lapse and conformal functions, as 
defined above, provides values of $\al(x_i,y_j,z_k)$ and 
$\psi(x_i,y,z_k)$ which satisfy the slicing and Hamiltonian constraint
equations to some degree of precision.  However, apart from 
considerations of the extent that the initial setup {\em does} 
represent two distinct and separated stars, the issues of whether 
or not there is substantial overlap of $\phi_0^{(1)}$ and $\phi_0^{(2)}$,
and whether or not superposition of the metric functions approximately
holds, are not very important.  In particular, the scalar field 
variables are, in 
principle, freely specifiable, so an overlapping configuration provides
initial data that is no less mathematically valid than one in which 
there is negligible overlap.  Furthermore, once we have fixed the 
scalar field values at $t=0$, 
we {\em always} solve the elliptic equations PDEs for the metric functions
to determine their initial values.  Ultimately then, the values obtained 
using the superposition 
formulae~(\ref{eq:id-al-sup}) and~(\ref{eq:id-psi-sup}) 
are only used to provide initial
estimates for the iterative multigrid method that we use for the 
elliptic system (see Chap.~\ref{numerical}).

\section{Boson Stars: Applying Approximate Lorentz Boosts} \label{sec:id-boost}

The procedure described in the previous section allows us to 
initialize our 3D dynamical code with data representing one or two stars 
that are initially at rest in the $(x,y,z)$ coordinate system.  In order 
to simulate scenarios such as a binary system in which the stars 
are in mutual orbit about one another, or, in general, where there 
is to be any motion of one or both stars at $t=0$, the scalar field 
initial conditions must be adjusted to provide the stars with 
initial velocities.  

In this section we describe how this is accomplished 
through the use of Lorentz transformations, applied to both 
the scalar field and metric variables.   We emphasize at this point
that the algorithm that we describe below does {\em not}, in general,
provide initial conditions (even for a single star) that correspond 
to a pure boost.  This is a result of an incompatibility between our 
straightforward approach to applying the Lorentz transformation and
the requirement that our spatial metric be conformally flat.  However,
at least at this juncture in our research, we do not consider this 
to be a major shortcoming.  What is most important is that we be 
able to produce initial conditions such that the stars propagate through
the solution domain, retaining their overall structure, at least 
approximately, as they move.  As will be seen in Chap.~\ref{results},
our current method certainly succeeds in this respect.

In the calculations that we report in this thesis 
(again, in Chap.~\ref{results}),
we have restricted attention to configurations in which one or 
two boson stars are boosted in the $x$ direction.  We therefore
adopt an abbreviated notation which suppresses the functional 
dependence of the unknowns on $y$ and $z$. Thus, for example,
$\phi(t,x)$ is shorthand for $\phi(t,x,y,z)$.  We let $(t,x)$ be 
the coordinates in the ``lab'' frame (i.e.~the coordinate system
in which we ultimately solve the PDEs), and denote the 
coordinates of the rest frame of any boosted star by $(t',x')$.

We now focus attention on the case of a single boson star.
Then if the star is boosted with speed $v > 0$ in the positive
$x$ direction, and keeping in mind that we have adopted units 
in which the speed of light, $c$, is unity, the Lorentz 
transformation is 
\bea 
x&=&\ga(x'+vt')\label{eq:lorentz_x} \, ,\\
t&=&\ga(t'+vx')\label{eq:lorentz_t} \, ,
\eea
where $\gamma$ is the usual Lorentz factor defined 
by $\gamma\equiv(1-v^2)^{-1/2}$.
The inverse Lorentz transformation is:
\bea 
x'&=&\ga(x-vt)\label{eq:lorentz_inv_xp} \, ,\\
t'&=&\ga(t-vx)\label{eq:lorentz_inv_tp} \, .
\eea

Now, from the static ansatz~(\ref{eq:SS_ansatz}) we have 
the following equations for 
the real and imaginary components of the scalar field, $\phi_1$
and $\phi_2$, respectively:
\bea
\phi_1'(t',x')&=&\phi_0(x')\cos{(\om t')} ,\\
\phi_2'(t',x')&=&-\phi_0(x')\sin{(\om t')} .
\eea
The time derivatives of 
these quantities, which we denote by $\bar{\Pi}'_1(t',x')$ and 
$\bar{\Pi}'_2(t',x')$, respectively, are then given by
\bea
\bar{\Pi}'_1(t',x')&\equiv&\frac{\pa \phi'_1(t',x')}{\pa t'}
   =-\om\phi_0(x')\sin{(\om t')},\\
\bar{\Pi}'_2(t',x')&\equiv&\frac{\pa \phi'_2(t',x')}{\pa t'}
=-\om\phi_0(x')\cos{(\om t')}.
\eea
As a scalar field is invariant under any coordinate transformation, the 
values of $\phi_1$ and $\phi_2$ are determined by
\bea
\phi_1(t,x)&\equiv& \phi_1'(t',x')= \phi_0 \lb \ga(x-vt)\rb \cos{\lhb \om \ga (t-vx)\rhb},
\label{eq:phi1_boost}
\\
\phi_2(t,x)&\equiv& \phi_2'(t',x')=-\phi_0 \lb \ga(x-vt)\rb  \sin{\lhb \om \ga (t-vx)\rhb},
\label{eq:phi2_boost}
\eea
where we have used the inverse Lorentz transformations given by
(\ref{eq:lorentz_inv_xp}) and (\ref{eq:lorentz_inv_tp}).
At the initial time, $t=0$, we thus have
\bea
\phi_1(0,x)&=& 
\phi_0(\ga x)\cos{\lb \om \ga vx\rb},
\label{eq:phi1_boost_id}
\\
\phi_2(0,x)&=&
\phi_0(\ga x)\sin{\lb \om \ga vx\rb}.
\label{eq:phi2_boost_id}
\eea
We remind the reader that we are suppressing the $y$ and
$z$ dependence of the unknowns, so that 
$\phi_0 ( \ga x )\equiv \phi_0(\gamma x,y,z)$.  This value
can be computed from the ODE solution, $\phi_0(r(R_J))$, via interpolation
to $r_{\gamma x}\equiv \sqrt{\gamma^2 (x-x_0)^2 + (y-y_0)^2 + (z-z_0)^2}$,
where $(x_0,y_0,z_0)$ are the lab frame coordinates of the point at 
which the star is centred.

The time derivatives, ${\bar \Pi}'_1$ and ${\bar \Pi}'_2$, are also 
scalar fields and thus invariant under 
coordinate transformations as well. However, in order to write 
down expressions for them in the lab frame, we need to go 
through the Lorentz transformation in more detail. 
We first note that the chain rule can 
be used to express the time derivatives in the star frame's in terms of 
derivatives in the lab frame:
\bea
\bar{\Pi}_1(t,x)&\equiv&\bar{\Pi}'_1(t',x')=
\fr{\pa \phi_1'}{\pa t'}(t',x')=\fr{\pa \phi_1}{\pa t'}(t,x)=
\fr{\pa \phi_1}{\pa x}(t,x) \fr{\pa x}{\pa t'} +
\fr{\pa \phi_1}{\pa t}(t,x) \fr{\pa t}{\pa t'},
\label{eq:Pi1_bar}
\\
\bar{\Pi}_2(t,x)&\equiv&\bar{\Pi}'_2(t',x')=
\fr{\pa \phi_2'}{\pa t'}(t',x')=\fr{\pa \phi_2}{\pa t'}(t,x)=
\fr{\pa \phi_2}{\pa x}(t,x) \fr{\pa x}{\pa t'} +
\fr{\pa \phi_2}{\pa t}(t,x) \fr{\pa t}{\pa t'}.
\label{eq:Pi2_bar}
\eea
The spatial derivatives appearing in the above expressions 
above can be calculated using equations~(\ref{eq:phi1_boost}) and 
(\ref{eq:phi2_boost}):
\bea
\fr{\pa \phi_1}{\pa x}(t,x)&=&\fr{\pa \phi_0}{\pa x}(\ga(x-vt)) \cos{\lhb \om \ga (t-vx)\rhb}
\nonumber
\\
& & +\om\ga v \phi_0(\ga(x-vt)) \sin{\lhb \om \ga (t-vx)\rhb} \, ,
\label{eq:dphi1dx}
\\
\fr{\pa \phi_2}{\pa x}(t,x)&=&-\fr{\pa \phi_0}{\pa x}(\ga(x-vt)) \sin{\lhb \om \ga (t-vx)\rhb}
\nonumber
\\
& & +\om\ga v \phi_0(\ga(x-vt)) \cos{\lhb \om \ga (t-vx)\rhb} \, .
\label{eq:dphi2dx}
\eea
Additionally, we have 
\beq
\fr{\pa \phi_0}{\pa t}(\ga(x-vt))=\fr{\pa \phi_0}{\pa x'}(x') \fr{\pa x'}{\pa t}=
-\ga v \left. \fr{\pa \phi_0}{\pa x'}(x') \right|_{x'=\ga(x-vt)},
\eeq
so that 
\bea
\fr{\pa \phi_1}{\pa t}(t,x)&=&\fr{\pa \phi_0}{\pa t}(\ga(x-vt))\cos{\lhb \om \ga (t-vx)\rhb}
-\om\ga  \phi_0(\ga(x-vt))\sin{\lhb \om \ga (t-vx)\rhb}
\nonumber 
\\
&=&-\ga v \left. \fr{\pa \phi_0}{\pa x'}(x') \right|_{x'=\ga(x-vt)}
\cos{\lhb \om \ga (t-vx)\rhb}
\nonumber
\\
& & -\om\ga  \phi_0(\ga(x-vt))\sin{\lhb \om \ga (t-vx)\rhb}  \, ,
\label{eq:dphi1dt}
\eea
and
\bea
\fr{\pa \phi_2}{\pa t}(t,x)&=&-\fr{\pa \phi_0}{\pa t}(\ga(x-vt))\sin{\lhb \om \ga (t-vx)\rhb}
-\om\ga  \phi_0(\ga(x-vt))\cos{\lhb \om \ga (t-vx)\rhb}
\nonumber 
\\
&=&\ga v \left. \fr{\pa \phi_0}{\pa x'}(x') \right|_{x'=\ga(x-vt)}
\sin{\lhb \om \ga (t-vx)\rhb}
\nonumber
\\
& & -\om\ga  \phi_0(\ga(x-vt))\cos{\lhb \om \ga (t-vx)\rhb} 
\label{eq:dphi2dt}\, .
\eea
Using~(\ref{eq:dphi1dx}), (\ref{eq:dphi2dx}), 
(\ref{eq:dphi1dt}) and (\ref{eq:dphi2dt}) in 
(\ref{eq:Pi1_bar}) and (\ref{eq:Pi2_bar}) yields
\bea
\bar{\Pi}_1(t,x) &=&\ga v \lhb \fr{\pa \phi_0}{\pa x}(\ga(x-vt)) 
- \ga \fr{\pa \phi_0}{\pa x'}(x')  \rhb 
\cos{\lhb \om \ga (t-vx)\rhb} 
\nonumber
\\
& & - \om \phi_0(\ga(x-vt))\sin{\lhb \om \ga (t-vx)\rhb}
\label{eq:Pi1_bar1},
\\
\bar{\Pi}_2(t,x) &=&\ga v \lhb  \ga \fr{\pa \phi_0}{\pa x'}(x')
 -\fr{\pa \phi_0}{\pa x}(\ga(x-vt))  \rhb 
\sin{\lhb \om \ga (t-vx)\rhb} 
\nonumber
\\
& & - \om \phi_0(\ga(x-vt))\cos{\lhb \om \ga (t-vx)\rhb}
\label{eq:Pi2_bar1},
\eea
where the derivatives with respect to $x'$ are to be evaluated at 
$x'=\ga(x-vt)$. At $t=0$, the solution in the lab frame is then
\bea
\bar{\Pi}_1(0,x) &=&\ga v \lhb \fr{\pa \phi_0}{\pa x}(\ga x) 
- \ga \left. \fr{\pa \phi_0}{\pa x'}(x') \right|_{x'=\ga x}  \rhb 
\cos(\om \ga vx) + \om \phi_0(\ga x)\sin(\om \ga vx)
\label{eq:Pi1_bar1_id},
\\
\bar{\Pi}_2(0,x) &=&\ga v \lhb \fr{\pa \phi_0}{\pa x}(\ga x)
 -\ga \left. \fr{\pa \phi_0}{\pa x'}(x') \right|_{x'=\ga x}  \rhb 
\sin(\om \ga vx) - \om \phi_0(\ga x)\cos(\om \ga vx)
\label{eq:Pi2_bar1_id}\, .
\eea

In order to completely determine the initial data---if only
to provide good initial estimates for the elliptic solver at 
$t=0$---we also need to consider the effect of the Lorentz boost
on the metric components in the lab frame.
In the rest frame of the star we have 
\beq
g'_{\mu\nu}(t',x')=
\lb
 \begin{array}{c c c c}
-\al'^2(t',x') & 0 & 0 & 0\\
 0 & \psi'^4(t',x') & 0 & 0\\
 0 & 0 & \psi'^4(t',x') & 0\\
 0 & 0 & 0 & \psi'^4(t',x')
 \end{array} \rb \, .
\eeq
We now adopt another notation, in which we use a tilde on lab-frame
metric components that have had Lorentz transformations applied to 
them.  This distingushes them from the {\em actual} lab-frame
metric variables that appear in the our CFA model, in view
of our above observation that the simple-minded boost procedure we use is 
incompatible with conformal flatness.
The metric components in the lab frame are then given by
\beq \label{eq:metric_transf}
\tilde{g}_{\la\de}(t,x)=\fr{\pa x'^{\mu}}{\pa x^{\la}} 
                \fr{\pa x'^{\nu}}{\pa x^{\de}} g'_{\mu\nu}(t',x') \, ,
\eeq
where, from the general $3+1$ form we also have 
\beq
\tilde{g}_{\la\de}(t,x)=
\lb
 \begin{array}{c c}
-\tilde{\al}^2(t,x)+\tilde{\ga}^{ij} \tilde{\bt}_i \tilde{\bt}_j & \tilde{\bt}_k(t,x)\\
 \tilde{\bt}_l(t,x) & \tilde{\ga}_{lk}(t,x) 
 \end{array} \rb \, .
\eeq
The Lorentz boost is given by 
\beq
\La^{\mu}{}_{\la}= \fr{\pa x'^{\mu}}{\pa x^{\la}} =
\lb
 \begin{array}{c c c c}
 \ga & -\ga v & 0 & 0\\
 -\ga v & \ga & 0 & 0\\
 0 & 0 &  1 & 0\\
 0 & 0 & 0 & 1
 \end{array} \rb .
\eeq
Thus, for example, the $\tilde{g}_{0x}$ component of the lab frame metric is
given by 
\beq
\tilde{g}_{0x}\equiv \tilde{\bt}_x(t,x) = \La^{\mu}{}_{0} \, \La^{\nu}{}_{x} \, g'_{\mu\nu}=
\lb \La^{0}{}_{0} \, g'_{0\nu} + \La^{x}{}_{0} \, g'_{x\nu} \rb \La^{\nu}{}_{x}
\, ,
\eeq
which yields
\beq
\label{eq:id-bt-cov-x}
\tilde{\bt}_x(t,x) = \ga^2 v \lhb \al'^2(t',x') - \psi'^4(t',x') \rhb \, .
\eeq
Similarly, it is possible to show that the spatial metric in the lab
frame is 
related to the metric in the star's rest frame by:
\beq
\label{eq:id_lab_spatial_metric}
\tilde{\ga}_{ij}=
\lb
 \begin{array}{c c c}
\ga^2 \lhb\psi'^4(t',x')-v^2\al'^2(t',x') \rhb & 0 & 0\\
 0 & \psi'^4(t',x') & 0\\
 0 & 0 & \psi'^4(t',x') 
 \end{array} \rb,
\eeq
As promised,
this lab frame 3-metric is {\em not} conformally flat.  However,
as the tilde notation emphasises, it is also 
{\em not} the spatial metric with which the
3D code is eventually initialized, and we emphasize that there is 
no inconsistency being introduced by the current method for 
determining initial data.  In this regard,
the metric~(\ref{eq:id_lab_spatial_metric}) is perhaps best viewed 
as an intermediary which aids in the computation of other quantities
(such as the scalar field momenta, $\Pi_1$ and $\Pi_2$) to
produce data {\em approximately} describing a boosted star.  A similar
comment applies to the other lab frame geometric quantities---namely 
the lapse, $\tilde{\alpha}$, and the non-vanishing shift 
component $\tilde{\beta}^x$---that 
are defined in this stage of the algorithm.
They too should be viewed 
as ``provisional'' values that are used primarily to set the 
scalar field quantities at $t=0$ so that something close 
to a boosted boson star results. On the other hand, the true initial values 
for the geometric quantities are {\em always} determined---once 
the scalar field variables are given---by solving the 
governing set of elliptic 
PDEs~(\ref{eq:lapse_cartesian})--(\ref{eq:shift_cartesian_z}),
in which the scalar field quantities 
play the role of sources.

Keeping this in mind then, we continue by 
using~(\ref{eq:id_lab_spatial_metric}) and~(\ref{eq:id-bt-cov-x}) to find
\beq
\tilde{\bt}^x(t,x) = \fr{ v \lhb \al'^2(t',x') - \psi'^4(t',x') \rhb}
                { \lhb\psi'^4(t',x')-v^2\al'^2(t',x')\rhb }\, .
\eeq
However, since the metric components in the star frame are all time
independent, the above equation becomes
\beq
\tilde{\bt}^x(t,x) = \fr{ v \lhb \al'^2(x') - \psi'^4(x') \rhb}
                { \lhb\psi'^4(x')-v^2\al'^2(x')\rhb } \, ,
\eeq
and restricting to the initial time $t=0$, we have 
\beq \label{eq:btx_id}
\tilde{\bt}^x(0,x) = \fr{ v \lhb \al'^2(\ga x) - \psi'^4(\ga x) \rhb}
                { \lhb\psi'^4(\ga x)-v^2\al'^2(\ga x)\rhb } \, .
\eeq

Using a derivation analogous to that just used to determine $\tilde{\bt}^x(0,x)$, we
can show that the initial values of the lapse in the lab frame are 
given by
\beq \label{eq:lps_id}
\tilde{\al}(0,x) = \fr{\al'(\ga x) \psi'^2(\ga x)}{\ga \sqrt{\psi'^4(\ga x)-v^2\al'^2(\ga x) }}\, ,
\eeq
while the conformal factor is simply
\beq \label{eq:psi_id}
\tilde{\psi}(0,x)=\psi'(\ga x)\, .
\eeq

Finally, using~(\ref{eq:phi1_boost_id}), 
(\ref{eq:phi2_boost_id}), (\ref{eq:Pi1_bar1_id}), (\ref{eq:Pi2_bar1_id}), 
(\ref{eq:btx_id}), (\ref{eq:lps_id}) and (\ref{eq:psi_id}), the initial 
values for the scalar field conjugate momenta are given by
\bea
\Pi_1(0,x)&=&\fr{\tilde{\psi}^6(0,x)}{\tilde{\al}^2(0,x)}\lhb \bar{\Pi}_1(0,x) - \tilde{\bt}^x(0,x) 
\fr{\pa \phi_1}{\pa x} (0,x)\rhb , 
\label{eq:Pi1_id} 
\\
\Pi_2(0,x)&=&\fr{\tilde{\psi}^6(0,x)}{\tilde{\al}^2(0,x)}\lhb \bar{\Pi}_2(0,x) - \tilde{\bt}^x(0,x) 
\fr{\pa \phi_2}{\pa x} (0,x)\rhb .
\label{eq:Pi2_id} 
\eea

To summarize, equations~(\ref{eq:phi1_boost_id}), 
(\ref{eq:phi2_boost_id}), (\ref{eq:Pi1_id}) and (\ref{eq:Pi2_id}) 
provide initial values for the scalar field components and 
their conjugate momenta in the lab reference frame. 
On the other hand, equations~(\ref{eq:btx_id}), 
(\ref{eq:lps_id}) and (\ref{eq:psi_id}) are used to provide initial 
{\em estimates} for the values of $\beta^x$, $\alpha$ and $\psi$ that 
are ultimately computed by solving the elliptic equations at $t=0$.
However, they are also used to construct the $t=0$ values of the 
scalar field conjugate momenta as given by equations~(\ref{eq:Pi1_id}) and 
(\ref{eq:Pi2_id}). 

As was the case in the previous section, for configurations involving 
two stars, the above calculations are performed for each star 
individually.  We then set values for the fields using the 
following superposition formulae:
\bea
\phi_1(0,x)&=&\phi_1^{(1)}(0,x)+\phi_1^{(2)}(0,x),\label{eq:phi1_id_2stars}\\
\phi_2(0,x)&=&\phi_2^{(1)}(0,x)+\phi_2^{(2)}(0,x),\label{eq:phi2_id_2stars}\\
\Pi_1(0,x)&=&\Pi_1^{(1)}(0,x)+\Pi_1^{(2)}(0,x),\label{eq:pi1_id_2stars}\\
\Pi_2(0,x)&=&\Pi_2^{(1)}(0,x)+\Pi_2^{(2)}(0,x),\label{eq:pi2_id_2stars}\\
\tilde{\bt}^x(0,x)&=&\tilde{\bt}^{x\,(1)}(0,x)+\tilde{\bt}^{x\,(2)}(0,x),\label{eq:btx_id_2stars}\\
\tilde{\al}(0,x)&=&\tilde{\al}^{(1)}(0,x)+\tilde{\al}^{(2)}(0,x)-1,\label{eq:lps_id_2stars}\\
\tilde{\psi}(0,x)&=&\tilde{\psi}^{(1)}(0,x)+\tilde{\psi}^{(2)}(0,x)-1\label{eq:psi_id_2stars}\, .
\eea
Here the superscripts $(1)$ and $(2)$ again refer 
to the solutions computed for the two individual stars.
For the scalar field variables, $\phi_1$, 
$\phi_2$, $\Pi_1$ and 
$\Pi_2$,~(\ref{eq:phi1_id_2stars})--(\ref{eq:pi2_id_2stars})
are used to set the actual initial values.  Again, 
however,~(\ref{eq:btx_id_2stars})--(\ref{eq:psi_id_2stars}) are used 
only to provide initial estimates for the multigrid elliptic solver which
computes the actual $t=0$ values of $\beta^x$, $\alpha$ and $\psi$.

We end this section by noting that it is probably possible 
to refine the algorithm described here---by iterating the 
steps of computing the conjugate momenta using expressions such
as~(\ref{eq:Pi1_id}) and (\ref{eq:Pi2_id}), and of solving 
the elliptic equations for the metric variables---in order 
to determine initial data that more closely approximates 
a purely boosted boson star.  This however, is another matter 
that will require further investigation: as stated earlier, for 
the purposes of the computations described in Chap.~\ref{results},
the procedure detailed here appears adequate.

 \resetcounters
\def\calH{{\mathcal H}}
\def\calT{{\mathcal T}}
\def\calR{{\mathcal R}}
\def\calG{{\mathcal G}}
\def\H#1#2{{\mathcal H}^{#1}{}_{#2}}

\chap{Numerical Techniques} \lab{numerical}

This chapter discusses the key numerical techniques that were 
used to find approximate solutions
of the partial differential equations (PDEs) that comprise our model. 
Furthermore, the methods that were 
employed to assess the basic correctness of the numerical implementation
and the accuracy of the generated solutions are presented.
Although some issues are covered here in more detail
than others, for completeness we have tried to describe all of the 
main numerical approaches and algorithms that factored into the 
construction of our code.  In this regard, however, we note that we have 
relegated some of the more basic and well-known concepts and techniques
to Appendices \ref{ap:stencil} and \ref{ap:ngs}.
Portions of 
this chapter closely follow the presentation 
in \cite{matt_mexico:2006} and the interested reader 
is referred to those notes (as well as references contained therein)
for examples, as well as more detailed discussions of some of the 
approaches we have used.
In addition, for the uninitiated, the venerable text due to Burden and 
Faires \cite{burden} provides an excellent general introduction to the field
of numerical analysis.

A crucial step in the construction of numerical solutions of PDEs 
is the choice of an
appropriate discretization approach. 
The one adopted here is 
finite difference approximation (FDA), and, fundamentally, involves
replacement of the differential operators appearing in the PDEs with
suitable finite difference operators. While the differential operators are
applied to functions defined on a particular subdomain, $\Om$, of the
continuum---which in our case is 
$\mathbb{R} \times \mathbb{R}^3$---in practice, the discrete operators are 
applied to functions defined 
on a discrete set of points, $\Om^h$, often referred to as a grid, 
or mesh.~\footnote{We note, however, that it can be useful to view the 
discrete operators as acting on functions defined on the continuum, 
particularly when considering error analysis of the type discussed,
for example, in \ref{subsec:IRV}.}
Sec.~\ref{sec:FDA} introduces the
main concepts we use in the finite difference 
discretization of our PDEs.
Based on the FDA concepts and assumptions discussed in 
Sec.~\ref{sec:FDA}, Sec.~\ref{sec:tools} then continues 
with a description of the analysis tools that were used to evaluate the 
fidelity of out numerical solutions, as presented in detail in 
Chap.~\ref{results}. 

As discussed previously, our model
consists of a coupled system of $4$ 
hyperbolic (first order in time) and $5$ 
elliptic (second order in space) PDEs. Since the equations belong to two
different classes, it is natural to expect that different 
techniques will be needed to efficiently obtain the full numerical
solutions.
In brief, we treat the hyperbolic equations 
using a second order (in the mesh spacing, $h$)  Crank-Nicholson approximation that 
is solved iteratively, while the elliptic equations, also discretized to 
$O(h^2)$, are solved using a multigrid technique. 
Here we note that, as illustrated by the pseudo-code of the overall 
flow of our code in Fig.~\ref{RNPL} (Sec.~\ref{sec:num_code}), 
at each stage of the 
Crank-Nicholson iteration that performs the basic time-step-advance of the 
hyperbolics, we re-solve the elliptics.
A review of basic relaxation techniques, including the point-wise 
Newton-Gauss-Seidel method used in the Crank-Nicholson iteration 
is given App.~\ref{ap:ngs},
while the multigrid method is discussed in some detail in Sec.~\ref{sec:MG}.

\section[Discretization of Partial Differential Equations] 
{Discretization of Partial Differential Equations: Finite Difference 
Approximation} \label{sec:FDA}

Two fundamental assumptions underlying a successful finite difference 
approximation (FDA) of a set of PDEs are:
\begin{enumerate}
\item Given suitable boundary and/or initial conditions, there is a 
unique solution of the system of PDEs to be discretized. 
\item This unique solution is smooth.  
\end{enumerate}
As argued in Chap.~\ref{formalism}, any
solution of the coupled hyperbolic-elliptic system of equations constituting
the initial-boundary value problem to be solved in this thesis is expected 
to be smooth provided the prescribed initial data is smooth. 
Also, although no general proof exists, there is no reason {\em a priori} to 
expect solutions of our system (again given appropriate initial and boundary
conditions) not to be unique.
Therefore the choice 
of finite difference approximation as the discretization approach becomes a 
natural one for the problem at hand.

The first step in the discretization of PDEs 
using finite differencing involves the definition 
of a grid (or mesh) over the solution domain
$\Om \subset \mathbb{R}^n$.  (We note that for the time being we will restrict
our attention to discretization of the {\em spatial} domain, so that $\Om$ is a
spatial volume: we will return to the issue of temporal discretization below.)
Loosely speaking, the grid is defined as a discrete set of points from 
the continuum domain that satisfy
some criteria.  These criteria can be as diverse as the physical problems 
that are governed by systems of PDEs. In general, though, 
the choice of grid points is governed by the geometry of
the domain, (and hence the geometries of the domain boundaries),
as well as the coordinate system in which the PDEs are expressed.
For domains with very complex structure, for example, it is common to use 
so-called unstructured grids that adapt themselves to the geometry. 
Grids used for simulations of car crashes or airplane aerodynamics 
are frequently of this kind.  Considering another example,
a physical problem with approximate spherical symmetry will often be formulated
in a coordinate system adapted to the symmetry, and it will therefore be 
natural to use a curvilinear grid which reflects that symmetry. 
For the case of the differential equations defining our model,
the coordinate system in which the equations are written is the 
familiar Cartesian (or rectangular) one.
Thus it is only logical that the grid adopted in the 
discretization also be Cartesian. In addition, the simplicity of our solution
domain, combined with the fact that the objects that we will be studying 
will 1) tend to have comparable extents in each of the three coordinate 
directions, and 2) tend to move through the computational domain,
suggests the use of the simplest grid possible: a uniform Cartesian grid. 
For our 3D spatial domain $\Om \subset \mathbb{R}^3$ this 
consists of an 
ordered set of points $(x_i,y_j,z_k)\equiv \Om^h \subset \Om$ such that 
the separation between adjacent points in any of the coordinate directions 
is some constant, $h$, known as the grid (or mesh) spacing, or discretization
scale.  Specifically we have
\beq
x_{i+1}-x_i = h, \qquad y_{j+1}-y_j = h, \qquad z_{k+1}-z_k = h \, .
\eeq
For any FDA the various grid spacings that may appear in the definition
of the mesh are the fundamental 
control parameters for the approximation: in particular one hopes to recover 
the continuum solution in the limit that all of the grid spacings tend to 0.  
We emphasize that in this
thesis we will exclusively use discretizations (including that of 
the time variable) that are characterized by the {\em single} discretization 
scale, $h$.

The specific grid points comprising our uniform Cartesian mesh are 
readily defined from the coordinate ranges of the continuum domain.
Thus, for
\bea
x_{\rm min}\le &x& \le x_{\rm max}, \\
y_{\rm min}\le &y& \le y_{\rm max}, \\
z_{\rm min}\le &z& \le z_{\rm max}, 
\eea
the grid points are 
\bea
x_i &=& x_{\rm min} + (i-1)\,h, \qquad \qquad  i=1 \dots n_{x},\\
y_j &=& y_{\rm min} + (j-1)\,h, \qquad \qquad  j=1 \dots n_{y},\\
z_k &=& z_{\rm min} + (k-1)\,h, \qquad \qquad  k=1 \dots n_{z},
\eea
such that
\begin{alignat}{2}
x_1 &= x_{\rm min} \qquad& \textrm{and} \qquad  x_{n_x} &= x_{\rm max}, \\
y_1 &= y_{\rm min} \qquad& \textrm{and} \qquad  y_{n_y} &= y_{\rm max}, \\
z_1 &= z_{\rm min} \qquad& \textrm{and} \qquad  z_{n_z} &= z_{\rm max}.
\end{alignat}
There are therefore
$N_g = n_x \times n_y \times n_z$ points in the mesh,
and we further note that there is an implicit assumption that each of the 
ranges $x_{\rm max}-x_{\rm min}$, $y_{\rm max}-y_{\rm min}$ 
and $z_{\rm max}-z_{\rm min}$ is evenly divisible by $h$.

\begin{figure}[th!]
\begin{center}
\input{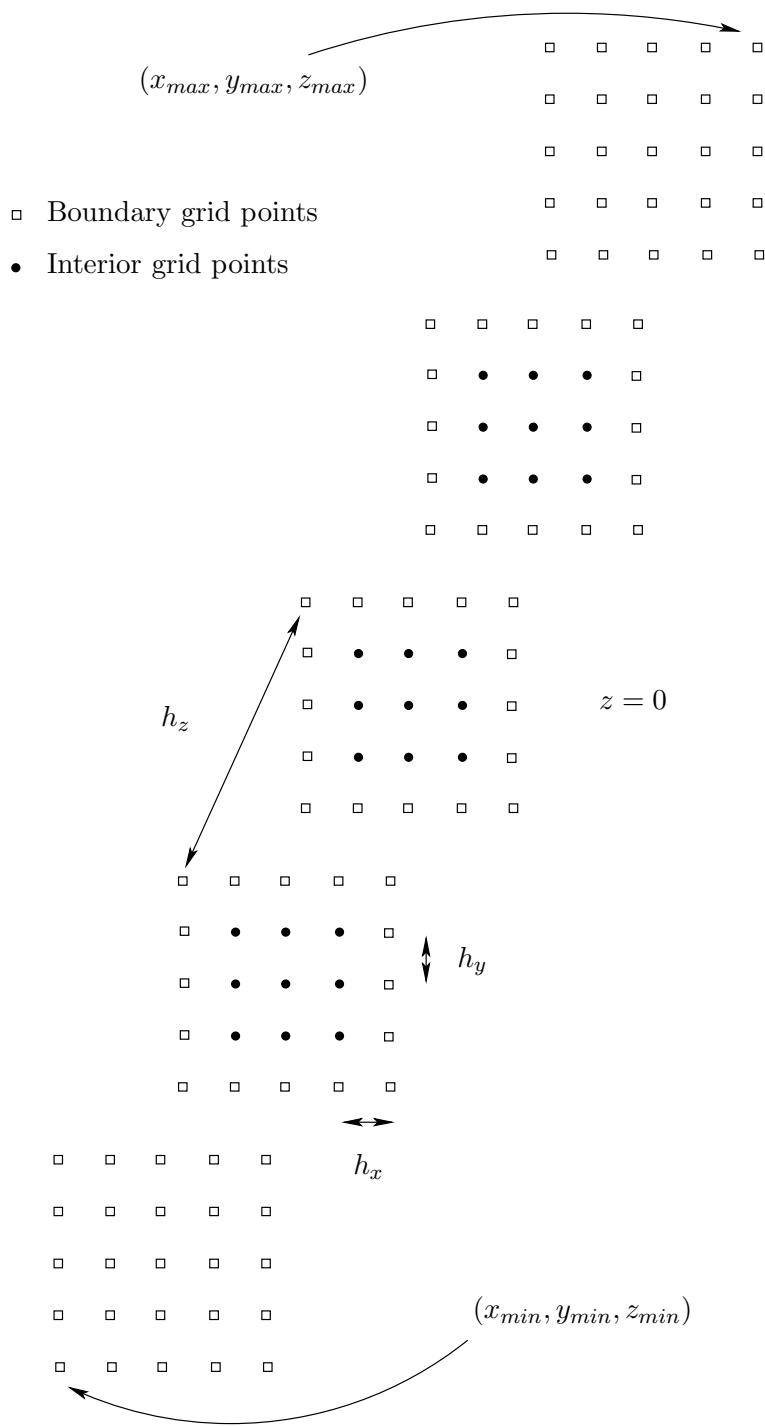}
\caption[3D Grid Sample.]
{Illustration of sample 3D finite difference grid of the type used
throughout this thesis. The figure shows a 3D Cartesian grid drawn in 
perspective. This
particular grid is composed of $5\times 5 \times 5$ grid points covering
the domain 
$x_{\rm min}\le x \le x_{\rm max}$,
$y_{\rm min}\le y \le y_{\rm max}$,
$z_{\rm min}\le z \le z_{\rm max}$.
The solid circles represent interior grid points,
while the hollow squares denote boundary grid points. For clarity,
the grid spacing in the $z$ direction is drawn out of scale, 
and we emphasize that
throughout this manuscript we take $h_x=h_y=h_z\equiv h$.
\label{grid3d}}
\end{center}
\end{figure}

Once the grid $\Om^h$ has been defined, we can introduce grid functions
$u^h:\Om^h \rightarrow \mathbb{R}$ on it. For the 3D Cartesian grid defined
above, the grid function $u^h$ is simply equivalent to the $N_g$ values
$u_{ijk}$ defined on the grid points $(x_i,y_j,z_k)$---i.e. $u_{ijk}$ 
denotes a discrete approximation to the continuum value 
$u(x_i,y_j,z_k)$. We also observe that the way that we have defined our
discrete unknowns is often called a {\em vertex centred} approach,
since the unknowns are defined at the grid 
points {\em per se}, instead of, for example, at the centre of a cubic cell 
composed of $8$ neighbouring grid points. Discretizations based on this 
latter approach, and which {\em have} frequently been used in numerical 
relativity---especially in studies involving sets of conservation laws 
such as hydrodynamics---are known as {\em cell centred}.
Finally, Fig.~\ref{grid3d} illustrates a sample 3D rectangular grid of 
the type used throughout this thesis.

\afterpage{\clearpage}

\begin{figure}[h!tb]
\begin{center}
\input{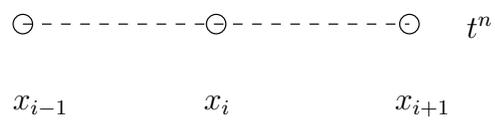}
\caption[Stencil for Second Spatial Derivative Operator]
{Illustration of stencil for a finite difference approximation of 
the second spatial derivative operator.
\label{stencilO2}}
\end{center}
\end{figure}

As already mentioned in the introductory section of this chapter,
the derivative operators appearing in the system of PDEs governing our 
model are approximated by finite difference 
operators. It is conventional to characterize any such 
operator by its {\em difference stencil} or {\em difference star}. 
This involves specification of the set of neighbouring 
grid points that appear in the definition of the difference operator. 
For example, if a real function of one variable, 
$u(x)$, is restricted to its values, $u_i$, defined on a uniform 1D grid, 
$\Om^h=\{x_i \, | \, x_{i+1}-x_i=h\}$, then
the second derivative of $u$ at $x=x_i$ can be approximated with the 
following
combination of grid function values:
\beq
\fr{d^2}{dx^2} u(x) \approx \fr{u_{i-1}-2u_i +u_{i+1}}{h^2}.
\eeq
The stencil of this finite difference operator (see Fig.~\ref{stencilO2}) 
has an overall multiplicative factor,
$1/h^2$, and is weighted by the ordered set of integers $(1,-2,1)$ 
corresponding to the coefficients that multiply the grid function
values in the numerator of the approximation.
In general, the particular grid function values, as well as the 
corresponding weights, that go into the definition of any finite difference
operator depend on both the order of the differential operator
being approximated and the level of accuracy (as a function of the 
mesh spacing) desired in the approximation.
App.~\ref{ap:stencil} 
tabulates all the FDAs used in the 
discretization of our model equations.
In addition, App.~\ref{ap:pdefda} describes a set of Maple procedures
we have written with an aim
to automate 1) the derivation of finite difference operators
of any order and 2) the subsequent discretization of PDEs, 
of any {\em differential} order, using these operators. The appendix includes several 
examples that illustrate the usage of these procedures. 

We now turn our attention to some additional basic concepts which are very 
useful in the analysis of FDAs.

\subsection{Basic Concepts Related to FDAs}\label{sec:basic-FDA}

Consider a system of partial differential equations described abstractly by
\beq \label{eq:luf}
L u - f = 0,
\eeq
where, in the most general case, 
$L$ is a set of $m$ differential operators acting on
a vector, $u$, of $m$ unknown dependent functions: $u=(u_1,\dots,u_m)$.
Each of the $u_i$ is a function of the $n$ independent 
variables (including time, as appropriate); 
that is, $u_i = u_i(x_1,\dots,x_n)$ for $i=1\dots m$. Similarly, $f$ 
is a vector of $m$ prescribed 
functions---often known as {\em sources}---with $f_i = f_i(x_1,\dots,x_n)$ 
in general.  For the remainder of this section, and strictly for the 
sake of presentation, we will specialize to the case $m=1$ (i.e.~for 
a single unknown function), while emphasizing that all of the discussion
is equally valid for the general-$m$ case.

We can write any finite difference approximation of~(\ref{eq:luf}) as 
\beq \label{eq:lufh}
L^h u^h - f^h = 0,
\eeq
where the operator $L^h$ is an FDA of the differential operator, $L$, $u^h$ is 
the discrete solution and $f^h$ are the values of the 
source function $f$ restricted to the grid points. The superscript $h$ notation 
emphasizes that $h$ is the fundamental control parameter of the discretization,
and we observe that~(\ref{eq:lufh}) will in general constitute 
a set of algebraic equations---possibly nonlinear---for the discrete 
unknowns, $u^h$.

In order to quantify how much the discrete solution, $u^h$, 
deviates from the continuum solution, $u$, for any specific value of $h$, 
it is natural to define the \emph{solution error}, $e^h$, as
\beq
e^h \equiv u^h - u .
\eeq
Clearly, in most cases of interest, the solution error {\em will} depend
on the (finite) value of $h$. 
Thus, a key element in the analysis of finite difference approximation schemes 
is the investigation of the 
relation between $e^h$ and $h$ as $h\to 0$. Does the discrete solution 
approach the continuum one? If so, at 
what rate? Roughly speaking, numerical analysts judge the quality of the 
FDA by how fast the error goes to zero as a 
function of $h$. This motivates the introduction of the concept of 
\emph{convergence} of the approximate solution. The approximation is said 
to converge if and only if
\beq
\lim_{h \rightarrow 0}{u^h} = u \, ,
\eeq
or, equivalently,
\beq
\lim_{h \rightarrow 0}{e^h} = 0 \, .
\eeq
The {\em order} of convergence (not to be confused with the differential
order of the system being approximated) measures the rate at which the 
error converges to zero. If
\beq
\lim_{h \rightarrow 0}{e^h} = O(h^p)\, ,
\eeq
then the discrete solution is said to converge to the continuum 
one with order $p$, or to be $p$-th order accurate. 
For example, if $p=1$, the solution is said to converge linearly (first-order
accurate), while 
if $p=2$, the solution converges quadratically (second-order accurate).

In principle, the discrete solution, $u^h$, of the algebraic system of 
equations~(\ref{eq:lufh}) can be calculated exactly (assuming that a 
solution exists).
In the context of numerical computation, ``exact'' typically
means ``as accurate as the particular representation of floating-point 
numbers being used allows''; this is often referred to as ``at the level
of machine precision''. 
Let us momentarily restrict our attention to the case that the 
system~(\ref{eq:lufh}) is linear in the unknowns $u^h$.  Then, again in 
principle (and ignoring issues concerning the conditioning of the 
linear system), we could employ standard numerical linear algebra software 
that implements some variant of Gaussian elimination to 
{\em directly} compute $u^h$ to machine precision.  Moreover, for the case 
that (\ref{eq:lufh}) is nonlinear in the $u^h$, we could employ a so-called
global Newton iteration: at each stage this would require the solution of 
a linear system, which again could be achieved via Gaussian elimination.
However, especially for the algebraic systems that result from the finite 
difference discretization of time-independent PDEs with dependence on 
multiple spatial variables (e.g. 2D and 3D systems), 
the ``exact'' calculation of $u^h$ in this manner is often 
prohibitively expensive computationally.  In particular, 
for such problems, the amount of computational work that must be expended 
to compute the solution {\em per unknown} tends to increase as the mesh is 
refined.  The reason for this increase in work per unknown as $h\to0$ 
can be traced to the phenomenon of {\em fill in} that occurs during the 
Gaussian elimination procedure, which is typically based on the so 
called $\mathbf{LU}$ decomposition, wherein a matrix, ${\bf A}$ is factored as 
${\mathbf A}={\mathbf L}{\mathbf U}$, where ${\mathbf L}$ and ${\mathbf U}$
are, respectively, upper- and lower-triangular matrices.  Even with 
special orderings of the unknowns (such as that provided
by nested dissection~\cite{george_liu}), the matrices that 
result from finite difference discretization of elliptic operators in 
2- and 3-D---though sparse---have bandwidths that increase as $h\to0$. This 
leads to an increasing density of non-zero elements appearing in 
${\bf L}$ and ${\bf U}$, which in turn results in the increasing cost per 
unknown as the mesh is refined. 

Given this situation, it often turns out to be more computationally 
efficient to calculate an {\em approximate} solution of the linear system
arising from an FDA of an elliptic equation
(or a linearization thereof)
through an iterative method.  In this case, an ``exact'' solution is only 
obtained in the limit of an infinite number of iterations 
(assuming the iteration converges), but in practice the iteration
can be terminated when some convergence criterion is achieved.  
We note 
that iterative techniques form the basis of the multigrid method that we
use to solve the elliptic PDEs in our model, and which we discuss in 
Sec.~\ref{sec:MG}. 
Additionally, they are also 
used {\em directly} to solve our time-implicit (Crank-Nicholson) 
discretization of the hyperbolic equations, as is described 
in more detail in Sec.~\ref{sec:CN}.

As the name suggests, an iterative method involves the 
computation of a sequence of approximations of the solution unknown,
$u^h$, and we will generically denote any of these approximations 
as $\tilde{u}^h$.  It should be noted that implicit in the use of 
an iterative technique is the fact that $\tilde{u}^h$ must be 
{\em initialized} in some fashion.

Given any iterate, $\tilde{u}^h$, an important 
quantity is the {\em residual},
$r^h$, associated with $\tilde{u}^h$, and defined by
\beq \label{eq:residual}
r^h \equiv L^h \tilde{u}^h -f^h .
\eeq
Thus, the residual quantifies the amount by which $\tilde{u}^h$ fails 
to satisfy the FDA~(\ref{eq:lufh}),
and the solution of~(\ref{eq:lufh}) by an iterative process 
is then equivalent to driving the residual to $0$. 

One last quantity that is very useful in the analysis of FDAs is the 
{\em truncation error}, defined by
\beq \label{eq:truncation}
\tau^h \equiv L^h u - f^h,
\eeq
where we note that $u$ is the {\em continuum solution of the 
differential system}~(\ref{eq:luf}).
Note that from Eq.~(\ref{eq:luf}) we have $f^h = Lu$,  where the 
right hand side of this last expression is understood to be evaluated on 
the discrete mesh.  We can thus rewrite~(\ref{eq:truncation}) as
\beq
(L^h - L) u = \tau^h ,
\eeq
and from this form we see that the truncation error directly measures 
the deviation between the actions of the finite difference and continuum 
operators on the continuum solution.

A finite difference approximation is said to be \emph{consistent} (with 
the underlying PDE) if and only if the truncation error 
goes to zero as $h$ tends to zero:
\beq
\lim_{h\rightarrow 0}{\tau^h} = 0 \, .
\eeq
Clearly, consistency is a necessary condition for convergence of the FDA.
Finally, the FDA is said to be $p$-th order accurate if:
\beq
\lim_{h\rightarrow 0}{\tau^h} = O(h^p) ,
\eeq
and where $p$ is strictly positive integer.

\subsection{Richardson Expansions} \label{richardson}

We note that for any given FDA, $L^h$, of some differential operator, $L$,
the functional form of the truncation error, $\tau^h$, can be explicitly
computed, typically using Taylor series expansion.
On the other hand, it is not 
immediately clear what, if anything, we can say about the functional 
form of the actual solution error, $e^h = u^h -u$.  However, this 
was, in fact, addressed almost a century ago in a landmark paper by 
L.~F.~Richardson~\cite{richardson:1911}.  
Richardson posited (stated without proof) that provided a finite 
difference scheme was $O(h^2)$ and centred, so that the
truncation error, $\tau^h$, had the 
form 
\beq
\label{eq:rich-tauh}
\tau^h = h^2 \tau_2(t,x,y,z) + h^4 \tau_4(t,x,y,z)
+ h^6 \tau_6( t,x,y,z) + \dots
\eeq
then the solution error, $e^h$, would have a similar expansion
\beq
\label{eq:rich-exp}
e^h = h^2 e_2(t,x,y,z) + h^4 e_4(t,x,y,z) + h^6 e_6(t,x,y,z) + \dots 
\eeq
Here, a key observation is that the functions $e_2$, $e_4$, $e_6$, $\dots$
appearing in the expansion of the solution error have {\em no $h$-dependence}.
This seemingly innocuous observation has far-reaching consequences for the 
analysis of the error in finite difference calculations, and we will refer 
to an asymptotic ($h\to0$) expansion of the form~(\ref{eq:rich-exp}) as 
a {\em Richardson expansion}.  Note that one immediate consequence 
of~(\ref{eq:rich-exp})---if it is indeed true for the particular 
second-order, centred FDA under consideration---is that an $O(h^2)$ 
truncation error implies an $O(h^2)$ solution error.  

In addition, we can postulate the existence of Richardson expansions in
schemes where non-centred difference approximations are employed: in this 
case, the expansion will include terms proportional to $h^p$, 
where $p$ can be both even {\em and} 
odd. 
Similarly in situations where the approximation is $O(h^p)$ accurate for 
$p>2$ the expansion will begin with a term $h^p\,e_p(t,x,y,z)$,
and will contain terms that have strictly even powers of $h$, or even and
odd powers of the mesh scale, depending on the 
type of finite differences used.

In all cases, should a Richardson expansion exist, we can 
argue that \emph{consistency} leads to \emph{convergence}:
\beq
\lim_{h\rightarrow 0}{\tau^h} = O(h^p)  \qquad \Rightarrow \qquad 
\lim_{h \rightarrow 0}{e^h} = O(h^p) \, .
\eeq
In general it is not feasible to \emph{prove} the existence of 
a Richardson expansion for a given FDA of a set of PDEs: such a proof
will be at least as difficult as proving global existence and uniqueness
of the PDEs themselves.
However, in simple cases where both the PDE and the discretization 
are amenable to closed form analysis it may be possible to provide
a proof---see page $111$ of reference \cite{gustafsson} 
for an example. Most importantly for us, in practice one can 
always establish the
existence of the expansion (non-rigourously) by construction, 
i.e.~through examination of the $h$-dependence of the numerical 
solutions themselves. 
In particular, if the discrete solution converges at the expected rate 
as $h\to0$, then one can usually be quite certain that a Richardson
expansion exists.  We will return to this point in 
Sec.~\ref{subsec:conv_factor}.

\subsection{The Crank-Nicholson Discretization Scheme} \label{sec:CN}

To conclude this section we consider a specific type of
FDA that can be applied to time dependent PDEs, and which was used in the 
finite-differencing of the hyperbolic equations appearing in our model.
We illustrate the technique using a very simple PDE, namely the 
one dimensional advection equation.
However, as discussed at the end of this sub-section, the 
method can be generalized
to virtually any system of PDEs that are first-order in time.

The 1D dimensional advection equation is perhaps the simplest
example of a hyperbolic PDE.  Posing it as an initial-boundary-value 
problem on a finite domain we have:
\bea \label{eq:adv_eq}
\fr{\pa u(t,x)}{\pa t}&=&\fr{\pa u(t,x)}{\pa x} 
\qquad -x_{\rm min} < x < x_{\rm max} \quad , \quad t \geq 0 ,\\
u(0,x)&=&u_0(x) ,\\
\label{eq:adv_eq_bdy}
u(t,x_{\rm max})&=&0 .
\eea
where the last condition can be interpreted as meaning that no disturbances
are entering the domain from the right.  Note that there is {\em no} 
boundary condition at $x=x_{\rm min}$, since the only characteristics 
in the problem are given by $t = -x + {\rm const}$, so that at $x=x_{\rm min}$,
the single characteristic field is strictly outgoing.  (Since it is largely 
irrelevant to the current discussion, and to keep the following
presentation straightforward, we will not stipulate any details concerning
the update of the 
grid function value at $x=x_{\rm min}$.)

One rule-of-thumb that is often used when constructing FDAs is to try to 
keep the stencils as centred as possible (i.e.~the geometric structure
of the stencil is to be kept symmetric about the grid point at which the 
approximation is applied).  Use of centred schemes typically results
in several benefits, including:
\begin{enumerate}
\item For given accuracy (e.g.~$O(h^2)$), a minimization of the number 
      of unknowns appearing in the stencil.
\item Conversely, for a given number of unknowns in the stencil, 
      a maximization of the accuracy of the scheme (e.g.~$O(h^2)$ vs 
      $O(h)$).
\item Symmetry of the matrices, $L^h$, that operate on the grid 
      function, $u^h$, and that result from the discretization (or a 
      linearization of the FDA).
\end{enumerate}
Concerning the last point, the properties of symmetric matrices are 
well known, and can often be used to prove existence and uniqueness 
of solutions of the discrete equations.
In addition, stability theorems that establish convergence of 
iterative techniques for solving the linear systems resulting 
from finite-difference approximation are generally easier to 
establish for the case of symmetric matrices.  Here and throughout 
this thesis, then, we use centred difference approximations (in time 
as well as in space) whenever possible.

Returning to the advection equation, we introduce a discrete domain 
$(t^n,x_i)$ as follows:
\bea
t^n&\equiv&n \De t, \qquad n = 0, 1,2, \dots \\
x_i&\equiv&x_{\rm min}+(i-1) h, \qquad i = 1,2,\dots n_x ,\\
u^n_i&\equiv&u^h(t^n,x_i) ,\\
\De x&\equiv&  h = \fr{(x_{\rm max}-x_{\rm min})}{(n_x - 1)}, \\
\De t&=&\lambda h ,
\eea
where $\lambda$ is known as the Courant factor.  Here, and throughout 
the thesis, we will assume that for any specific sequence
of calculations in which the mesh resolution is changed, $\lambda$ is 
held fixed.  This means that the overall difference scheme will always 
be characterized by the {\em single} mesh spacing, $h$.
 
\begin{figure}[h!]
\begin{center}
\input{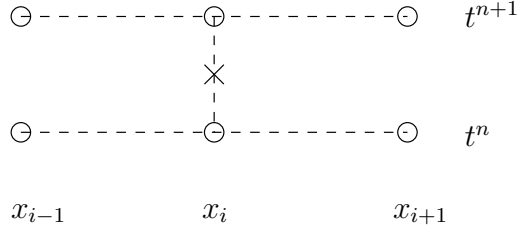}
\caption[Crank-Nicholson Stencil.]
{Crank-Nicholson Stencil. The fictitious grid point, $(t^{n+1/2},x_i)$,
about which the scheme is centred, is marked with the $\times$ symbol.
\label{stencilCN}}
\end{center}
\end{figure}

The Crank-Nicholson discretization scheme is a two-level (i.e. involves 
unknowns at two discrete instants of time, $t^n$ and $t^{n+1}$), 
$O(h^2)$ method, which, as shown in Fig.~\ref{stencilCN},  is centred in 
both space and time about the fictitious grid point, $(t^{n+1/2},x_i)$.
One obvious advantage of this method is that it uses only 
two time levels.  Related to this is the fact that among schemes which
are $O(h^2)$ accurate in time, the actual magnitudes of the 
temporal contributions to 
the truncation and solution errors tend to be minimized.
Finally, another appealing feature of Crank-Nicholson
discretization is that it tends to minimize the appearance
of instabilities that frequently
arise in the finite difference approximation of hyperbolic equations.
As applied to the advection equation, and using the standard $O(h^2)$ 
approximation of the first spatial derivative, the Crank-Nicholson scheme
is 
\beq \label{eq:adv_cn}
\fr{u^{n+1}_i-u^n_i}{ \De t} =  
    \mu_t \lb \fr{u^n_{i+1}-u^n_{i-1}}{2 \De x} \rb, \quad \quad
    2 \le i \le n_x -1 \,\, .
\eeq
Here $\mu_t$ is an $O(\De t^2$) ($O(h^2)$) time-averaging operator. Its action 
on any grid function is defined by
\beq
\mu_t u^n_i = \ha \lb u^{n+1}_i + u^n_i \rb .
\eeq
Inserting the above definition into
(\ref{eq:adv_cn}) yields the following explicit expression for the 
Crank-Nicholson scheme as applied to the advection equation:
\beq
\fr{u^{n+1}_i-u^n_i}{ \De t} =  
\ha \lhb \fr{u^{n+1}_{i+1}-u^{n+1}_{i-1}}{2 \De x} + 
\fr{u^n_{i+1}-u^n_{i-1}}{2 \De x} \rhb, \quad \quad 2 \le i \le n_x -1 \,\, .
\eeq
These last equations can be rewritten as
\beq \label{eq:adv_cn_matr}
-\fr{\la}{4} u^{n+1}_{i+1}+u^{n+1}_i+\fr{\la}{4}u^{n+1}_{i-1} = 
u^n_i +\fr{\la}{4} \lb u^n_{i+1}-u^n_{i-1} \rb , 
\quad \quad 2 \le i \le n_x -1 \,\, .
\eeq
The above equations, 
supplemented with the boundary condition $u^{n+1}_{n_x}=0$ (that 
follows from~(\ref{eq:adv_eq_bdy})), and some auxiliary condition of 
the form 
$u^{n+1}_1 = U(u^{n+1}_i, \,\, i=2\ldots n_x;\,\, u^n_i,\,\,i=1\ldots n_x)$,
constitute a linear system for the unknowns 
$u^{n+1}_i$, $i=1,2,\dots,n_x$ (provided that $U$ is linear in 
$u^{n+1}_i$, $i=2\ldots n_x$).  That is, we can write the update in 
the form 
\beq
{\mathbf A} {\mathbf u^{n+1}} = {\mathbf b} .
\eeq
where ${\mathbf A}$ is an $n_x\times n_x$ matrix, and ${\mathbf u^{n+1}}$ 
and ${\mathbf b}$ are length-$n_x$ column vectors.

We note that ${\mathbf A}$ is {\em not} diagonal---that is, there is 
coupling between the individual advanced-time unknowns, $u^{n+1}_i$, 
and hence our discretization is an example of what is known as an
\emph{implicit} scheme. (In contrast, an {\em explicit} scheme would 
be one for which it would be possible to write, for arbitrary $n_x$, explicit
expressions for the $u^{n+1}_i$.)  Now, assuming that the update for 
$u^{n+1}_1$ is of the form
$u^{n+1}_1 = U(u^{n+1}_2;\,\, u^n_i,\,\,i=1\ldots n_x)$,
then $\mathbf A$ is actually a 
tridiagonal matrix. That is, only the elements of its diagonal, 
as well as the the immediate upper
and lower diagonals are non-vanishing.  
We note in passing that these tridiagonal systems can be 
solved very efficiently---with $O(n_x)$ operations---using 
specialized solvers such as those available in the widely used 
LAPACK linear algebra package~\cite{lapack}.

The evolution equations used in this dissertation were cast as a system of 
first order PDEs in $(t,x,y,z)$, with each individual PDE assuming 
the following general form:
\beq
\fr{\pa u(t,x,y,z)}{\pa t} = F(x,y,z,u,u_x,u_{xx}) \, .
\eeq
Each evolution equation was then discretized using an $O(h^2)$ Crank-Nicholson
scheme. As explained above in the context of the
advection equation, and emphasizing that {\em all} mesh spacings,
$\De x$, $\De y$, $\De y$ and $\De t$ are $O(h^2)$, 
this involves the following approximation for 
the time derivative:
\beq
\fr{\pa u(t,x,y,z)}{\pa t} \approx 
\fr{u^{n+1}_{ijk}-u^n_{ijk}}{\De t}  =  
\left.\fr{\pa u(t,x,y,z)}{\pa t}\right\vert^{n+1/2}_i + O(h^2) \, .
\eeq
Again, as was the case for the advection equation, the spatial part 
of the PDEs are replaced by approximations that 
average between the advanced and current time levels
using the time averaging operator, $\mu_t$. Specifically, for a general
right hand side, $F$, we have
\beq
F(x,y,z,u,u_x,u_{xx}) \approx 
\ha \lb \hat F^{n+1}_{ijk} + \hat F^n_{ijk} \rb  =
\left.F(x,y,z,u,u_x,u_{xx})\right\vert^{n+1/2}_i + O(h^2) \, ,
\eeq
where $\hat F$ denotes an approximation to $F$ in which $u_x$ and $u_{xx}$ 
have been replaced by the usual $O(h^2)$, centred finite-difference
formulae.
The result of the Crank-Nicholson discretization, as applied to our
system of $4$ hyperbolic PDEs, 
Eqs.~\ref{eq:phidot_adm_cart}--\ref{eq:pidot_adm_cart}, is a non-linear system
of algebraic equations to be solved for the advanced-time 
unknowns, $u^{n+1}_{ijk}$.  Here we remind the reader that $u$ 
represents any of the hyperbolic variables appearing in our model;
that is $u$ is any of $\phi_1$, $\phi_2$ $\Pi_1$ or $\Pi_2$.

Fortunately there are efficient iterative methods to solve this type 
of algebraic system of equations.  In particular, we chose to apply 
point-wise iterative Newton-Gauss-Seidel 
relaxation.~\footnote{The reader can refer to App.~\ref{ap:ngs} for a 
brief overview of the method or to 
Varga~\cite{Varga:MIA} for a more comprehensive
treatment.}  
We also note that the relaxation was applied to each of the 4 hyperbolic
unknowns individually (the {\em decoupled} approach), rather 
than {\em collectively}, which would have involved the solution
of a $4\times4$ system at each grid point.

We should also remark that discretizations of hyperbolic systems of 
equations are notorious for being susceptible to numerical 
instability~\cite{gustafsson_et_al,richtmyer_morton}.
However, the implicit Crank-Nicholson discretization 
scheme is well known for being an
\emph{unconditionally} stable scheme---i.e.~stable for arbitrary
values of the Courant factor---for large classes of time 
dependent problems.  Although we have no mathematical
proof of stability for the scheme applied to our hyperbolic equations,
we have not encountered any issues with numerical stability in the 
calculations described in this thesis.

\newpage
\section{Convergence Testing of Finite Difference Solutions} \label{sec:tools}

This section introduces two techniques that we employ to establish that
our finite difference solutions {\em do} converge to continuum solutions 
of the PDEs governing our model.   The first technique, which is now 
almost universally employed in numerical relativity work (as well as in
other fields), is a basic convergence test that, for fixed initial data,
examines the behaviour of the finite difference solution as the discretization
scale, $h$, is varied.  In our case, this is done by defining a quantity,
which we call the convergence factor, $Q^h(t)$, and which is easily 
computed from a series of calculations in which successive discretization
scales are typically related by a factor of 2.  When the empirically measured 
$Q^h(t)$ behaves in the expected fashion, we establish that the finite 
difference solution is converging to {\em some} continuum solution, but there
is then still the possibility that it may not actually be a solution of 
the original PDEs.  This could easily happen if the overall finite difference 
approximation was {\em not} consistent with the model PDEs, but {\em was}
consistent with some other set of PDEs (an omission or accidental
modification of some term in any of the original PDEs would generally lead 
to precisely such a situation).   We thus use a second technique, which is 
perhaps not so commonly used in our field, which is termed independent 
residual evaluation, wherein we essentially directly verify that our 
discrete solutions {\em do} approach the desired continuum solutions as 
$h\to0$.

\subsection{Convergence Factor} \label{subsec:conv_factor}

As suggested above, we perform convergence tests in 
a very straightforward manner: we fix initial data, and then 
calculate numerical solutions using at least three different 
discretization scales (which we will often refer to as {\em levels}
of discretization). Successive levels are characterized by mesh 
spacings that are in a $2:1$ ratio.  Thus, at a minimum---and again
emphasizing that we fix the initial data, as well as all parameters defining 
the numerical solution, {\em except} for $h$---we calculate
$u^h$, $u^{2h}$ and $u^{4h}$, which are finite difference
solutions generated with mesh spacings $h$, $2h$ and $4h$, respectively. 
In addition, and again for the sake of simplicity and convenience, we 
ensure that each grid in the sequence ``aligns'' with the next coarsest 
grid.  That is, the grid points of the level-$h$ mesh are a subset of 
those of the level-$2h$ mesh, which in turn are a subset of those of
the level-$4h$ mesh.  We then define our convergence factor, $Q^h(t)$,
as follows
\beq
Q^h(t) \equiv \fr{\Vert u^{4h}-u^{2h}\Vert}{\Vert u^{2h}-u^h\Vert}\label{eq:qfactor_def} ,
\eeq 
where $\Vert \cdot \Vert$ is any appropriate spatial norm, such as 
$l_2$ ($\Vert \cdot \Vert_2$) or 
$l_\infty$ ($\Vert \cdot \Vert_\infty$),~\footnote{We note that, 
in contrast to what is often done in 
numerical analysis, we define the $l_2$ norm here to include a normalization
by an appropriate power of the number, $n$, of grid 
points---i.e. by $1/\sqrt{n}$.
When dealing with functions defined on different meshes, this is 
the natural and convenient approach: for example, grid functions defined 
on meshes with different resolutions will then tend to have approximately
the same norm.}
\beq
\Vert u^h\Vert_{2} = 
                \fr{1}{\sqrt{n}} \lhb \sum_{i=1}^n |u^h_i|^2 \rhb^{1/2}\, ,
\eeq
\beq
\Vert u^h\Vert_{\infty} = \max_{1 \leq i \leq n} |u^h_i| \, ,
\eeq
and where it is to be understood that the subtraction of 
individual grid function values implicit in the expressions
$u^{2h}-u^h$ and $u^{4h}-u^{h}$ occurs only at the set of 
grid points common to the two meshes.
Now, assuming that 
\begin{enumerate}
\item our FDA is $O(h^2)$ and completely centred, so that the 
      truncation error, $\tau^h$, contains no
      terms proportional to $h^{p_{\rm o}}$, where $p_{\rm o}$ is 
      an odd integer,
\item any finite difference solution, $u^h$ has a Richardson expansion
      of the form~(\ref{eq:rich-exp}),
      
\end{enumerate}
then the discrete solutions $u^h$, $u^{2h}$ and $u^{4h}$ can be 
expanded as
\bea
u^h &=& u + h^2 e_2 + h^{4} e_{4} + \dots \\
u^{2h} &=& u + (2h)^2 e_2 + (2h)^{4} e_{4} + \dots \\
u^{4h} &=& u + (4h)^2 e_2 + (4h)^{4} e_{4} + \dots 
\eea
Substitution of the above expressions into~(\ref{eq:qfactor_def}) yields
\bea
	Q^h(t) &=& 
		\frac{\left\Vert \lb u + (4h)^2 e_2 + (4h)^{4} e_{4} + O(h^6)\rb - 
             \lb  u + (2h)^2 e_2 + (2h)^{4} e_{4} + O(h^6)\rb\right\Vert}
            {\left\Vert \lb u + (2h)^2 e_2 + (2h)^{4} e_{4} + O(h^6)\rb - 
             \lb u + h^2 e_2 + h^{4} e_{4} + O(h^6)\rb\right\Vert} 
       \\ \nonumber
          &\approx&
		\frac{\left\Vert 12 h^2 e_2 + 240 h^4 e_4 \right\Vert}
		     {\left\Vert 3 h^2 e_2 + 15 h^4 e_4 \right\Vert} \, .
\eea
Thus, as the mesh spacing tends to 0, we have
\beq
\lim_{h\to0} Q^h(t) 
	= \lim_{h\to0} 
		\frac{\left\Vert 12 h^2 e_2 + 240 h^4 e_4 \right\Vert}
			  {\left\Vert 3 h^2 e_2 + 15 h^4 e_4 \right\Vert}
   = \lim_{h\to0} \frac{12 h^2 \left\Vert e_2 \right\Vert}
			  {3 h^2 \left\Vert e_2 \right\Vert} = 4 \, .
\label{eq:qfactor2} 
\eeq

Note that in deriving~(\ref{eq:qfactor2}) we have assumed that the terms 
in the (asymptotic) Richardson expansion are monotonically decreasing.  This 
will usually be the case when the mesh scale, $h$, is significantly smaller
than the typical scale over which $u$ varies.  Conversely, if $h$ is 
of the order of the variation scale of $u$, one cannot reasonably expect 
that $Q^h(t)$ will be close to 4.

Finally, as we have already mentioned, one needs a {\em minimum} of 3 levels 
of discretization to compute $Q^h(t)$.  In practice, it is best to use 
as many levels as possible in testing for convergence.~\footnote{Available 
computational resources tend to be the limiting factor in this respect: 
for 3+1 dimensional computations such as that performed in this thesis, a 
calculation with mesh size $h/2$ requires about $2^4=16$ as much computer 
time as one with mesh size $h$.}  As finer and finer discretization scales 
are used, we expect $Q^h(t)$ to get closer and closer to the constant 4, and 
deviations from this anticipated behaviour can be used as a powerful 
diagnostic for detecting subtle mistakes in the FDA (e.g.~where certain 
terms that are relatively small have only been discretized to $O(h)$ 
accuracy, or perhaps not even consistently with the PDEs).

\subsection{Independent Residual Evaluation} \label{subsec:IRV}

As mentioned previously, although a convergence test of the form detailed 
above can provide strong evidence that a discrete solution is 
converging to {\em something} as $h\to0$, it does {\em not} directly 
establish that the limiting solution satisfies the original set 
of PDEs.  We therefore introduce a second technique, known as independent
residual evaluation~\footnote{This technique was first introduced by Choptuik in
his studies of critical phenomena in gravitational collapse~\cite{choptuik:1994},
where its use was crucial to the validation of new and unexpected results.
}, that aims to remedy this shortcoming. 
Importantly,
the method can be used to test the correctness of an arbitrary FDA of 
an arbitrary set of PDEs, as well as the actual computer code that 
solves the FDA.  We feel that is difficult to overemphasize the 
value of using this technique for testing the {\em complete} process
of discretizing a set of PDEs, and then solving for the discrete 
solutions.
Particularly for complicated PDEs in multiple 
dimensions, the FDAs used, and the algorithms needed to solve them
can be extremely complex: there are thus many places where mistakes 
can be made, and it will not always be obvious from the numerical
solutions that something has gone awry.  Independent residual evaluation
has the potential to detect virtually any and all errors that have been 
made, assuming only the following:
\begin{enumerate}
\item The PDEs that we start from are correct.
\item The discrete solutions, $u^h$, that are computed have Richardson
      expansions.
\end{enumerate}
We also note that one of the greatest advantages of independent residual
evaluation is that it does {\em not} require the existence of any
exact or previously-computed solution of the PDEs.  

We illustrate the technique by again considering a general set of 
PDEs written in the form introduced 
in~Sec.~\ref{sec:basic-FDA}:~\footnote{As in 
Sec.~\ref{sec:basic-FDA}, and for clarity of exposition, we suppress
the $(t,x,y,z)$ dependence of continuum functions
($u$, $f$, $\tau_2$ etc.) as well as their  discrete counterparts ($u^h$,
$f^h$ etc.).
In addition, we will proceed by assuming that 
$L$ and any finite difference approximations of $L$, as well
as the operators $E_2$  and ${\hat E}_2$ appearing in (\ref{eq:Lhuh-exp}) and 
(\ref{eq:Lhatuh-exp}), respectively, are linear;
however, the technique is equally applicable to nonlinear equations.}
\beq
L u - f = 0 \, .
\eeq
The continuum equation is discretized as 
\beq
\label{Lhuh}
L^h u^h - f^h =0 \, ,
\eeq
with the presumption being that the discretization is correct.  Thus,
assuming that $L^h$ is a second order approximation of $L$, we will
have
\beq
\label{eq:Lhuh-exp}
L^h = L + h^2 E_2 + O(h^4) \, ,
\eeq
where $E_2$ is a differential operator, whose form can be explicitly 
computed, and that will be of higher degree
than $L$ (e.g.~if $L$ involves second derivatives, then $E_2$ will generally 
involve fourth derivatives, given that the FDA is $O(h^2)$).
We now assume that we have computed numerically an approximate discrete
solution, ${\tilde u}^h$, and have shown that it is converging at 
second order 
to {\em some} continuum function, ${\bar u}$ (e.g.~so that $Q^h(t) \approx
4$).  We can therefore be confident that ${\tilde u}^h$ has a Richardson 
expansion of the form
\beq
\tilde{u}^h = \bar{u} + h^2 {\bar e}_2 + O(h^4) \, .
\eeq
Furthermore, by construction (i.e. by virtue of the actual numerical
calculation), we assert that the residual, $r^h$, associated with
$\tilde{u}^h$ has a magnitude less than some 
convergence tolerance, $\ep$, where $\ep$ can generally be much smaller 
than the typical magnitude of the truncation error, $\tau^h$.  That
is, we have
\beq
\left\Vert L^h \tilde{u}^h - f^h \right\Vert = \left\Vert r^h\right\Vert < \ep 
\ll \Vert \tau_h \Vert\, .
\eeq
 
Relative to the discretization~(\ref{Lhuh}), then, we have established 
that we have a convergent solution, which limits to ${\bar u}$.
What we have {\em not} established, however, is whether ${\bar u}\equiv u$.
In order to do so, we need to simultaneously establish that $L^h$ {\em is}
a consistent approximation of $L$, and that our implementation 
correctly solves the algebraic equations~(\ref{Lhuh}).

We thus consider an independent (distinct) discretization of the PDE:
\beq
\label{eq:indres}
\hat{L}^h \hat{u}^h - f^h = 0\, .
\eeq
This new discrete operator $\hat{L}^h$ can be expanded in the same manner
as $L^h$ was:
\beq 
\label{eq:Lhatuh-exp}
\hat{L}^h = L + h^2 {\hat E}_2 + O(h^4)  \, 
\eeq 
where ${\hat E}_2$ is another higher-order differential operator that 
will be {\em not} be the same as the operator $E_2$ appearing in~(\ref{eq:Lhuh-exp}).
Note that we assume here that ${\hat L}^h$ is also an $O(h^2)$ approximation
to $L$, but this is not essential (e.g. an $O(h)$ approximation could also
be adopted).
Given this second discretization, the process of independent residual 
evaluation consists simply of applying the left hand side of~(\ref{eq:indres})
to our putative numerical solution, ${\tilde u}^h$.
Defining $I^h$ to be the independent residual we have
\begin{eqnarray}
I^h\equiv \hat{L}^h \tilde{u}^h - f^h &=& \lhb L + h^2 {\hat E}_2 + O(h^4)\rhb 
                      \lhb \bar{u} + h^2 {\bar e}_2 + O(h^4) \rhb - f^h \\
 &=& L \bar{u} - f^h + h^2 \lb {\hat E}_2 {\bar u} + L {\bar e}_2 \rb +
    O(h^4) \\
 &=& L \bar{u} - f^h + O(h^2) \, .
\end{eqnarray}
Here we have again used the (inessential) assumption that 
$L$ and $E_2$, are both linear.

Now, if the computed continuum solution, ${\bar u}$ is {\em not} a solution of the 
original PDE we will generically have $\bar{u}(t,x,y,z) = u(t,x,y,z) + 
e_0(t,x,y,z)$, where $e_0=O(1)$, and as $h\rightarrow 0$ we will find
\beq I^h\equiv \hat{L}^h \tilde{u}^h - f^h = L u - f^h + L e_0 + O(h^2) = 
L e_0 + O(h^2) \, .
\eeq
Thus, unless $L e_0 \equiv 0$, which is extremely improbable, then as $h\to0$,
we will find that $I^h$ will converge to some non-zero {\em function},  
given by $L e_0$.  In other words we will have $I^h = O(1)$.   Conversely,
if ${\bar u}$ and $u$ {\em are} the same function, we will have
\beq
\label{eq:indres1}
I^h\equiv \hat{L}^h \tilde{u}^h - f^h = L u - f^h + O(h^2) = O(h^2) ,
\eeq
with the crucial observation being that $I^h \to 0$ as $h\to0$, i.e. that 
$I^h$ really is ``residual'' quantity with respect to both the independent
discretization, and the continuum PDE.  If we compute $I^h$ in a 
typical sequence of calculations aimed establishing convergence (i.e. with 
all problem parameters, save $h$, fixed), and observe behaviour as given
by~(\ref{eq:indres1}) then we have provided very strong evidence 
that the approximate discrete solution $\tilde{u}^h$ {\em is} converging 
to the true continuum solution $u$ of the PDE.

In order for the technique of independent residual evaluation to be 
effective, it is vital that the second discretization, ${\hat L}^h$,
{\em is} consistent with the original PDE operator $L$.  Otherwise, 
measurement of an $I^h$ which is $O(1)$ could signal an inconsistency 
in ${\hat L}^h$, instead of some problem with the principal discretization,
$L^h$, or in the solution of the algebraic equations that result
from that discretization.  In
this regard, one should note the following.  First, there is no need 
to {\em solve} the system~(\ref{eq:indres}) for ${\hat u}^h$: rather, the 
independent discretization ${\hat L}^h$ is simply {\em applied} to 
the computed solution ${\tilde u}^h$.  Intimately related to this 
observation is the fact that for the case of time dependent systems, 
one does not need to worry about the stability of ${\hat L}^h$.  In
addition, as already noted, there is no need for ${\hat L}^h$
to be of the same order of accuracy as $L^h$. For example, if $L^h$ is 
$O(h^2)$ accurate, but ${\hat L}^h$ is only $O(h)$, then if ${\tilde u}^h$ 
is converging to the true continuum solution, $u$, we will find that $I^h$ is
$O(h)$ (i.e. still a residual quantity). However, if convergence is to some 
function ${\bar u} \ne u$, then $I^h$ will still be $O(1)$.

All of this means that one has lot of latitude in how ${\hat L}^h$ is 
constructed, and, perhaps more importantly, one can easily use 
symbolic computing software to generate ${\hat L}$ from a high-level 
description of the PDEs that is relatively easy to check for 
correctness.  This significantly decreases the probability that a
$I^h$ that is measured to be $O(1)$ is due to a mistake in the derivation or 
implementation of ${\hat L}$.

For example, in the construction of the independent residual
operator for our set of equations,  we used $O(h)$ 
forward difference approximations for time derivatives, 
and $O(h^2)$ forward difference approximations for the spatial 
derivatives.\footnote{For grid points close to the outer boundary 
of the computational domain where the forward difference formulae 
could not be applied, we used $O(h^2)$ backwards difference expressions.}
This is to be contrasted to the $O(h^2)$ approximations 
used for all derivatives in the construction of the basic difference 
scheme itself (Crank-Nicholson for the time derivatives, standard 
$O(h^2)$ centred approximations for the spatial derivatives).
In addition, the Maple procedure described in~App.~\ref{ap:pdefda} was used 
to generate the actual code to evaluate $I^h$.
The results from our application of independent residual 
evaluation to series of 
calculations for several distinct initial data sets are described 
in the next chapter, along with the results from measurement of the 
basic convergence factor $Q^h(t)$ discussed in Sec~\ref{subsec:conv_factor}.

\newpage
\section{Multigrid Techniques} \label{sec:MG}

\subsection{Introduction} \label{sec:MG-intro}

As stated in the previous section, iterative techniques
such as point-wise Newton-Gauss-Seidel (NGS), provide
efficient methods for solving the systems of nonlinear algebraic 
equations that result from our time-implicit discretization of the 
hyperbolic PDEs of our model (those that govern the scalar 
field).  Specifically, we can generally expect convergence to the 
level of intrinsic discretization error
using a number of sweeps through the mesh which is independent of the 
mesh scale, $h$. Thus, in the absence of any elliptic PDEs, the operation count 
for solving the hyperbolics would be linear in the number of 
discrete unknowns.~\footnote{Indeed, as noted by 
Teukolsky~\cite{Teukolsky:1999rm}, for many purely
hyperbolic systems of PDEs, one can expect the so-called ``iterative 
Crank-Nicholson'' method to converge to the level of the solution
error in precisely
2 iterations.  However, in our case, due to the additional coupling with 
the elliptic equations, this observation does not apply.}

However, as is well known, relaxation methods---even when they are 
accelerated using the successive-overrelaxation (SOR) technique---are 
{\em not} very efficient for solving the systems that result from 
finite-difference discretization of elliptic PDEs.  In particular, the amount 
of computational work {\em per unknown} needed to solve such systems
generally increases as $h\to0$. 

For example, consider what one might call the Poisson equation in one spatial 
dimension:
\beq
\label{eq:1dpoisson}
u(x)_{xx}= f(x) \, ,
\eeq
and, neglecting boundary conditions, discretize this equation on a uniform 
mesh using the usual centred, $O(h^2)$ FDA of $u_{xx}$:
\beq
\fr{u_{i-1}-2u_i+u_{i+1}}{h^2} - f_i = 0 \, .
\eeq
As described in~App.~\ref{ap:ngs}, assuming that the NGS iteration 
visits the grid points in so-called lexicographic order (i.e.~$i=2, 3, \cdots
n_x-1$), the $i$-th component of the {\em running residual} vector is 
given by
\beq
[{\bf r}^{(k)}_i]_i = \fr{\tilde{u}^{(k)}_{i-1}-2\tilde{u}^{(k-1)}_i
+\tilde{u}^{(k-1)}_{i+1}}{h^2} - f_i \, .
\eeq
Each relaxation sweep (iteration) is supposed to bring the approximate solution 
$\tilde{u}^{(k)}_i$ closer to the exact discrete solution $u_i$. The 
convergence of this process can also be viewed in terms of driving the
running residuals to 0. 
In fact, in the analysis of the convergence of relaxation 
methods~\cite{Varga:MIA},
the effect of the iteration on the running 
residual vector is often described in terms of the action
of the {\em residual amplification matrix}, 
$\mathbf{A}$. Assuming a linear set of discrete equations~\footnote{Again, 
the analysis can be extended to the nonlinear case, when a linearization 
method such as Newton iteration is used.} this matrix
maps the residual vector at iteration $k$, $\mathbf{r}^{(k)}$, to the 
corresponding vector at iteration $k+1$, $\mathbf{r}^{(k+1)}$:
\beq
\mathbf{r}^{(k+1)} = \mathbf{A} \mathbf{r}^{(k)} .
\eeq
Clearly, $\mathbf{A}$ must be a contraction map in order for the iteration
to converge, and, ideally, the spectral radius, $\rho$,  of the 
amplification matrix would always be bounded away from unity---i.e. $\rho(\mathbf{A}) < 1$---to
{\em ensure} rapid convergence.
Unfortunately, for the case of NGS applied to the simple, but representative,
set of equations defined by~(\ref{eq:1dpoisson}), this is not the case.  Indeed, it can be shown that the spectral radius 
of the amplification matrix, $\mathbf{A}_{\rm NGS}$ satisfies
\beq
\label{eq:rhongs}
\lim_{h\to0} \rho(\mathbf{A}_{\rm NGS}) = 1 - O(h^2) \, .
\eeq
Operationally, this means that one must perform $O(h^{-2}) = O(n_x^2)$ 
relaxation sweeps in order to achieve convergence, so that, for 1D problems,
the overall operation count is $O(n_x^3)$, rather than the optimal $O(n_x)$.
Furthermore, the type of behaviour given by~(\ref{eq:rhongs}) is seen 
when NGS is applied to virtually any finite differenced elliptic system,
in any number of spatial dimensions.  Although the use of overrelaxation (SOR) 
can significantly improve convergence, one still generally has 
\beq
\label{eq:rhosor}
\lim_{h\to0} \rho(\mathbf{A}_{\rm SOR}) = 1 - O(h) \, ,
\eeq
and the work per unknown needed to achieve convergence still increases as 
the mesh is refined.~\footnote{Furthermore, one can only attain~(\ref{eq:rhosor})
when the choice of relaxation parameter is precisely optimal, and determining
an optimal value is non-trivial in practice, particularly for coupled nonlinear elliptic 
systems such as ours.}
In short, especially for cases such as ours, where an elliptic system must 
be solved at each discrete time step, relaxation techniques do not provide
a viable route for solving discretized elliptic equations.

On the other hand---and again using the simple toy problem~(\ref{eq:1dpoisson})
for illustrative purposes---by expanding the residual vector in terms of the 
eigenvectors of the amplification matrix, it is possible to show that the 
asymptotic convergence rate~(\ref{eq:rhongs}) is dominated by the 
eigenvectors with wavelengths that are long compared to the mesh scale, 
$h$. That is, each relaxation sweep damps long-wavelength components of the 
running residual by factors given by the associated 
eigenvalues, $\sigma^{\rm long}_i$, where $\sigma^{\rm long}_i = 1 - O(h^2)$.
Conversely, for short-wavelength modes---which we will define as 
modes having wavelengths, $\lambda_i$, in the range 
$2h \le \lambda_i \le 4h$---one finds that the corresponding eigenvalues are 
bounded away from unity, and, importantly, {\em are independent of $h$}.
In other words, as the amplification matrix 
is repeatedly applied to the residual vector, 
the short wavelength (high frequency) components of the residual
are rapidly annihilated, while the long wavelength (low 
frequency) components are very slowly damped. 
In short, relaxation methods such 
as Gauss-Seidel tend to be very good {\em smoothers} when applied to 
finite difference discretizations of elliptic PDEs, and that this smoothing 
property applies both to the (running) residual vector, $\mathbf r^{(k)}$ 
and the 
deviation between the approximate solution after the $k$-th relaxation sweep, 
$\mathbf {\tilde u}^{k}$, and the exact solution, $\mathbf{\tilde u}^h$ of the 
discrete equations.

Fortunately there is an extremely efficient methodology for solving 
finite difference approximations of elliptic equations. 
This is the \emph{multigrid} technique~\cite{brandt:1977,trottenberg}, which 
was largely developed by A.~Brandt in the late 1970's and early 1980's.
In terms of computational complexity, the most striking feature of multigrid
is that, in many cases, it is able to provide solutions of the discrete 
equations with $O(N)$ computational work and $O(N)$ storage, where $N$ is the 
total number of discrete unknowns.~\footnote{The multigrid method is itself iterative, 
so by solving the discrete equations we again mean ``to the level of the intrinsic 
solution error, $u^h - u$'', where $u$ is the continuum solution and $u^h$ is the 
exact solution of the discrete system.} Importantly, this computationally-optimal
performance can be achieved for quite general nonlinear systems of elliptic PDEs,
such as the one encountered in our model.  It should be noted that from the fact 
that multigrid can provide solutions with $O(N)$ work one can immediately
deduce that its convergence rate in such instances must be $h$-independent. 

The classic multigrid method (i.e.~the approach due to Brandt) 
builds on two fundamental observations. 
The first of these we have just made: relaxation methods, such 
as (Newton)-Gauss-Seidel, can be used to efficiently {\em smooth} both
the residuals and solution errors on any particular grid.  
In multigrid then, one uses relaxation not to {\em solve} the 
discrete set of equations, but only to smooth the residuals and solution 
errors.~\footnote{Again, when we speak of ``solution errors'' in the 
context of an iterative method such as multigrid, we generally mean the errors
in the approximate solution, ${\tilde u}^h$, 
at any stage of the iteration, 
relative to the exact solution, $u^h$, of the difference equations.  
However, assuming that the true solution error, $u^h - u$ is smooth, 
smoothing ${\tilde u}^h - u^h$
is clearly equivalent to smoothing ${\tilde u}^h - u$.}

The second observation is that once the discrete elliptic problem has 
been smoothed, it can be well represented on an coarser grid, having,
for example, a discretization scale $2h$.  Especially for multidimensional
equations, the work need to solve the coarse-grid problem will be a fraction
of that required for the fine-grid solution.  Moreover, we can then 
apply these ideas recursively: relaxation on the $2h$ grid quickly annihilates
modes in the residuals and errors that are high-frequency (short-wavelength)
with respect to $2h$, and then we can pose a version of the problem on
an even coarser grid having mesh spacing, $4h$, and so on.  Eventually, 
this smoothing and coarsening process on grids with ever increasing mesh
spacings leads to a grid with so few points that {\em solution} of the 
discrete equations using relaxation (or even a direct technique) requires
a negligible amount of computational work.  The process of working on 
coarser and coarser grids is then reversed: smooth corrections to grid 
functions on finer grids are interpolated from the coarse grids.  This
process of interpolation introduces new high frequency error components 
in the fine-grid functions, but these can be effectively damped with 
a few more relaxation sweeps.  Once the algorithm returns to the original, 
finest mesh, the relaxation sweeps performed after the interpolation
of the correction from the second-finest mesh typically results in a 
solution in which the error has been reduced by a substantial amount. 

In the next subsection we will discuss these basic ideas in somewhat
more detail for the case of a particular multigrid method, known
as the Full Approximation Storage (FAS) scheme,  which 
is suitable for the treatment of nonlinear problems.

\subsection{The FAS Algorithm}

Again adopting a notation in which the dependence of functions on the relevant 
set of independent variables (e.g. $(x,y,z)$) is implicit, we write a general 
nonlinear system of elliptic PDEs as 
\beq \label{eq:Nuf}
N[u] = f \, .
\eeq
Here $N$ denotes the set of nonlinear differential operators acting 
on the solution vector, $u$, while $f$ is a vector of source functions which is
independent of $u$ or any of its spatial derivatives.
We then consider some finite difference approximation of~(\ref{eq:Nuf}) 
which, as usual, is characterized by a {\em single} discretization scale
$h$.   This yields a nonlinear set of algebraic equations that we 
write as 
\beq \label{eq:Nuf_FDA}
N^h[u^h] = f^h \, .
\eeq
The solution of~(\ref{eq:Nuf_FDA}) is expected to be computed through some 
iterative process. Each 
step of the iteration defines the residual vector, $r^h$, corresponding to the 
current approximation, $\tilde{u}^h$, of the discrete solution, $u^h$:
\beq \label{eq:res_u}
r^h \equiv N^h[\tilde{u}^h] - f^h \, .
\eeq
The goal of the iterative process is to drive the residual to zero or,
equivalently, to drive the solution error, $v^h$, to zero, where $v^h$ is 
defined by
\beq \label{eq:exact_sol}
u^h = \tilde{u}^h + v^h \, .
\eeq
 
As mentioned in the last section, relaxation methods such as Newton-Gauss-Seidel
tend to be excellent smoothers. The FAS multigrid method takes advantage of 
this fact to (implicitly) smooth the correction $v^h$ on each member of 
a {\em hierarchy} of grids. Each grid in the hierarchy is labelled by a parameter $l$ 
(the {\em level} parameter), 
where $l=1$ is the coarsest level and $l=l_{\rm max}$ is the finest.
From considerations of both total computational cost, as well as ease 
of implementation, the grids in the hierarchy usually satisfy $h_{l+1} = h_l/2$,
and we have adopted this choice in our current work.
Starting on the finest mesh, i.e.~on level $l_{\rm max}$, we perform a 
few relaxation sweeps using a method such as NGS.~\footnote{We note that,
as discussed in Sec.~\ref{sec:space_compact}, {\em point-wise} Newton-Gauss-Seidel is not always 
an effective smoother for elliptic discretizations, but it is for our system,
provided that we work in the standard (i.e. non-compactified) Cartesian 
coordinates.}
Provided that these sweeps {\em do} smooth both $r^h$ and $v^h$ on the 
fine grid, we can then proceed to define a coarse grid problem. 
This is done by setting up an appropriate equation for the correction (solution
error), $v^h$, on the coarse grid.

In order to define this equation, first note that if $N^h$ were {\em linear},
we could apply it to both sides of~(\ref{eq:exact_sol}) to get
\beq
N^h[u^h] = N^h[\tilde{u}^h + v^h]= N^h[\tilde{u}^h]+N^h[v^h]=r^h+f^h+N^h[v^h] \, .
\eeq 
Using~(\ref{eq:Nuf_FDA}) this could be further simplified to 
\beq \label{eq:Nvr}
N^h[v^h]=-r^h \, .
\eeq
This last equation actually forms the basis of a multigrid algorithm 
known as the \emph{linear correction scheme} (LCS).
However, since $N^h$ is a nonlinear operator, we cannot proceed
along this route.

Nonetheless, Eq.~(\ref{eq:Nvr}) 
{\em does} suggest a similar treatment for the nonlinear case. Subtracting the 
definition of the residual, Eq.~(\ref{eq:res_u}), from the basic difference 
equation~(\ref{eq:Nuf_FDA}) yields
\beq \label{eq:FAS}
N^h[u^h]- N^h[\tilde{u}^h] = N^h[\tilde{u}^h+v^h]- N^h[\tilde{u}^h] = -r^h \, .
\eeq
Thus, rather than having an explicit equation for the correction $v^h$,
(\ref{eq:FAS}) is to be viewed as an equation 
for the ``full approximation'', $u^h=\tilde{u}^h+v^h$, and hence the 
name ``Full Approximation Scheme''.

From the smoothing assumption, all terms appearing in~(\ref{eq:FAS})
are smooth on the scale, $h$, of the finer grid.  One can then sensibly pose a 
coarse grid form of~(\ref{eq:FAS}) as follows:
\beq \label{eq:CCG}
N^{2h}[u^{2h}]- N^{2h}[I^{2h}_h \tilde{u}^h] = - I^{2h}_h r^h ,
\eeq
Here, $u^{2h}$ is the unknown which is to be computed on the coarse grid (note that 
it does {\em not}, in general, satisfy $N^{2h} u^{2h} = f^{2h}$), and 
$I^{2h}_h$---known as a {\em restriction} operator---transfers a fine grid function
to the coarse grid.  Once $u^{2h}$ has been determined by solving~(\ref{eq:CCG}), 
the fine grid unknown, $u^h$,  is updated using
\beq \label{eq:uupdate}
\tilde{u}^h = \tilde{u}^h + I^h_{2h} \lb u^{2h} - I^{2h}_h \tilde{u}^h \rb .
\eeq
Here, $I^h_{2h}$---known as a {\em prolongation} operator transfers a coarse 
grid function to the fine grid.  In practice, $I^h_{2h}$ generally performs 
polynomial interpolation of an order that can depend on: the differential order
of the elliptic system, the order of accuracy of the FDA, and 
the specific smoother being used.
As Brandt~\cite{brandt:1977} emphasizes, 
Eq.~(\ref{eq:uupdate}) is to be preferred over the more obvious
\beq \label{eq:uupdate-naive}
\tilde{u}^h = I^h_{2h} u^{2h} 
\eeq
since the former retains (useful) high frequency information already computed 
in $u^h$, whereas the latter does not.

Eqs.~(\ref{eq:CCG}) and (\ref{eq:uupdate}) constitute the core of 
the FAS algorithm.  Our quick derivation of the FAS scheme
also corresponds to a specific point of view, wherein the multigrid method
is seen as a solver that uses a hierarchy of coarser grids to accelerate the 
convergence of error components which have long wavelengths on the fine grid.

Again as stressed by Brandt~\cite{brandt:1977},  there is a useful ``dual''
interpretation of the FAS algorithm in which the fine grids are used 
to provide correction terms to coarse grid systems, essentially allowing 
unknowns on coarse grids to be determined to the same accuracy---relative 
to the continuum solution---as the unknowns on the finest grid.  It is 
instructive to quickly work through the alternate derivation of~(\ref{eq:CCG})
from this vantage point, and to do so we must introduce the concept of 
{\em relative truncation error}.

Recall from Sec.\ref{sec:basic-FDA}, Eq.~(\ref{eq:truncation}), 
that the truncation error,
$\tau^h$, of a finite difference scheme is defined in terms of the action of 
the discrete operator on the continuum solution:
\beq \label{eq:tau}
\tau^h \equiv N^h[u] - f^h \, .
\eeq
For the purposes of the current development, it is convenient to use 
the FDA (\ref{eq:Nuf_FDA}) in the form $f^h = N^h[u^h]$ to rewrite 
the above equation as 
\beq \label{eq:tau1}
\tau^h =  N^h[u] - N^h[u^h] \, .
\eeq
Now, if it was possible to know the exact value of the truncation error 
in advance, then by adding it to the right hand side of (\ref{eq:Nuf_FDA})
we would have an equation
\beq
N^h[u^h]=f^h + \tau^h ,
	\eeq
whose solution, $u^h$, would be identical to 
the restriction of the continuum solution, $u$, of the PDE~(\ref{eq:Nuf}) 
to the mesh points.
Unfortunately {\em a priori} knowledge of $\tau^h$ is, of course, equivalent 
to {\em a priori} knowledge of $u$, so at first glance, this observation does not 
seem very useful.

However, suppose that it is possible to compute some 
approximation, $\tilde{\tau}^h$ of $\tau^h$.  Then, by adding $\tilde{\tau}^h$
to the right hand side of Eq.~(\ref{eq:Nuf_FDA}), we would get an equation
\beq
N^h[u^h_{\star}]=f^h + \tilde{\tau}^h ,
\eeq
for a new grid function, $u^h_\star$, which should be more accurate than
the solution $u^h$ of~(\ref{eq:Nuf_FDA}).  That is, we should find
$\Vert u^h_{\star} - u\Vert < \Vert u^h - u\Vert$, where $\Vert \cdot \Vert$ denotes some 
norm.

In analogy to the definition of the truncation error, $\tau^h$, we now 
define the relative truncation error, $\tau^{2h}_h$, which involves 
quantities defined on two adjacent levels within the multigrid
hierarchy, having discretization scales $h$ and $2h$ respectively:
\beq \label{eq:tau_rel}
\tau^{2h}_h \equiv N^{2h}[I^{2h}_h u^h] - I^{2h}_h \lb N^h[u^h] \rb .
\eeq
Again, $I^{2h}_h$ is a fine-to-coarse transfer operator (restriction operator).
Once more using Eq.~(\ref{eq:Nuf_FDA}) in the form $f^h = N^h[u^h]$, and assuming
that $I^{2h}_h f^h = f^{2h}$, this definition can be rewritten as
\beq \label{eq:CCG_tau}
N^{2h}[u^{2h}] = f^{2h} + \tau^{2h}_h .
\eeq
Thus, completely parallelling our previous interpretation of $\tau^h$, 
the relative truncation error, $\tau^{2h}_h$, can be viewed
as the correction that must be added to the source term of the coarse 
grid difference equations in order that the 
coarse grid solution actually coincide with the (restricted) 
fine grid solution. Once more, we are unable to compute $\tau^{2h}_h$ precisely
unless we have the exact solutions, $u^h$ and $u^{2h}$, of the two discrete systems
in hand.  However, during the multigrid solution process, we can certainly calculate
an approximation, ${\tilde \tau}^{2h}_h$, of $\tau^{2h}_h$ using the current
estimate, ${\tilde u}^h$, of the fine grid unknown:
\beq \label{eq:tautilde_rel}
{\tilde \tau}^{2h}_h \equiv N^{2h}[I^{2h}_h {\tilde u}^h] - I^{2h}_h \lb N^h[{\tilde u}^h] \rb .
\eeq

Then, replacing $\tau^{2h}_h$ with ${\tilde \tau}^{2h}_h$ in~(\ref{eq:CCG_tau}), we have
derived the ``dual'' equation for the coarse grid unknown, $u^{2h}$ (i.e. the full
approximation): 
\beq \label{eq:CCG_tautilde}
N^{2h}[u^{2h}] = f^{2h} + {\tilde \tau}^{2h}_h .
\eeq
 
Now from~(\ref{eq:CCG}), and again assuming that $I^{2h}_h f^h = f^{2h}$ and
that $I^{2h}_h$ is linear (which in practice it invariably is), we have
\begin{eqnarray}
N^{2h}[u^{2h}] &=&  N^{2h}[I^{2h}_h \tilde{u}^h] - I^{2h}_h r^h  \\
               &=& N^{2h}[I^{2h}_h \tilde{u}^h] - 
                     I^{2h}_h \left( N^h {\tilde u}^h - f^h\right) \\
               &=& f^{2h} + {\tilde \tau}^{2h}_h
\end{eqnarray}
which is precisely Eq.~(\ref{eq:CCG_tautilde}).  Thus the two viewpoints 
lead to the same coarse grid correction equations, and therefore are indeed 
equivalent.

As summarized in~Sec.~\ref{sec:MG-intro}, the FAS coarse grid correction
equations are applied recursively, with relaxation sweeps on level $l$
being followed by the initiation of a coarse grid correction on level $l-1$.
When the coarsest grid ($l=1$) is reached, $u^{h_1}$ can generally be 
computed very inexpensively using relaxation or perhaps some other 
method (such as a direct solution of the difference equations, 
using a global Newton iteration). 
Once $u^{h_1}$ has been calculated, another sequence---consisting of fine grid 
function updates (using prolongation) followed by additional
smoothing sweeps---is performed for $l=2, ... , l_{\rm max}$ 
This entire process of ``working down'' the hierarchy from the finest to 
coarsest grid, then ``back up'' to the finest grid, is known as a $V$-cycle,
due to the $V$-shape that results from a standard pictorial representation
of the algorithm in which the vertical direction encodes the discretization
level (with $h$ increasing downwards), while horizontal displacement represents
successive stages of the procedure.  A pseudo-code version of the basic 
FAS algorithm described above, and which was used in this thesis, is shown
in~Fig.~\ref{MG_FAS}.

To conclude this chapter we reiterate that the elliptic PDEs appearing 
in our model, Eqs.~\ref{eq:phidot_adm_cart}--\ref{eq:shift_cartesian_z} 
were discretized using a centred $O(h^2)$ FDA
(see App.~\ref{ap:stencil} for the specific difference operators used).
In addition, we used \emph{red-black} ordering for the NGS relaxation
sweeps, 
\emph{full-weighted} restriction for $I^{2h}_{h}$  and \emph{trilinear} 
interpolation for $I^{h}_{2h}$.  The interested reader is directed 
to~\cite{brandt:1977,trottenberg} for further discussion of these 
technicalities.  Finally, although we {\em did} experiment
with collective relaxation, due to an original concern that 
the mixed second derivatives in some of the equations might be problematic,
we ultimately found that the decoupled approach produced faster
running code.

\begin{figure}
\centerline{\fbox{\parbox{12.5cm}{
\input{MG_FAS.txt}
}}}
\caption[A Pseudo-Code Representation of the FAS $V$-Cycle Multigrid Algorithm.]
{A pseudo-code representation of the FAS $V$-cycle multigrid algorithm used in this
thesis.}
\label{MG_FAS}
\end{figure}

 \resetcounters
\def\calH{{\mathcal H}}
\def\calT{{\mathcal T}}
\def\calR{{\mathcal R}}
\def\calG{{\mathcal G}}
\def\H#1#2{{\mathcal H}^{#1}{}_{#2}}
\def\shape{{\tt shape}}
\def\bbox{{\tt bbox}}
\def\vnfeIV{{\tt vnfe4}}
\def\vnfeV{{\tt vnfe5}}

\chap{Code Validation and Results} \lab{results}

This chapter presents some results from the numerical solution of 
the system of PDEs derived in Chap.~\ref{formalism} and summarized in
Sec.~\ref{sec:EOM_overview}. The results discussed here are restricted
to the case of boson stars having a potential that only includes 
a mass term, so that $U(|\phi|^2)=m^2|\phi|^2$.   \footnote{As mentioned 
in Chap.~\ref{formalism}, these configurations are sometimes referred 
to as {\em mini boson stars}, since for any plausible particle mass, the 
total gravitating mass of the star is very small. Consequently, such
objects are thought to be highly unlikely to be of any astrophysical
significance, especially as dark matter candidates.}
As discussed in Chap.~\ref{initialdata}, the mass parameter $m$ can be 
chosen arbitrarily as part of our overall specification of a system
of units, and we have performed all of our calculations with $m=1$.

Sec.~\ref{sec:num_code} provides some details concerning our numerical 
code as well as the nature of the various initial data configurations 
that we have considered.  This is followed by three sections devoted 
to validation and error analysis of the code.  Specifically, we 
report the results of convergence tests and independent residual 
evaluation, as defined and discussed in Ch.~\ref{numerical}, 
for representative calculations involving:
\begin{enumerate}
\item A generic initial configuration of the scalar field (Sec.~\ref{sec:gen_ID}).
\item Single boson stars at rest in the computational domain (Sec.~\ref{sec:SS_ID}).
\item A single boson star moving through the computational domain (Sec.~\ref{sec:boost_ID}).
\end{enumerate}
The overall success of these convergence tests---most notably the 
convergence of the independent residuals---is considered the 
strongest evidence we have that our code {\em does} correctly compute
solutions of our model that converge to the continuum limit as 
the basic mesh scale, $h$, approaches 0.
In addition, we provide evidence that our numerical results 
conserve ADM mass and Noether charge to the expected order of accuracy.

The final two sections of the chapter focus on the dynamics of two boson
stars, and the calculations discussed therein constitute the 
major new computational results of this thesis.
Sec.~\ref{sec:headon} is concerned with the simulation of a 
head-on collision between two boson stars, in which we observe 
``solitonic'' behaviour that has previously been seen in other 
studies of self-gravitating scalar field configurations
\cite{dale,ChoiBEC,cwlai:phd,Choptuik:2009ww}.  
Sec.~\ref{sec:orbit}
then considers the interesting case of the simulation of 
boson star binaries for three different choices of initial data
parameters, each of which produces a distinct end state:
\begin{enumerate}
\item Long-lived orbital motion.
\item Merger of the stars that results in a conjectured rotating and
      pulsating boson star.
\item Merger of the stars that leads to the conjectured formation of
      a black hole.
\end{enumerate}
Finally, 
Sec.~\ref{sec:discussions} summarizes our main results and outlines
some possible directions for improvements and additional developments 
of this project.

\section{Summary of the Numerical Code} \label{sec:num_code}

The set of PDEs governing our model---as derived and discussed in
Ch.~\ref{formalism}----was discretized using the finite difference 
techniques described in Ch.~\ref{numerical}. Specifically, an $O(h^2)$ 
Crank-Nicholson scheme was applied to:
\begin{enumerate}
\item The 2 first-order-in-time equations that
      govern the time evolution of the real and imaginary components
      of the complex scalar field, $\phi_1(t,x,y,z)$ and $\phi_2(t,x,y,z)$,
      respectively.
\item The 2 first-order-in-time equations for the corresponding
      conjugate momenta, $\Pi_1(t,x,y,z)$ and $\Pi_2(t,x,y,z)$.
\end{enumerate}
Additionally, $O(h^2)$
centred FDAs were used to discretize the following elliptic PDEs 
appearing in the model:
\begin{enumerate}
\item The slicing condition for the lapse function, $\al(t,x,y,z)$.
\item The Hamiltonian condition for the conformal factor, $\psi(t,x,y,z)$.
\item The momentum constraints for the three components of the 
      shift vector, $\bt^x(t,x,y,z)$, $\bt^y(t,x,y,z)$ and
      $\bt^z(t,x,y,z)$.
\end{enumerate}
We reemphasize that in generating the results described below,
Dirichlet boundary conditions were imposed on all 
variables (see Sec.~\ref{sec:bc-dirichlet}).

The numerical code that was written to solve the finite difference equations 
(FDEs) originating from the discretization sketched above consisted 
of separate components for the hyperbolic and elliptic variables.
For the hyperbolic unknowns, the Crank-Nicholson FDEs were generated and 
solved using RNPL (Rapid Numerical Prototyping 
Language)~\cite{marsa:phd,RNPL_web}.  RNPL takes a high-level specification
of difference equations written in a natural operator form, and 
produces routines that employ point-wise Newton-Gauss-Seidel iteration
to compute advanced-time unknowns.  As discussed in Chap.~\ref{numerical}
this iteration typically converges rapidly for time-implicit discretizations
of wave equations, and our experience was consistent with that 
observation.  For the case of the FDEs governing the discrete elliptic 
unknowns we wrote FORTRAN 77 routines that implemented 
an FAS multigrid algorithm, as also described in the previous chapter.

A pseudo-code description of the overall code flow is given 
in~Fig.~\ref{RNPL}.
Execution of the program begins with a call to the initial data 
solver denoted~{\tt IVP\_solver} in the 
figure.~\footnote{We note that our code can handle 
essentially arbitrary initial configurations for the scalar field.  Here
we focus discussion on the case of most interest, in 
which the initial data represents one or more boson stars.}
In turn, this solver makes use of a set of routines that generate 
solutions representing the static spherically symmetric boson stars that
were discussed in detail in Chap.~\ref{initialdata}.  
The interested reader is referred to App.~\ref{ap:bsidpa} for 
documentation of the highest level routine, {\tt bsidpa}, that can 
be used to determine spherically symmetric boson star profiles 
for arbitrary polynomial self-interaction potentials. 

The solver for a single boson star returns a set of spherically
symmetric functions, including $\alpha(R)$, $a(R)$ $\phi(R)$ and 
$\Phi(R)$, where $R$ is areal radius.  These functions~\footnote{Refer to
Sec.~\ref{sec:id-boost} for details concerning the 
computation of the conjugate momenta, $\Pi_1$ and $\Pi_2$,
from the spherically symmetric solution for a single boson star.}
must then be transformed to the isotropic radial coordinate, $r$, and 
interpolated to the Cartesian computational domain as 
described in Secs.~\ref{sec:id-ansatz} and \ref{sec:id-1d3d}, respectively.  In
addition, if the star is to be boosted at the initial time, 
the transformations detailed in Sec.~\ref{sec:id-boost} are applied.

In instances where the evolution is to describe the dynamics of 
a boson star binary, the above process is carried out for each 
star, and the resulting functions from the computations for 
each individual star are added to produce the data for the binary. 
Although not necessary mathematically---since the initial values
of the scalar field variables are essentially unconstrained, other 
than requirements of smoothness and sufficiently rapid decay as 
$r\to\infty$---when we set up binaries in this manner we try to 
ensure that the stars are sufficiently well separated that the 
initial configuration really {\em does} describe two isolated objects.

Once the scalar field variables have been fixed at $t=0$, the 
multigrid solver is invoked to determine the initial values of 
the elliptic variables, $\al(0,x,y,z)$, $\psi(0,x,y,z)$ and 
$\bt^x(0,x,y,z)$.  Here we note that $\bt^x(0,x,y,z)$ is only 
non-zero when a boost has been applied (either to a single 
star, or to both stars in the binary case).  Furthermore, since 
we only apply boosts in the $x$ direction we always have 
$\bt^y(0,x,y,z)=\bt^z(0,x,y,z)=0$.  In performing the multigrid 
solution we start from initial estimates given by the 
values of the geometric variables computed from the boson star 
solver---and post-processed in the manner described in Sec.~\ref{sec:id-boost}.

Once the initial data has been determined, the code enters the 
main time-stepping loop.  After initialization of the advanced 
values {\tt f(t+dt)} and {\tt g(t+dt)} to those from the 
previous step {\tt f(t)} and {\tt g(t)}, each time step proceeds by 
a sub-iteration in which:
\begin{enumerate}
\item The advanced values, {\tt g(t+dt)}, of the geometry variables
      are updated using a single multigrid FAS $V$-cycle with 
      the scalar field values {\tt f(t+dt)} acting as sources.
\item The advanced values, {\tt f(t+dt)}, of the scalar field variables 
      are updated using a single point-wise Newton-Gauss-Seidel relaxation
      sweep, with the values {\tt g(t)}, {\tt g(t+dt)} and {\tt f(t)} 
      acting as sources.
\item The residuals of the finite difference equations for the 
      advanced values of the scalar field are computed.
\end{enumerate}
This sub-iteration continues until the $l_2$ norm of the scalar field 
residuals is below some specified tolerance, which was set to
$10^{-7}$ for the calculations described below.
Once the sub-iteration has converged, the advanced ({\tt t+dt}) values are
relabelled as current values (pseudo-routine {\tt swap\_levels}).
Invocation of the {\tt analysis} routine then effects calculation
of quantities such as the independent residuals, the ADM mass and 
Noether charges, as well as the periodic output of grid function values
to disk.  This completes one pass of the main loop: the time-stepping
procedure is then repeated 
until the specified final integration time, $t_{\rm max}$, is
reached.

\begin{figure}
\centerline{\fbox{\parbox{12.5cm}{
\input{RNPL.txt}
}}}
\caption[A Pseudo-Code Representation of the Numerical Code.]
{A pseudo-code representation of the numerical code.  See the 
text for details concerning the overall program flow.
\label{RNPL}}
\end{figure}

In order to facilitate the presentation of the numerical results that 
follow, we now introduce some convenient nomenclature and notation.
Recalling that we use a single, uniform grid (mesh) to perform 
our computations, the actual number of grid points associated with any 
of our simulations can be represented by a vector we call the 
\shape\ of the mesh. \shape\ is defined by
\beq
\shape = [N_x,\,N_y,\,N_z] ,
\eeq
where $N_x$, $N_y$ and $N_z$ are the number of grid points in the 
$x$, $y$ and $z$, coordinate directions, respectively.
The total number of grid points in the 
computational domain is thus simply given by $N_x N_y N_z$.
In addition, when performing computations on a series of meshes 
with grid spacings in which each finer scale is $1/2$ that of 
the previous grid, it is useful to introduce the notion of the
discretization {\em level}.  Let $N=\min(N_x,N_y,N_z)$.
Then we demand that there be an integer $l$ such that 
\beq
N= 2^l+1 \, ,
\eeq
where $l$ is precisely what we call the level of the calculation.  
In many cases we run our code
with $N_x=N_y=N_z$.  In instances where this is not true, then tacit 
in this definition is the assumption that we will compute with some minimum 
level $l_{\rm min}$, with corresponding $N_{\rm min} = 2^{l_{\rm min}} +1$.
Further assuming that $N=N_x$, the coarsest grid will then be 
characterized by 
\beq
\shape_{l_{\rm min}} = [N_{\rm min},\,N_{y0},\,N_{z0}] \, ,
\eeq
with $N_{y0} \ge N$, $N_{z0} \ge N$. Finer grids will then have
\beq
\shape_{l_{\rm min}+i} = [2^i(N-1)+1,\,2^i(N_{y0}-1)+1,\,2^i(N_{z0}-1)+1]\, .
\eeq

Another definition concerns the coordinates spanned by a particular
grid; i.e.~the limits of the computational domain.  
As discussed in~\ref{sec:FDA} this domain is given by 
\bea
x_{\rm min}\le &x& \le x_{\rm max}\,, \\
y_{\rm min}\le &y& \le y_{\rm max}\,, \\
z_{\rm min}\le &z& \le z_{\rm max} \,.
\eea
We thus introduce the notion of a \emph{bounding box vector}, denoted 
by \bbox, and defined by
\beq
\bbox = [x_{min},x_{max},y_{min},y_{max},z_{min},z_{max}] \,.
\eeq

We conclude this section with some brief remarks concerning the 
computing requirements for our calculations, as well as the nature 
of some of the figures we use to illustrate key results.

First, the simulations presented here were run on single nodes of one 
of two clusters located at UBC, which are known as \vnfeIV\ and \vnfeV. 
Each node of \vnfeV\ has four $2.4$ GHz Dual-Core 
AMD(R) Opteron(TM) processors with a total of $4$ GB of memory.
Nodes on \vnfeIV\ have two 
$2.4$ GHz Intel(R) Xeon(TM) CPUs and a total of $2$ GB of 
memory. The most extensive calculations were run on \vnfeV,
lasted approximately $260$ hours, and used about $700$ MB of memory. 
Larger simulations (up to $1$ GB), although possible, were deemed
impractical in terms of providing results on the timescale of a 
few days at most.

Second, many of the figures that follow are surface plots of functions,
and for these plots the following should be noted:
\begin{enumerate}
\item The ``grid lines'' in each direction---which are included 
      as a visual aid---are generally only a subset 
      of the total that are available: i.e.~the number of grid lines does 
      not reflect the true resolution of the computation.
\item Unless otherwise stated, the vertical displacement of the 
      surface corresponds to the value of the 
      function being plotted.
\end{enumerate}

\newpage

\section{Generic Initial Data}\label{sec:gen_ID}

A straightforward way to test our code for convergence and consistency
is to choose suitably generic initial data which can be specified
in closed form, and then evaluate the results of its time evolution. 
The initial configuration for the results discussed in this section
was based on the following generalized gaussian profile for one 
component of the complex scalar field:
\beq \label{eq:gaussian}
\phi(x,y,z)=\phi_0 \exp{\lhb -\lb \fr{x-x_0}{\de_x} \rb^2 
                             -\lb \fr{y-y_0}{\de_y} \rb^2 
                             -\lb \fr{z-z_0}{\de_z} \rb^2 \rhb}  \, .
\eeq
Here, $\phi_0$, $x_0$, $y_0$, $z_0$, $\de_x$, $\de_y$ and $\de_z$ are 
adjustable parameters.
The freely specifiable variables $\{\phi_1,\phi_2,\Pi_1,\Pi_2\}$ were
then initialized as follows: 
\begin{itemize}
\item $\phi_1(0,x,y,z) = \phi(x,y,z)$, with $(x_0,y_0,z_0)=(4,3,2)$\,,
		$\phi_0 = 0.05$ and $(\de_x,\de_y,\de_z)=(4.5,4,4)$\,,
	\item $\Pi_2(0,x,y,z)=-3\phi_1(0,x,y,z)$\,,
	\item $\phi_2(0,x,y,z)=\Pi_1(0,x,y,z) = 0$\,.
\end{itemize}
This produces a off-origin, slightly non-spherical, gaussian lump
of field that exhibits non-trivial dynamics, but whose evolution
also mimics, to some extent, that of a distorted stable boson star.
We would like to emphasize that the presence of symmetries in 
the initial data could lead to a failure in detecting certain implementation
errors.
Thus, in particular, we designed this initial configuration so
that it did {\em not} have any of the reflection symmetries
$x\to-x$, $y\to-y$, and $z\to-z$.  We should note, however, that
the other initial datasets we subsequently describe \emph{do} 
have at least one of these symmetries, and thus in principle, should 
lead to time evolutions that also possess the symmetry.~\footnote{This 
assumes that there are no dynamically unstable modes in the 
configurations considered that would break the symmetries, and that could 
be excited by truncation error and/or roundoff error effects.}
We are aware that we could
have exploited this fact to reduce the cost of many of our computations.
Nonetheless, we did not pursue this option since we wanted to keep 
our code general, and also wanted to avoid the complications of 
implementing symmetry conditions in the multigrid solver at this stage.

The bounding box for the experiments described here was 
$\bbox=[-11,19,-12,18,-13,17]$, and three different grids with 
resolutions in a $1:2:4$ ratio were used for the convergence tests 
and independent residual evaluations. 
The grid with the finest resolution (level $7$), had $\shape=[129,129,129]$,
while the coarsest grid (level $5$) had $\shape=[33,33,33]$. 
The Courant factor adopted here---and for all simulations discussed in this 
chapter---was $\la=0.4$.

\begin{figure}[p]
\begin{center}
\epsfxsize=15.0cm
\ifthenelse{\equal{\highQ}{true}} {
\epsffile{figs/gaussianasy/gaussianasy.eps}
}{
\epsffile{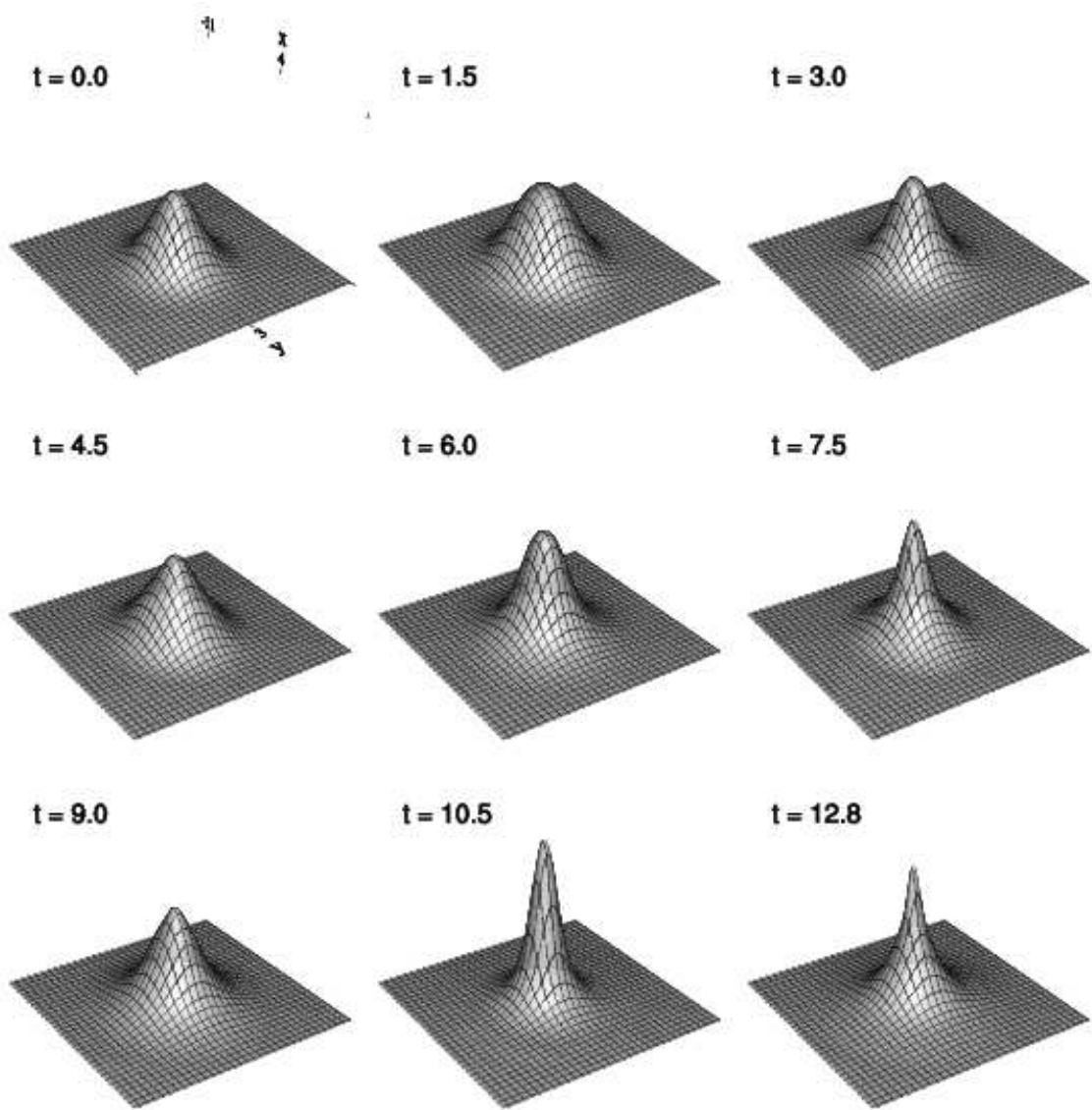}
}
\caption
[Time Evolution of a Generic Gaussian Profile.]
{Time Evolution of a Generic Gaussian Profile. 
This figure displays a time-series of $z=2$ cuts of the scalar field modulus $|\phi(t,x,y,z)|$ 
from a short evolution (roughly one dynamical time) of the initial 
configuration defined by Eq.~\ref{eq:gaussian}. 
Here the value of $\phi_0$ appearing in~(\ref{eq:gaussian}) is 
$0.05$, the range of the plotted data is $0\le |\phi(t,x,y,z)| \le 0.12$, 
and the 
solution was computed using a grid with ${\tt shape} = [129,129,129]$ (level
7).
Note that the solution oscillates, but is apparently gravitationally
bound, as there is no evidence for dispersal of the scalar field 
}
\label{gaussian}
\end{center}
\label{fig:generic}
\end{figure}

Fig.~\ref{gaussian} shows a time-series of $z=2$ cuts of the 
scalar field from a short evolution (roughly one 
dynamical time) of the initial configuration defined above, and where 
the finest resolution grid was used.  Here, and for other 
plots of this type shown subsequently, evolution proceeds left-to-right and 
top-to-bottom.  The solution oscillates, but is apparently gravitationally
bound, as there is no evidence for dispersal of the scalar field. 
However, we are not so much interested in the evolution of this data 
for its physical content, as for what it can tell us about the convergence
properties of our code.

\begin{figure}
\begin{center}
\epsfxsize=18.0cm
\epsffile{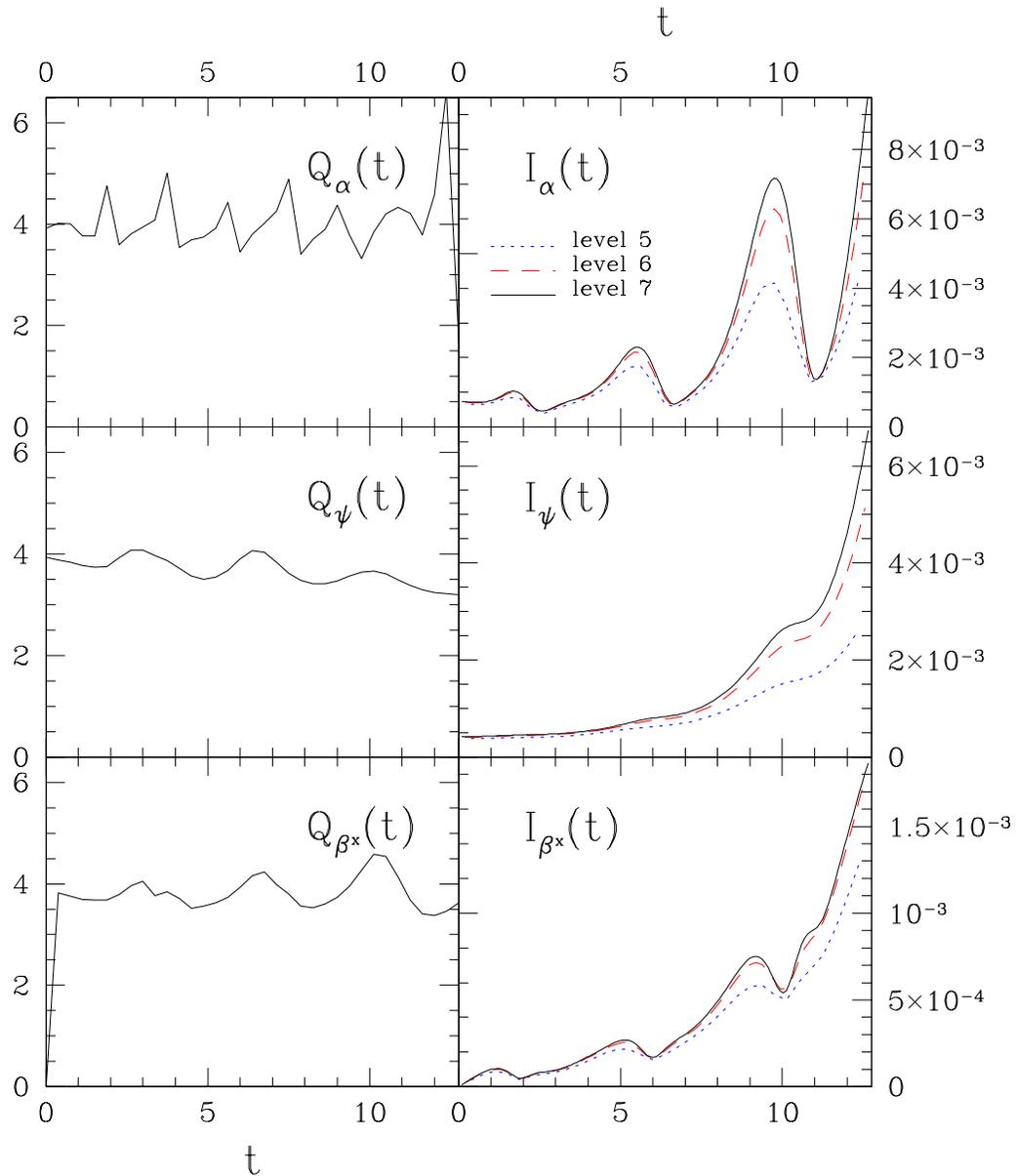}
\end{center}
\caption
[Generic Initial Data: $Q^h(t)$ and $\Vert I(t)\Vert_2$ for $\al$, $\psi$ and $\bt^x$.]
{Left panels: plots of the convergence factor $Q^h(t)$, as a function
of time, $t$, 
for the geometric variables $\al$, $\psi$ and $\bt^x$, 
using data from the calculation described in the caption
of Fig.~\ref{fig:generic}.
The convergence factor remains close to the value $4$ that is
expected for our second-order scheme.
Note that $\bt^x(0)=0$ at all resolutions, and we have thus 
defined $Q^h_{\beta^x}(0)= 0/0 \equiv0$.
Right panels: $l_2$ norms of the rescaled independent 
residuals (see Eq.~\ref{eq:indres_resc_geometry}), 
$\Vert I(t)\Vert_2$, 
for the same set of variables. 
As the level of refinement increases, the plots of the 
rescaled $I(t)$ become increasingly coincident, providing strong
evidence of convergence of our code to the continuum solution.
}
\label{QI-l-ps-btx}
\end{figure}

\begin{figure}
\begin{center}
\epsfxsize=18.0cm
\epsffile{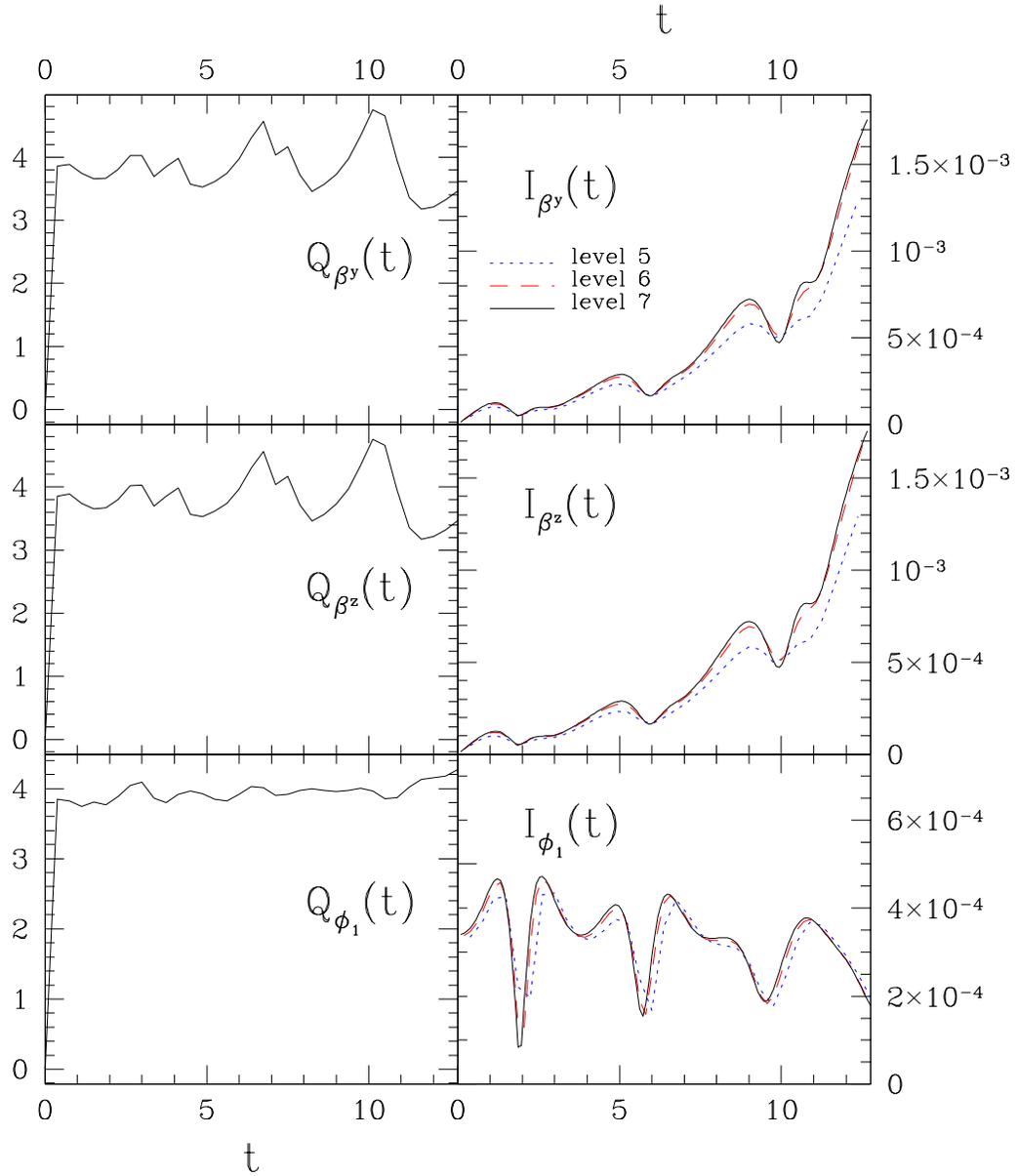}
\end{center}
\caption
[Generic Initial Data: $Q^h(t)$ and $\Vert I(t)\Vert_2$ for $\bt^y$, $\bt^z$ and $\phi_1$.]
{Left panels: Plots of the convergence factor $Q^h(t)$ 
for $\bt^y$, $\bt^z$ and $\phi_1$,
using data from the calculation described in the caption
of Fig.~\ref{fig:generic}.
Again, all convergence factors are close to the expected value of $4$.
Right panels: $l_2$ norms of scaled independent 
residuals, $\Vert I(t)\Vert_2$, 
for the same set of variables, and where we have defined 
$Q^h_{\bt^y}(0)=0/0\equiv0$ and
$Q^h_{\bt_z}(0)=0/0\equiv0$.
Scaling factors for the two shift vector components and the 
scalar field are given by~(\ref{eq:indres_resc_geometry}) and
(\ref{eq:indres_resc_matter}) respectively. 
We see clear evidence for the convergence of the independent
residuals at their expected rates ($O(h^2)$ for $\bt^y$ and $\bt^z$,
$O(h)$ for $\phi_1$).
}
\label{QI-btyz-ph1}
\end{figure}

\begin{figure}
\begin{center}
\epsfxsize=18.0cm
\epsffile{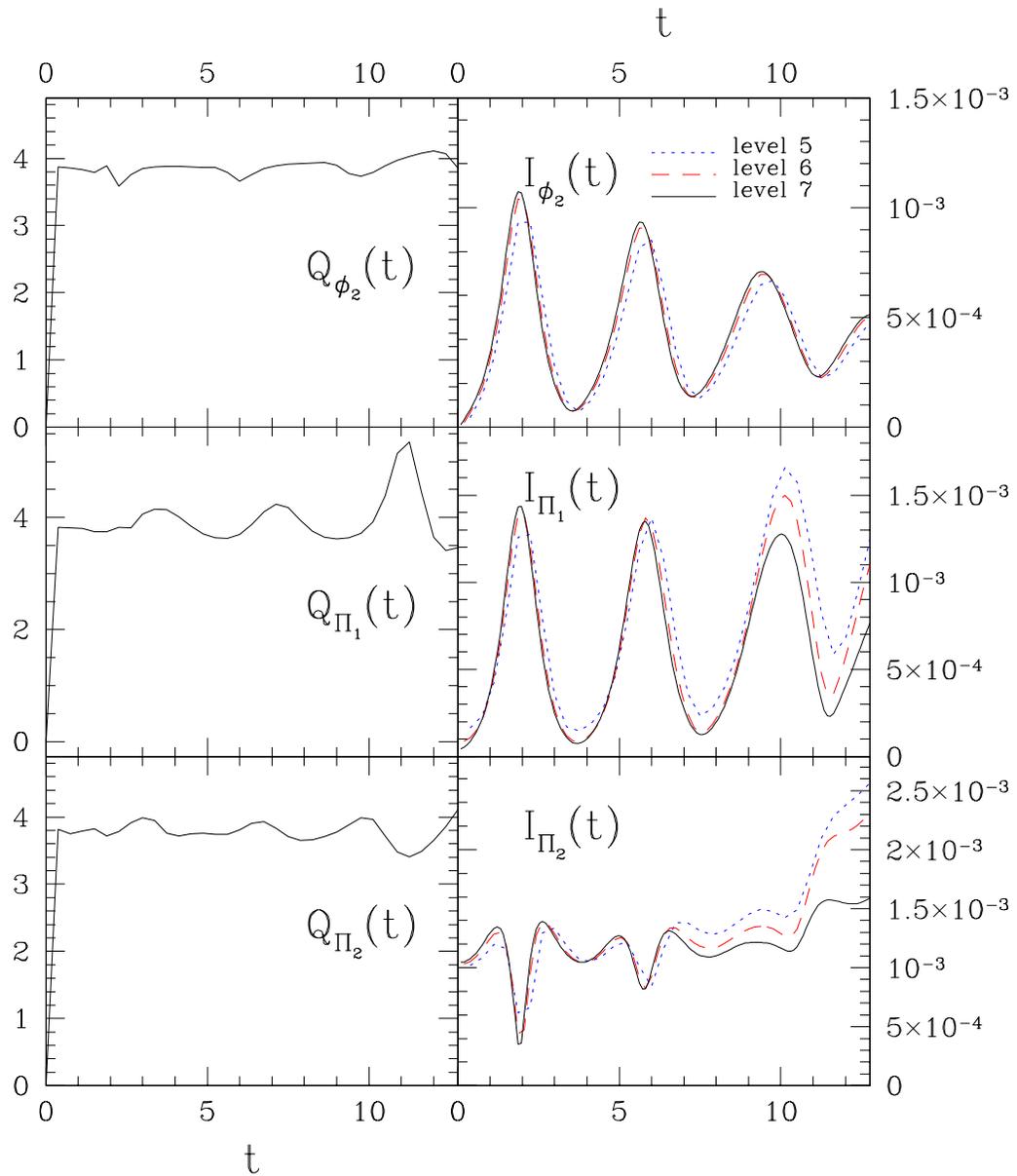}
\end{center}
\caption
[Generic Initial Data: $Q^h(t)$ and $\Vert I(t)\Vert_2$ for $\phi_2$, $\Pi_1$ and $\Pi_2$.]
{Left panels: Plots of the convergence factor $Q^h(t)$ 
as a function of time for 
the scalar field variables $\phi_2$, $\Pi_1$ and $\Pi_2$,
using data from the calculation described in the caption
of Fig.~\ref{fig:generic}.
Right panels: $l_2$ norms of scaled independent residuals, 
$\Vert I(t)\Vert_2$, 
for the same set of variables. 
The residuals were rescaled using~(\ref{eq:indres_resc_matter}) and
clearly are converging as $O(h)$, as expected.
}
\label{QI-ph2-pi12}
\end{figure}

In that regard, we first note that all grid functions remain quite 
smooth during the calculations.  Second, as can be seen in the left 
panels of Figs.~\ref{QI-l-ps-btx}, \ref{QI-btyz-ph1} and \ref{QI-ph2-pi12},
the convergence factors $Q^h(t)$ (defined by \ref{eq:qfactor_def}) for the dynamic 
variables provide clear evidence of second-order convergence.  
That is, $Q^h(t)$ for all of the variables remains close to the value 4 
that is expected for our second-order scheme in the limit that the 
mesh spacings go to 0 (according to \ref{eq:qfactor2}).
Third, and perhaps most importantly, the right panels of these figures 
show that the independent residuals are also converging.  
Here we note that in order to more easily assess the convergence 
rates of the independent residuals from the graphs,  we have 
rescaled them so that, should they be converging at the expected 
order, the plots from computations at different discretization levels
will be roughly coincident.  Thus, for the geometric variables, 
where we expect second order convergence, the $l_2$-norms of the 
independent residuals 
were rescaled in the following way:
\beq \label{eq:indres_resc_geometry}
\Vert I^l(t)\Vert_2 \to 4^{(l-5)} \Vert I^l(t)\Vert_2 ,
\eeq
where $l$ refers to the level of discretization: $5$, $6$ or $7$
in this case. The independent residuals for the scalar
field quantities were rescaled in similar fashion.  However,
since we expect first-order convergence in this instance, the 
scaling factor is modified accordingly:
\beq \label{eq:indres_resc_matter}
\Vert I^l(t)\Vert_2\to 2^{(l-5)}\Vert I^l(t)\Vert_2 \, .
\eeq
We also observe that, unless otherwise specified, all of the subsequent
plots of independent residuals in this chapter show rescaled values as well.

Thus, Figs.~\ref{QI-l-ps-btx} and \ref{QI-btyz-ph1} provide good 
evidence that the rescaled independent residuals for all of
the geometric variables are converging as $O(h^2)$, as expected.
We also note that the 
functional form of $I^h(t)$ for the metric variables is roughly 
resolution-independent, so that we apparently {\em do} have
\beq
\lim_{h\to0} I^h=I_2(t) \, h^2 \qquad \textrm{and not} 
	\qquad \lim_{h\to0} I^h=I_2(t,h) \, h^2 .
\eeq
 
Similar remarks can be made concerning the independent residuals for 
the variables
$\phi_1$, $\phi_2$, $\Pi_1$, and $\Pi_2$, except that the convergence of the 
$I^h(t)$ in this case is linear in $h$.  This is to be expected since,
as also discussed in Chap.~{\ref{numerical}}, the independent discretization 
adopted for approximate time derivatives was a first-order forward 
difference.

\begin{figure}
\centering
\includegraphics[width=7.4cm]{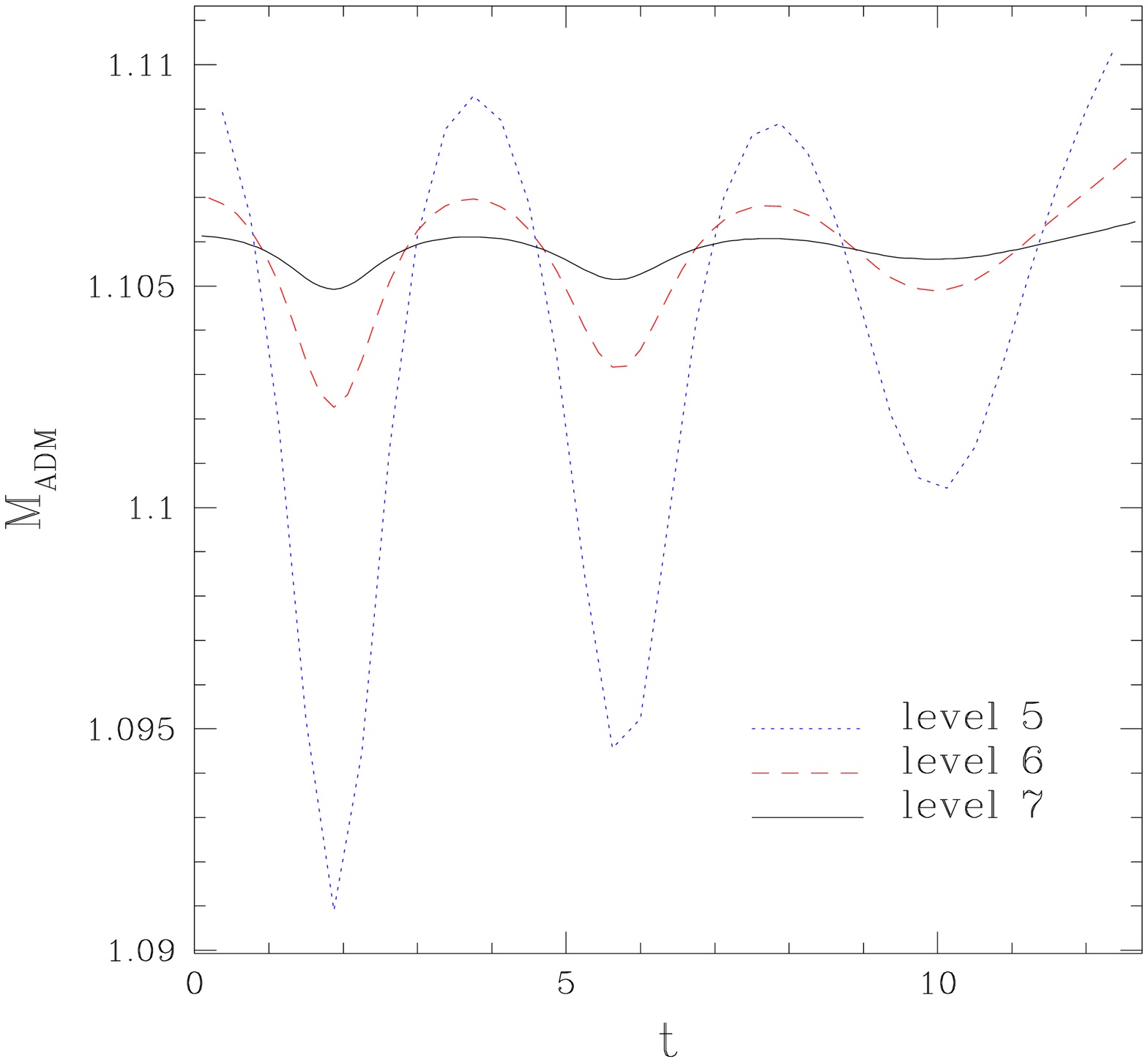}
\includegraphics[width=7.4cm]{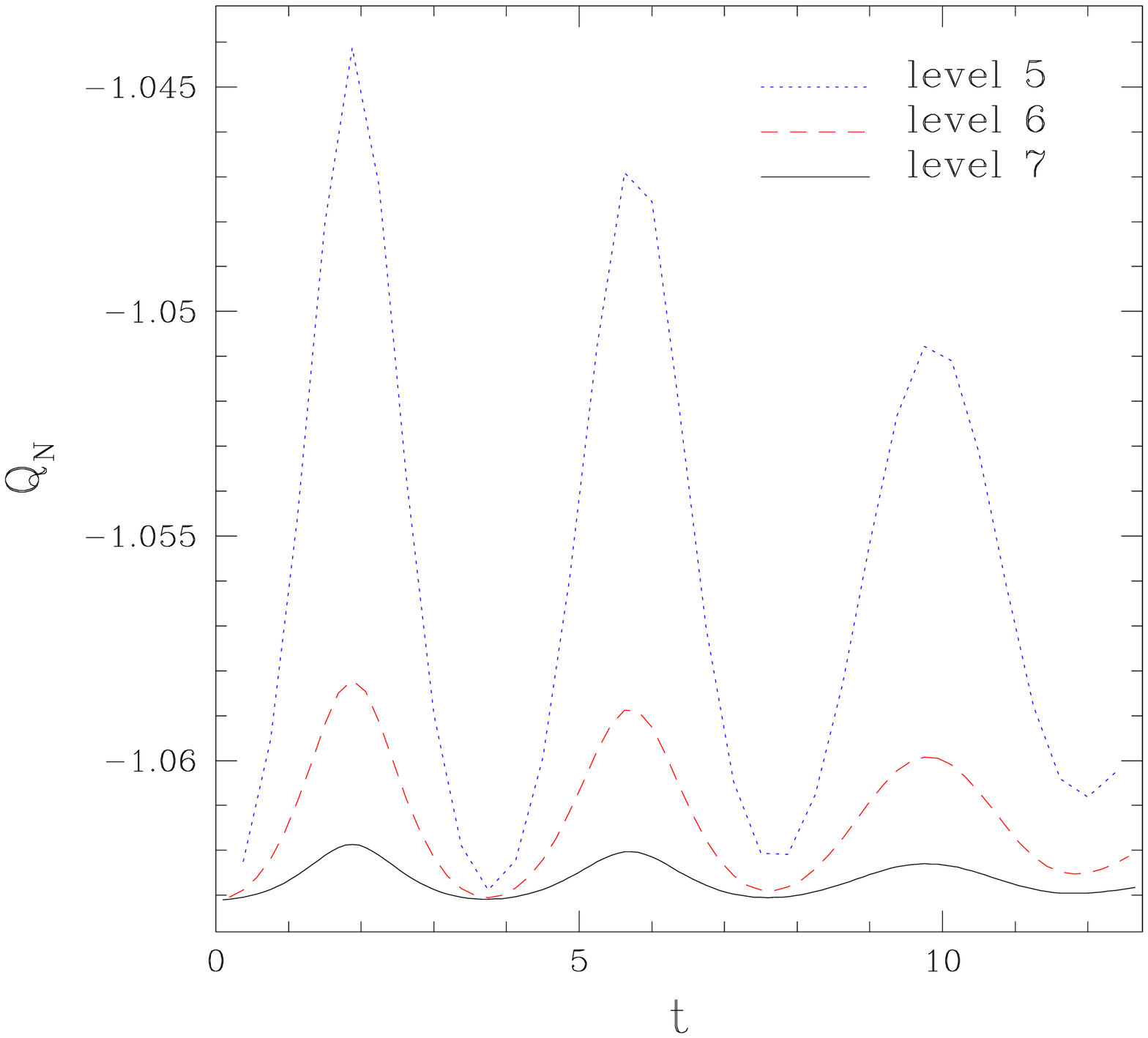}
\caption
[ADM Mass $M_{\rm ADM}(t)$ and Noether Charge $Q_{N}(t)$ for the Generic Initial Data.]
{ADM mass $M_{\rm ADM}(t)$ and Noether charge $Q_{N}(t)$
from the calculation described in the caption
of Fig.~\ref{fig:generic}.
Both plots indicate a trend to conservation of the 
respective quantity, with a convergence rate of $O(h^2)$, as clearly
illustrated in Fig.~\ref{dmass_noether_gauss} below.
}
\label{mass_noether_gauss}
\end{figure}

\begin{figure}
\centering
\includegraphics[width=7.4cm]{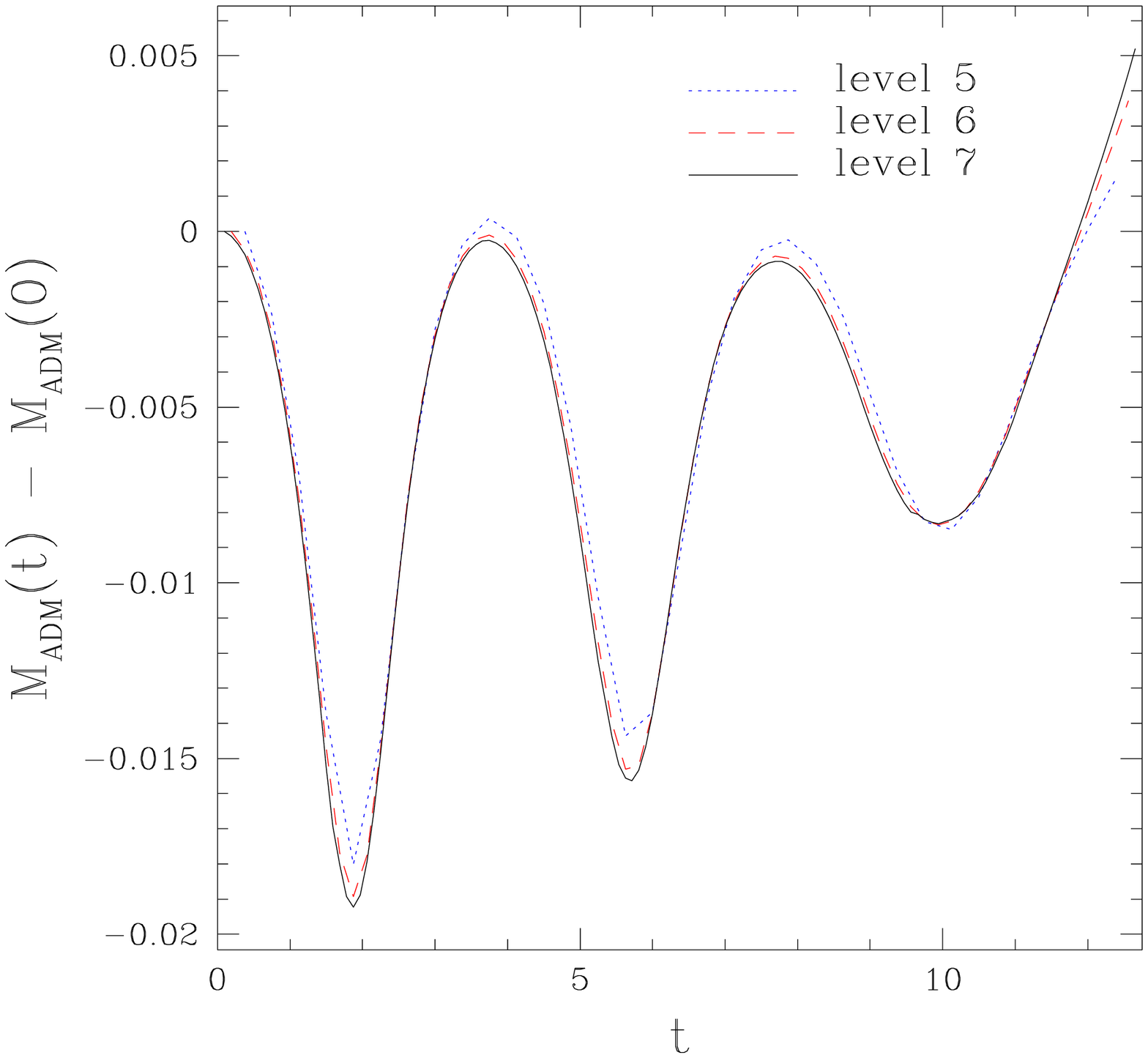}
\includegraphics[width=7.4cm]{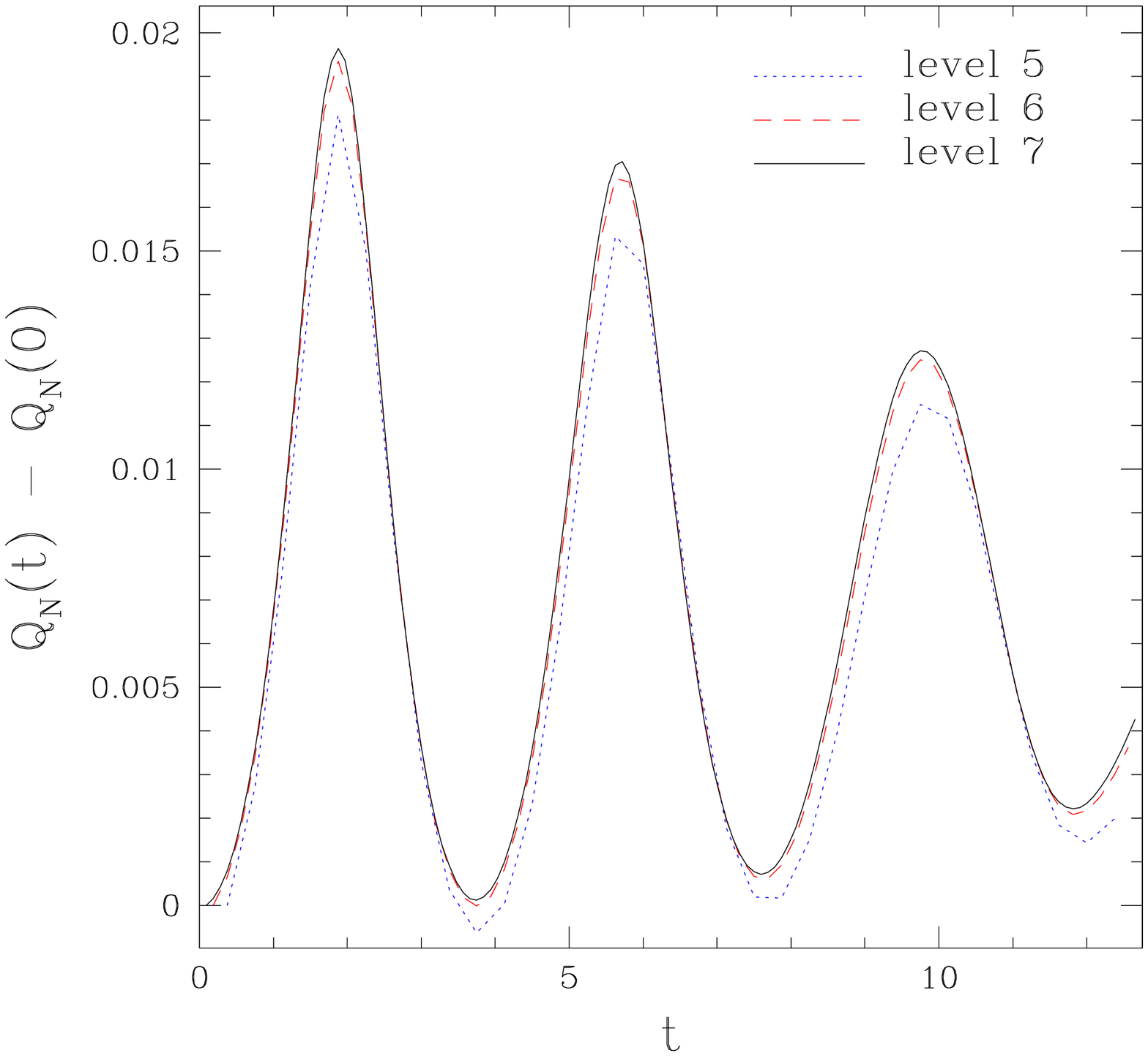}
\caption
[Rescaled ADM Mass and Noether Charge Deviations 
for Generic Initial Data.]
{Rescaled deviations in ADM mass and Noether charge, as functions of time, $t$, 
from the calculation described in the caption
of Fig.~\ref{fig:generic}.
The deviations were rescaled using~(\ref{eq:DM_ADM}) and~(\ref{eq:DQ_N}),
and the near-coincidence of the curves is strong evidence of ``convergence
to conservation''.
}
\label{dmass_noether_gauss}
\end{figure}

Fig.~\ref{mass_noether_gauss} shows the ADM mass, $M_{\rm ADM}$, and 
the Noether charge, $Q_{N}$, as a function of time for levels $5$, $6$ and
$7$. As the resolution increases, $M_{\rm ADM}(t)$ and $Q_{N}(t)$ 
tend to constant functions, indicating that the continuum conservation
laws are being recovered as the mesh spacing tends to 0.\footnote{
We reemphasize, however, that we know of no proof that $M_{\rm ADM}(t)$ {\em 
should} be conserved for other than spherically symmetric cases.
}
In order to better examine the rate of ``convergence to conservation''
of these values,
Fig.~\ref{dmass_noether_gauss} plots {\em deviations} of the 
ADM mass $M_{\rm ADM}(t)$ and the Noether charge $Q_{N}(t)$ from
their initial 
values, $M_{\rm ADM}(0)$ and $Q_{N}(0)$, as a function of time.
Additionally, parallelling what we did for the independent residuals,
the deviations computed at different levels of discretization have
been rescaled assuming an $O(h^2)$ convergence rate:
\bea
\De M^l_{\rm ADM}(t) = 4^{l-5} 
\lb M^l_{\rm ADM}(t) - M^l_{\rm ADM}(0) \rb ,
\label{eq:DM_ADM}\\
\De Q^l_{N}(t) = 4^{l-5} 
\lb Q^l_{N}(t) - Q^l_{N}(0) \rb ,
\label{eq:DQ_N}
\eea 
for $l = 5$, $6$ and $7$.
Again, the fact that the plots of the rescaled deviations are nearly coincident 
provides strong evidence of second order convergence (to conservation,
as a function of time) of both $Q_N$ and $M_{\rm ADM}$.

Before proceeding to describe our next set of numerical experiments, 
we make a brief digression to emphasize a key assumption that underpins the 
efficacy of finite differencing.  To keep the discussion simple, we 
consider the case of a scalar function of one independent variable, $u(x)$,
and assume that we have discretized the continuum domain with a uniform
grid with mesh spacing $h$.
Then when any differential operator, $L$, acting on $u$ is approximated 
with a finite difference operator, $L^h$, the error in the approximation
will be of the form
\beq
	L \left[ u(x) \right] - L^h \left[ u(x) \right] =
		h^p E_p \left[ u(x) \right] + \cdots
\eeq
where the positive integer $p$ is the order of the approximation, and 
$E_p$ is a differential operator of {\em higher (differential) degree}
than $L$.  For example, the usual $O(h^2)$ centred approximation to 
$du/dx\equiv u'(x)$ can be written as 
\beq
\fr{u_{i+1}-u_{i-1}}{2h} \equiv \fr{u(x+h)-u(x-h)}{2h} = u'(x) + \fr{h^2}{12} u'''(\xi) ,
\eeq
where $\xi \in [x-h,x+h]$ and $ h^2/12 u'''(\xi)$ is the error in
the approximation.  Note, however, that this result is valid {\em only}
if $u'''(x)$ exists and is continuous on the $[x-h,x+h]$.

This simple example highlights the importance of {\em smoothness} of 
both the continuum {\em and} discrete solutions in the context of 
finite differencing. Specifically, for the discrete case, ``smooth'' naturally
means ``smooth on the scale of the mesh'', and when grids that have 
$h$ comparable to the scale of variation of the solution unknowns are 
used, one cannot expect ``good'' results from convergence tests, 
including independent residual evaluation.  The reader should keep
this point in mind, especially in later sections of this chapter where 
the coarsest mesh spacings used {\em are} comparable to the scales on 
which the solution is changing. Moreover, as we will see in the next section, 
(discrete) non-smoothness related to issues such as the treatment of 
boundary conditions can also easily spoil the convergence of our 
finite difference scheme.

\newpage
\section{Static Spherically Symmetric Initial Data - One Star}\label{sec:SS_ID}

In the previous section we used a generic type of initial data to 
establish the correctness and convergence of our numerical code. 
The results of our convergence tests (including those of the conserved
quantities), and independent residual evaluation provide strong evidence
that the discrete equations of motion are consistent, have been implemented
correctly, and that the numerical solution obtained converges to the 
continuum solution.  In fact, we consider the generic initial data test 
to be sufficiently comprehensive to establish the overall validity 
of the code.  Nevertheless, in this section and the next
we continue with code tests, albeit with slightly different goals.

If there are no problems with the discrete equations of motion, or 
with convergence of sufficiently smooth solutions, then any deficiencies 
identified in the numerical results can likely be traced to characteristics 
of the initial data, or the resolution(s) used in the calculations. 
With this in mind, there were two key motivations for the tests performed
in this section.

First, it is natural to use the spherically-symmetric static configurations
that describe single boson stars to simultaneously test 1) the evolution code,
and 2) our solutions of the ODEs that determine the static solutions.
Recall from Sec.~\ref{sec:id-ansatz} that 
these solutions are computed from the following ansatz for the complex scalar 
field:
\beq
  \phi(t,R) = \phi_0(R)\,e^{-i\omega t},
\eeq
where $R$ is the areal radius.
It is important to note that even though the stars are characterized by a 
time {\em independent} gravitational field, $\phi(t,r)$ is still 
time {\em dependent}
due to the $e^{-i\omega t}$ factor in the above expression.  We should 
thus expect that the metric variables computed from such initial 
data should exhibit only approximate time independence, but that the
time independence should become more exact as the discretization scale,
$h$, tends to 0.

Second, as also discussed in Chap.~\ref{initialdata}, the boson stars 
computed using the above ansatz form a one-parameter family, 
where $\phi_0(0)$ is a convenient choice for the family parameter.  
Different members 
of this family have different ADM masses, as well as different overall
strengths of the gravitational field (as quantified, for example, by
$\max_R 2m(R)/R$).  In this context it seems reasonable to expect that for 
fixed discretization parameters stronger gravitational fields should 
lead to larger numerical errors.

We thus consider the numerical evolution of 
two distinct boson stars using our code. Both are on the stable branch 
(see Fig.~\ref{Madm1d}) and have central scalar field
values $\phi_0\equiv\phi_0(0)=0.03$ and 
$\phi_0\equiv\phi_0(0)=0.06$, respectively.  
Here, and in the following, we adopt a notation such that 
$S_{\phi_0(0)}$ denotes the boson star having a central field modulus
$\phi_0(0)$.  With this nomenclature, the stars that we will study 
are $S_{0.03}$ and $S_{0.06}$. 
$S_{0.03}$ is less massive, and thus
less compact than $S_{0.06}$, as can be inferred from Fig.~\ref{bs1d}. 
To evolve each of these configurations, we choose a
set of grid parameters that is slightly different than that adopted in the 
previous section: namely $\bbox = [-15,15,-15,15,-15,15]$, with 
the coarsest and finest of the three grids used
given by $\shape=[33,33,33]$ and $\shape=[129,129,129]$, respectively.

\begin{figure}
\begin{center}
\epsfxsize=15.0cm
\ifthenelse{\equal{\highQ}{true}} {
\epsffile{figs/sphsym/sph003.eps}
}{
\epsffile{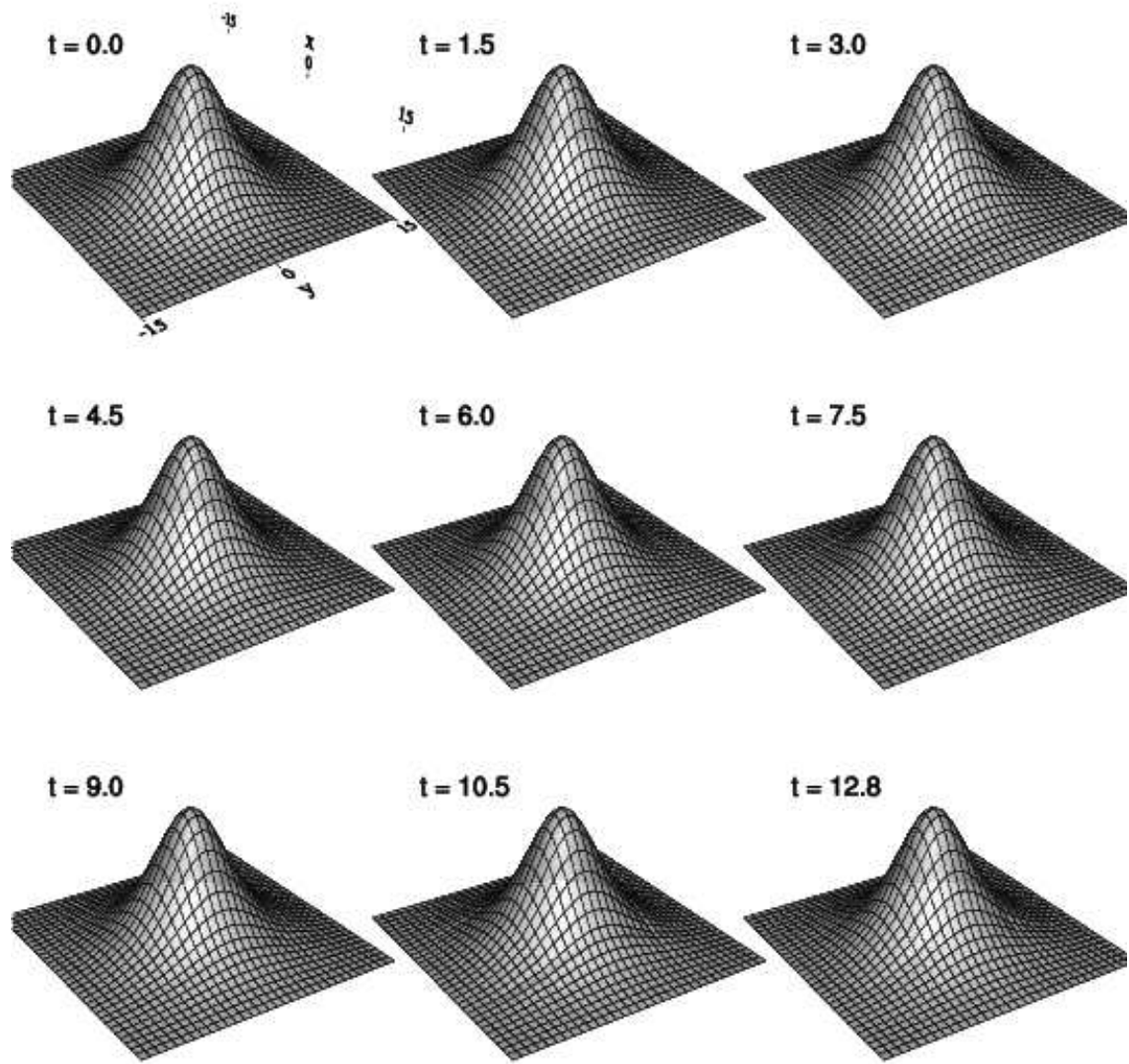}
}
\caption
[Time Evolution of the Static Boson Star, $S_{0.03}$.]
{Snapshots of the time evolution of $|\phi(t,x,y,0)|$ for 
the static boson star, $S_{0.03}$, 
as computed on the finest grid (${\tt shape} = [129,129,129]$).  
At least to the naked eye, the evolution appears time independent. 
}
\label{sph003}
\end{center}
\end{figure}

\begin{figure}
\begin{center}
\epsfxsize=15.0cm
\ifthenelse{\equal{\highQ}{true}} {
\epsffile{figs/sphsym/sph006.eps}
}{
\epsffile{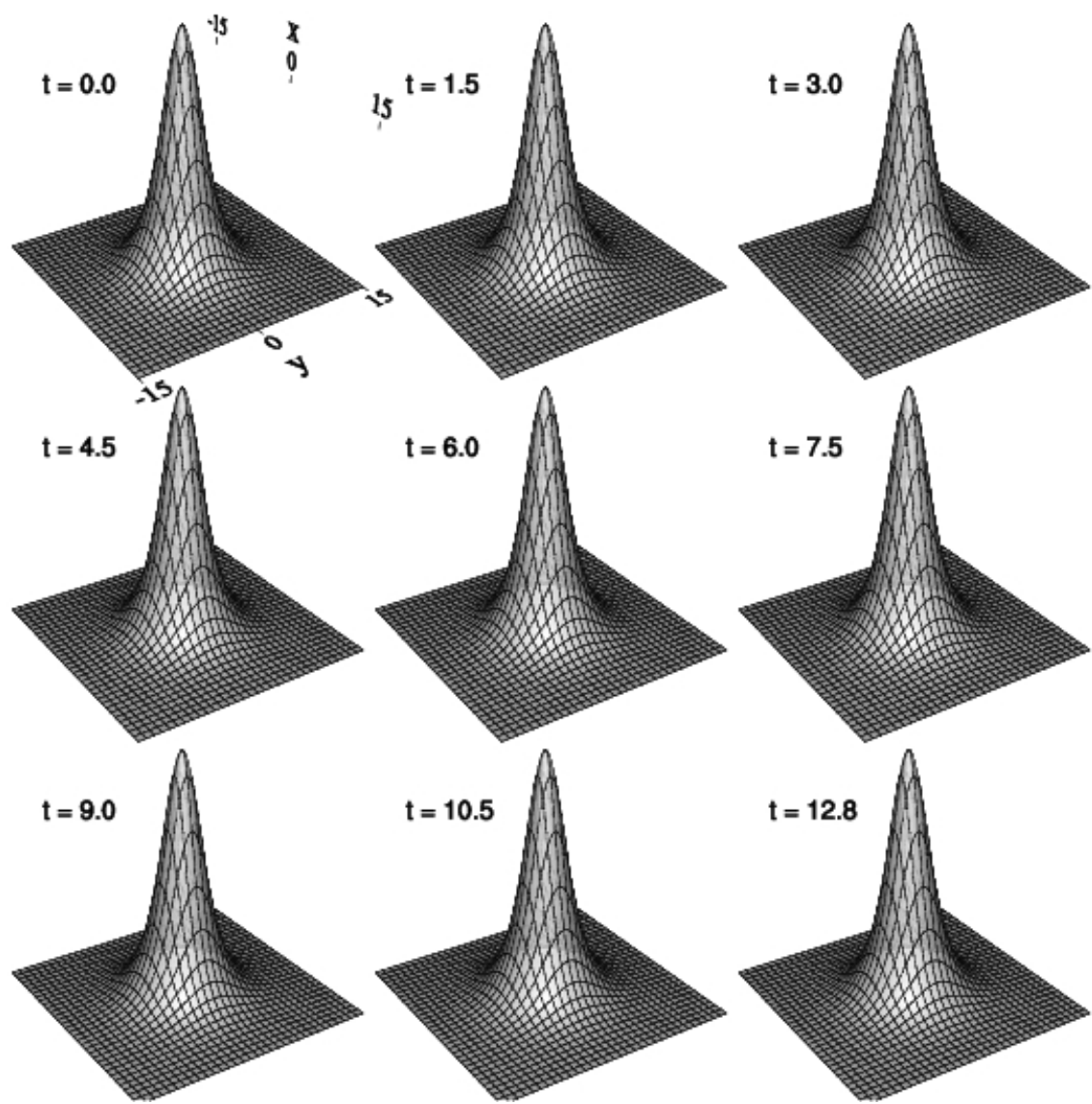}
}
\caption
[Time Evolution of the Static Boson Star, $S_{0.06}$.]
{
Snapshots of the time evolution of $|\phi(t,x,y,0)|$ for
the static boson star, $S_{0.06}$,
as computed on the finest grid (${\tt shape} = [129,129,129]$).
As for the data displayed in Fig.~\ref{sph003}, the evolution appears time independent.
}
\label{sph006}
\end{center}
\end{figure}

Figs.~\ref{sph003} and \ref{sph006} show some snapshots from the 
time evolution of $S_{0.03}$ and $S_{0.06}$, respectively, as computed 
on the finest grid.  At least to the naked eye these figures 
seem to indicate that that computed geometries are time independent 
in both cases.  It is thus tempting to conclude that with respect to 
the first purpose of this particular experiment, the code has passed
the test.

\begin{figure}
\begin{center}
\epsfxsize=16.0cm
\epsffile{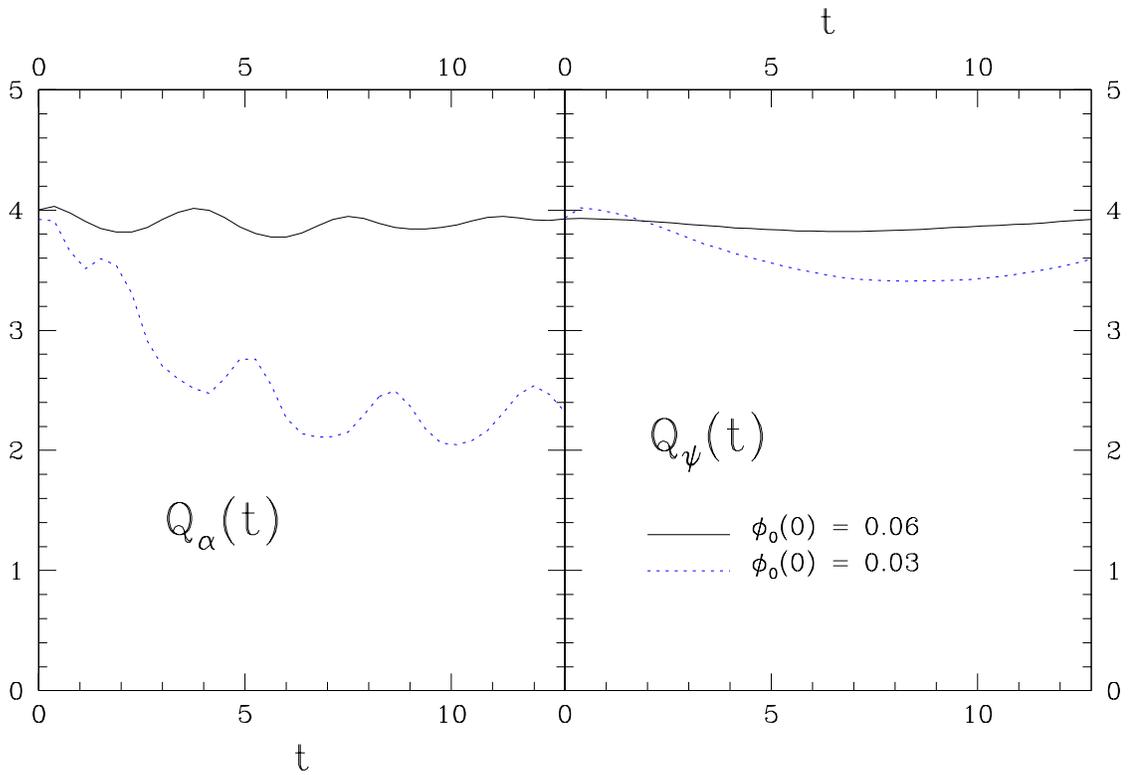}
\end{center}
\caption
[$S_{0.03}$ and $S_{0.06}$ Convergence Factors $Q^h(t)$ for $\alpha$ and $\psi$.]
{Convergence factors $Q^h(t)$ of $\alpha$ and $\psi$ for evolutions of the
two static configurations $S_{0.03}$ (solid black curves) and
$S_{0.06}$ (blue dashed curves).
The superior convergence of the $S_{0.06}$ data at the resolutions 
used (${\tt shape} = [33,33,33]$, $[65,65,65]$ and $[129,129.129]$)
is apparent.
}
\label{Q-l-psi}
\end{figure}

\begin{figure}
\begin{center}
\epsfxsize=16.0cm
\epsffile{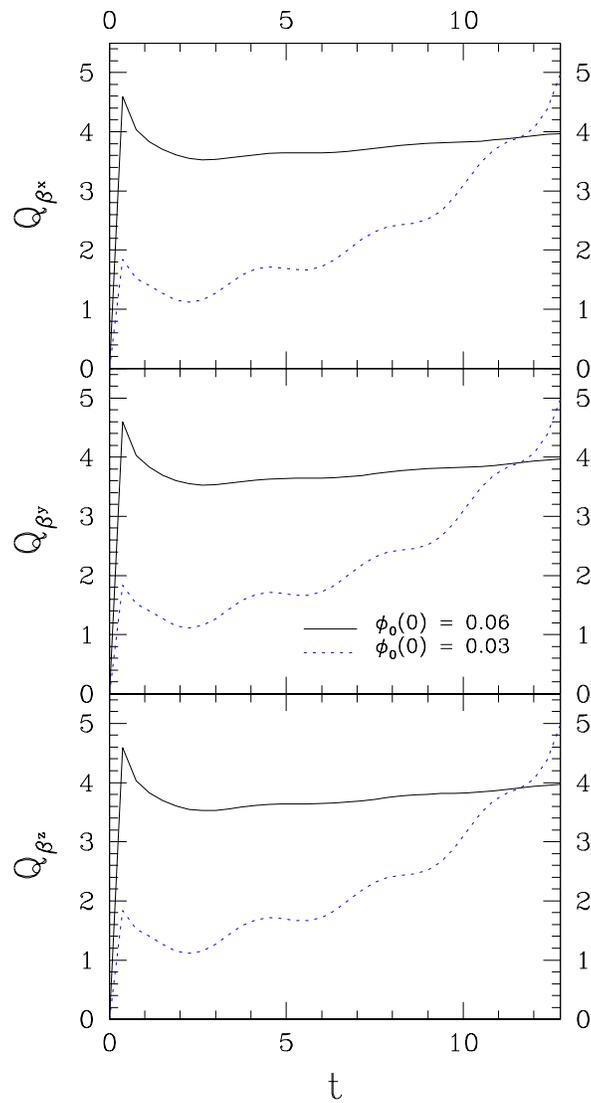}
\end{center}
\caption
[$S_{0.03}$ and $S_{0.06}$ Convergence Factors $Q^h(t)$ for $\bt^i$.]
{Convergence factors $Q^h(t)$ of $\bt^k$ for evolutions of the
two static configurations $S_{0.03}$ (solid black curves) and
$S_{0.06}$ (blue dashed curves).  As in Fig.~\ref{Q-l-psi}, 
the superior convergence of the $S_{0.06}$ is clear.
}
\label{Q-bti}
\end{figure}

\begin{figure}
\begin{center}
\epsfxsize=16.0cm
\epsffile{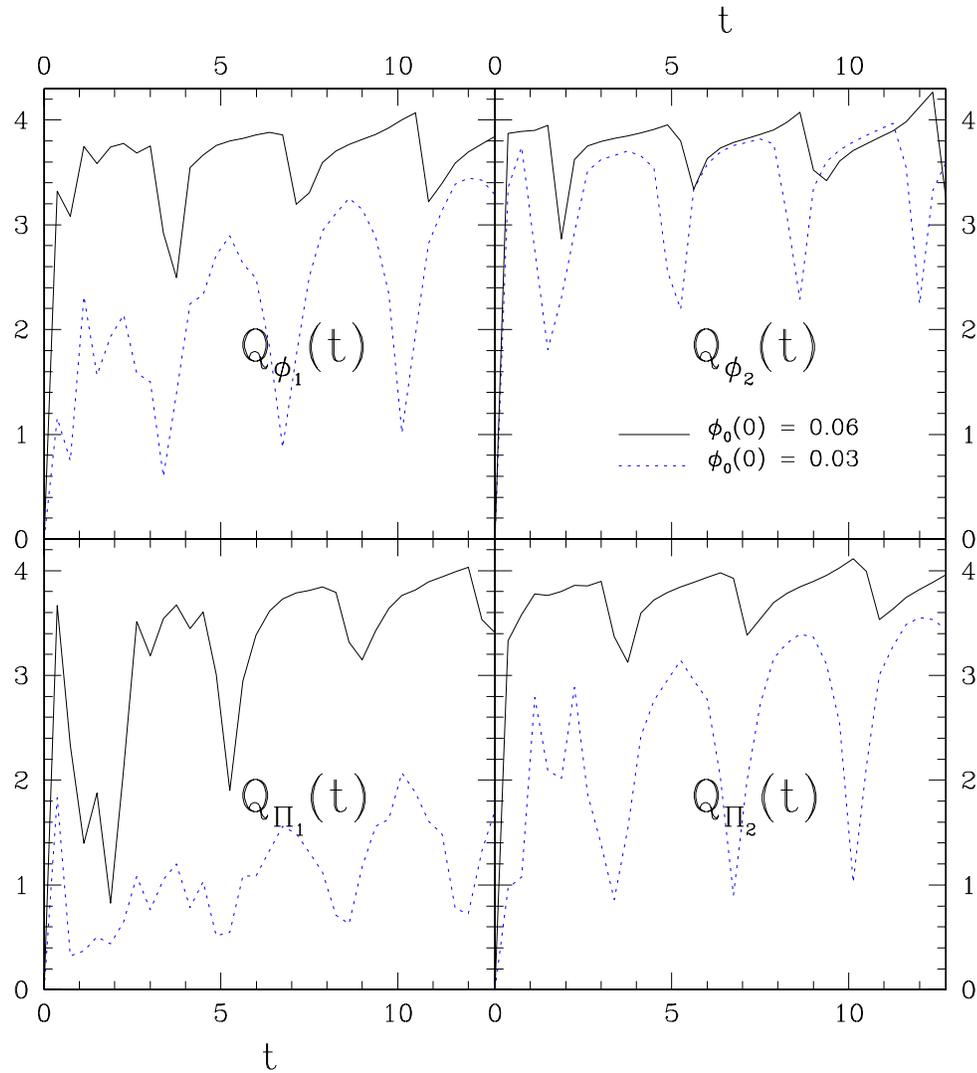}
\end{center}
\caption
[$S_{0.03}$ and $S_{0.06}$ Convergence Factors $Q^h(t)$ for $\phi_A$ and $\Pi_A$.]
{Convergence factors $Q^h(t)$ of $\phi_A$ and $\Pi_A$ 
for evolutions of the
two static configurations $S_{0.03}$ (solid black curves) and
$S_{0.06}$ (blue dashed curves).  As in the previous two figures, 
better convergence is observed in the $S_{0.06}$
data.   There is, however, a noticeable ``glitch'' in the $S_{0.06}$ 
results for $Q_{\Pi_1}(t)$ in the interval $1\le t \le 3$.  Although
we do not completely understand this behaviour at the current time, 
we suspect that spurious reflections from the boundaries may be
to blame.
}
\label{Q-phi-pi}
\end{figure}

However, as Figs.~\ref{Q-l-psi}, \ref{Q-bti} and \ref{Q-phi-pi}
show, the measured convergence rates for $S_{0.03}$ quantities are 
significantly poorer than
those from the $S_{0.06}$ calculation. This runs counter to the 
expectation that stronger
gravitational fields and more compact scalar field configurations should 
lead to higher discretization error for a fixed mesh size.  
As argued above, we are then led to suspect that the convergence 
problem is related to the nature of the specific data that defines 
$S_{0.03}$.

\begin{figure}
\begin{center}
\epsfxsize=18.0cm
\epsffile{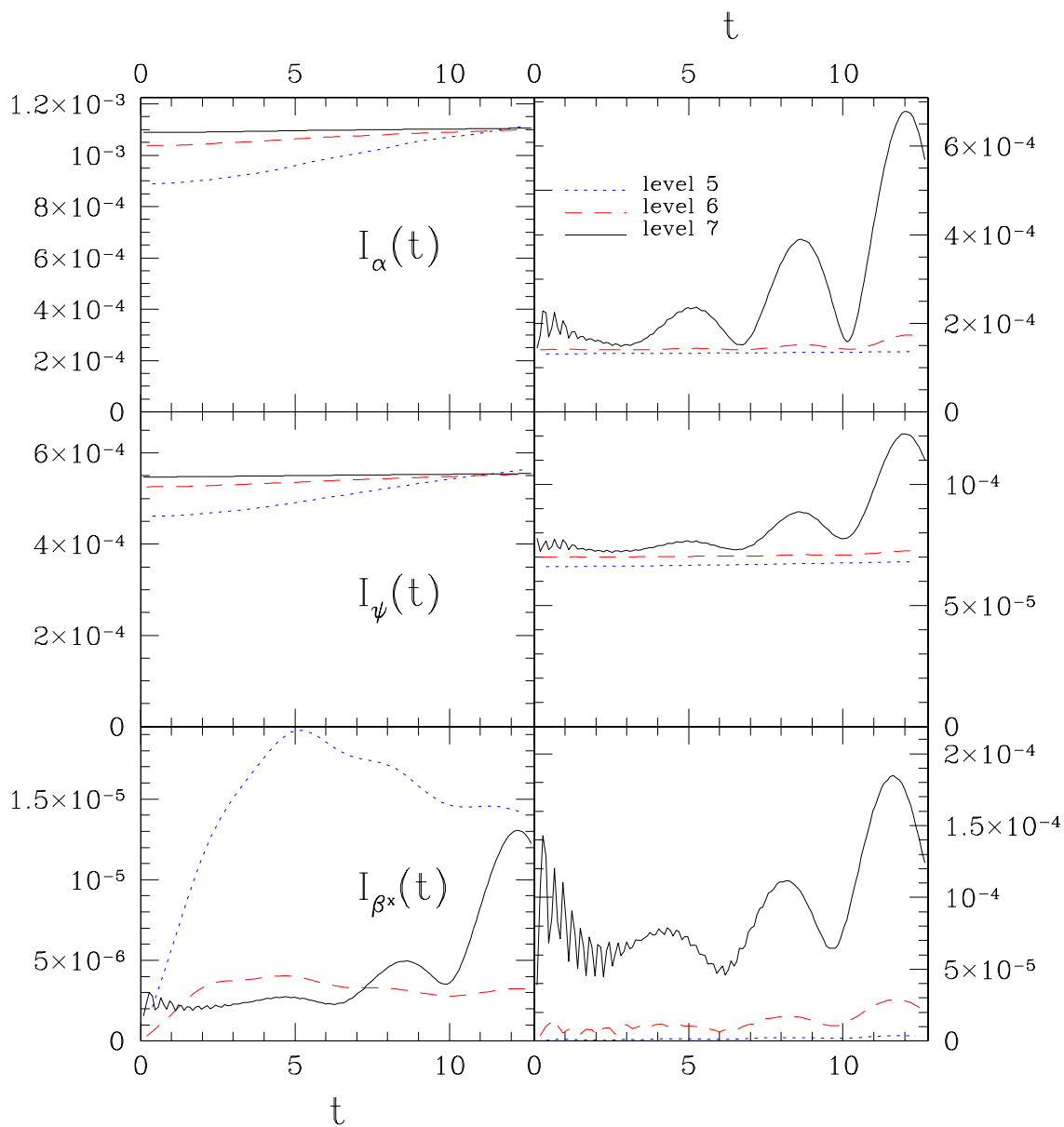}
\end{center}
\caption
[$S_{0.06}$ and $S_{0.03}$ Independent Residual $l_2$-Norms, 
$\Vert I(t)\Vert_2$, for $\al$, $\psi$ and $\bt^x$.]
{Rescaled independent residuals, $\Vert I(t)\Vert_2$, for
$\al$, $\psi$ and $\bt^x$ from evolutions of the two static
configurations $S_{0.03}$ (left panels) and $S_{0.06}$ (right panels).
As previously, the residuals were rescaled 
using~(\ref{eq:indres_resc_geometry}) so that coincidence of curves 
computed at different levels of discretization corresponds to
convergence of the residuals at the expected $O(h^2)$ rate.  
Apart from the residuals associated 
with $\bt^x$ (as well as $\bt^y$ and $\bt^z$, see Fig.~\ref{I-btyz-ph1}),
there is strong evidence that $I(t)$ is $O(h^2)$ for the geometric
quantities.  However, as discussed in the text, 
spurious boundary effects adversely impact the convergence of
the independent residuals for the $S_{0.03}$ case. 
}
\label{I-l-ps-btx}
\end{figure}

\begin{figure}
\begin{center}
\epsfxsize=18.0cm
\epsffile{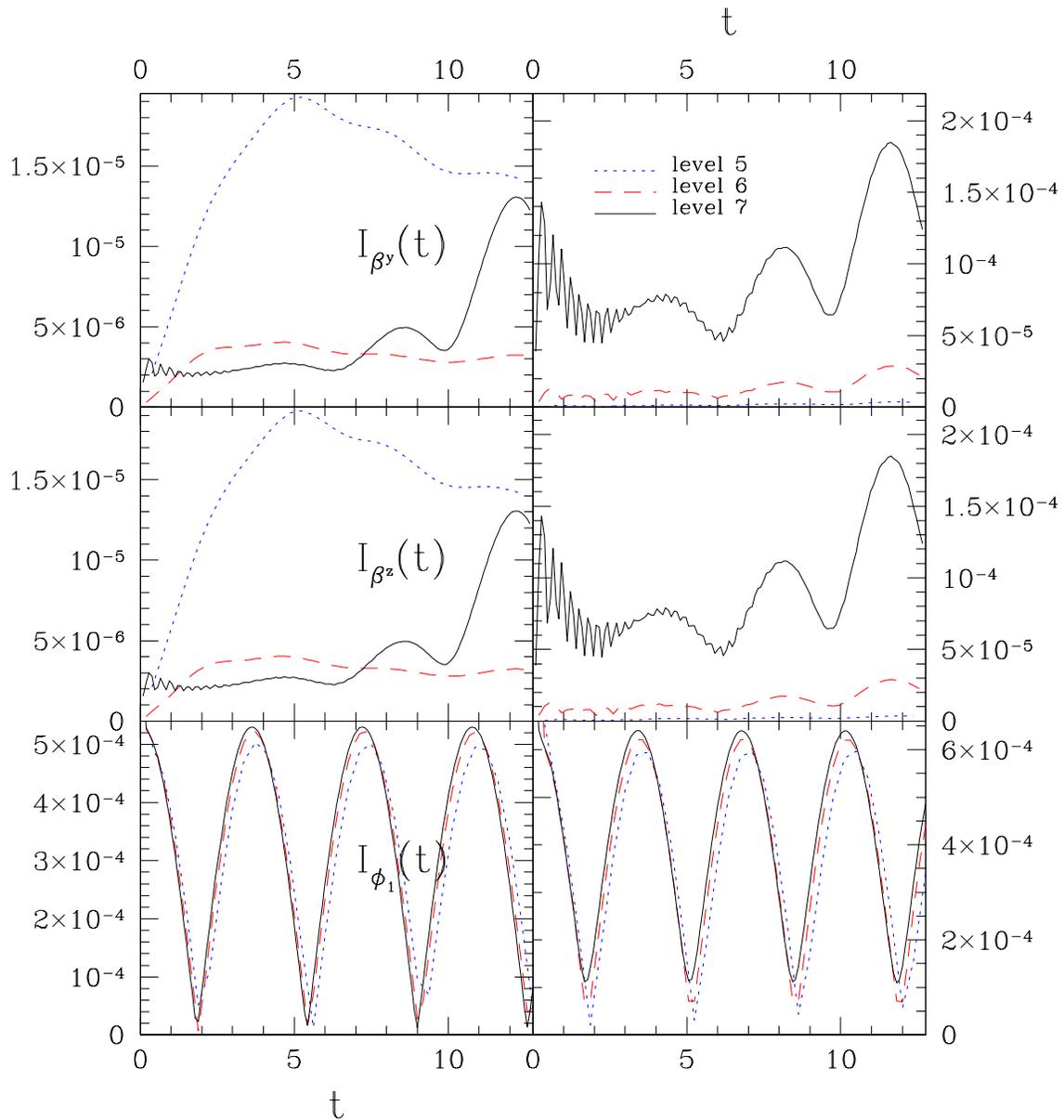}
\end{center}
\caption
[$S_{0.06}$ and $S_{0.03}$ Independent Residual $l_2$-Norms, 
$\Vert I(t)\Vert_2$, for $\bt^y$, $\bt^z$ and $\phi_1$.]
{
Rescaled independent residuals, $\Vert I(t)\Vert_2$, for
$\bt^y$, $\bt^z$ and $\phi_1$
from evolutions of the two static
configurations $S_{0.03}$ (left panels) and $S_{0.06}$ (right panels).
Residuals were rescaled
using~(\ref{eq:indres_resc_geometry}) for $\bt^y$ and $\bt^z$, 
and~(\ref{eq:indres_resc_matter}) for $\phi_1$.  Although there 
is clear evidence for the expected $O(h)$ convergence of $I_{\phi_1}(t)$ 
for both calculations, for the case of the shift vector components 
(and as in the previous figure), the independent residuals are 
clearly {\em not} $O(h^2)$ quantities.  However, as discussed in the 
text, the continuum static solutions have $\bt^i(t,x,y,z)\equiv0$, 
and Fig.~\ref{l2norm-bti} provides clear evidence that for both 
calculations, all components of the shift vector {\em do} converge 
to 0 as $O(h^2)$.  Combined with the results from the test that used 
generic initial data (Sec.~\ref{sec:gen_ID}), 
this suggests that the anomalous behaviour 
of $I_{\beta^i}(t)$ can be traced to the fact that the $\bt^i$
vanish in the continuum limit, and, possibly to boundary effects. 
However, additional investigation is needed to provide a completely 
satisfactory explanation of the observed results.
}
\label{I-btyz-ph1}
\end{figure}

\begin{figure}
\begin{center}
\epsfxsize=18.0cm
\epsffile{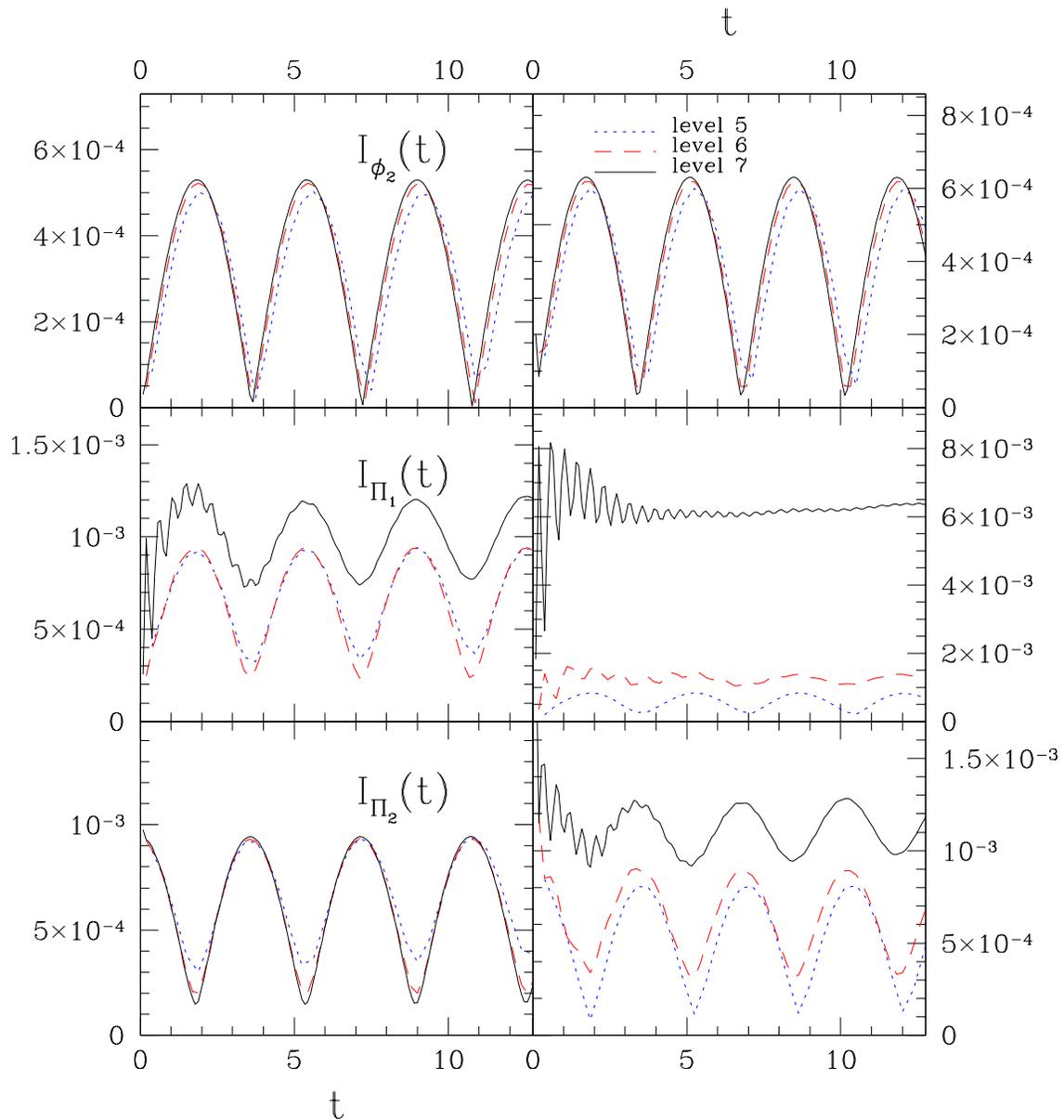}
\end{center}
\caption
[$S_{0.06}$ and $S_{0.03}$ Independent Residual $l_2$-Norms, 
$\Vert I(t)\Vert_2$, for $\phi_2$, $\Pi_1$ and $\Pi_2$.]
{
Rescaled independent residuals, $\Vert I(t)\Vert_2$,
for $\phi_2$, $\Pi_1$ and $\Pi_2$
from evolutions of the two static
configurations $S_{0.03}$ (left panels) and $S_{0.06}$ (right panels).
Residuals were rescaled
using~(\ref{eq:indres_resc_matter}) for $\phi_1$. 
Except for $I_{\Pi_1}(t)$, the expected $O(h)$ convergence of the
residuals is observed for the $S_{0.06}$ data.
As previously noted in the caption of Fig.~\ref{I-l-ps-btx}, and 
as discussed in detail in the text, boundary effects produce
a deterioration of the convergence of the $S_{0.03}$ residuals.
We suspect the same effect is at play in the case of 
$I_{\Pi_1}(t)$ from the $S_{0.06}$ calculations.
}
\label{I-ph2-pis}
\end{figure}

The defect in the $S_{0.03}$ computations, relative to the 
$S_{0.06}$ results, is even more apparent when 
one examines the rescaled independent residuals.
These are shown in Figs.~\ref{I-l-ps-btx}, 
\ref{I-btyz-ph1} and \ref{I-ph2-pis}, which plot $\Vert I(t)\Vert_2$
for all dynamical variables and at all three resolutions used, 
and where the rescaling has been done using~(\ref{eq:indres_resc_geometry}) 
and~(\ref{eq:indres_resc_matter}) as appropriate.
In each panel, the left and right plots correspond to the 
$S_{0.06}$ and $S_{0.03}$ calculations respectively.
Particularly worrisome is the fact that for $S_{0.03}$ the independent 
residuals for the shift vector components and $\Pi_1$ are actually 
{\em increasing} as the resolution decreases (this may not be immediately
obvious due to the rescaling, but is in fact the case).
On the other hand,
consistent with the measured values of $Q^h(t)$, all independent 
residuals for $S_{0.06}$---with the exception of $I_{\Pi_1}$ and 
$I_{\bt^i}$---are
converging as expected.~\footnote{We do not yet understand the difference
in the behaviour of $I_{\Pi_1}$ relative to the independent residuals
for the other matter variables, but note that there is also 
a large ``glitch'' in $Q_{\Pi_1}$ at early times (see Fig.~\ref{Q-phi-pi}) 
that is almost certainly related.}

\begin{figure}
\begin{center}
\epsfxsize=16.0cm
\epsffile{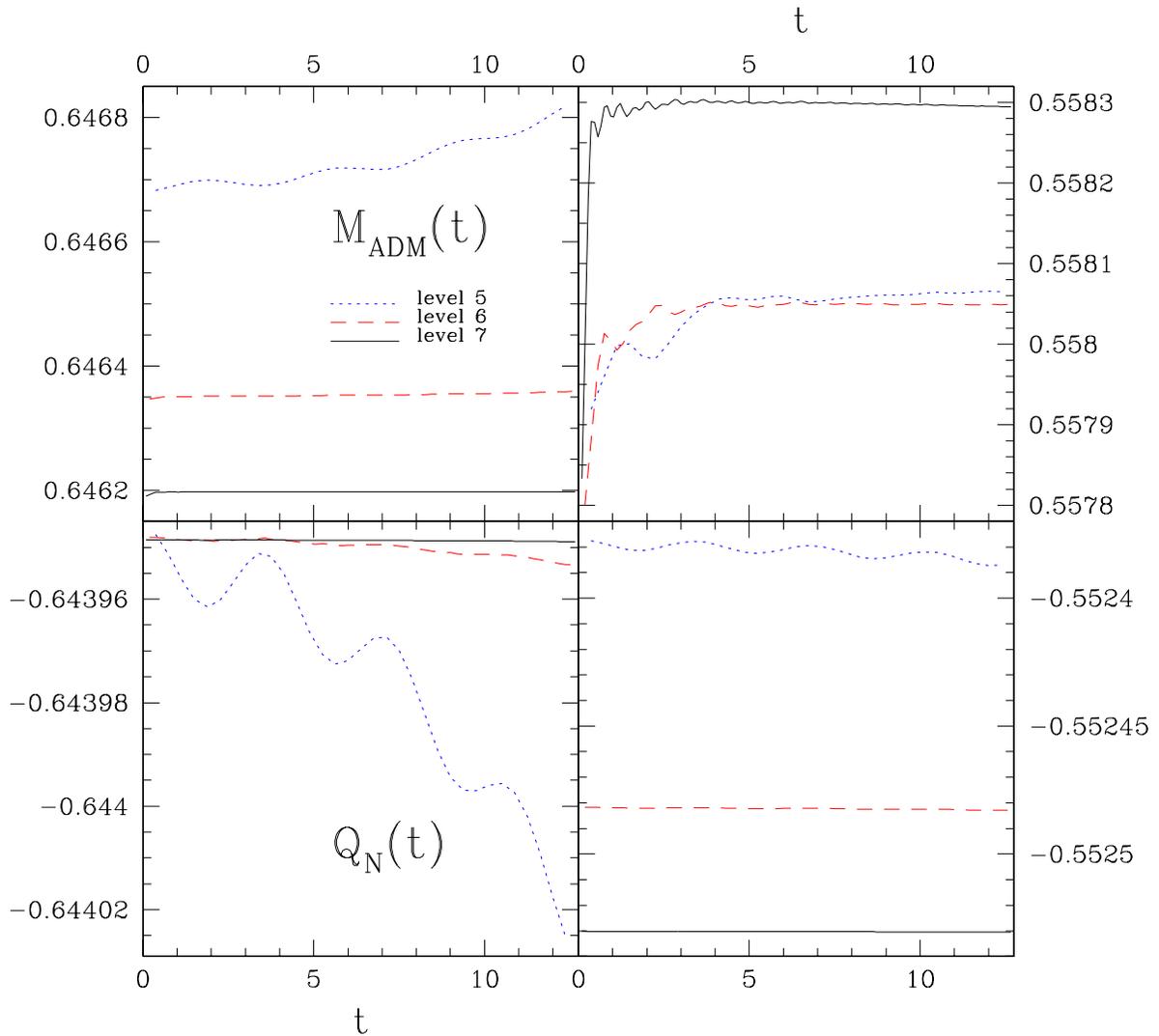}
\end{center}
\caption
[$M_{\rm ADM}(t)$ and $Q_{N}(t)$ for $S_{0.06}$ and $S_{0.03}$.]
{Plots of the ADM mass $M_{\rm ADM}(t)$ and 
Noether charge $Q_{N}(t)$ for $S_{0.06}$ (left) and $S_{0.03}$ (right),
and for the three resolutions used in the calculations.
``Convergence to conservation'' is observed 
in both quantities for $S_{0.06}$, but not for $S_{0.03}$.
}
\label{M-NoetherSS}
\end{figure}

\begin{figure}
\begin{center}
\epsfxsize=16.0cm
\epsffile{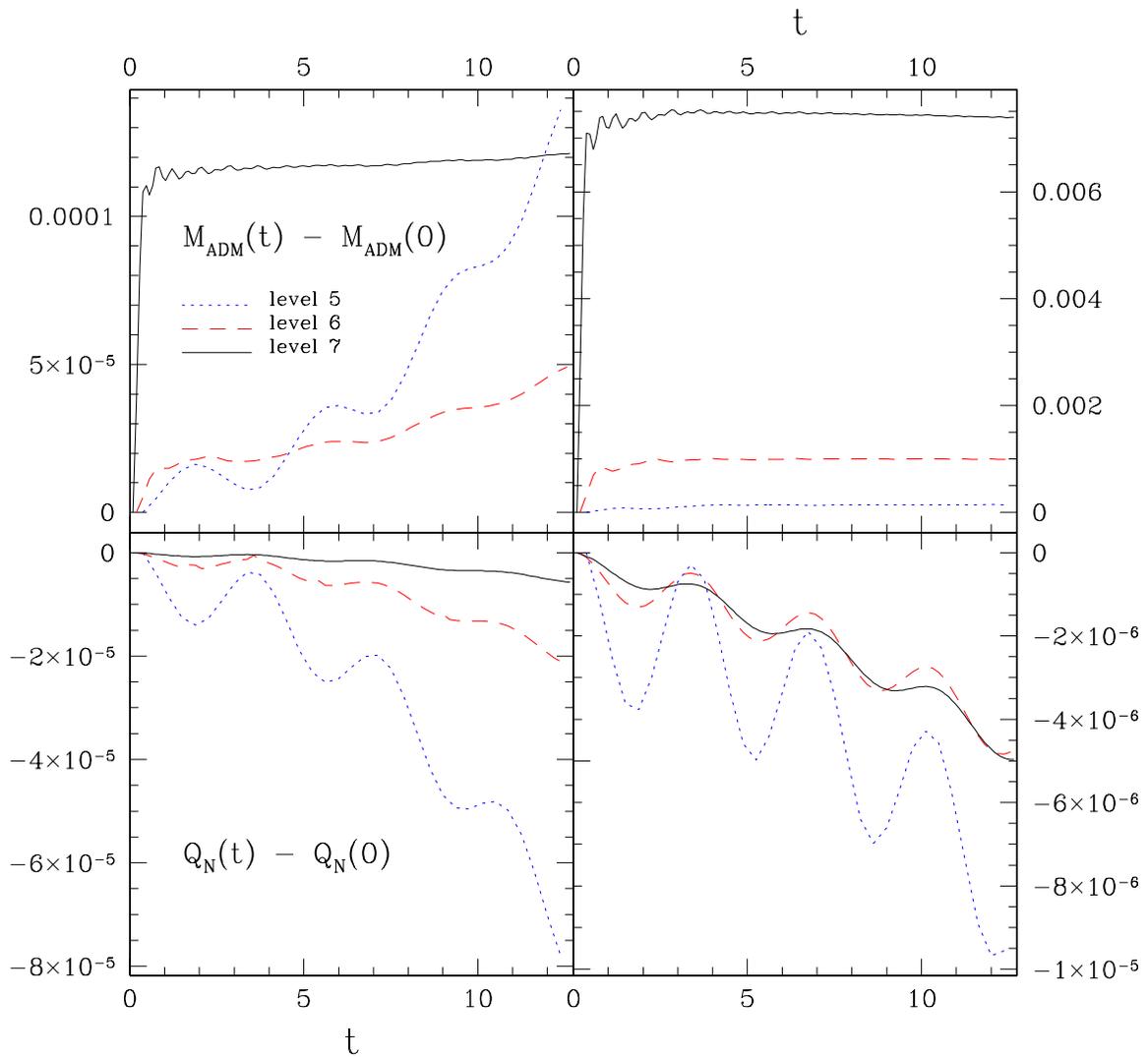}
\end{center}
\caption
[Rescaled Deviations of ADM Mass and Noether Charge for $S_{0.06}$ and $S_{0.03}$.]
{Plots of the rescaled deviations of the ADM mass $M_{\rm ADM}(t)$ and
Noether charge $Q_{N}(t)$ for $S_{0.06}$ (left) and $S_{0.03}$ (right),
and for the three resolutions used in the calculations.
$O(h^2)$ ``convergence to conservation'' is evident for the $S_{0.06}$ data,
while for $S_{0.03}$, the convergence rate is indeterminate at best.
}
\label{dM-NoetherSS}
\end{figure}

Finally, although Fig.~\ref{M-NoetherSS} provides some evidence
that there is convergence to conservation of mass and Noether charge
for both sets of calculations, it is clear that the rate 
of convergence for mass conservation in the $S_{0.03}$ is 
indeterminate at best, and that even for the $S_{0.06}$ computations
it is not definitively second order, 
as we can see from Fig.~\ref{dM-NoetherSS}.

We have already mentioned that these results run counter to 
our intuition that numerical calculations of 
$S_{0.06}$, with its stronger gravitational
fields and steeper scalar field gradients, should be less
accurate than those of $S_{0.03}$, for any given resolution.
Moreover, since $S_{0.03}$ {\em is} less centrally condensed, 
the observed lack of convergence cannot be due to any 
inherent non-smoothness of the solution relative to the scales
of the grids being used.
What has been omitted from the analysis thus far is 
consideration of the boundary conditions. 
As discussed in Sec.~\ref{subsec:bdy_cond}, Dirichlet boundary 
conditions were used for all of the calculations reported in 
this thesis: this means that the outer boundary values for all
dynamical variables remain unchanged relative to their initial
values as the evolution proceeds.  As we argued in that section,
Dirichlet conditions can provide a reasonable approximation to 
the true boundary conditions---which are to be imposed at 
spatial infinity---so long as the boundaries of the 
computational domain are sufficiently far removed from the region 
containing matter.  This leads us to suspect that for $S_{0.03}$
the computational domain defined by $\bbox = [-15,15,-15,15,-15,15]$
is simply not large enough. 

\begin{figure}
\begin{center}
\epsfxsize=15.0cm
\ifthenelse{\equal{\highQ}{true}} {
\epsffile{figs/sphsym/sph003zoom.eps}
}{
\epsffile{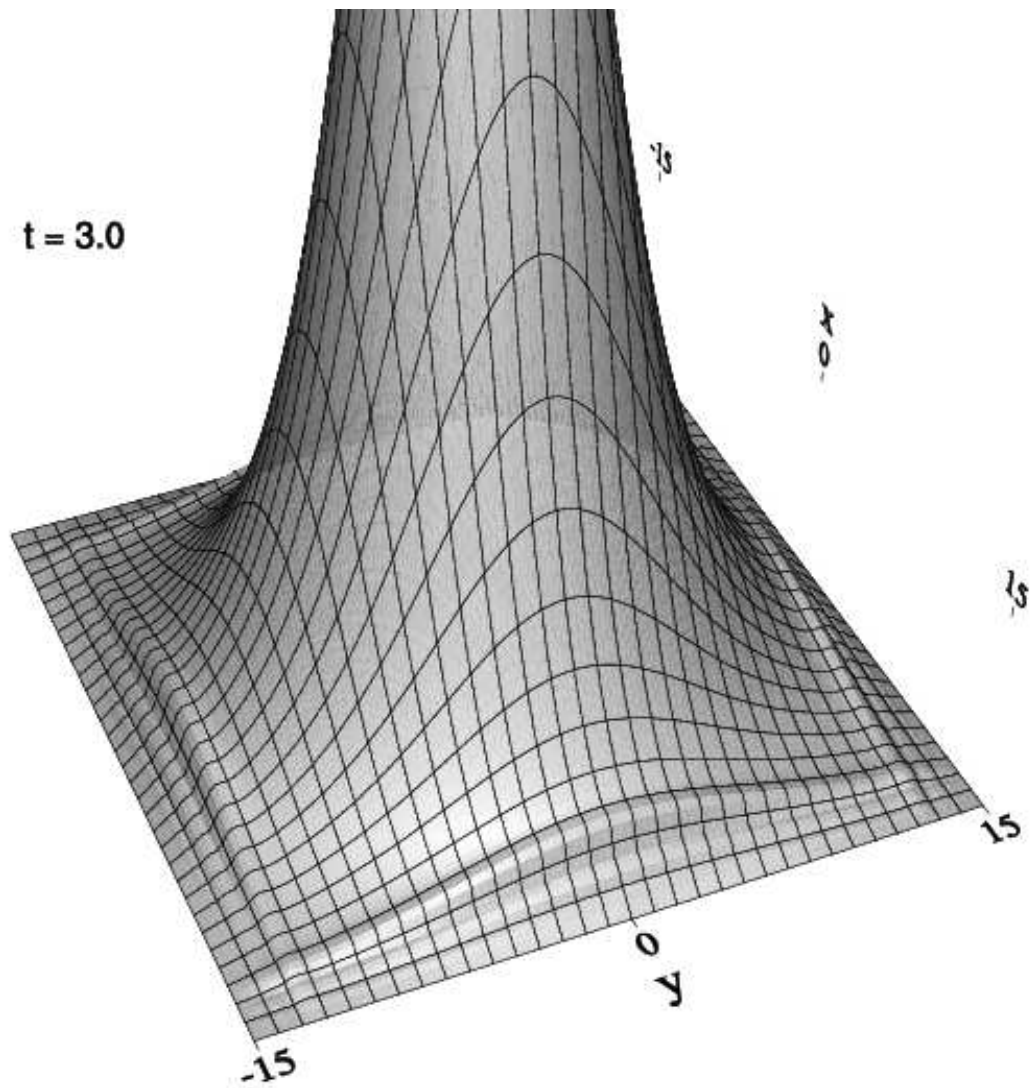}
}
\caption
[Magnified View of $|\phi(3.0,x,y,0)$| from $S_{0.03}$ Evolution.]
{Magnified view of $|\phi(3.0,x,y,0)|$ (i.e. $t=3.0$) from the 
$S_{0.03}$ evolution.  The jump in the solution is estimated to be of 
the order of $1\%$ of $\phi_0\equiv|\phi(0,0,0,0)|.$
}
\label{sph003zoom}
\end{center}
\end{figure}

Indeed, a closer examination of the $t=3.0$ frame of Fig.~\ref{sph003}, 
as shown in Fig.~\ref{sph003zoom}, reveals a discontinuous wave of 
scalar matter that originates at the boundaries of the numerical domain 
and then propagates inwards. The jump in the solution is estimated to be 
of the order of $1\%$ of the central value, $\phi_0$. 
Clearly, this relatively small, but spurious, boundary effect has 
a dramatic impact on the measured convergence of the solution.
As emphasized at the end of the previous section, 
when computing convergence factors and 
independent residuals, we always assume that the continuum
solution is smooth, and that our discrete approximations are 
similarly smooth (on the scale of the mesh).  Here we see 
vividly what can happen to calculated measures of convergence when
this assumption is violated.

The observed propagating discontinuity in the $S_{0.03}$ 
calculation can be immediately connected to the fact that 
the star has a relatively long tail which extends beyond $r=15$. 
This in turn means that there is significant dynamical behaviour 
in the scalar field variables near the boundaries of the computational
domain---in particular at the next-to extremal grid points.  
Since Dirichlet conditions are used, the time variation in the 
function values at these grid points generates the discontinuity which
then travels inwards, eventually contaminating the entire solution.

\begin{figure}
\begin{center}
\epsfxsize=15.0cm
\ifthenelse{\equal{\highQ}{true}} {
\epsffile{figs/sphsym/sph006zoom.eps}
}{
\epsffile{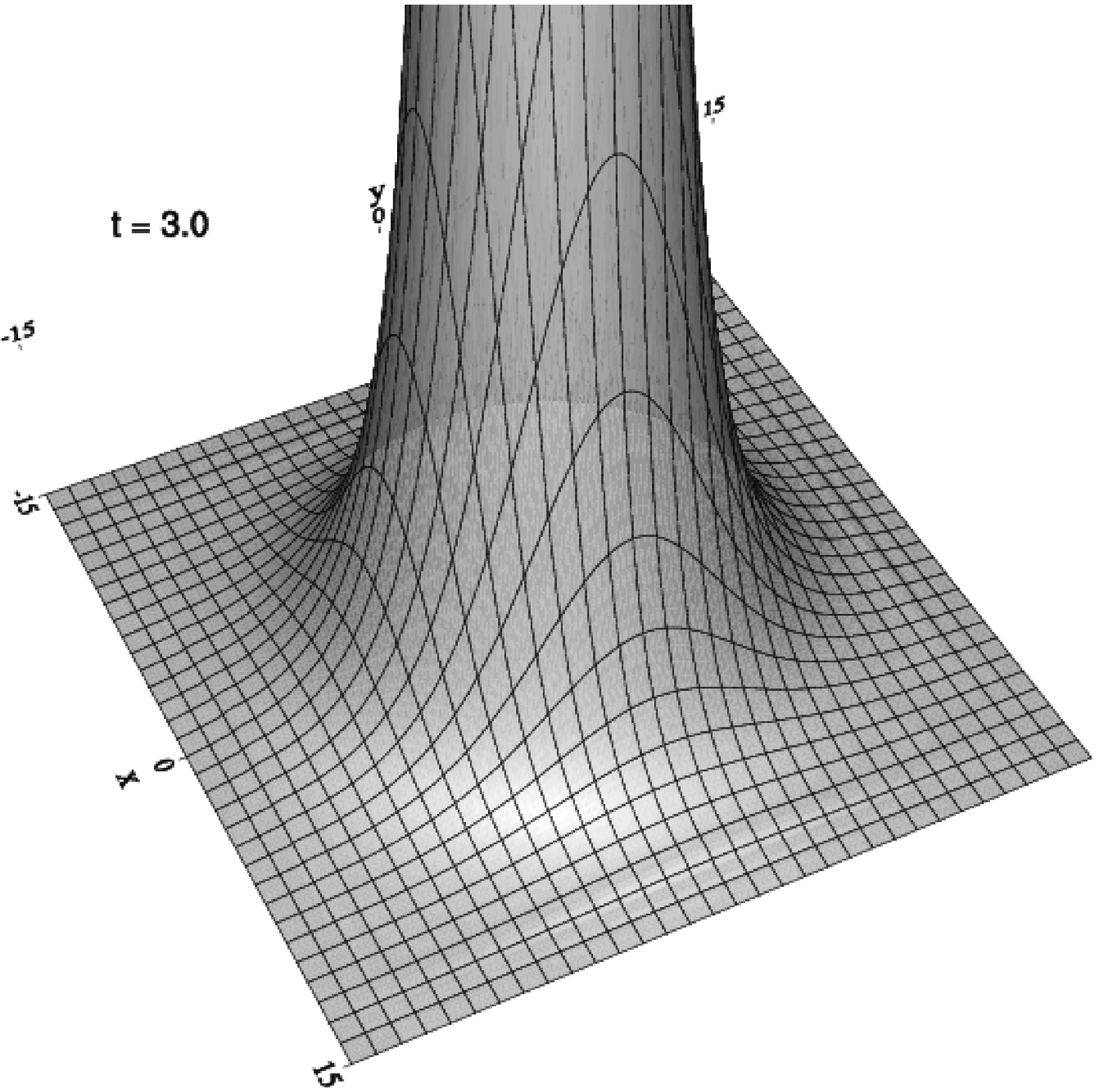}
}
\caption
[Magnified View of $|\phi(3.0,x,y,0)$| from $S_{0.06}$ Evolution.]
{
Magnified view of $|\phi(3.0,x,y,0)|$ from the $S_{0.06}$ evolution. 
Here it virtually impossible to visually detect the discontinuity in 
the scalar field near the computational boundaries.
}
\label{sph006zoom}
\end{center}
\end{figure}

For $S_{0.06}$ this effect is much less pronounced,
as can be seen from Fig.~\ref{sph006zoom}.
Thus, although there is without doubt some level of discontinuous 
behaviour at and near the boundaries in this case, the resulting 
impact on the solution is well below the level of truncation
error at the resolutions used.  In other words, we can consider 
$S_{0.06}$ as providing another example of an initial dataset that
is ``sufficiently smooth'', given the computational domain, 
discretization parameters and boundary conditions. 

To conclude this section we return to the issue of the anomalous
behaviour of the independent residuals, $I_{\bt^i}$, that are 
associated with the shift vector components 
(see Figs.~\ref{I-l-ps-btx} and \ref{I-btyz-ph1}).
Here we argue that the interpretation of the measured values 
of $I_{\bt^i}$ requires special attention for 
the case of static initial data.  

\begin{figure}
\begin{center}
\epsfxsize=18.0cm
\epsffile{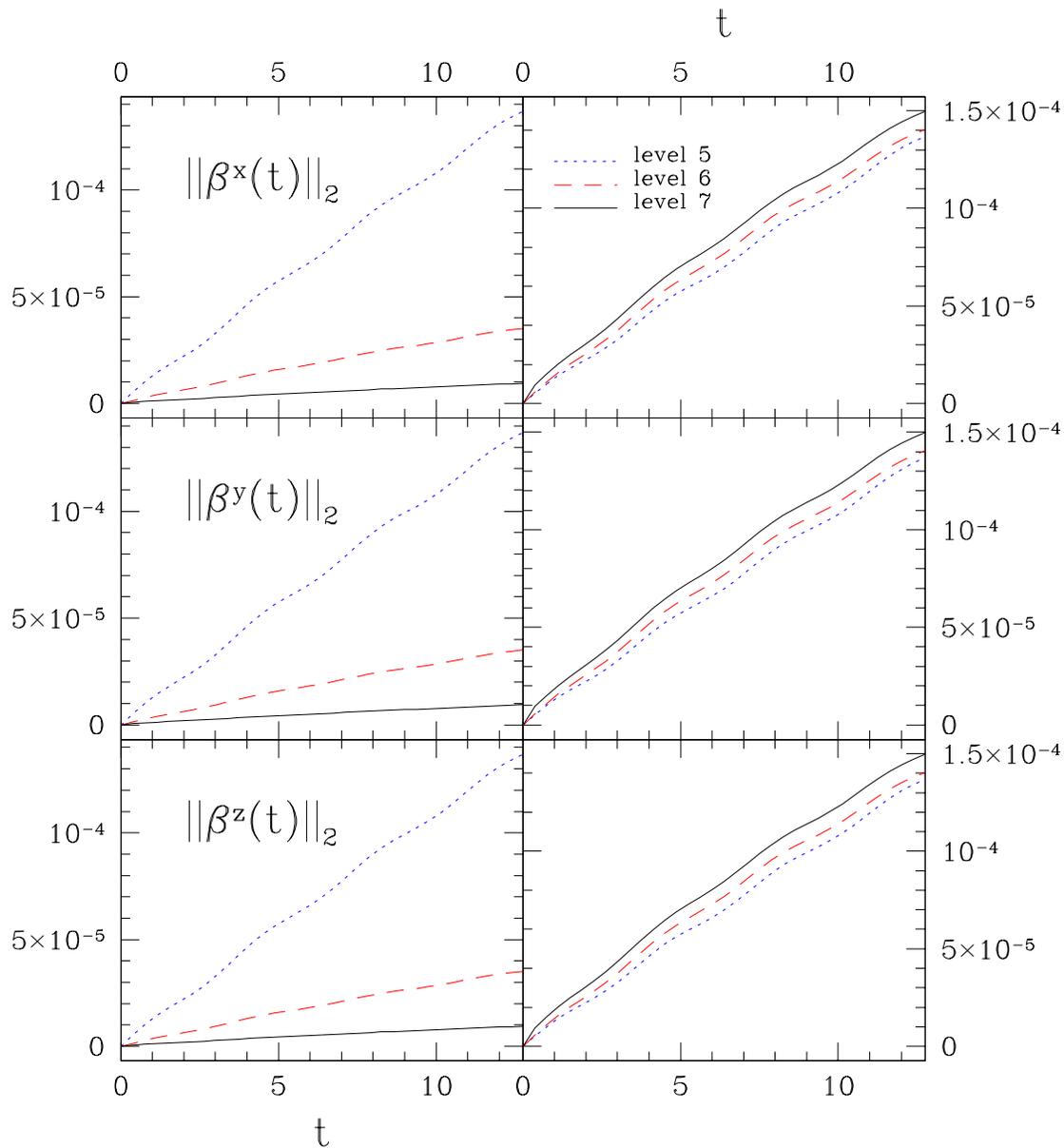}
\end{center}
\caption
[$\Vert \bt^x(t) \Vert_2$, $\Vert \bt^y(t) \Vert_2$ and $\Vert \bt^z(t) \Vert_2$
for $S_{0.06}$.]
{Left panels: Spatial $l_2$-norms of the shift vector components, 
$\Vert \bt^x(t) \Vert_2$, $\Vert \bt^y(t) \Vert_2$ 
and $\Vert \bt^z(t) \Vert_2$ from the $S_{0.06}$ calculation. 
Right panels: Values of these $l_2$ norms rescaled 
using~(\ref{eq:indres_resc_geometry}).  These plots provide strong 
evidence that despite the issues with the independent residuals
$I_{\bt^i}(t)$ displayed in previous figures, and discussed in the 
text, all of the shift vector components are converging to the continuum
limit, $\bt^i(t,x,y,z)\equiv0$, as $O(h^2)$.
}
\label{l2norm-bti}
\end{figure}

As discussed in Sec.~\ref{sec:id-ansatz}, a static 
spherically symmetric spacetime which has a time coordinate adapted 
to the time symmetry---as ours is---implies that $\bt(t,x^i)\equiv0$.
Moreover, the finite difference approximations we use for the elliptic
equations that govern $\bt^i$, in combination
with the multigrid algorithm that we use to solve the resulting 
algebraic equations, guarantees that $[\bt^i]^n_{ijk}\equiv0$ satisfies 
the discretized elliptic equations, provided that $[J^i]^n_{ijk}\equiv0$.
Here the $[J^i]^n_{ijk}$ are the grid values (at discrete time $t^n$) 
of the components of the 3-momentum,
as defined by~(\ref{eq:JA_adm1_cart}).  Now, at the initial
time $t=t^1=0$ we {\em do} have
$[J^i]^1_{ijk}\equiv0$, so that $[\bt^i]^1_{ijk}\equiv0$ as well.
However, the subsequent evolution on 
the Cartesian grid $(x_i,y_j,z_k)$ 
generates non-sphericities in the scalar field 
variables, which leads to non-zero values for $[J^i]^n_{ijk}$ for 
any $t^n > 0$. This in turn results
in $[\bt^i]^n_{ijk} \ne 0$ for any discrete time $t^n\ne0$.
This fact is clearly illustrated in the plots on 
the left hand side of Fig.~\ref{l2norm-bti}, which
show $\Vert \bt^i(t) \Vert_2$ from the 
$S_{0.06}$ computations performed at discretizations levels $5$,
$6$ and $7$.  At any given resolution there is clearly 
approximately linear growth in the norms of all of the shift 
vector components.  However, the plots on the right hand side 
of the figure---which display values of the norms that have
been rescaled by factors $4^{l-5}$---show that all of the components 
are converging to 0 as $O(h^2)$.  Thus, although we do not yet 
understand why the independent residuals for the shift 
components fail to scale as expected for static data, it seems clear
that the failure is not related to a convergence problem of the 
$\bt^i$ themselves.

\newpage

\section{Single Boosted Boson Star} \label{sec:boost_ID}

In the previous section we tested our code using spherically-symmetric
initial datasets representing single boson stars.  Modulo the problems 
identified when using stars not sufficiently compact with 
respect to the size of the computational domain, we observed satisfactory 
convergence of the numerical results to static continuum solutions, which
in addition to providing a further test of the PDE code, also established 
that our procedure for computing the boson star initial data was working.
However, there is one significant piece of our overall process for
setting up initial data for boson stars that has not yet been tested.
To evolve configurations describing stars with non-zero 
initial velocities---as is done in the next two sections---then 
once a spherically symmetric solution for a star has been generated 
and interpolated to the 3D numerical domain (Sec.~\ref{sec:id-1d3d}), 
an (approximate) Lorentz boost is applied (Sec.~\ref{sec:id-boost}). 
This section thus reports the results of the same battery of tests used
in the previous section, but now applied to the special case of 
a boosted spherically symmetric solution.

\begin{figure}
\begin{center}
\epsfxsize=15.0cm
\ifthenelse{\equal{\highQ}{true}} {
\epsffile{figs/sphsymBoost/sph006boost.eps}
}{
\epsffile{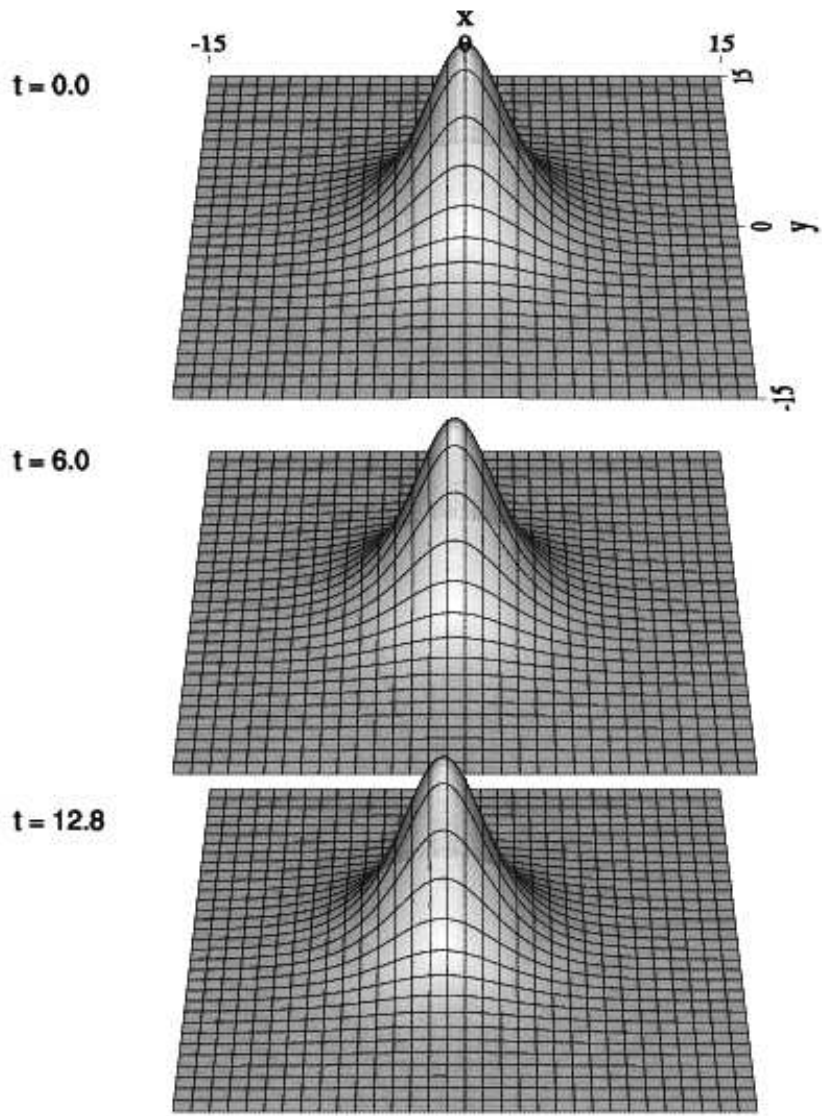}
}
\caption
[Time Evolution of a Single Boosted $S_{0.06}$ Star.]
{Time evolution of a single $S_{0.06}$ star which has been 
boosted in the $x$-direction with a velocity parameter, $v_x=-0.1$
(see Sec.~\ref{sec:id-boost} for a full discussion of the algorithm used to 
determine the boosted data).
Plotted are snapshots of $\vert \phi(t,x,y,0)\vert$ for $t = 0.0$, $6.0$
and $12.8$. 
Careful inspection of 
these frames reveals that the star is slowly moving in the $-x$ direction.  
A rough numerical estimate of the rate of change of the star's coordinate
position with respect to coordinate time yields 
$\De x/\De t \approx -0.1$, consistent with expectations.
}
\label{sph006boost}
\end{center}
\end{figure}

We perform the tests on the star $S_{0.06}$
and maintain the same grid parameters employed in the previous section:
$\bbox = [-15,15,-15,15,-15,15]$, using three levels of resolution
ranging from $\shape = [33,33,33]$ to $\shape=[129,129,129]$.  The 
star is boosted in the $x$-direction with a boost parameter $v_x=-0.1$.
Fig.~\ref{sph006boost} displays some snapshots of 
$\vert\phi(t,x,y,0)\vert$ 
from an evolution computed on the finest grid.  Careful study of the 
sequence shows that the star is slowly (relative to the time scale of the 
sequence) moving across the computational domain from right to left 
(i.e.~in the $-x$ direction).
A rough numerical estimate of the rate of change of the star's coordinate 
position with respect to coordinate time yields $\De x/\De t \approx -0.1$,
consistent with expectations.

\begin{figure}
\begin{center}
\epsfxsize=18.0cm
\epsffile{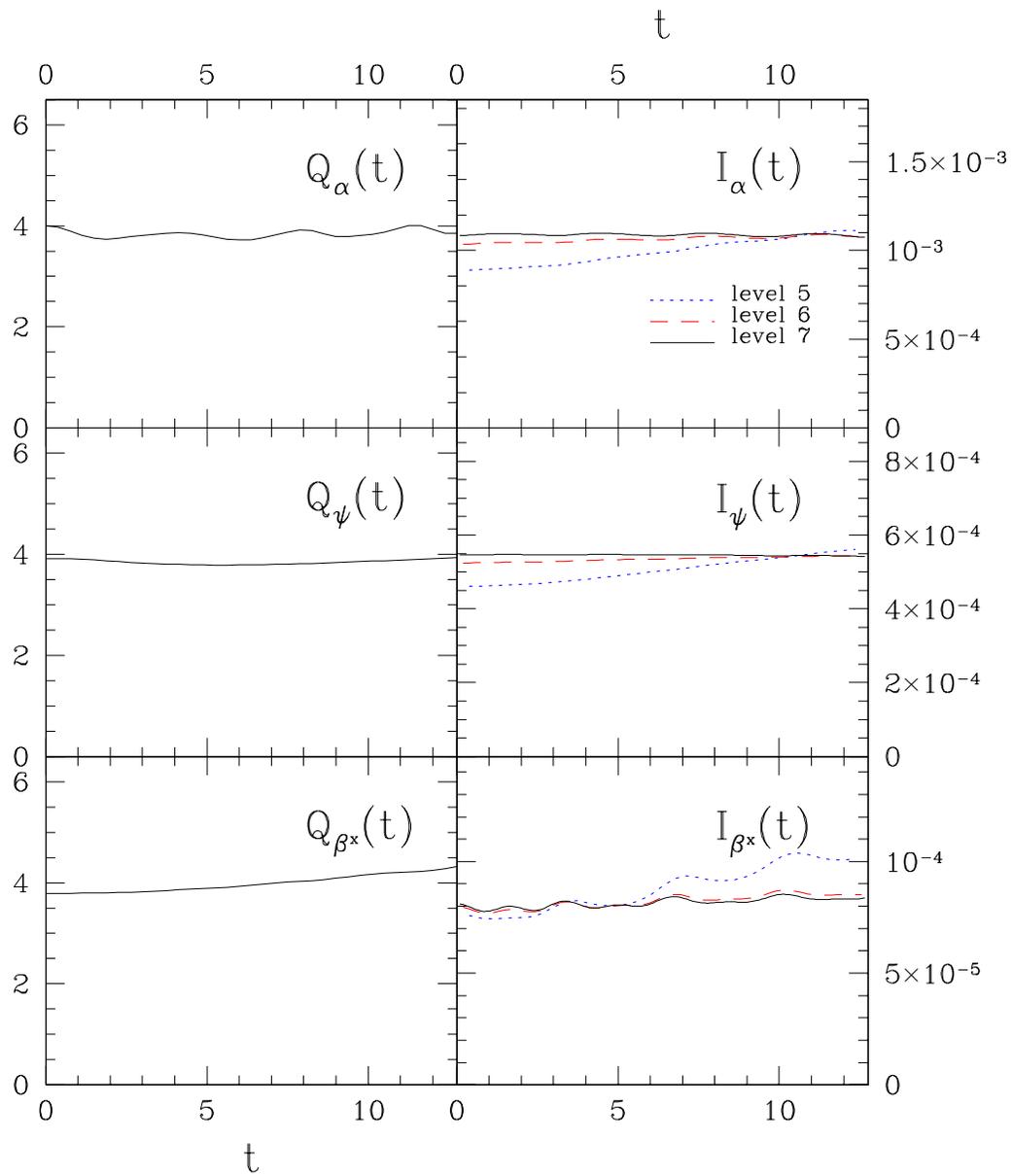}
\end{center}
\caption
[Single Boosted $S_{0.06}$ Star: 
$Q^h(t)$ and $\Vert I(t)\Vert_2$ for $\al$, $\psi$ and $\bt^x$.]
{
$Q^h(t)$ (left) and rescaled $\Vert I(t)\Vert_2$ (right) for $\al$, $\psi$ and 
$\bt^x$ from the boosted $S_{0.06}$ experiment.
Values of the independent residuals norms $\Vert I(t)\Vert_2$ have been 
rescaled using~(\ref{eq:indres_resc_geometry}). 
These plots provide strong evidence of $O(h^2)$ convergence for all
three of the geometric variables.
}
\label{boost-QI-l-ps-btx}
\end{figure}

\begin{figure}
\begin{center}
\epsfxsize=18.0cm
\epsffile{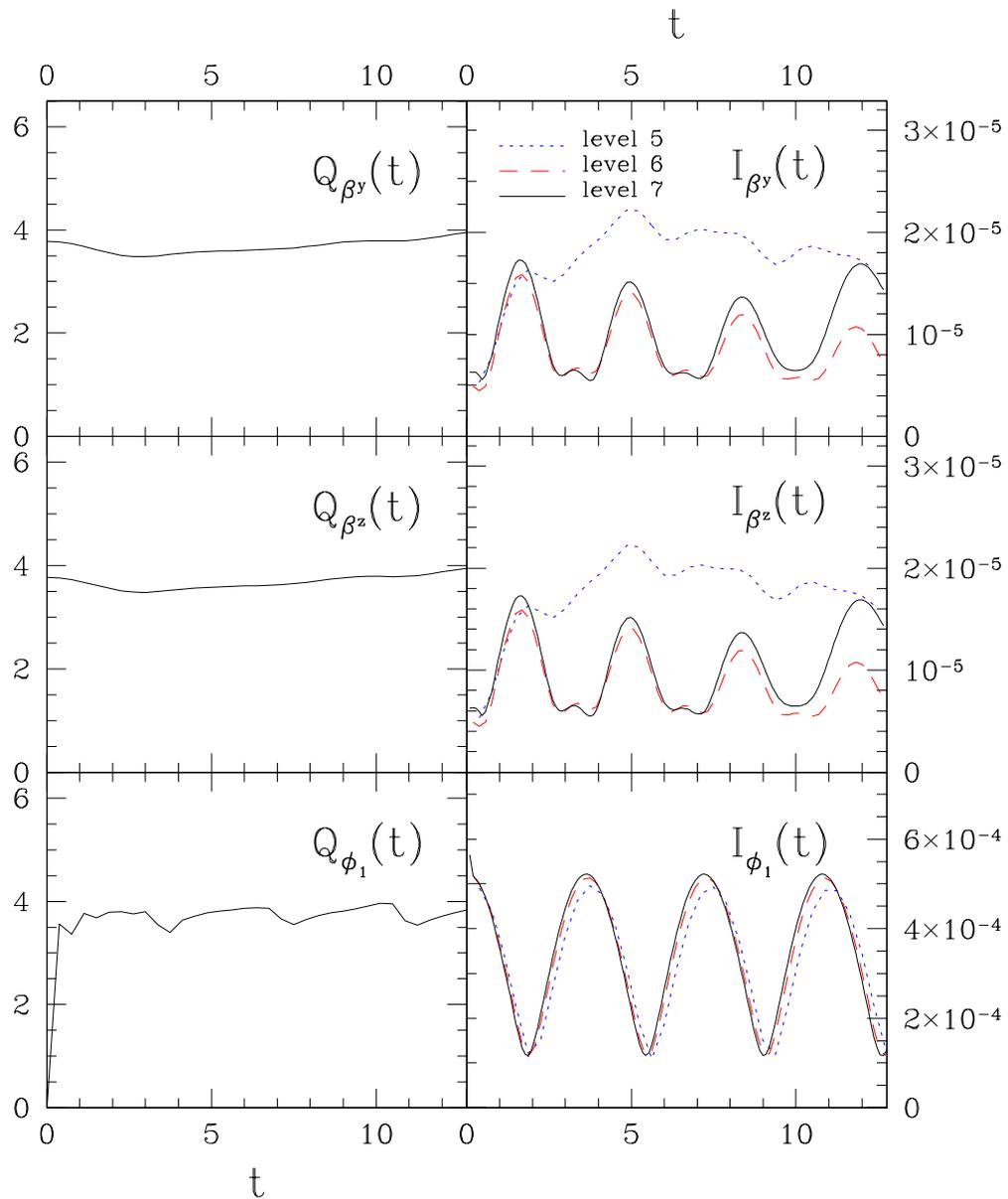}
\end{center}
\caption
[Single Boosted $S_{0.06}$ Star: 
$Q^h(t)$ and $\Vert I(t)\Vert_2$ for $\bt^y$, $\bt^z$ and $\phi_1$.]
{
$Q^h(t)$ (left) and rescaled $\Vert I(t)\Vert_2$ (right) for 
$\bt^y$, $\bt^z$ and $\phi_1$ from the boosted $S_{0.06}$ calculations.
Values of the independent residual norms $\Vert I(t)\Vert_2$ have been
rescaled using~(\ref{eq:indres_resc_geometry}) for $I_{\bt^y}(t)$ and 
$I_{\bt^z}(t)$ 
and~(\ref{eq:indres_resc_matter}) for $I_{\phi_1}(t)$.
These plots provide strong evidence of $O(h^2)$ convergence for all
of $\bt^y$, $\bt^z$ and $\phi_1$.
}
\label{boost-QI-btyz-ph1}
\end{figure}

\begin{figure}
\begin{center}
\epsfxsize=18.0cm
\epsffile{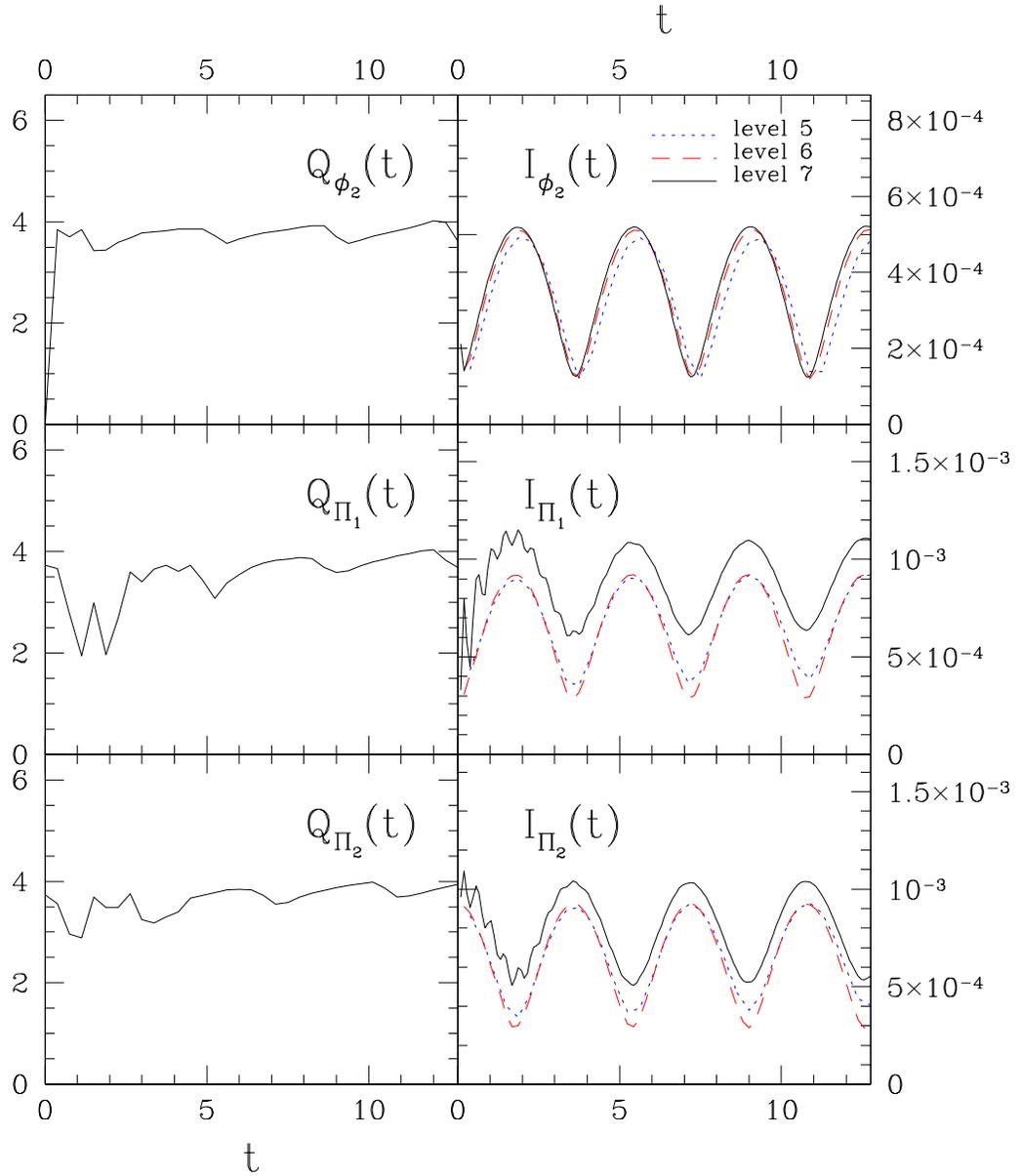}
\end{center}
\caption
[Single Boosted $S_{0.06}$ Star: 
$Q^h(t)$ and $\Vert I(t)\Vert_2$ for $\phi_2$, $\Pi_1$ and $\Pi_2$.]
{
$Q^h(t)$ (left), and rescaled $\Vert I(t)\Vert_2$ (right) for
$\bt^y$, $\bt^z$ and $\phi_1$ from the boosted $S_{0.06}$ calculations.
Values of the independent residual norms $\Vert I(t)\Vert_2$ have been
rescaled using~(\ref{eq:indres_resc_matter}).
Here we observe some irregularities in the norms of $I_{\Pi_1}(t)$ 
and $I_{\Pi_2}(t)$ that we suspect may again be due to our use 
of Dirichlet conditions on a computational domain whose extent is 
comparable to the size of the star.   Nonetheless, these plots still
provide strong evidence for $O(h^2)$ convergence of the scalar 
field variables.
}
\label{boost-QI-ph2-pi12}
\end{figure}

\begin{figure}
\centering
\includegraphics[width=7.4cm]{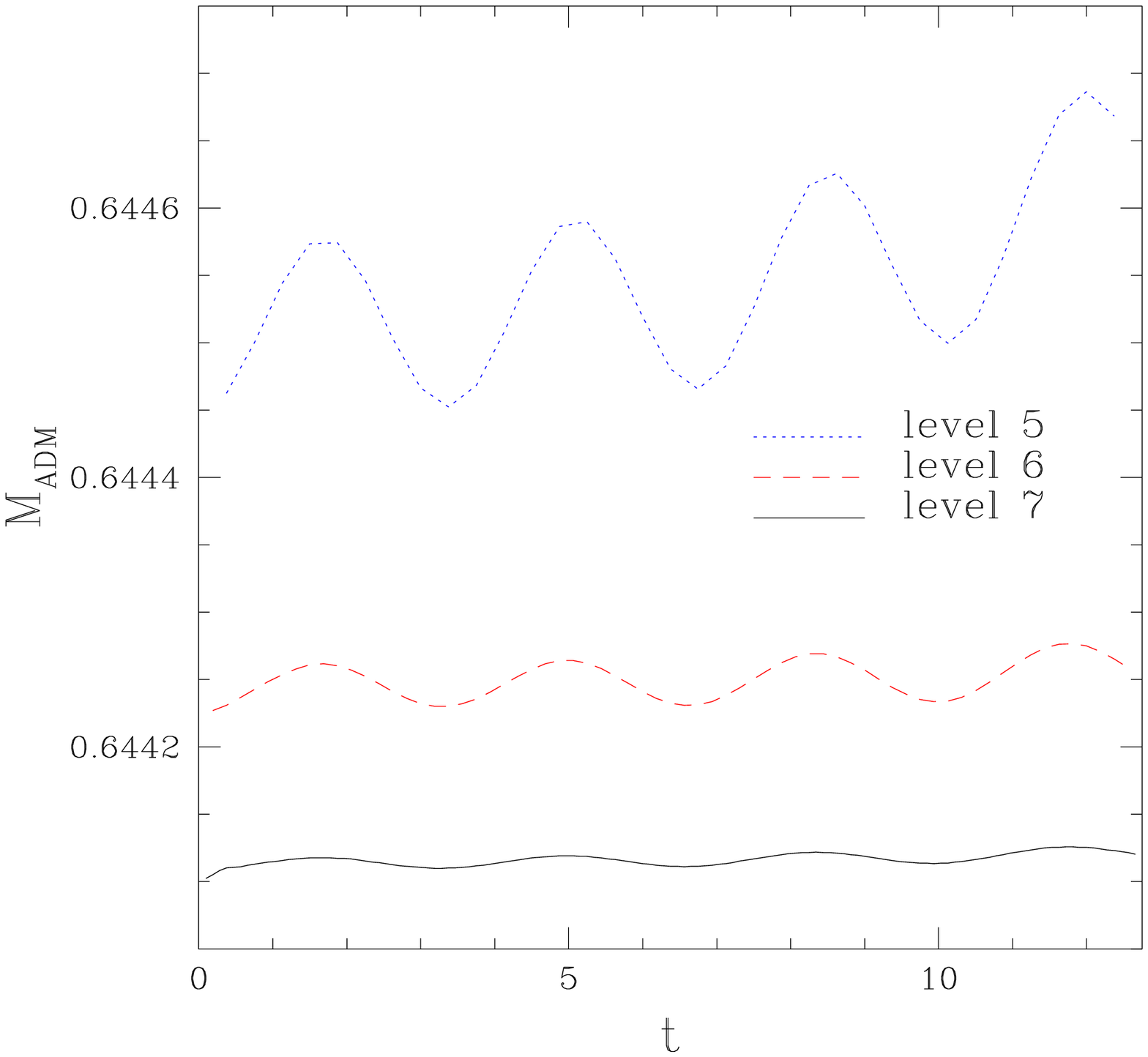}
\includegraphics[width=7.4cm]{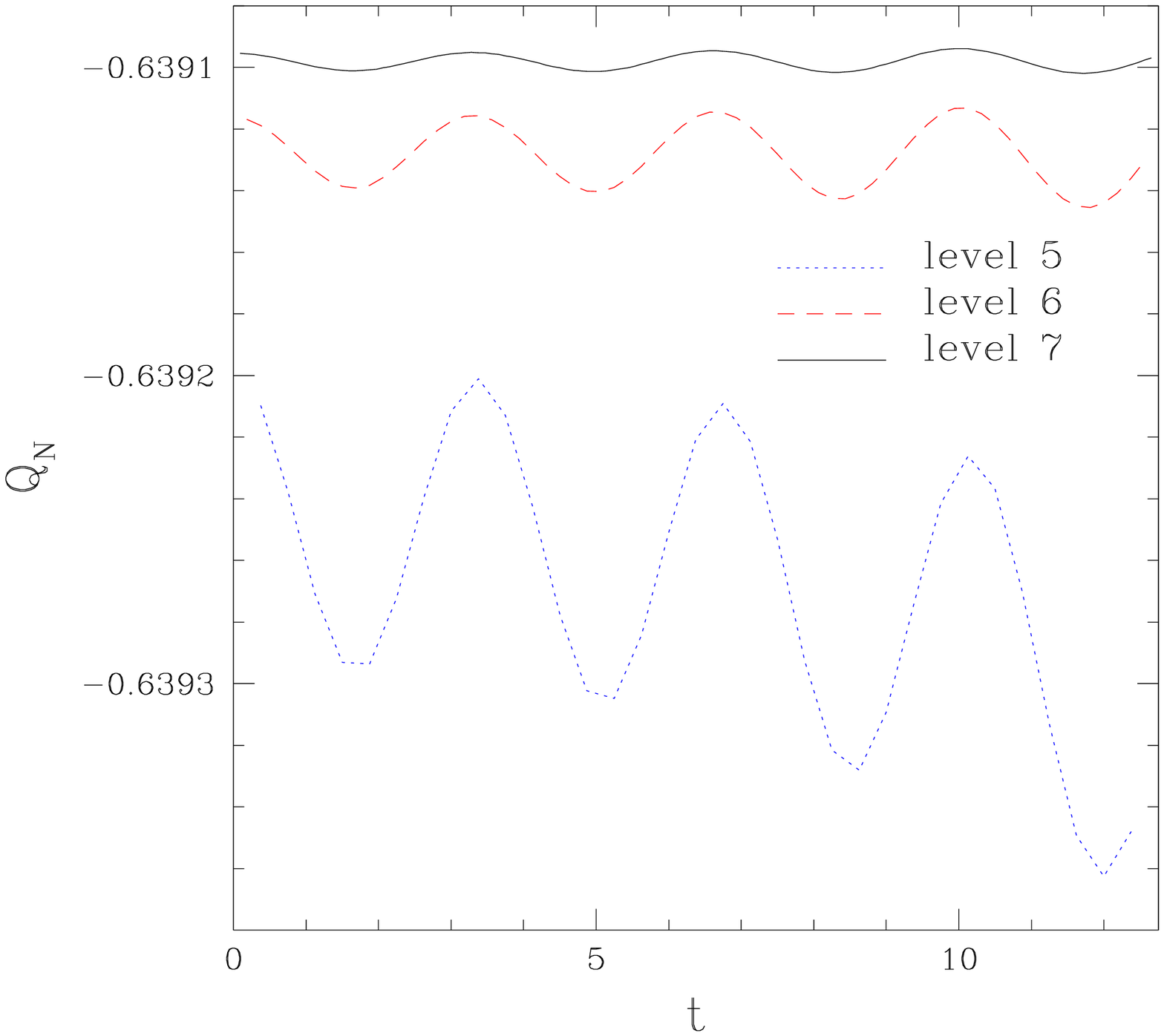}
\caption
[ADM Mass $M_{\rm ADM}(t)$ and Noether Charge $Q_{N}(t)$ for Boosted $S_{0.06}$ Star.]
{ADM mass $M_{\rm ADM}(t)$ (left) and Noether charge $Q_{N}(t)$ (right) 
for the 
boosted $S_{0.06}$ calculation.
These plots suggest that both the ADM mass and Noether charge 
are conserved as $h\to0$. 
}
\label{massBoost}
\end{figure}

\begin{figure}
\centering
\includegraphics[width=7.4cm]{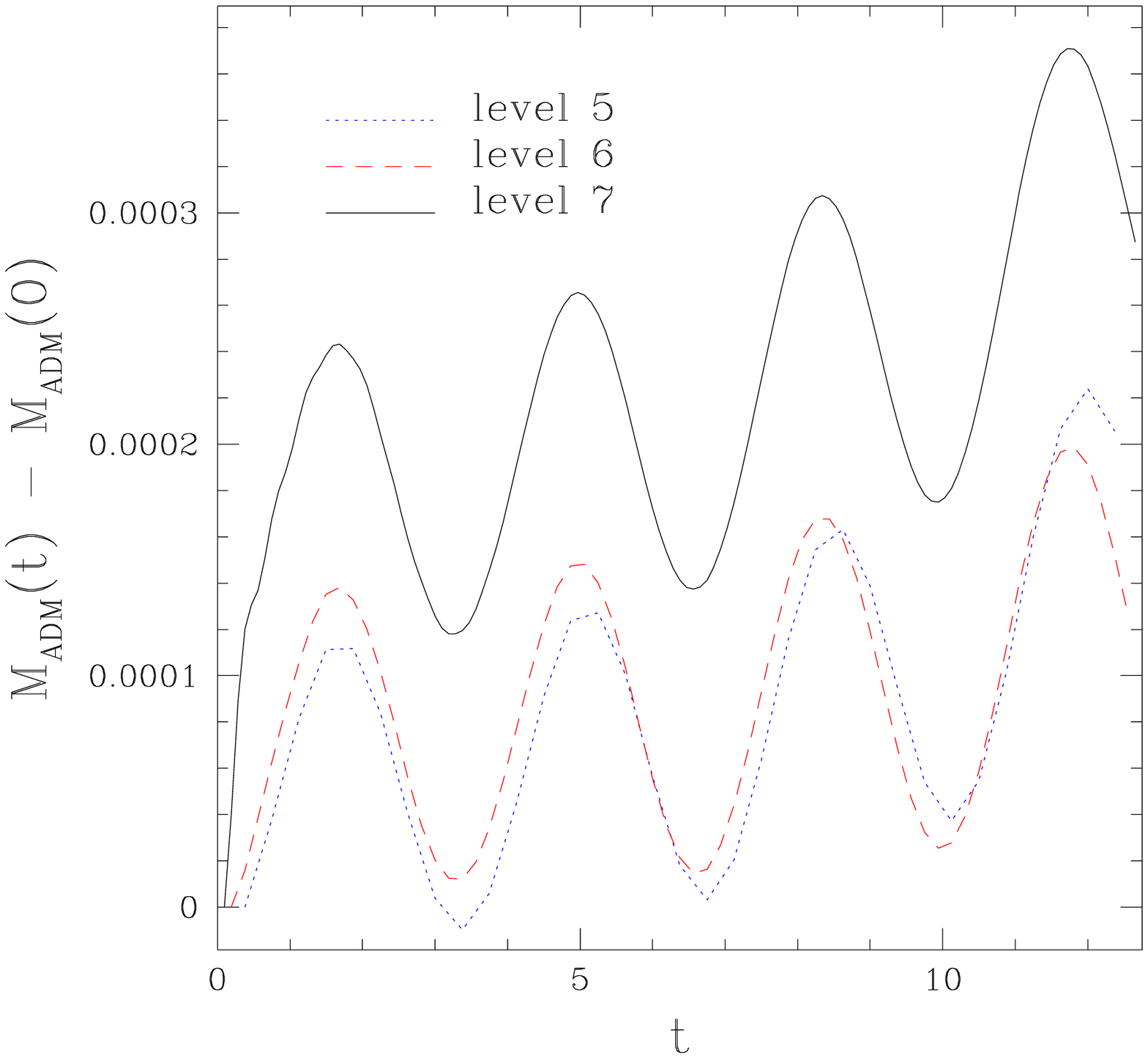}
\includegraphics[width=7.4cm]{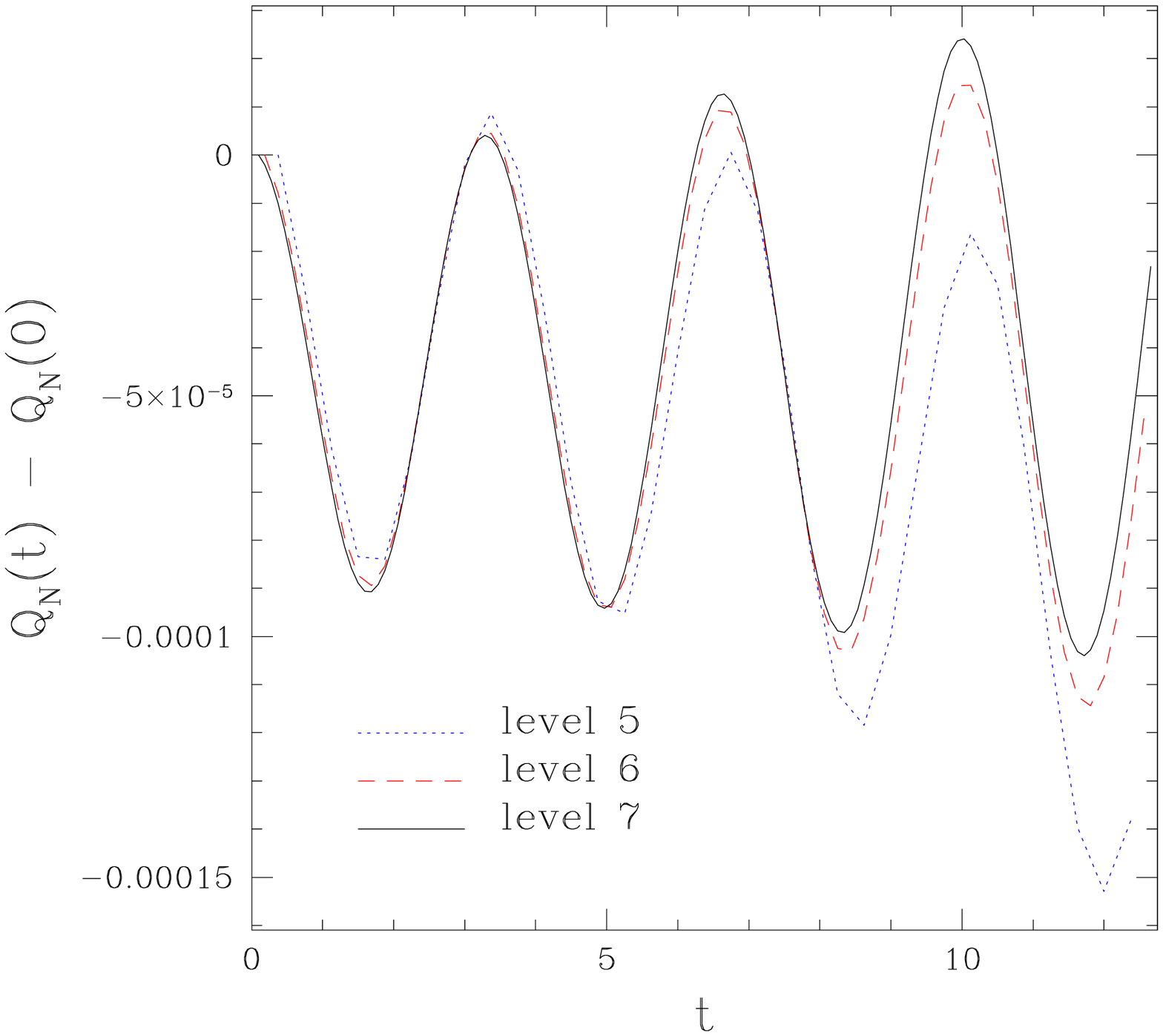}
\caption
[ADM Mass and Noether Charge Deviations 
for Single Boosted $S_{0.06}$ Star.]
{Rescaled deviations of ADM mass (left) and Noether charge (right), 
relative to their initial values, 
for the boosted $S_{0.06}$ calculation.
The deviations were rescaled using~(\ref{eq:DM_ADM}) and (\ref{eq:DQ_N}). 
Although it seems evident that the computed Noether charge is converging
to a conserved value as $O(h^2)$, the convergence rate for $M_{\rm ADM}$
cannot be determined with certainty using the available data.
}
\label{dmassBoost}
\end{figure}

Convergence factors, $Q^h(t)$, and independent residual norms, 
$\Vert I(t)\Vert_2$, are shown in the left and right panels, respectively,
of Figs.~\ref{boost-QI-l-ps-btx}, \ref{boost-QI-btyz-ph1}
and \ref{boost-QI-ph2-pi12}.  Once again, these results provide strong
evidence that the numerical solution is converging to a continuum 
solution of the PDEs governing our model, and at the expected $O(h^2)$ 
rate.  Additionally, Fig.~\ref{massBoost} and Fig.~\ref{dmassBoost} indicate 
that both the
ADM mass and the Noether charge are being conserved as $h\to0$, and that 
the deviations from conservation are also $O(h^2)$.

We hope that the results from the extensive and comprehensive set 
of tests that we have described thus far in this chapter will have 
convinced the reader of the following:
\begin{itemize}
\item  That we have consistently discretized the 
   equations of motion for our model to $O(h^2)$ in both space and 
   time.
\item That our implementation of the resulting finite difference approximation 
	is correct, and produces convergent (and thus, implicitly, stable) 
	results.
\item That our procedures for computing and boosting single boson 
   stars are also correct.
\item That convergence testing will reveal situations where the 
   numerical solution has developed significant non-smoothness on the 
   scale of the mesh.
\end{itemize}

We now proceed to a discussion of the key numerical calculations in this 
thesis: those that involve two boson stars in interaction with one 
another.

\newpage
\section{Head-on Collision of Two Boson Stars} \label{sec:headon}

In this section we report the results from an numerical experiment
describing the head-on collision of two boson stars which
are initially boosted towards one another.
We note that head-on collisions of self-gravitating boson stars have been 
investigated previously by other researchers. The prior
studies include work by: Choi~\cite{dale,ChoiBEC} who implemented 
an axisymmetric Newtonian code,~\footnote{This required the
solution of the Schr\"odinger equation for the complex scalar field, 
and the Poisson equation for the Newtonian gravitational field.}
Balakrishna \cite{balakrishna:phd}, using a 3D fully relativistic code,
Lai~$\cite{cwlai:phd}$ who used a fully general relativistic, but 
axisymmetric code, Palenzuela {\em et al}~\cite{Palenzuela:2006wp},
using a 3D general-relativistic code, and most recently 
Choptuik and Pretorius~\cite{Choptuik:2009ww} who used an axisymmetric 
general-relativistic code. One particularly interesting 
effect that Choi first demonstrated for the Newtonian case is so-called
``solitonic behaviour'', wherein upon collision the stars interpenetrate 
and effectively pass through one another, emerging from the interaction
relatively unscathed.  
Subsequently, this behaviour has also been seen in the general relativistic 
calculations reported in~$\cite{cwlai:phd}$ and~\cite{Choptuik:2009ww}.
Here we show that the solitonic effect is also present within the 
context of our current model, which in some loose sense is an 
intermediary between the purely Newtonian and fully general relativistic 
calculations.

The experiment was prepared as follows.  First, we adopted a discrete 
domain defined by $\bbox=[-50,50,-25,25,-25,25]$.  Note that this 
yields a computational volume with an $x:y:z$ aspect ratio of 
$2:1:1$, and that the coordinate extent in the $y$ and $z$ directions is 
roughly double that used in the calculations reported in previous 
sections.  This last point is important since the star configuration
that we used in the collision is relatively extended, so we 
needed to be careful to ensure that our calculations were not significantly
affected by the boundary-discontinuity issue described in Sec.~\ref{sec:SS_ID}.

Next we generated initial 
data for the star $S_{0.02}$ (i.e.~a star with central field modulus 
$\phi_0=0.02$) and the star $S_{0.03}$.  
Both solutions were then interpolated
onto the 3D numerical domain with centres at $(25,0,0)$ and $(-25,0,0)$,
respectively.  The stars were then given boosts in the $x$ direction
having equal magnitudes but opposite senses: $v^1_x =-0.4$ and $v^2_x= 0.4$.
This of course results in an initial configuration in which the 
stars are already approaching one another with a significant relative
velocity.

\begin{figure}
\begin{center}
\epsfxsize=15.0cm
\ifthenelse{\equal{\highQ}{true}} {
\epsffile{figs/headondiffM/headondiffM1.eps} 
}{
\epsffile{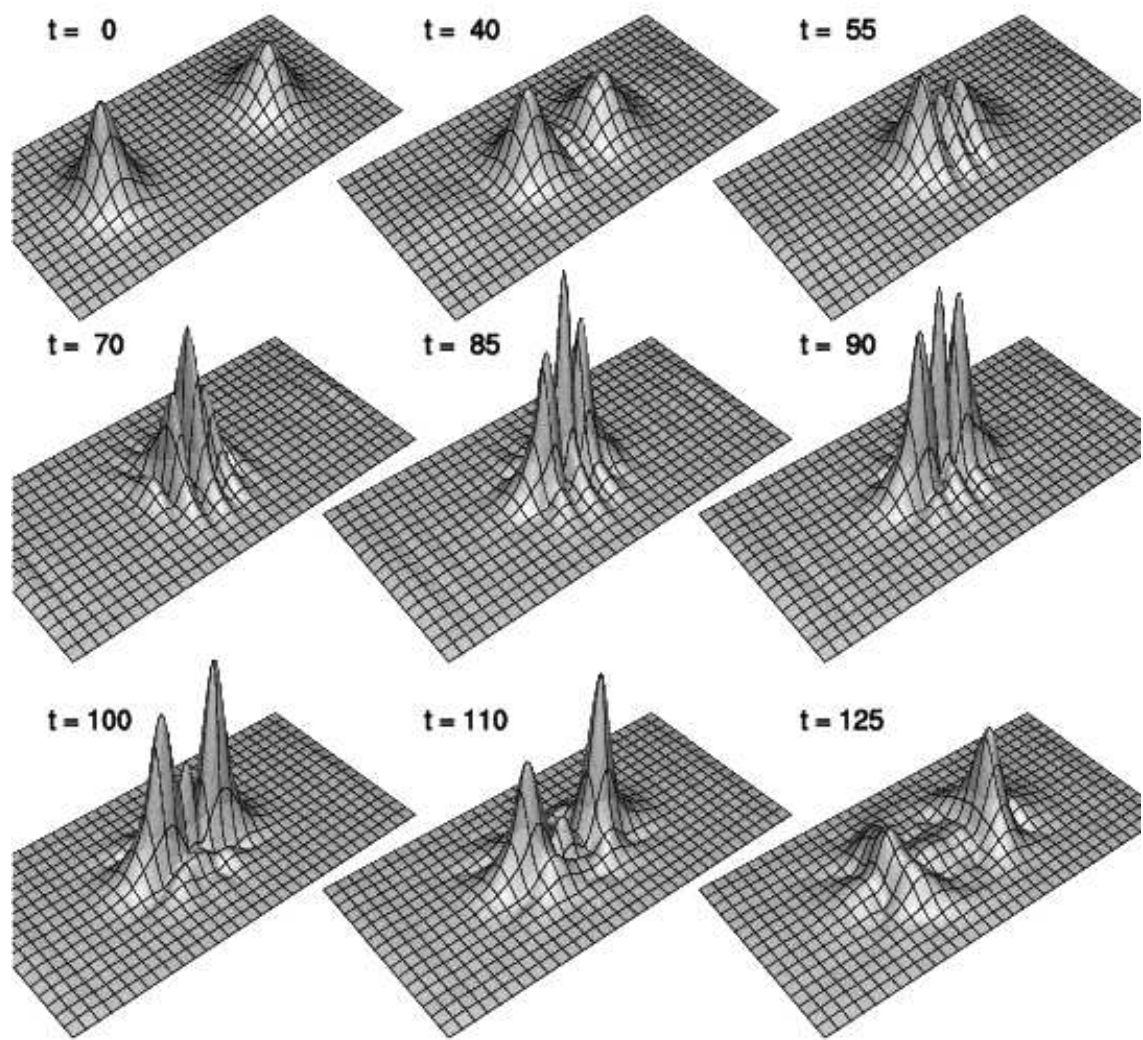} 
}
\caption
[Head-on Collision of Boson Stars Solitonic Dynamics.]
{Head-on Collision of Boson Stars: Solitonic Dynamics.
This figure shows a series of snapshots from the 
time evolution of $\vert\phi(t,x,y,0)\vert$, for 
$0\le t \le 125$.
The initial configuration describes two different boson stars,
$S^1_{0.02}$ and $S^2_{0.03}$, centred 
at $(25,0,0)$ and $(-25,0,0)$, respectively, and boosted with
velocity parameters $v^1_x =-0.4$ and $v^2_x= 0.4$.
At time $t \approx 40$, the stars are interacting substantially, 
and by $t=55$ it is virtually impossible to identify the individual 
stars.  During this interaction epoch ``interference patterns'' 
that are characteristic of boson star collisions of this type 
are evident. 
By $t \approx 100$ the period of interaction has essentially concluded
and distinct star-like configurations can again be identified.  The stars
emerge with their shape roughly preserved---thus exhibiting ``solitonic''
dynamics---and continue to propagate in the directions of their 
respective original boosts.
}
\label{soliton}
\end{center}
\end{figure}

The computations performed in this section were made using three different
grid resolutions. Let the grids be denoted by $G_1$, $G_2$, and 
$G_3$, where $G_1$ is the coarsest and $G_3$ is the finest mesh, the 
associated \shape\ parameters are given by:
\bea
G_1: \qquad \shape_1&=& [129,65,65],\\
G_2: \qquad \shape_2&=& [161,81,81],\\
G_3: \qquad \shape_3&=& [193,97,97].
\eea
Fig.~\ref{soliton} shows the time evolution of $\vert\phi(t,x,y,0)\vert)$,
$0\le t \le 125$ as calculated on the grid $G_3$, using
the initial data described above.  Significant direct interaction of 
the stars begins at $t\approx40$, and by $t=55$ it is essentially 
impossible to identify two distinct objects.  The period of strong
interaction---during which ``interference patterns'' are clearly 
visible---persists until $t\approx100$.   The stars then emerge from 
the collision with their initial shapes roughly preserved,
and continue to propagate until the end of the 
calculation.

\begin{figure}
\begin{center}
\epsfxsize=16.0cm
\epsffile{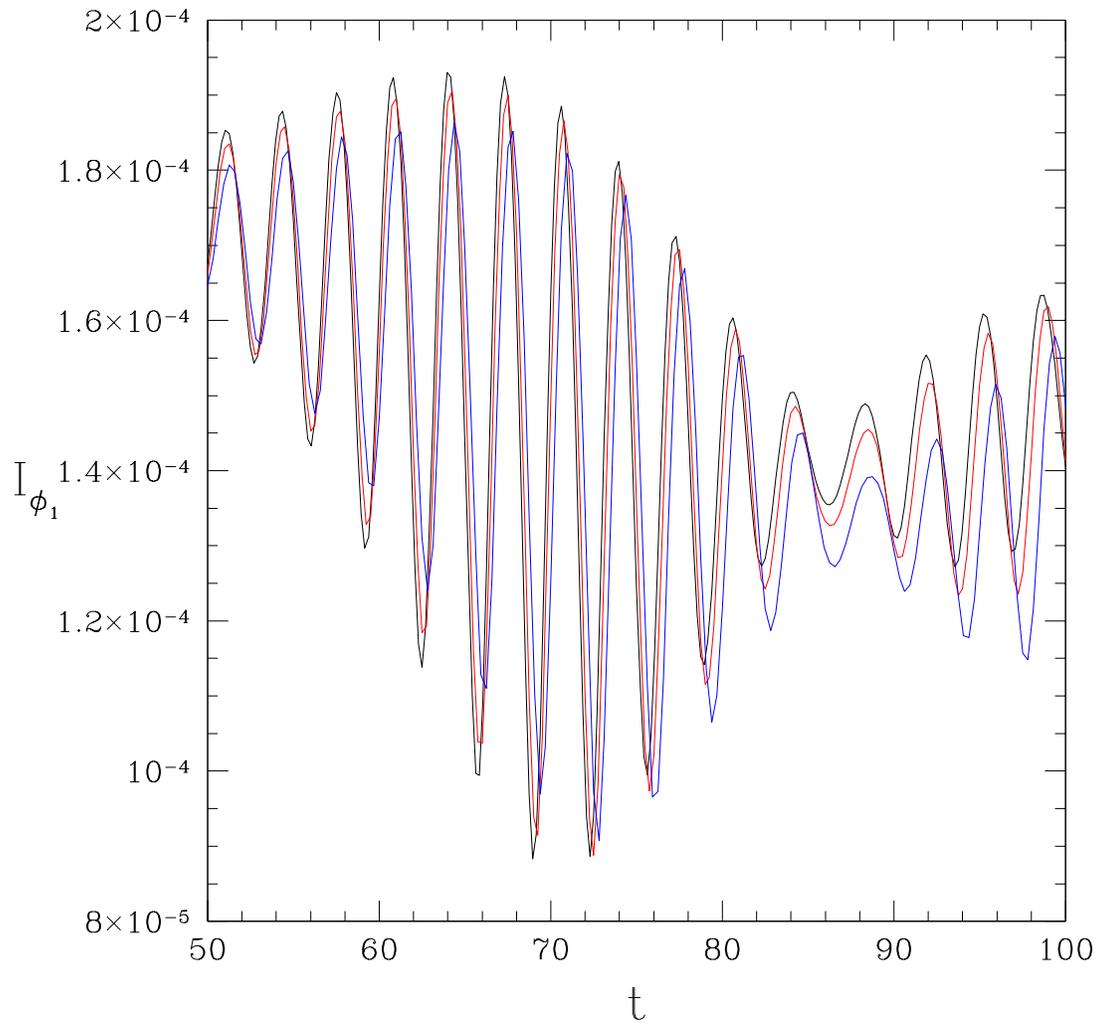}
\end{center}
\caption
[Head-on Collision of Boson Stars: $\Vert I_{\phi_1}(t)\Vert_2$.] 
{
This figure shows the rescaled $l_2$ norms of the independent 
residuals, $\Vert I_{\phi_1}(t) \Vert_2$, from the calculation of the head-on collision
of $S^1_{0.02}$ and $S^2_{0.03}$.
The plot focuses on the time interval $50\leq t \leq 100$, when
the interaction between the stars is strongest.
The blue, red and black curves correspond to runs using 
grids $G_1$, $G_2$ and $G_3$ (see text), which had mesh scales 
$h_1$, $h_2$ and $h_3$, respectively.  The red and black values 
have been rescaled by factors of $h_1/h_2$ and $h_1/h_3$ respectively,
and the near-coincidence of the curves shows that the independent 
residuals are $O(h)$, as expected.
\label{indphi1headon}}
\end{figure}

During the period of interaction, the scalar field modulus reaches 
a maximum value $|\phi(t,x,y,z)|$ $\simeq0.059$. 
In addition, the stars come out of the collision with
an estimated average coordinate velocity $\De x/\De t \sim 0.24$.
The rescaled $l_2$ norms of the independent residual, $I_{\phi_1}(t)$,
for the scalar field component, $\phi_1$, are plotted in 
Fig.~\ref{indphi1headon}.
They provide validation of the computation since they are converging 
at the expected $O(h)$ rate.

Given the previous observations of the solitonic nature of head-on 
boson stars in fully general relativistic cases, the results of this 
experiment can be interpreted as providing qualitative evidence that 
the CFA is capturing at least some of the essential physics described 
by a solution of the full Einstein-Klein-Gordon equations.   However,
we must again note that Newtonian calculations also yield the same type of
behaviour.  Thus, without some sort of  direct comparison with fully 
general relativistic results, our computations should be viewed 
as providing a relatively weak validation of the suitability of 
the CFA as a replacement for the Einstein equations in 
the treatment of head-on collisions of boson stars.

\begin{figure}
\begin{center}
\epsfxsize=16.0cm
\epsffile{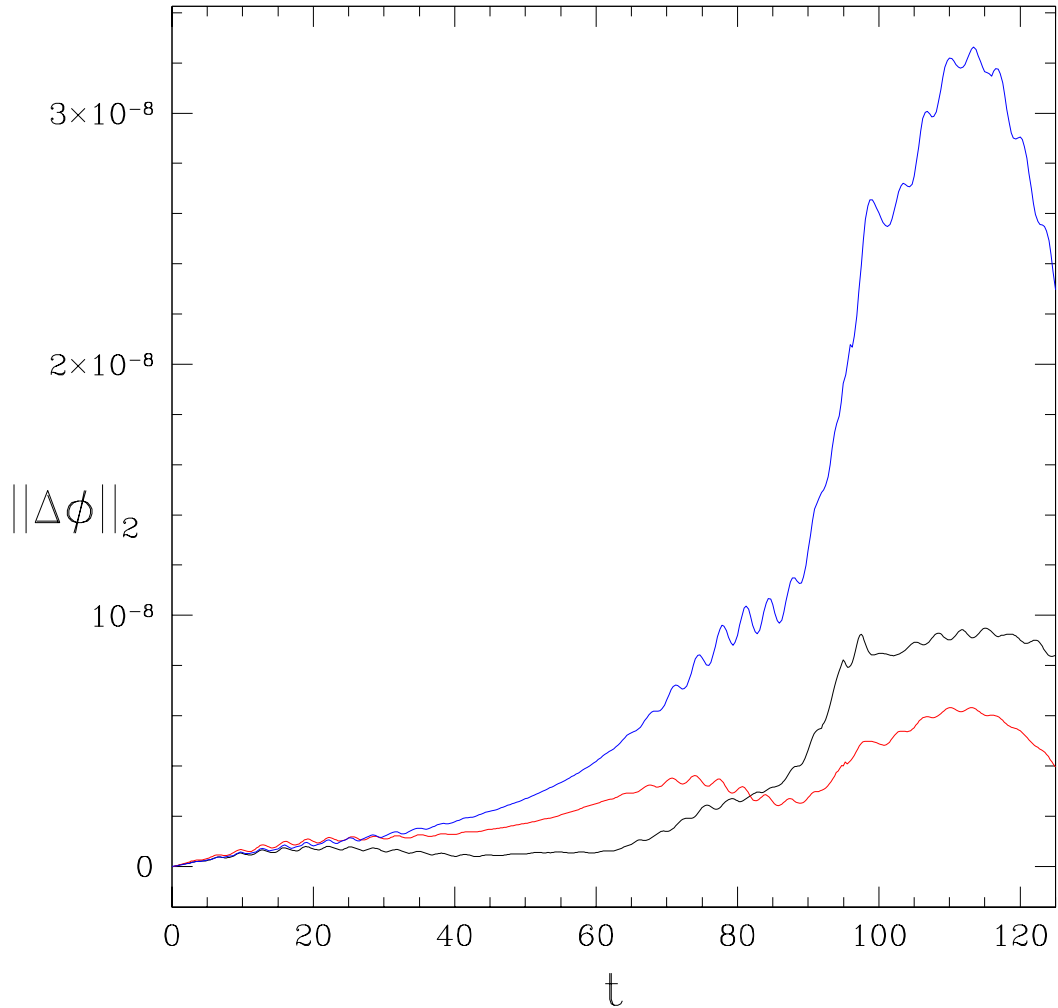}
\end{center}
\caption
[Head-on Collision of Boson Stars: $y\to-y$ Symmetry Assessment.] 
{Head-on Collision of Boson Stars: $y\to-y$ Symmetry Assessment.
This figure plots a measure of the extent to which our code
breaks the $y\to-y$ symmetry that is expected in the continuum
limit for the head-on collision of $S^1_{0.02}$ and $S^2_{0.03}$.
Specifically, we define $\De \phi(t,x)\equiv \vert \phi(t,x,5,0) \vert - 
\vert \phi(t,x,-5,0) \vert $ and then plot $\Vert \De \phi(t,x) \Vert_2$,
where the norm is taken over the remaining spatial coordinate, $x$.
Blue, red and black curves correspond to calculations 
on $G_1$ (coarsest resolution), $G_2$ and $G_3$ (finest resolution), 
respectively.  The maximum value of
$\vert \phi(t,x,5,0) \vert$ attained during the computation is 
about $10^{-2}$.
We thus conclude that our code preserves the $y\to-y$ symmetry 
to about a part in $10^6$ for this particular set of calculations.
}
\label{dphisym}
\end{figure}

Finally, we note that the initial conditions used
in this section have
$y\to-y$ and $z\to-z$ symmetries.
As we have mentioned previously, these were {\em not} exploited in 
this (or any other) calculation.
However, in this case we \emph{have} estimated how well the 
reflection symmetry $y\to-y$ is maintained during
the evolution.  Results of this test are shown in 
Fig.~\ref{dphisym}, which suggest that, at any of the resolutions 
used, the maximum deviation from exact symmetry in a typical dynamical 
variable---$\vert\phi\vert$ in this instance---was about
$1$ part in $10^6$.

\newpage
\section{Orbital Dynamics of Two Boson Stars}\label{sec:orbit}

In this section we summarize what we consider to be the most interesting
and important calculations that we have performed with our code to
date.
We focus on dynamics 
that involves two identical boson stars---$S_{0.02}$---but 
now set initial data so that, at least initially, some form of orbital 
motion results.
The stars continue to be centred in the $z=0$ plane, but at $t=0$ have
centres on the $y$ axis which are equidistant from the origin.  We 
then give the stars initial boosts in the $x$ direction which are of equal 
magnitude, $v_x$,
but of opposite sign.  
By adjusting the size of $v_x$, we can produce 
evolutions that lead to one of three distinct end states, as already 
summarized in the introductory section of this chapter.
Specifically, we have run 
calculations for $v_x = 0$,
$0.03$, $0.05$, $0.07$, $0.08$, $0.09$, $0.1$ and $0.11$. 
From this sequence of computations, and 
again noting that we used stars with $\phi_0=0.02$, we have identified 
approximate ranges for $v_x$ associated with the three types of 
behaviour:
\begin{enumerate}
\item Long-lived orbital motion: $0.09\le v_x \le 0.11$.
\item Merger of the stars that results in a conjectured rotating and
      pulsating boson star: $0.07\le v_x <0.09$.
\item Merger of the stars that leads to the conjectured formation of
      a black hole: $0\le v_x <0.07$.
\end{enumerate}

Following a description of the computational setup used for
our parameter survey (in $v_x$), we will proceed to discuss 
representative examples from each class in some detail.

All the calculations documented in this section were performed on
a computational domain defined by $\bbox=[-60,60,-60,60,-60,60]$, and 
all were made using four different grid resolutions. 
Denoting the grids $G_1$, $G_2$, $G_3$ and $G_4$, where $G_1$ is the 
coarsest mesh and $G_4$ is the finest, the corresponding \shape\
parameters are given by:
\bea
G_1: \qquad \shape_1&=& [65,65,65],\\
G_2: \qquad \shape_2&=& [81,81,81],\\
G_3: \qquad \shape_3&=& [97,97,97],\\
G_4: \qquad \shape_4&=& [113,113,113].
\eea
As usual, there is a $z\to-z$ symmetry in the computations, and the 
plots that are displayed were generated from $z=0$ cuts of the full
3D solutions.  The two identical stars, $S^1_{0.02}$ and $S^2_{0.02}$,
are initially centred at $C^1=(0,20,0)$ and $C^2=(0,-20,0)$, and 
given boosts in the $x$ direction with $v^1_x = -v_x$ and $v^2_x=v_x$.
We emphasize that $v_x$ is being treated as a control parameter 
for our experiments: in particular, the {\em only} difference in 
parameter settings for the three cases described below is the value of $v_x$ that is
used.

\subsection{Case 1: $v_x=0.09$---Long Lived Orbital Motion}\label{sec:orbit-case1}

\begin{figure}
\begin{center}
\epsfxsize=15.0cm
\ifthenelse{\equal{\highQ}{true}} {
\epsffile{figs/dynamics/orbitv09.eps}
}{
\epsffile{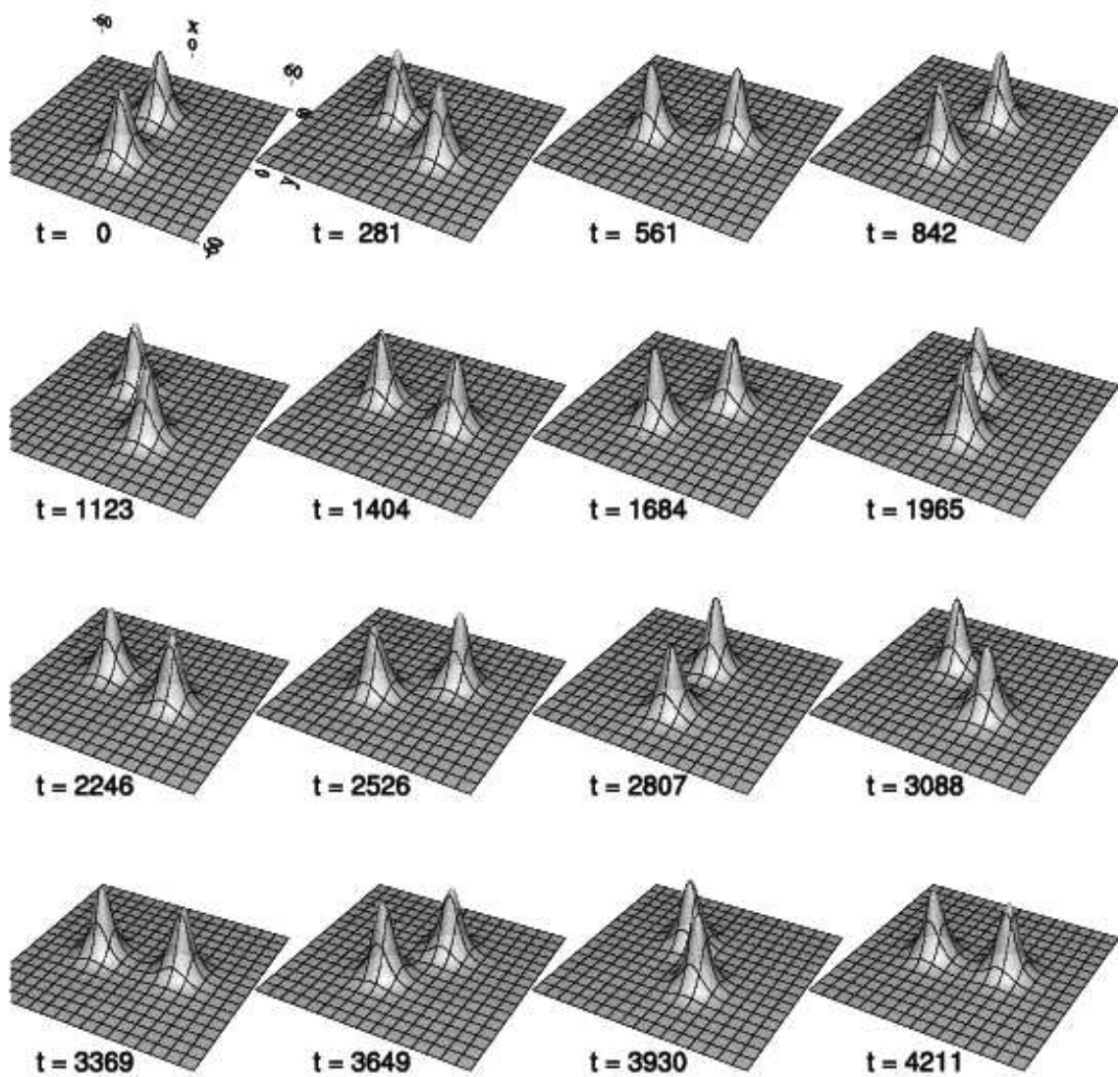}
}
\caption
[Long Lived Orbital Motion: Time Evolution of $\vert \phi(t,x,y,0) \vert$.]
{
This figure shows a series of snapshots of $\vert \phi(t,x,y,0) \vert$ 
for $0 \le t \le 4211$ from a calculation describing long lived orbital 
motion of two identical boson stars.  The two stars, 
$S^1_{0.02}$ and $S^2_{0.02}$, are initially centred at
$C^1=(0,20,0)$ and $C^2=(0,-20,0)$ and boosted with parameters
$v_x = - v^{(1)}_x = v^{(2)}_x=0.09$.
During this period of time that is displayed, the stars execute about 
$2\frac14$ orbits, and the maximum value of $\vert \phi(t,x,y,0) \vert$ remains
approximately constant during the evolution.  
}
\label{orbitv09}
\end{center}
\end{figure}

\begin{figure}
\begin{center}
\epsfxsize=16.0cm
\epsffile{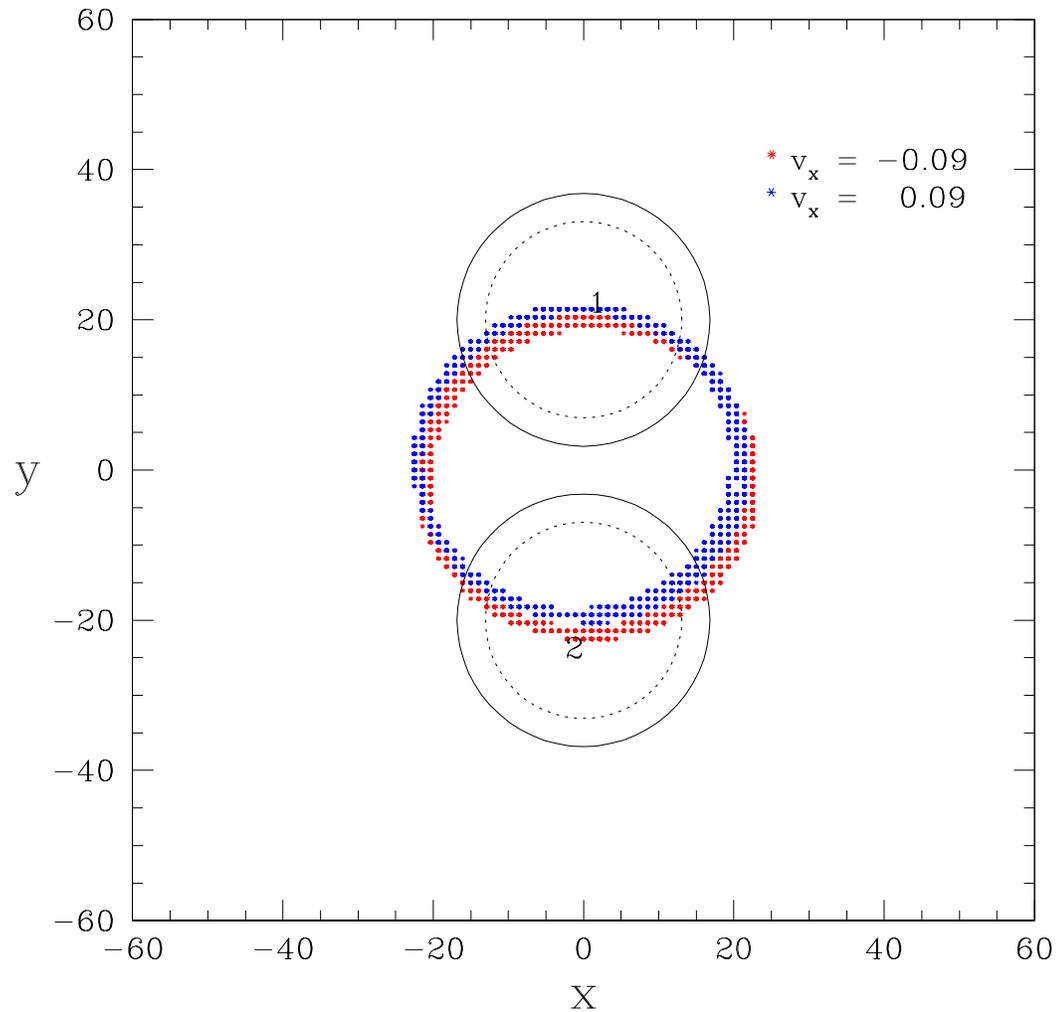}
\end{center}
\caption
[Long Lived Orbital Motion: Trajectories.]
{
Long lived orbital motion: stellar trajectories.
This figure shows trajectories of the two
stars, $S^1_{0.02}$ and $S^2_{0.02}$, as they orbit one another during 
the coordinate time interval $0\leq t \leq 4500$.  
Each red or blue dot represents the local maximum of 
$\vert \phi(t,x,y,0) \vert$ at a particular instant of time.  
The orbit is slightly eccentric and is apparently precessing with time.  
In this and other figures of this type, in order to provide some 
sense of the size of the stars  we have also plotted circles 
that approximately delimit the stellar surfaces.  Specifically,
the solid and dashed lines have radii $R_{99}$ (containing 
99\% of $M_{\rm ADM}$) and $R_{95}$ (containing
95\% of $M_{\rm ADM}$), respectively.
}
\label{traj2d-v09}
\end{figure}

\begin{figure}
\begin{center}
\epsfxsize=16.0cm
\epsffile{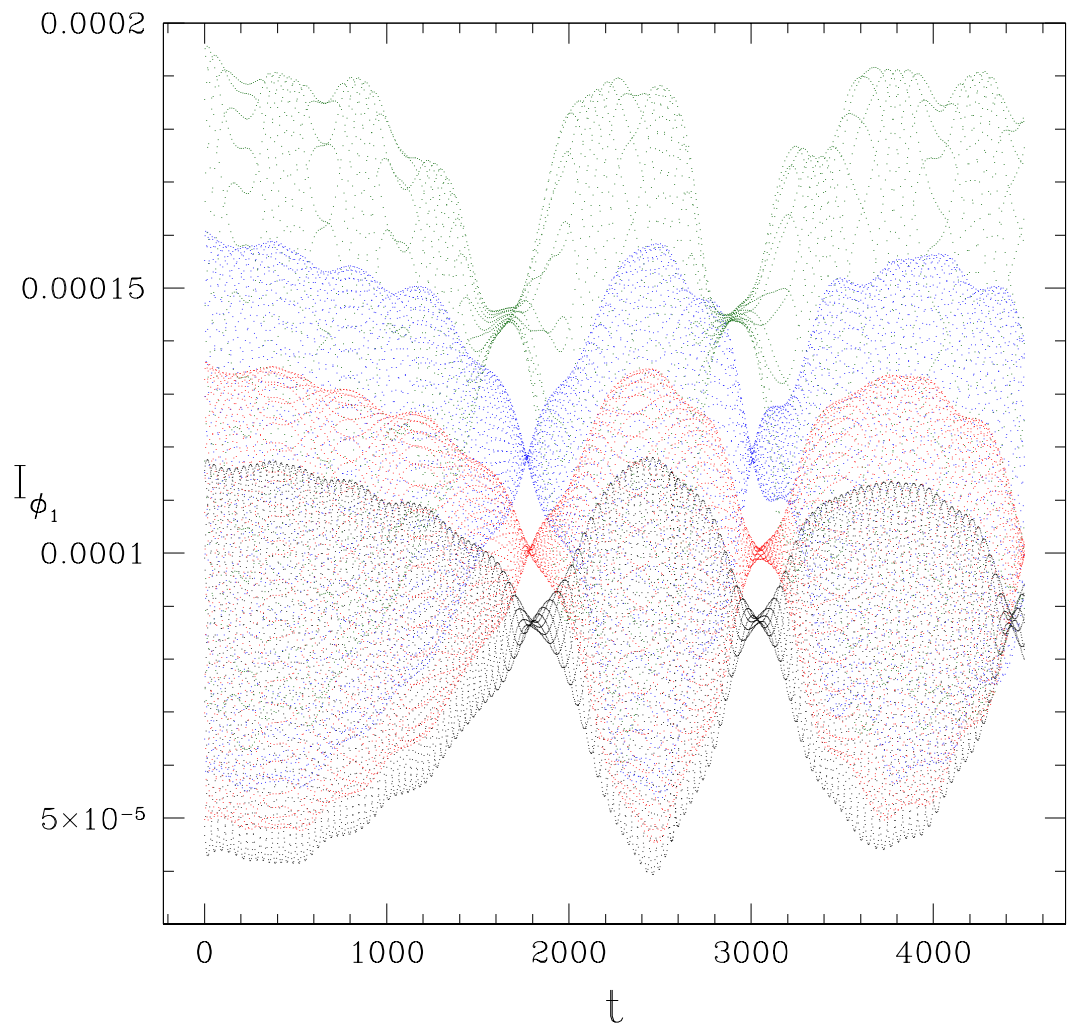}
\end{center}
\caption
[Long Lived Orbital Motion: $\Vert I_{\phi_1}(t)\Vert_2$.]
{Long Lived Orbital Motion: $\Vert I_{\phi_1}(t)\Vert_2$.
This figure shows the $l_2$-norm of the independent residuals
$I_{\phi_1}(t)$ for $0\le t \le 4500$. During this interval
of coordinate time
the stars perform about $2\frac14$ orbits.
Residuals plotted with green, blue, red and black dots were computed 
using grids $G_1$ (coarsest resolution), $G_2$, $G_3$ and 
$G_4$ (finest resolution), respectively,
with specific grid parameters defined in the text.  
Note that these residuals have {\em not} 
been rescaled.  The plots provide strong evidence of convergence of the 
independent residuals as $h\to0$.
}
\label{indphi1v09}
\end{figure}

\begin{figure}
\begin{center}
\epsfxsize=16.0cm
\epsffile{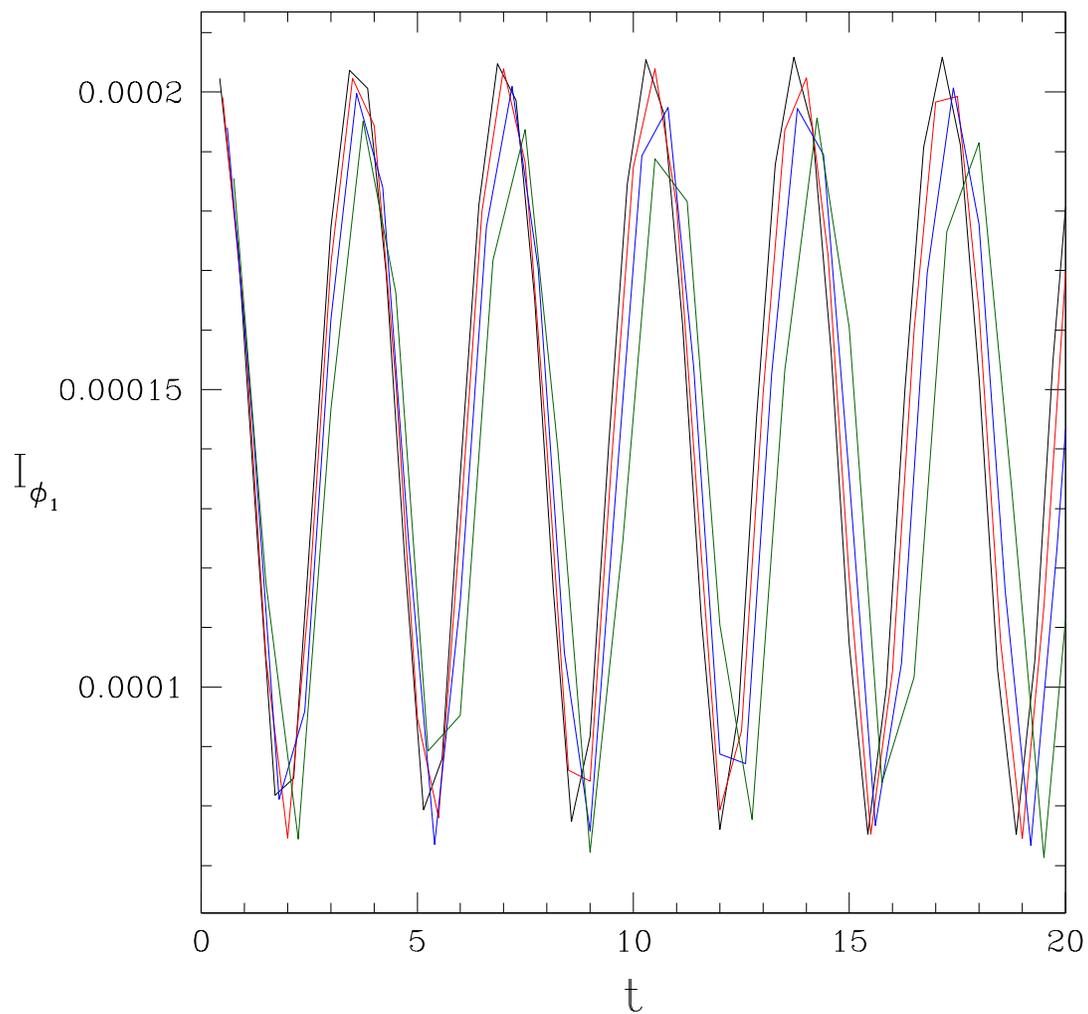}
\end{center}
\caption
[Long Lived Orbital Motion: Rescaled $\Vert I_{\phi_1}(t)\Vert_2$.]
{Long Lived Orbital Motion: Rescaled $\Vert I_{\phi_1}(t)\Vert_2$.
This figure shows the same independent residuals displayed in
Fig.~\ref{indphi1v09}, but on the shorter time interval $0\leq t \leq 20$,
and with scaling factors $h_1/h_2$, $h_1/h_3$, and $h_1/h_4$ multiplying
the values computed on $G_2$ (blue), $G_3$ (red) and $G_4$ (black), 
respectively.  The expected $O(h)$ convergence of the residuals is 
apparent.
}
\label{indphi1v09short}
\end{figure}

\begin{figure}
\begin{center}
\epsfxsize=16.0cm
\epsffile{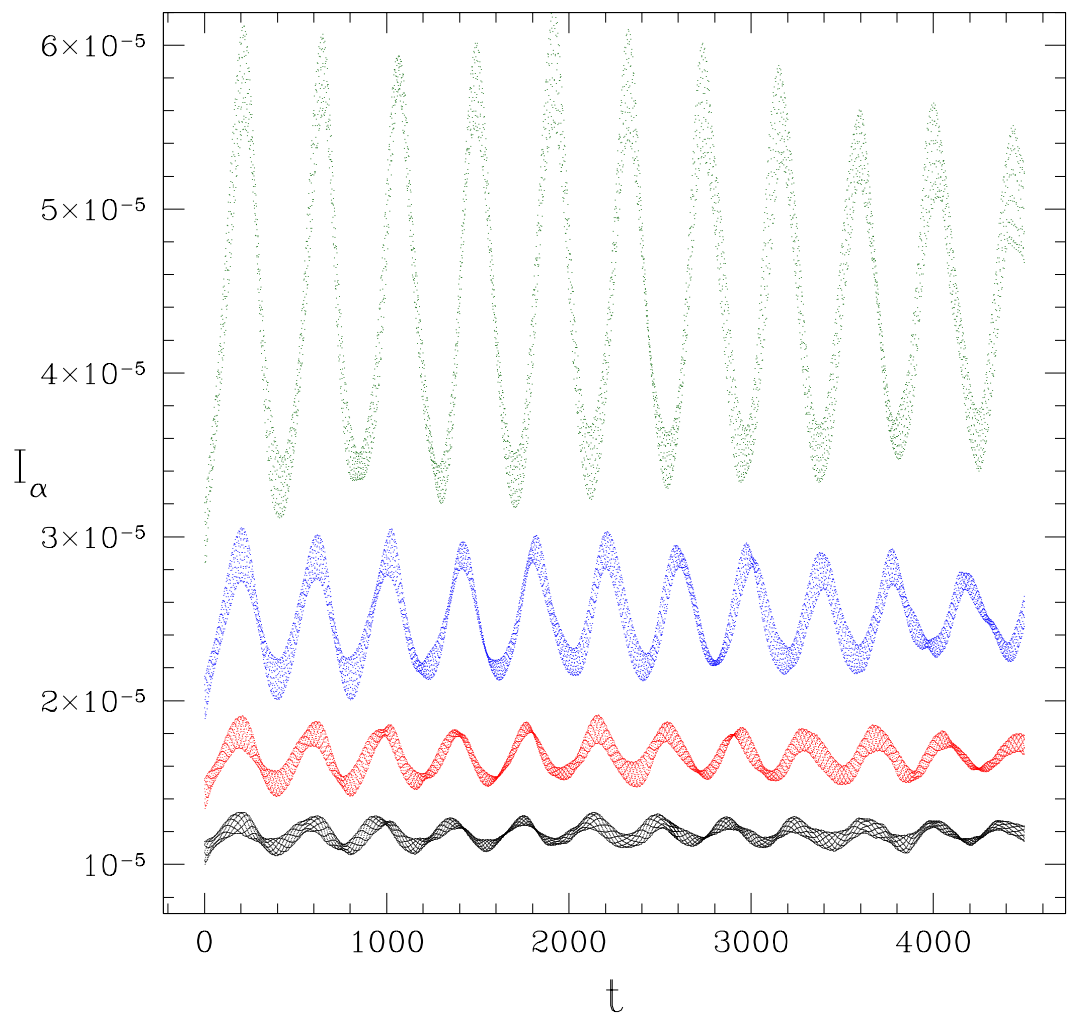}
\end{center}
\caption
[Long Lived Orbital Motion: $\Vert I_{\alpha}(t)\Vert_2$.]
{
Long Lived Orbital Motion: $\Vert I_{\alpha}(t)\Vert_2$.
This figure shows the $l_2$-norm of the independent residuals
$I_{\alpha}(t)$ for $0\le t \le 4500$, and as computed on 
grids $G_1$ (green, coarsest resolution), $G_2$ (blue), 
$G_3$ (red), and $G_4$ (black, finest resolution).
As with Fig.~\ref{indphi1v09}, these residuals have {\em not} 
been rescaled.
Once more, the plots strongly suggest convergence of the
independent residuals as $h\to0$.
}
\label{indlpsv09}
\end{figure}

\begin{figure}
\begin{center}
\epsfxsize=16.0cm
\epsffile{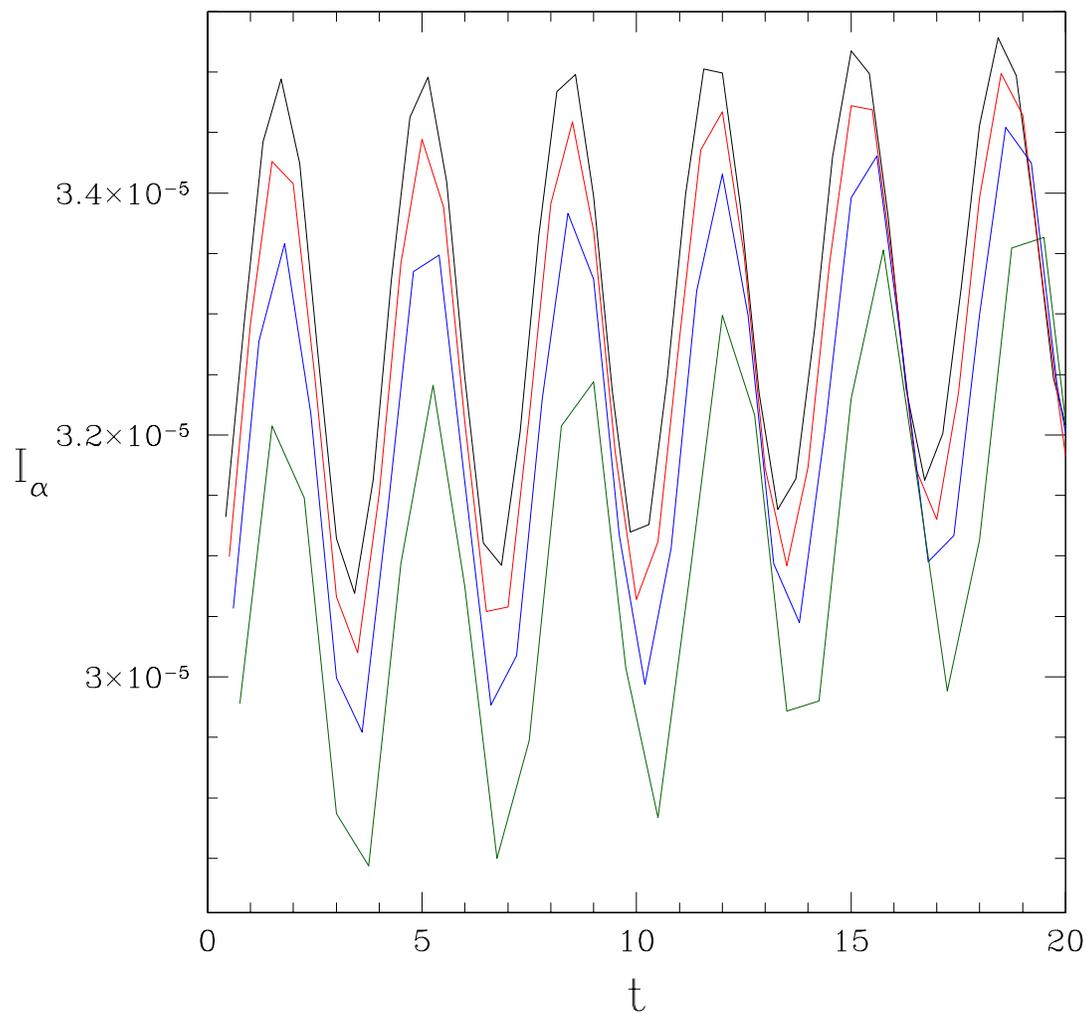}
\end{center}
\caption
[Long Lived Orbital Motion: Rescaled $\Vert I_{\alpha}(t)\Vert_2$.]
{Long lived Orbital Motion: Rescaled $\Vert I_{\alpha}(t)\Vert_2$.
This figure shows the same independent residuals displayed in
Fig.~\ref{indlpsv09}, but on the shorter time interval $0\leq t \leq 20$,
and with scaling factors $(h_1/h_2)^2$, 
$(h_1/h_3)^2$, and $(h_1/h_4)^2$ multiplying
the values computed on $G_2$ (blue), $G_3$ (red) and $G_4$ (black),
respectively.  The $O(h^2)$ convergence rate expected for 
all the residuals associated with the metric variables is
apparent.
}
\label{indlpsv09short}
\end{figure}

Results from this calculation are shown in 
Figs.~\ref{orbitv09}--\ref{indlpsv09short}.  Fig.~\ref{orbitv09} displays 
a series of snapshots of $\vert\phi(t,x,y,0)\vert$ for $0\le t\le4211$.
During this period of coordinate time the stars orbit one another about 
$2\frac14$ times.  The orbit is slightly eccentric, as can be seen from 
Fig.~\ref{traj2d-v09} which shows the orbital trajectory of the centre
of each star (defined to be the position of the local maximum of 
$\vert\phi(t,x,y,0)\vert$).  We note, however, that we have not yet made 
any attempt to estimate the eccentricity.  
Detailed examination of 
the trajectories suggests that the orbit is also precessing.  Since 
orbital precession is a purely general relativistic effect (assuming
that the stars remain spherical in their respective 
rest-frames) this qualitative result provides additional 
encouraging evidence that the CFA is capturing some of the key
physical effects that one would see in a solution of the full Einstein
equations.  It would therefore be very interesting to attempt to 
quantify the observed rate of precession and compare it to the value 
obtained (or estimated) from a general relativistic calculation.  This,
however, is something that remains to be done.

Plots of the $l_2$ norms of two of the independent residuals---$I_{\phi_1}(t)$
and $I_\alpha(t)$---from the computations performed on the 
four different grids are shown in 
Figs.~\ref{indphi1v09}--\ref{indlpsv09short}.
Here, and for all of the remaining figures in this 
chapter, the colours green, blue, red and black are used for quantities 
computed on grids $G_1$, $G_2$, $G_3$ and $G_4$ respectively. Additionally,
and in contrast to the corresponding plots in 
Secs.~\ref{sec:gen_ID}--\ref{sec:boost_ID}
we have {\em not} scaled the independent residuals in this section or 
the next, as the resulting graphs are difficult to interpret
(exceptions are Fig.~\ref{indphi1v09short} and  Fig.~\ref{indlpsv09short}).  
Nonetheless, a detailed quantitative examination of the residuals, such as 
that provided by Fig.~\ref{indphi1v09short} and Fig.~\ref{indlpsv09short}, 
reveals
that the independent residuals are converging at the rates expected.

\subsection{Case 2: $v_x=0.07$---Formation of Pulsating \& Rotating Boson Star}\label{sec:orbit-case2}

\begin{figure}
\begin{center}
\epsfxsize=15.0cm
\ifthenelse{\equal{\highQ}{true}} {
\epsffile{figs/dynamics/orbitv07.eps}
}{
\epsffile{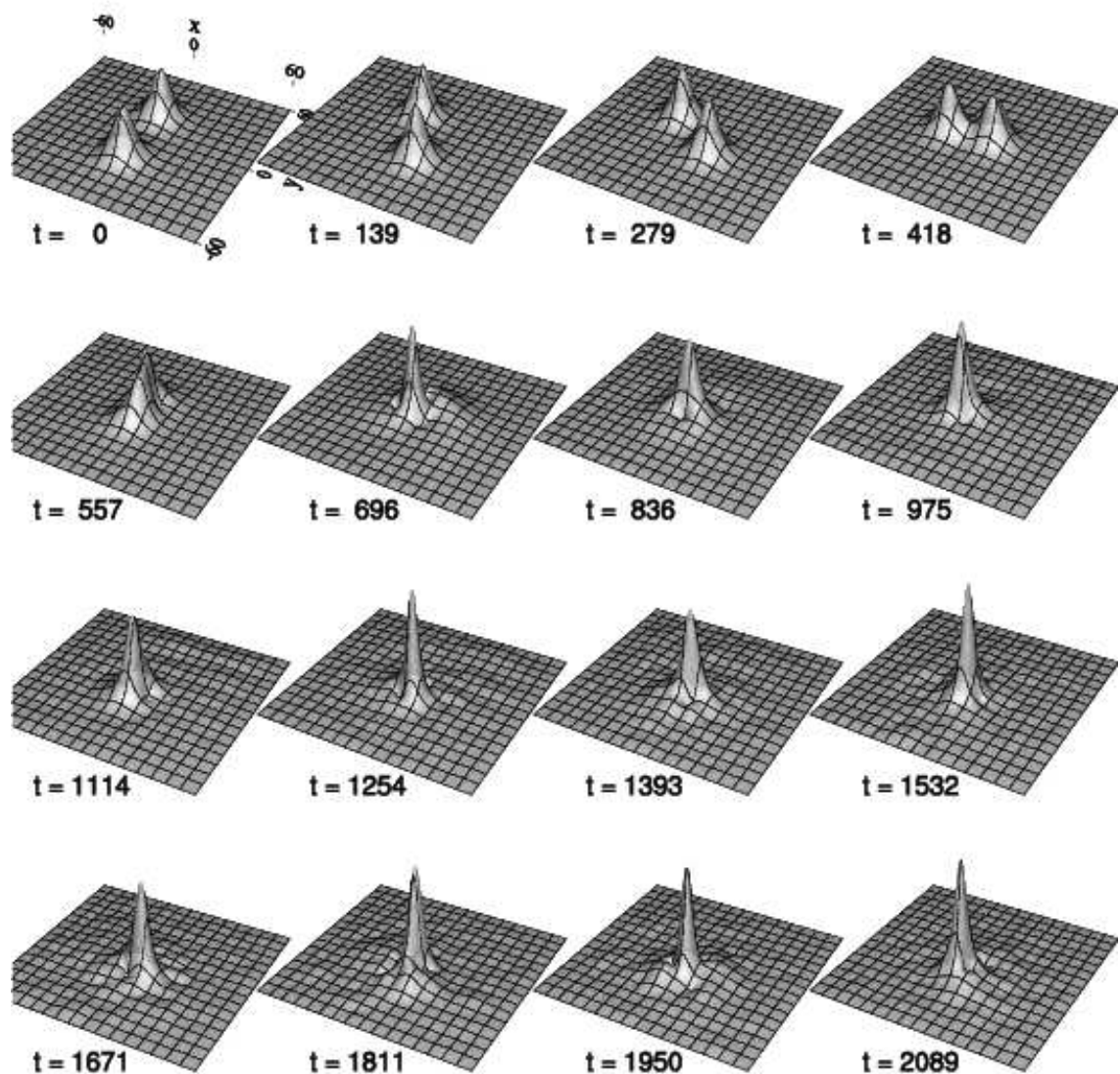}
}
\caption
[Formation of Pulsating and Rotating Boson Star: Time Evolution Snapshots.]
{
This figure displays a series of surface plots of $\vert \phi(t,x,y,0) \vert$ 
for $0 \le t \le 2089$, from a calculation in which two identical
stars, $S^1_{0.02}$ and $S^2_{0.02}$, are initially centred at 
$C^1=(0,20,0)$ and $C^2=(0,-20,0)$, and boosted with parameters
$v_x = - v^{(1)}_x = v^{(2)}_x=0.07$.  
The stars subsequently execute somewhat less than half of an orbit before they
graze one another.  A quick ``plunge'' phase follows, leading
to the formation of a (conjectured) pulsating and rotating boson star
characterized by a central field modulus $\phi_0\simeq0.05$.
}
\label{orbitv07}
\end{center}
\end{figure}

\begin{figure}
\begin{center}
\epsfxsize=16.0cm
\epsffile{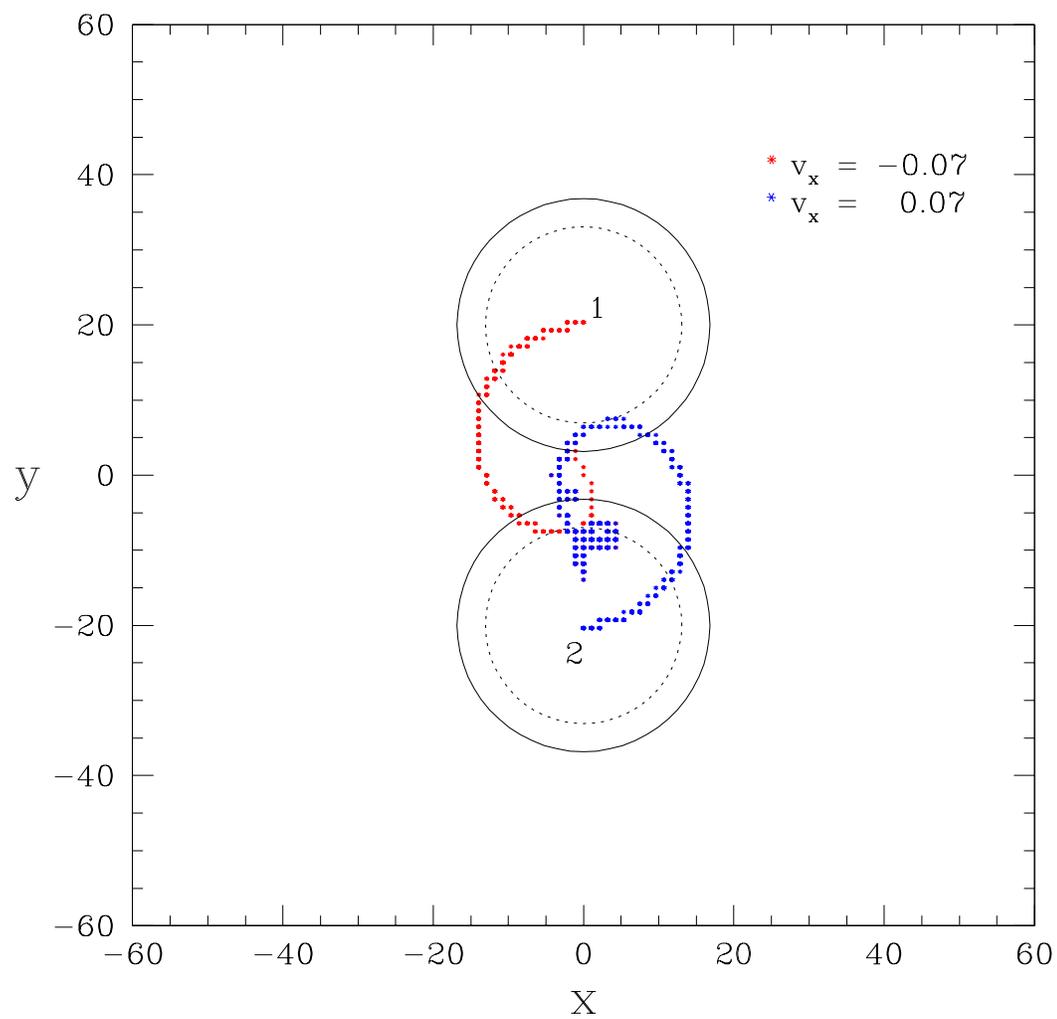}
\end{center}
\caption
[Formation of Pulsating and Rotating Boson Star: Trajectories.]
{Formation of Pulsating and Rotating Boson Star: Trajectories.
This figure shows the trajectories of the stars
$S^1_{0.02}$ and $S^2_{0.02}$ for the computation described in
the caption of Fig.~\ref{orbitv07}.
Each red or blue dot plots the position a local maximum of 
$\vert \phi(t,x,y,0) \vert$ at a particular instant of time.
Here the tracks
clearly reveal a short period of orbital motion which is followed by a 
rapid ``plunge'' and merger of the stars.
}
\label{traj2d-v07}
\end{figure}

\begin{figure}
\begin{center}
\epsfxsize=16.0cm
\epsffile{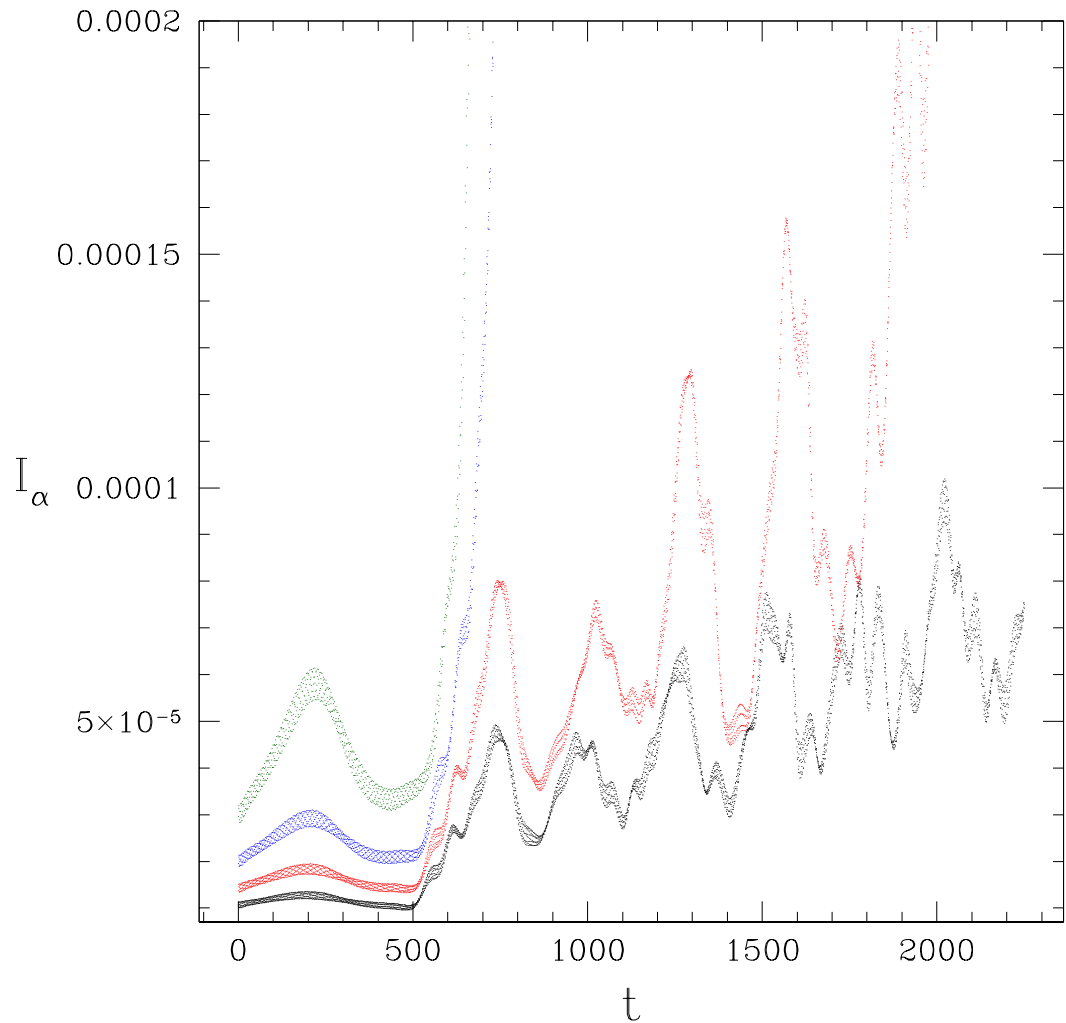}
\end{center}
\caption
[Formation of Pulsating and Rotating Boson Star: $\Vert I_{\alpha}(t)\Vert_2$.]
{Formation of Pulsating and Rotating Boson Star: $\Vert I_{\alpha}(t)\Vert_2$.
This figure shows the $l_2$-norm of the independent residual $I_{\alpha}(t)$ 
for the calculation described in Fig.~\ref{orbitv07}.  
We note that the residuals have {\em not} been rescaled in this case,
but that the values computed on grids $G_1$ 
(green, coarsest resolution), $G_2$ (blue), $G_3$ (red), and 
$G_4$ (black, finest resolution) \emph{do} seem to be 
converging to 0 as $h\to0$. 
However, it also clear that there is a qualitative change in the nature 
of the independent residuals computed on $G_1$ and $G_2$ once 
the collision has occurred at $t\sim500$.  This is a good indication
that the calculations on those relatively coarse meshes are not 
reliable for $t\gtrsim 500$.
}
\label{indlpsv07}
\end{figure}

Results from this calculation are summarized in 
Figs.~\ref{orbitv07}--\ref{indlpsv07}.  Fig.~\ref{orbitv07} shows 
a series of surface plots of $|\phi(t,x,y,0)|$ for $0\le t\le2089$, 
while Fig.~\ref{traj2d-v07} shows trajectories of the stars prior 
to their merger.  In this case orbital motion persists for 
a little less than one half of a full rotation before the stars 
graze one another.  This is followed by a rapid ``plunge'' phase which 
leads to the formation of what we conjecture to be essentially
a single spinning and pulsing boson star, with a central field 
modulus $\phi_0\simeq0.05$.

As usual, we monitor independent residuals to ensure that our 
calculations are reliable, and Fig.~\ref{indlpsv07} shows the $l_2$ norm 
of $I_\alpha(t)$. We first note that the plot shows that $I_\alpha$
is tending to 0 as the mesh spacing decreases.  However, it is also 
clear from the figure that the convergence deteriorates at
later times, especially once the stars have merged.  Indeed, the 
apparent divergence of $I_\alpha(t)$ on $G_1$ (green) and $G_2$ (blue)
for $t \gtrsim 500$ is a clear indication that the numerical solutions
computed on those grids are not trustworthy once the collision
has occurred.  In addition, the merger results in the ejection (radiation)
of a significant amount of scalar matter.  Some of this ejecta hits 
the outer computational boundary and, due to our use of Dirichlet
boundary conditions, is largely reflected back into the solution 
domain.  These spurious reflections eventually contaminate the entire
calculation, so that even though the solution may still exhibit 
convergence as measured via independent residuals, the physical 
interpretation at late times is dubious at best.

\subsection{Case 3: $v_x=0.05$---Formation of a Black Hole}\label{sec:orbit-case3}

\begin{figure}
\begin{center}
\epsfxsize=15.0cm
\ifthenelse{\equal{\highQ}{true}} {
\epsffile{figs/dynamics/orbitv05.eps}
}{
\epsffile{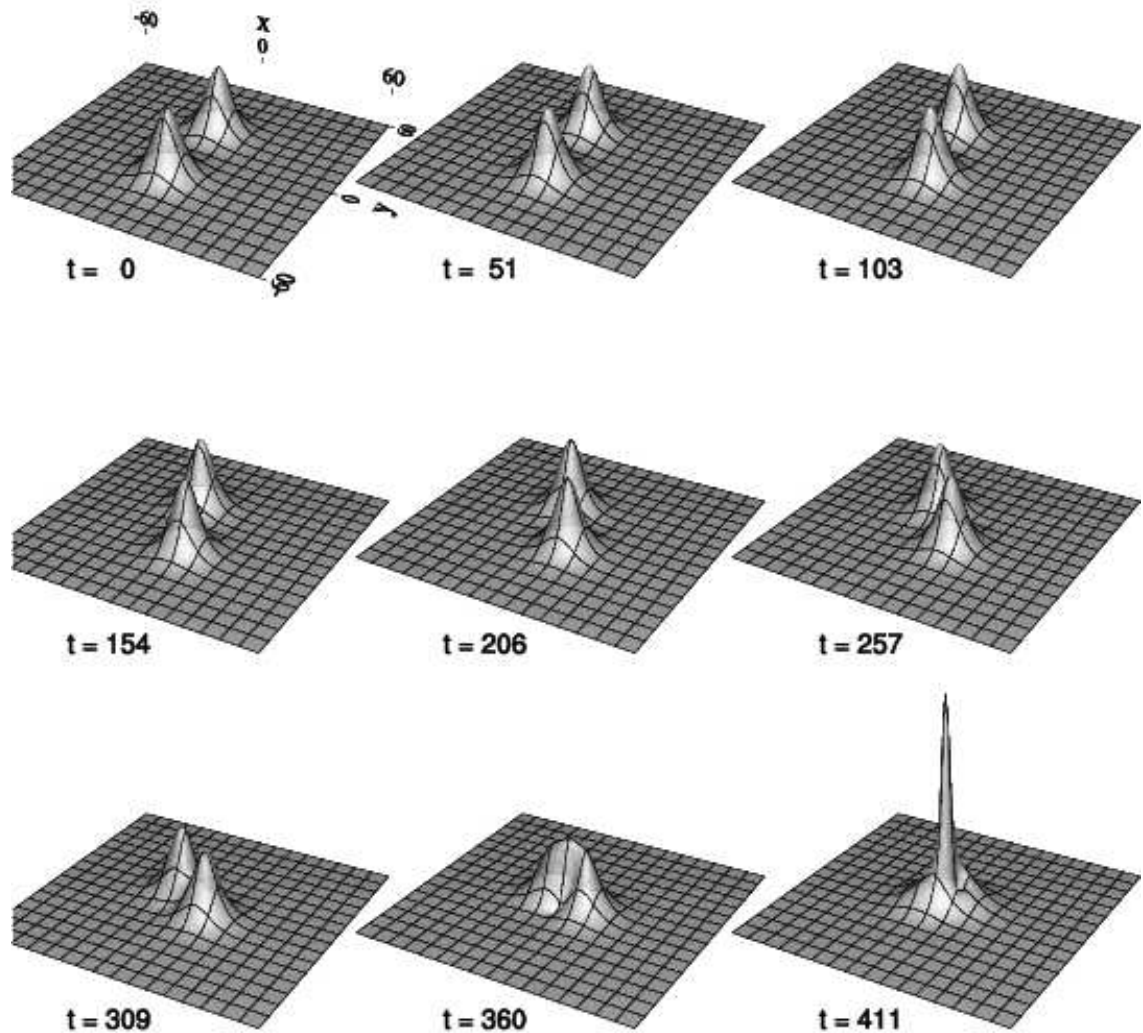}
}
\caption
[Formation of a Black Hole: Time Evolution Snapshots.]
{
This figure displays a series of surface plots of $\vert \phi(t,x,y,0) \vert$
for $0 \le t \le 411$, from a calculation in which two identical
stars, $S^1_{0.02}$ and $S^2_{0.02}$, are initially centred at
$C^1=(0,20,0)$ and $C^2=(0,-20,0)$, and boosted with parameters
$v_x = - v^{(1)}_x = v^{(2)}_x=0.05$.  As with all previous plots 
of this type, the specific results visualized were computed on the 
finest grid, $G^4$, which has ${\tt shape} =[113, 113, 113]$.
In this case, the stars rapidly approach each other, and, after a plunge 
and merger phase, apparently undergo gravitational collapse 
to form a single black hole.  At $t=411$ 
the maximum value of $\vert \phi(t,x,y,z) \vert$ (the height of 
the ``spike'' in the last frame) is $\simeq 0.06$, and even larger 
values---$\vert \phi(t,x,y,z) \vert \gtrsim 0.14$---are seen at later
times.  We note that as can be verified from~Fig.~\ref{Madm1d},
such large values are associated with boson stars on the {\em unstable}
branch.  It thus seems plausible that the evolution will lead to black
hole formation, but computations with considerably more resolution in
the central, strongly interacting region would be required to test 
this hypothesis.
}
\label{orbitv05}
\end{center}
\end{figure}

\begin{figure}
\begin{center}
\epsfxsize=16.0cm
\epsffile{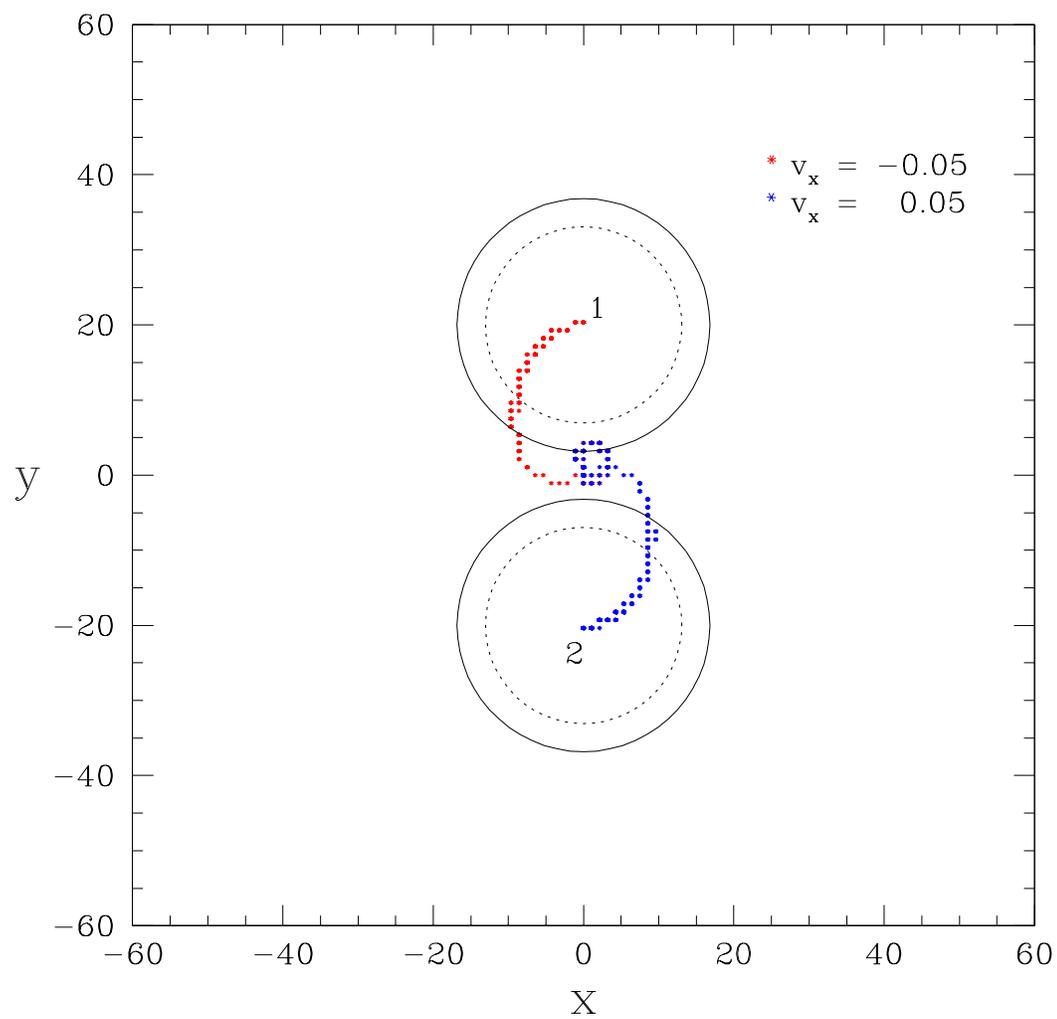}
\end{center}
\caption
[Formation of a Black Hole: Trajectories.]
{Formation of a Black Hole: Stellar Trajectories.
This figure shows the trajectories of the stars 
$S^1_{0.02}$ and $S^2_{0.02}$ from the calculation
described in the caption of Fig.~\ref{orbitv05}, and
during the coordinate time interval
$0\leq t \leq 1500$.  
Once more, each red and blue dot represents a local 
maximum of $\vert \phi(t,x,y,0) \vert$ at a particular instant of time.  
The resulting tracks show a rapid plunge of the stars towards one another,
resulting in a merger that we believe ends in black hole formation.
}
\label{traj2d-v05}
\end{figure}

\begin{figure}
\begin{center}
\epsfxsize=16.0cm
\epsffile{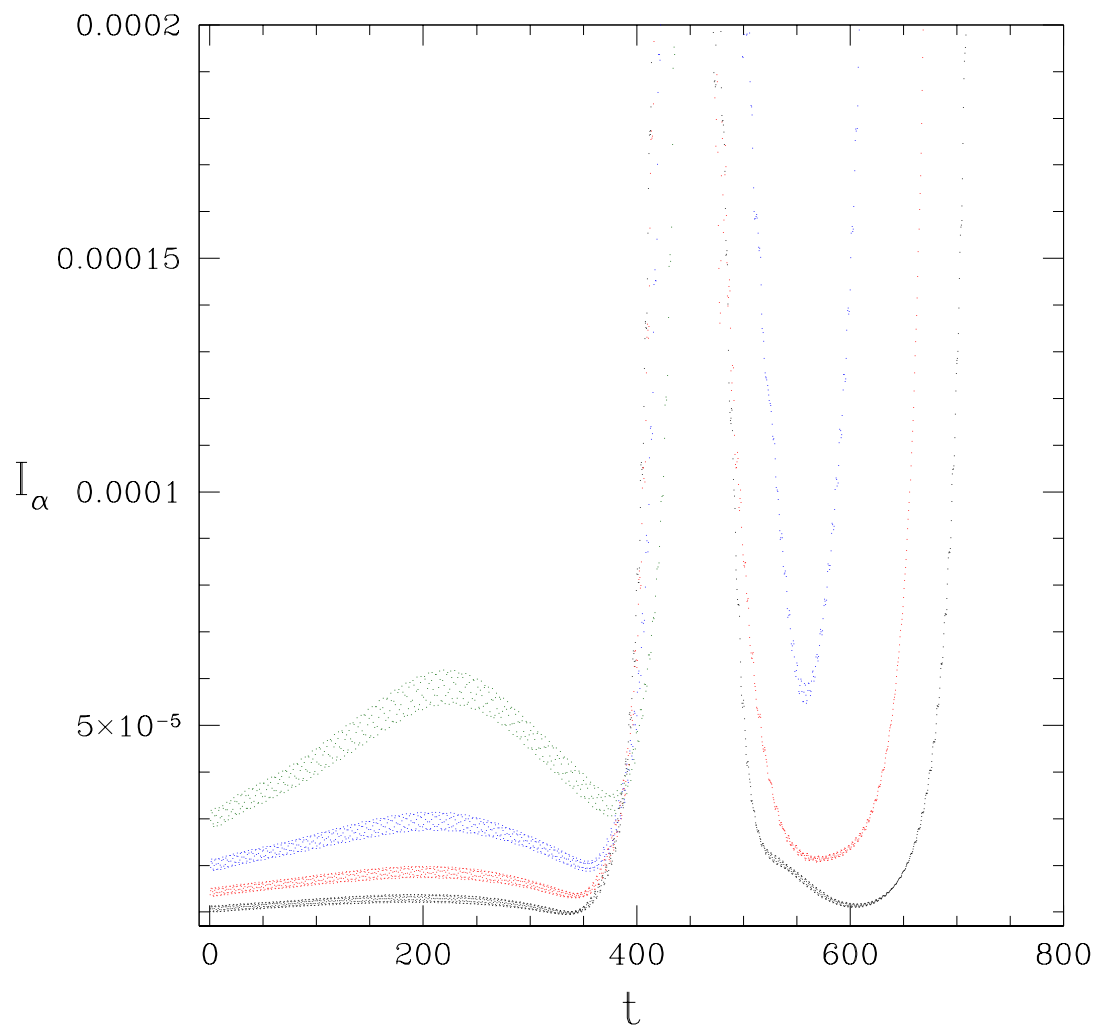}
\end{center}
\caption
[Formation of a Black Hole: $\Vert I_{\alpha}(t)\Vert_2$.]
{Formation of a Black Hole: $\Vert I_{\alpha}(t)\Vert_2$.
This figure shows the $l_2$-norm of the independent residuals $I_{\alpha}(t)$
for the calculation described in the caption of Fig.~\ref{orbitv05}.
The residuals have not been rescaled.
At early times, the residual values computed on grids $G_1$
(green, coarsest resolution), $G_2$ (blue), $G_3$ (red), and
$G_4$ (black, finest resolution) once more appear to be 
converging to 0 as $h\to0$,
However, for $t \gtrsim 400$, when we suspect that gravitational
collapse to a black hole would result if the finite difference resolution 
was high enough, we see an obvious deterioration in the convergence of 
the residuals computed
on {\em all} grids.
We interpret this as a clear indication that none of the computations are 
reliable for $t \gtrsim 400$.
}
\label{indlpsv05}
\end{figure}

Results from our last numerical experiment are shown in 
Figs.~\ref{orbitv05}--\ref{indlpsv05}:  here the initial boost
parameter is $v_x=0.05$.
As usual, we start with surface plots of 
$\vert\phi(t,x,y,0)\vert$, this time on the interval $0\le t\le411$.
Following a very brief period of orbital motion, the 
two stars quickly plunge towards one another and merge, as can clearly 
be seen in the trajectory plots shown in~Fig.~\ref{traj2d-v05}.
In this case we believe that the evolution results in the formation
of a single black hole.  This hypothesis could best be tested by looking 
for marginally trapped surfaces~\cite{wald} in our data---that is, 
surfaces with 
$S^2$ topology for which the divergence of outgoing null geodesics 
emanating from the surface vanishes.  If such a surface was located, 
then, assuming cosmic censorship~\cite{wald} holds, we could 
infer that the data described is a black hole.  However, at the current 
time we have not implemented this approach.  Instead, our conjecture
is based on the behaviour of the metric functions in the central
interaction region.  
For example, the lapse function attains 
extremely small values in that part of the solution domain, 
effectively freezing the evolution there.
This ``collapse of the lapse'' is a very 
well known feature of maximal slicing and is usually correlated
with the formation of a black hole.

However, we must emphasize that various indicators---such as the 
plot of the independent residuals for the lapse function shown
in Fig.~\ref{indlpsv05}---strongly suggest that the results for
this specific choice of $v_x$ are not trustworthy for times
$t\gtrsim380$, even on the finest grid used.  At later times 
the matter is 
highly centrally condensed, and, assuming that collapse to a black hole
{\em is} occurring, would become increasingly so as the evolution
proceeded.  Thus, only a substantial increase in resolution in the 
interaction region would allow us to provide a definitive answer 
concerning the end state of members of this class of initial data.

Finally, since code ``crashes''---due to floating point overflows,
for example---are rather frequent occurrences in numerical relativity calculations,
we should point out that our code did {\em not} ``crash'' at late
times for this configuration (or for any other simulation
reported in this thesis, for that matter).  Rather, we simply 
stopped the evolution at a coordinate time when, even on the finest 
grid, the solution was obviously very poorly resolved.

\newpage
\section{Discussions and Further Developments} \label{sec:discussions}

It is worth noting that the second and third experiments discussed
in the previous 
section are qualitatively consistent with some
of the results reported by 
Palenzuela~{\em et.~al.}~in~\cite{Palenzuela:2007dm}.  
Those authors have 
implemented a fully general relativistic 3D code for 
the Einstein-Klein-Gordon system, and have also 
performed calculations in which the initial data 
contains two identical spherically symmetric boson 
stars that have been given equal but opposite boosts.
One of the calculations discussed in~\cite{Palenzuela:2007dm} 
is quite similar in setup to our $v_x=0.07$ calculation 
described in Sec.~\ref{sec:orbit-case2}, in which the final state appears to 
be a pulsating and rotating boson star.
The specific stars Palenzuela and collaborators used 
in that case each had an ADM mass, $M_{\rm ADM}=0.5$, while the 
ones used here ($S_{0.02}$) had $M_{\rm ADM}\simeq0.475$. 
They also positioned their stars so that, at least in coordinate 
space, they were closer to one another at the initial time than ours were.
Defining $d$ to be the coordinate separation of the centres of 
the stars, they took $d=32$. On the other hand, we chose $d=40$ so that 
our stars---which have $R_{99}\sim17.25$---were essentially 
completely separated at $t=0$.
These differences in the initial configurations are relatively minor,
so we feel that a general comparison of our results with theirs 
is not unreasonable.  Specifically, they identified the end state 
of their evolution as a spinning ``bar'' configuration, accompanied
by dispersal of scalar matter, and this at least roughly agrees
with the interpretation of our end state as a spinning and pulsing
star.  
In addition, the range of values of the 
boost parameter, $v_x$, that led to black hole formation in their 
calculations, namely $0\le v_x<0.04$,  is in rough agreement with our 
findings, where we conjecture that black hole 
formation occurs for $0\le v_x <0.07$.

Palenzuela~{\em et. al.}~also report a very interesting case in which 
the stars merge into a transient single lump of scalar field from which 
two configurations of approximately the same form as the initial stars 
eventually emerge, and propagate away from one another.  
We have {\em not} observed any behaviour of this type in the calculations 
we have done with our code thus far.

Arguably, the most novel of the results presented here are those 
that describe long term orbital motion.  
To our knowledge, the calculations reported in Sec.~\ref{sec:orbit} 
are the first in which more than two orbits of a 
binary boson star have been simulated. 
Furthermore, we have noted that there is
strong evidence of orbital precession, which 
indicates that our model has captured some of the key 
physical effects predicted by general relativity.

We note that one of the original motivations for this work was 
to see whether the compression effect reported by Wilson and 
Mathews~\cite{Wilson:1995uh,Wilson:1996ty,Mathews:1997vw,Mathews:1997nm,matthews:1998, Marronetti:1998vm, marronetti_mathews_wilson:1999,Mathews:1999km,Wilson:2002yh} 
for the case of binary neutron star inspiral (and using 
the conformally flat approximation with maximal slicing) was 
present in our boson star analog of that setup.
In this regard our current results are inconclusive.
In one calculation, where the boost parameter was $v_x=0.08$, 
there {\em did} appear to be compression of the stars as they 
orbited.  However, this is another case where significantly
higher resolution is needed to provide a clear answer.  Moreover,
should something like the Wilson-Mathews effect be convincingly 
seen in our calculations, we would want to vary the scalar field
self-interaction potential, $U(|\phi_0|^2)$ to see what impact
that had on the results.  Here the key point is that the 
reported compression effect for the case of fluid stars depends
on the equation of state that is adopted.

We conclude this chapter with a few comments concerning 
future plans for improvements and extensions of our code.
The most pressing issue is clearly that of getting 
adequate finite-difference resolution for those 
instances where the boson stars interact strongly, and especially 
when black holes appear to be forming. This is  simply not possible 
given that our current code uses a single uniform mesh, and only 
runs on one processor.  We therefore plan to parallelize the 
code but, perhaps more importantly, to incorporate adaptive 
mesh refinement (see, for example, the classic paper by
Berger and Oliger~\cite{BO:1984}), 
so that the increased resolution can be concentrated where 
it is needed.  
In fact, using the AMRD/PAMR software infrastructure developed 
by Pretorius~\cite{AMRD_web,PAMR_web}, we have already developed 
a version of the 
code that should both run in parallel (with good scaling on up to 
100's of processors) and provide AMR.  However, this implementation
is still being debugged and tested.

A second component of our algorithm that obviously needs improvement
is the treatment of the boundary conditions.
As discussed in
Sec.~\ref{sec:space_compact} a (formally) exact implementation of 
the physical boundary conditions could be achieved
through compactification of the 
spatial domain. Although it is promising, this approach would 
require the development of a more sophisticated multigrid solver,
capable of efficiently solving highly anisotropic equations,
and doing so in parallel.  This will not be an easy task.
Given that we are incorporating AMR into our code---which should 
allow us to substantially extend the boundaries of the computational
domain---we feel that it would be most fruitful at this time 
to implement the strategy described 
in~Sec.~\ref{sec:sommerfeld}, which uses mixed (Robin) conditions based 
on known falloff behaviour for the metric variables, and 
Sommerfeld conditions for the scalar field unknowns.

We end by emphasizing that the experiments described above represent only 
a start of a thorough investigation of the multi-dimensional parameter space of 
boson star orbital dynamics.  We hope that once we have implemented the 
improvements just discussed, we will be able to use our code to 
survey this space at significantly lower computational cost relative 
to fully general relativistic calculations.   Ideally, this survey 
would identify regions of parameter space that warrant more detailed 
investigation using codes that solve the full Einstein equations.

\resetcounters

\def\calH{{\mathcal H}}
\def\calT{{\mathcal T}}
\def\calR{{\mathcal R}}
\def\calG{{\mathcal G}}
\def\H#1#2{{\mathcal H}^{#1}{}_{#2}}
\def\shape{{\tt shape}}
\def\bbox{{\tt bbox}}
\def\vnfeIV{{\tt vnfe4}}
\def\vnfeV{{\tt vnfe5}}

\chap{Conclusion} \lab{conclusion}

The main new work presented in this thesis involved the implementation of a
finite-difference code to solve a nonlinear system of elliptic-hyperbolic partial
differential equations in 3 space dimensions plus time. 
The PDEs describe the evolution of a self-gravitating massive complex 
scalar field, where the gravitational interaction is given by the conformally
flat approximation of the Einstein equations.
This code was thoroughly tested to establish overall correctness of
the implementation (in particular, consistency with the continuum
PDEs), and was shown to generate results that converge to the continuum
limit at the expected rate in the mesh spacing.
The implementation was then used to study interactions between
two boson stars.  

In simulations of head-on collisions, we observed the same type of 
solitonic behaviour that has been seen in previous computations using 
this type of matter~\cite{dale,ChoiBEC,cwlai:phd,Choptuik:2009ww}.
We also performed a preliminary survey of orbital dynamics of boson 
star binaries, finding evidence for three distinct end states:
a long-lived, precessing elliptic orbit, merger of the two stars into a 
final, more compact boson star, and merger that we conjecture 
produces a black hole. 
These results are consistent with the findings
of Palenzuela and collaborators~\cite{Palenzuela:2007dm}, who solved
the full Einstein-scalar system.
This concordance suggests that the CFA 
may be sufficiently accurate to capture at least some 
of the key features of general relativistic dynamics, even in 
fully 3D scenarios.

The major weakness of our current code is its limited resolution.
For many of the runs we have described, the resolution was sufficient
for us to be confident that our modeling was capturing the key
physics exhibited in the evolutions.  However, in other instances,
and especially in those cases where we suspect that black holes
are forming, the calculations are {\em not} reliable at late integration
times, and only a substantial increase in resolution will fix 
this flaw.  As we have already discussed, we are currently working 
on a parallel, adaptive version of the code that should 
provide this fix.

Determining conditions under which the CFA does or does 
not provide a good approximation of general relativity is another
research priority.
Ultimately, this pursuit may even be able to shed some light on the 
elusive issue of identifying the true dynamical 
(radiative) degrees of freedom of Einstein's theory.

 \bibliographystyle{hunsrt}
 \bibliography{phd}
\appendix
 \resetcounters

\def\bsidpa{{\tt bsidpa}}
\def\dft{{\tt dft}}
\def\lsoda{{\tt lsoda}}
\def\odepack{{\tt odepack}}
\def\wshoot{{\tt wshoot}}

\chapter[BSIDPA---Boson Star Initial Data Function]
{BSIDPA---Boson Star Initial Data Function in Polar Areal Coordinates}\label{ap:bsidpa}

\bsidpa\ is a freely available\footnote{\bsidpa\ and supporting routines are maintained 
in the UBC numerical relativity group's {\tt ftp}
repository \cite{mundim:bsidpa_site}.} FORTRAN 77 subroutine that
generates numerical solutions describing static, spherically symmetric boson 
stars as described in Chap.~\ref{initialdata} (the reader should refer
to that chapter for details concerning notation, definitions of 
functions used below, etc. which will not be repeated here).
The routine name is an acronym for \textbf{B}oson 
\textbf{S}tar \textbf{I}nitial \textbf{D}ata in 
\textbf{P}olar-\textbf{A}real coordinates. 
\bsidpa\ has been written so that the user can specify 
a general polynomial self-interaction potential, $U(\phi_0(R))$,
for the scalar field:
\beq \label{eq:si_pot}
U(\phi_0) = \sum_{i=1}^{p_n} p_i \phi_0^i \, ,
\eeq
where $p_n$ is the degree of the interaction polynomial, and the 
coefficients $p_i, \, i = 1, 2, \ldots p_n$ are user-supplied.
Note that $p_n=2$, with $p_1=0$ and $p_2=1$,  is the case 
considered in this thesis.  In other words, the potential
only has a mass term, which leads to static configurations 
that are sometimes called mini boson stars.
As detailed in Chap.~\ref{initialdata}, for any given potential,
the static boson stars form a one-parameter family, and it is convenient
to use the central value of $\phi_0(r)$---i.e.~$\phi_0(0)$---as the 
family parameter.  We also recall that the 
ODE system~(\ref{eq:aprime_ham_IVP})--(\ref{eq:Phiprime_IVP})
(hereafter often referred to as ``the ODEs''), which \bsidpa\
solves numerically using the \lsoda\ integrator from 
ODEPACK~\cite{odepack}, constitutes an 
eigenvalue problem, with eigenvalue (eigenfrequency) $\omega=\omega(\phi_0(0))$.
Thus, another key input to \bsidpa\ is the central field value, 
$\phi_0(0)$.  If the user knows the corresponding eigenfrequency, that 
can also be supplied.
However, if 
$\omega$ is not known for the given value of $\phi_0(0)$, \bsidpa\ will 
attempt to compute it using a bisection algorithm. The bisection 
is based on the 
observation that, generically, as $R\to\infty$ we have 
$\phi_0(R) \to \infty$ for $\omega_{\rm hi} < \omega$, while 
$\phi_0(R) \to -\infty$ for $\omega_{\rm lo} > \omega$~\footnote{
Observe that in Chap.~\ref{initialdata} we used a notation such that
$\omega_{-}\equiv\omega_{\rm hi}$ and $\omega_{+}\equiv\omega_{\rm lo}$.
}.  In this latter case, the user can either supply an initial 
bracket $[\omega_{\rm hi},\omega_{\rm lo}]$ such that 
$\omega_{\rm hi} \le \omega \le \omega_{\rm lo}$, or \bsidpa\ can
attempt to locate such a bracket automatically.

A typical invocation of \bsidpa\ is illustrated by the following 
fragment of FORTRAN 77 code:
\begin{verbatim}
      ...

      integer   sht, n, pn, pt, dft, tail_type
      real*8    phi0, rmax, lsoda_tol, w_tol, dr, whi, wlo, wshoot, wresca
      real*8    a(n), alpha(n), phi(n), pp(n), m(n), zr(n), r(n)
      real*8    p(pn)

      ...

      call bsidpa(a,alpha,phi,pp,m,zr,r,p,phi0,rmax,lsoda_tol,w_tol,dr,
     &            whi,wlo,wshoot,wresca,sht,n,pn,pt,dft,tail_type)

      ...
\end{verbatim}

The various arguments to the routine can be classified
as being either inputs or outputs.
Further, some input arguments are required, while others are ``optional'',
in the sense that a flag can be set so that they are 
assigned default values.~\footnote{Note, however, that since FORTRAN
77 has no provision for subprograms with varying-length argument lists,
any invocation of \bsidpa\ must have exactly 23 arguments}

\noindent {\em Input Arguments} 

The following input arguments 
must be supplied:

\begin{list}{}{\lsep}
\item {\tt phi0:} The central value of the scalar field, $\phi_0(0)$ 
\item {\tt rmax:} The range of the integration: i.e. the ODEs 
are integrated from $R=0$ to $R={\tt rmax}$.
\item {\tt pt:} The potential type. Currently, the only valid value 
for {\tt pt} is
	\begin{itemize}
		\item ${\tt pt}=1\,$:  Polynomial potential as given by~(\ref{eq:si_pot}).
	\end{itemize}
\item {\tt p(pn):} real*8 vector of length {\tt pn} defining
the coefficients $p_i$ of the polynomial potential.
\item {\tt n:} Number of grid points in the interval 
$0\leq R \leq {\tt rmax}$ at which to compute the solution of the ODEs.
Specifically, the solution will be computed at 
$R_j\equiv (j-1)\Delta_R \,, j = 1, 2 \ldots,\, {\tt n}$, where 
$\Delta_R \equiv {\tt rmax}/({\tt n} - 1)$.
\item {\tt sht:} 
Flag that controls how the eigenfrequency, \wshoot, is to 
be determined by \bsidpa.
Valid values are as follows:
\begin{itemize}
\item ${\tt sht}=0\,$: User supplies \wshoot. 
\item ${\tt sht}=1\,$: Routine uses a shooting technique and 
bisection to determine \wshoot\
to within the user-specified tolerance {\tt w\_tol} (see below). 
Also see documentation of {\tt dft} below which controls whether 
the routine or user is responsible for determining an initial 
bracket for \wshoot.
\end{itemize}

\item \wshoot: Solution eigenfrequency as defined above.  Note: 
for ${\tt sht}=0$ this is an input argument, while for ${\tt sht}=1$ 
it is an output argument.
\item {\tt dft:} Flag that controls use of default values for 
``optional'' arguments as follows:
\begin{itemize}
\item $\dft = 0\,$: Routine uses default values for 
{\tt lsoda\_tol}, {\tt w\_tol}, {\tt dr}, {\tt whi} and {\tt wlo}.
\item $\dft = 1\,$: Routine uses default values for 
{\tt dr}, {\tt whi} and {\tt wlo}; user must supply values 
for {\tt lsoda\_tol} and {\tt w\_tol}.
\item $\dft = 2\,$: User must supply values for 
{\tt lsoda\_tol}, {\tt w\_tol}, {\tt dr}, {\tt whi} and {\tt wlo}.
\end{itemize}
\end{list}

The ``optional'' input parameters 
as well as their
default values, are defined as follows:
\begin{list}{}{\lsep}
\item {\tt lsoda\_tol:} \lsoda\ tolerance parameter. \lsoda\ will
attempt to keep the local error in the solution of the ODEs below
this value.  Default: ${\tt lsoda\_tol} =10^{-10}$.
\item {\tt w\_tol:} Tolerance for computation of the eigenfrequency 
estimate, \wshoot.  The bisection/shooting algorithm stops 
when $|\omega_{\rm low}-\omega_{\rm hi}| \le  {\tt w\_tol}$.
Default: ${\tt w\_tol}=10^{-9}$
\item {\tt dr:} Solution output interval used by \bsidpa\ when the algorithm 
is determining a bracket for the eigenvalue \wshoot.  It is useful
to be able to set {\tt dr} independently of the value $\Delta_R$ defined 
from {\tt rmax} and {\tt n} (see above) since if the estimate \wshoot\
is sufficiently poor, \lsoda\ may fail on the interval 
$0 \le R \le \Delta_R$.  In this case, setting ${\tt dr} \ll \Delta_R$
may aid in the bracket-locating process. Default: ${\tt dr}=10^{-4}$.
\item {\tt [whi,wlo]:} Values defining the initial bracket for eigenfrequency
\wshoot, and which should thus satisfy ${\tt whi} \le \wshoot \le {\tt wlo}$.
More specifically, for $\omega={\tt whi}$ and {\tt wlo}, respectively
the solution $\phi_0(R)$ should tend to $+\infty$ and 
$-\infty$, respectively, for large $R$.
\bsidpa\ \emph{does} have an algorithm encoded to automatically
look for an appropriate bracket, but this must be used with caution.
Automatic bracketing becomes particularly difficult in cases
where eigenfrequencies corresponding to the ground state,
and one or more of the excited states, are close to one another.
That can happen, for example, when one or more of the potential 
coefficients, $p_i$, are large.
When performing parameter space surveys in such instances, it 
is recommended that a continuation method be used, whereby an
initial bracket for a solution with $\phi_0(0)+\delta\phi_0(0)$ is given by
$[\omega(\phi_0(0))-\delta\omega,\,\omega(\phi_0(0))+\delta\omega]$, and where
$\delta\phi_0(0)$ and $\delta\omega$ are adjusted as necessary to ensure
proper bracketing.
Default: ${\tt [whi,wlo]}= [1.00,1.01]$ (appropriate for a mini boson
star with $m=1$ ($p_1=0$, $p_2=1$) and $\phi_0(0)=0.01$).
\item {\tt tail\_type:} Flag for controlling what closed form expression
is used to define $\phi_0$ at large values of $R$, via a fitting
procedure to the numerical solution of the ODEs.
Currently, only ${\tt tail\_type}=1$ is implemented, which results in a 
fit to 
\begin{equation}
\label{eq:tail-exp-fit}
    T_1(R)= A \exp(-BR) \, ,
\end{equation}
where $A$ and $B$ are coefficients determined from the fit.
\end{list}

\noindent {\em Output Arguments} 

\bsidpa\ returns the following output arguments, all of which 
are vectors of length {\tt n}:
\begin{list}{}{\lsep}
\item {\tt r(n):} The areal coordinate, $R$.
\item {\tt a(n):} The computed radial metric function, $a(R)$.
\item {\tt alpha(n):} The computed lapse function, $\al(R)$, rescaled 
so that $\lim_{R\to\infty} \alpha(R) = 1$.
\item {\tt phi(n):} The computed scalar field modulus, $\phi_0(R)$.
\item {\tt pp(n):} The computed derivative of the 
scalar field modulus, $\Phi_0(R)$.
\item {\tt m(n):} The mass aspect function $m(R)\equiv R(1-a(R)^{-2})/2$.
\item {\tt zr(n):}  The ``compactness function'' $z(R)\equiv2m(R)/R$.
This function provides a measure of how gravitationally compact a 
given star is.
\item {\tt wresca:} Rescaled value of $\om$.

\end{list}

Tab.~\ref{table:phi0_w} lists eigenfrequencies calculated 
using \bsidpa\ for a sequence of values of the family parameter,
$\phi_0(0)$, for the case $U(\phi_0) = (\phi_0)^2$.
Various properties of these stars are given in 
Tab.~\ref{table:phi0_properties1}. These include the cutoff 
radius, $R_{\rm cutoff}$ 
, the derivative of $m(R)$ at the cutoff,
the ADM mass, $M_{\rm ADM}$, and the two measures of the stellar 
radius, $R_{99}$ and $R_{95}$, that were defined in Sec.~\ref{sec:id-ansatz}.
Note that the contribution to the total mass from the tail is negligible
in all cases.
Tab.~\ref{table:phi0_properties2} lists additional properties including
$\min_R \alpha(R)$, $\max_R a(R)$, $\max_R z(R)$, the radius, 
$R_{{\rm max} z}$, at which $z(R)$ attains its maximum, and the 
Schwarzschild radius, $R_{\rm S}\equiv 2 M_{\rm ADM}$, associated
with each star.
The values of $\max_R z(R)$ are particularly noteworthy, clearly 
showing
that the overall gravitational field of the stars gets stronger as 
$\phi_0(0)$ increases.

\begin{table}[htbp]
\begin{center}
\begin{tabular}[l]{| c | c | c |}
\hline
$\phi_0(0)$ & {\tt wshoot}         &    {\tt wresca}         \\
\hline
0.005 & 1.01173913657665  & 0.987921656870952  \\
0.006 & 1.01414229214191  & 0.985551416322374  \\
0.007 & 1.01656427949667  & 0.983196152260810  \\
0.008 & 1.01900526873767  & 0.980855788252950  \\
0.009 & 1.02146543309092  & 0.978530270851457  \\
0.01  & 1.02394494622946  & 0.976219519074950  \\
0.02  & 1.04984409445897  & 0.953910295230624  \\
0.03  & 1.07788555219769  & 0.933010708580807  \\
0.04  & 1.10828046126291  & 0.913467522154004  \\
0.05  & 1.14126635619905  & 0.895235350281201  \\
0.06  & 1.17711107464507  & 0.878276194485022  \\
0.07  & 1.21611731750891  & 0.862559036385476  \\
0.08  & 1.25862797105685  & 0.848059515066958  \\
0.09  & 1.30503232180141  & 0.834759601661149  \\ 
\hline
\end{tabular}

\caption[Central Scalar Field Values and Eigenfrequencies 
$\om$ for Mini Boson Stars.]{Central scalar field values and 
eigenfrequencies for mini boson stars (i.e. for boson stars
with interaction potential $U(\phi_0)=(\phi_0)^2$). 
Listed are bare and rescaled values of the eigenfrequency,
{\tt wshoot} and {\tt wresca}, respectively, for a sequence 
of central scalar field values, $\phi_0(0)$.
The calculations were performed using
${\tt rmax} = 100$, ${\tt n}=2^{16}+1$, ${\tt dr}=10^{-4}$,
and default values for all other parameters.
}
\label{table:phi0_w}
\end{center}
\end{table}

\begin{table}[htbp]
\begin{center}
\begin{tabular}[l]{| c | c | c | c | c | c | }
\hline
$\phi_0(0)$ & $R_{\rm cutoff}$ & $dm/dR|_{R_{\rm cutoff}}$  & $M_{ADM}$ & $R_{99}$ & $R_{95}$ \\
\hline
0.005&91.12&1.34$\times10^{-9}$&0.2646&35.78&28.15\\
0.006&81.42&1.13$\times10^{-9}$&0.2878&32.58&25.62\\
0.007&73.58&1.56$\times10^{-9}$&0.3086&30.09&23.65\\
0.008&71.63&0.84$\times10^{-9}$&0.3275&28.08&22.06\\
0.009&65.72&0.92$\times10^{-9}$&0.3449&26.40&20.74\\
0.01 &63.85&1.22$\times10^{-9}$&0.3609&24.99&19.63\\
0.02 &45.49&2.06$\times10^{-9}$&0.4751&17.26&13.51\\
0.03 &36.62&2.62$\times10^{-9}$&0.5424&13.77&10.74\\
0.04 &31.99&3.25$\times10^{-9}$&0.5844&11.67&9.065\\
0.05 &28.39&3.94$\times10^{-9}$&0.6103&10.22&7.909\\
0.06 &25.44&4.83$\times10^{-9}$&0.6251&9.148&7.048\\
0.07 &23.74&5.37$\times10^{-9}$&0.6319&8.311&6.375\\
0.08 &22.16&5.99$\times10^{-9}$&0.6327&7.637&5.830\\
0.09 &20.84&7.21$\times10^{-9}$&0.6290&7.081&5.379\\
\hline
\end{tabular}

\caption[Mini Boson Star Properties: Tail, Mass and Radius.]{ Mini Boson 
Star Properties: Tail, Mass and Radius. This table lists the location,
$R_{\rm cutoff}$, at which $\phi_0(R)$ was fit to~(\ref{eq:tail-exp-fit}),
$dm/dR|_{R_{\rm cutoff}}$, the ADM mass of the star, 
$M_{\rm ADM}$, and the areal radii containing 
$99\%M_{\rm ADM}$ and $95\%M_{\rm ADM}$. 
The computational parameters used to obtain this data are described
in the caption of Tab.~\ref{table:phi0_w}.
Note that the derivative of the mass aspect function at $R_{\rm cutoff}$ is 
always extremely small, implying that the contribution to $M_{\rm ADM}$ 
from the tail is negligible. 
}

\label{table:phi0_properties1}
\end{center}
\end{table}

\begin{table}[htbp]
\begin{center}
\begin{tabular}[l]{| c | c | c | c | c | c | }
\hline
$\phi_0(0)$ & $\min{\alpha(R)}$ & $\max{\psi(R)}$ & $\max{z(R)}$ & $R_{\max{z}}$ & $R_{S}$  \\
\hline
0.005&0.9764&1.0111&0.0207&19.03&0.529\\
0.006&0.9718&1.0134&0.0248&17.32&0.575\\
0.007&0.9672&1.0157&0.0288&15.98&0.617\\
0.008&0.9626&1.0180&0.0328&14.90&0.655\\
0.009&0.9580&1.0203&0.0367&14.00&0.690\\
0.01 &0.9534&1.0226&0.0406&13.24&0.722\\
0.02 &0.9086&1.0454&0.0780& 9.05&0.950\\
0.03 &0.8656&1.0678&0.1124& 7.14&1.085\\
0.04 &0.8242&1.0898&0.1441& 5.98&1.169\\
0.05 &0.7844&1.1114&0.1734& 5.17&1.220\\
0.06 &0.7461&1.1325&0.2004& 4.56&1.250\\
0.07 &0.7093&1.1532&0.2253& 4.08&1.264\\
0.08 &0.6738&1.1735&0.2483& 3.69&1.265\\
0.09 &0.6397&1.1933&0.2694& 3.36&1.258\\
\hline
\end{tabular}

\caption[Mini Boson Star Properties: Extrema of Metric Components.]{ 
Mini Boson Star Properties: Extrema of Metric Components. This 
table lists the minimum value of the lapse function, $\min{\alpha(R)}$, 
the maximum value of the conformal factor, $\max{\psi(R)}$ and the
maximum values of the compactness function $\max{z(R)}$, 
as well as the location,  
$R_{\max{z}}$, at which $z(R)$ attains its maximum.
For comparison purposes, the Schwarzschild radius,
$R_{S}\equiv 2 M_{\rm ADM}$, 
associated with each star, is also tabulated. 
The computational parameters used to obtain this data are described
in the caption of Tab.~\ref{table:phi0_w}.
}

\label{table:phi0_properties2}
\end{center}
\end{table}

 \resetcounters

\chapter{Finite Difference Approximations}\label{ap:stencil}
 
The technique of finite differencing 
is the most common approach to the discretization of time
dependent PDEs that has been used 
in numerical relativity to date.  The specific finite difference 
approximations (FDAs) that we employ in this thesis are all straightforward,
and the novice can consult any basic text on the subject, such
as Mitchell and Griffiths~\cite{mitchell_griffiths}, for details 
on how they can be derived using, for example, Taylor series 
expansion.  The main purpose of this appendix is to collect 
and explicitly display all of the FDAs that are used in 
the 3D finite difference code described in the body
of the thesis.
In addition, in Sec.~\ref{sec:ddrp} we briefly describe a technique which 
is useful for deriving FDAs with good regularity properties when 
one is working in curvilinear coordinates (spherical polar, 
cylindrical~etc.).

\section{Finite Difference Operators} \label{sec:stencil}
We first recall (Sec.~\ref{sec:FDA}) that we adopt 
Cartesian (rectangular) coordinates for our computations, and that 
our spatial solution domain is then defined by 
\bea
x_{\rm min}\le &x& \le x_{\rm max}, \\
y_{\rm min}\le &y& \le y_{\rm max}, \\
z_{\rm min}\le &z& \le z_{\rm max},
\eea
where 
$x_{\rm min}$, $x_{\rm max}$, $y_{\rm min}$, $y_{\rm max}$, $z_{\rm min}$ and
$z_{\rm max}$ are prescribed values.

We discretize this domain by introducing a finite difference grid (or mesh),
in which the same constant mesh spacing, $h$, is used in each of 
the three coordinate directions.  The discrete spatial
coordinates $(x_i, y_j, z_k)$
that label the grid points are given by
\bea
x_i &=& x_{\rm min} + (i-1)\,h, \qquad \qquad  i=1 \dots n_{x},\\
y_j &=& y_{\rm min} + (j-1)\,h, \qquad \qquad  j=1 \dots n_{y},\\
z_k &=& z_{\rm min} + (k-1)\,h, \qquad \qquad  k=1 \dots n_{z},
\eea
such that
\begin{alignat}{2}
x_1 &= x_{\rm min} \qquad& \textrm{and} \qquad  x_{n_x} &= x_{\rm max}, \\
y_1 &= y_{\rm min} \qquad& \textrm{and} \qquad  y_{n_y} &= y_{\rm max}, \\
z_1 &= z_{\rm min} \qquad& \textrm{and} \qquad  z_{n_z} &= z_{\rm max}.
\end{alignat}
Here there is an implicit assumption that each of the
ranges $x_{\rm max}-x_{\rm min}$, $y_{\rm max}-y_{\rm min}$
and $z_{\rm max}-z_{\rm min}$ is evenly divisible by $h$.

Furthermore, assuming that the calculation is performed on the time 
interval $0\le t \le t_{\rm max}$, we similarly discretize the
time coordinate by defining values, $t^n$, given by
\beq
t^n = (n-1)\,\lambda h, \qquad  t=1 \dots n_{t} \, .
\eeq
Here, $\lambda$---which is the ratio of the temporal and spatial
mesh spacings---is known as the Courant number. It will generally satisfy
$\lambda < 1$ from stability considerations, and will be held 
constant when we perform a sequence of computations in 
which $h$ is varied (when we are performing a convergence test,
for example).  Again, there is a tacit assumption that $\lambda h$ 
evenly divides $t_{\rm max}$.

For a generic solution unknown, $u(t,x,y,z)$, which  is 
to be approximated on the grid, we then introduce the standard finite 
difference notation 
\beq
	u^n_{i,j,k} \equiv u(t^n,x_i,y_j,z_k)\, ,
\eeq
for the so-called grid-function values, $u^n_{i,j,k}$.

Having dispensed with basic definitions, we observe that the 
finite difference approximations that we have used fall into 
two sets.  Members of the first set, which are listed in 
Tables~\ref{table:1st_center}--\ref{table:2nd_mixed_center},
were used in the actual discretization of the PDEs governing our 
model, and are all so-called centred approximations.  
The second set, enumerated in 
Tables~\ref{table:1st_forward}--\ref{table:2nd_mixed_bw_bw},
were used to compute independent residuals~(see Sec.~\ref{subsec:IRV}), 
and all of these are either forward or backward approximations.
In all cases, the approximations are second order in the mesh 
spacing, $h$.  This includes the FDA~(\ref{eq:dudt-fda}) for 
the first time derivative, $\pa u/\pa t$, which is used
in differencing the evolution equations~(\ref{eq:phidot_adm_cart}) 
and (\ref{eq:pidot_adm_cart}) using a Crank-Nicholson scheme,
and where the difference equations are centred at the ``fictitious''
grid point $(t^{n+1/2},x_i,y_j,z_k)\equiv(t^n+\lambda h/2,x_i,y_j,z_k)$.

\begin{table}[htp]
\begin{eqnarray}
\fr{\pa u}{\pa x} & \rightarrow  & \fr{u^n_{i+1,j,k}-u^n_{i-1,j,k}}{2h}  \\
\fr{\pa u}{\pa y} & \rightarrow  & \fr{u^n_{i,j+1,k}-u^n_{i,j-1,k}}{2h}  \\
\fr{\pa u}{\pa z} & \rightarrow  & \fr{u^n_{i,j,k+1}-u^n_{i,j,k-1}}{2h}  \\
\label{eq:dudt-fda}
\fr{\pa u}{\pa t} & \rightarrow  & \fr{u^{n+1}_{i,j,k}-u^n_{i,j,k}}{\la h}
\end{eqnarray}
\caption[Centred FDAs of First Order Derivatives.]{
Centred FDAs of First Order Derivatives.
}
\label{table:1st_center}
\end{table}

\begin{table}[htp]
\begin{eqnarray}
\mu_t u & \equiv  & \fr{u^{n+1}_{i,j,k}+u^n_{i,j,k}}{2}  
\end{eqnarray}
\caption[Time Averaging Operator.]{
Time Averaging Operator.
}
\label{table:time_average}
\end{table}

\begin{table}[htp]
\begin{eqnarray}
\fr{\pa^2 u}{\pa x^2} & \rightarrow  & \fr{u^n_{i+1,j,k}-2u^n_{i,j,k}+u^n_{i-1,j,k}}{h^2}  \\
\fr{\pa^2 u}{\pa y^2} & \rightarrow  & \fr{u^n_{i,j+1,k}-2u^n_{i,j,k}+u^n_{i,j-1,k}}{h^2}  \\
\fr{\pa^2 u}{\pa z^2} & \rightarrow  & \fr{u^n_{i,j,k+1}-2u^n_{i,j,k}+u^n_{i,j,k-1}}{h^2}  
\end{eqnarray}
\caption[Centred FDAs of Second Order Derivatives.]{
Centred FDAs of Second Order Derivatives.
}
\label{table:2nd_center}
\end{table}

\begin{table}[htp]
\begin{eqnarray}
\fr{\pa^2 u}{\pa x \pa y} & \rightarrow  & \fr{u^n_{i-1,j-1,k}
          -u^n_{i-1,j+1,k}-u^n_{i+1,j-1,k}+u^n_{i+1,j+1,k}}{4h^2}  \\
\fr{\pa^2 u}{\pa y \pa z} & \rightarrow  & \fr{u^n_{i,j-1,k-1}
          -u^n_{i,j+1,k-1}-u^n_{i,j-1,k+1}+u^n_{i,j+1,k+1}}{4h^2}  \\
\fr{\pa^2 u}{\pa x \pa z} & \rightarrow  & \fr{u^n_{i-1,j,k-1}
          -u^n_{i+1,j,k-1}-u^n_{i-1,j,k+1}+u^n_{i+1,j,k+1}}{4h^2}  
\end{eqnarray}
\caption[Centred FDAs of Mixed Second Order Derivatives.]{
Centred FDAs of Mixed Second Order Derivatives.
}
\label{table:2nd_mixed_center}
\end{table}

\begin{table}[htp]
\begin{eqnarray}
\fr{\pa u}{\pa x} & \rightarrow  & \fr{-3u^n_{i,j,k}
                      +4u^n_{i+1,j,k}- u^n_{i+2,j,k}}{2h}  \\
\fr{\pa u}{\pa y} & \rightarrow  & \fr{-3u^n_{i,j,k}
                      +4u^n_{i,j+1,k}- u^n_{i,j+2,k}}{2h}  \\
\fr{\pa u}{\pa z} & \rightarrow  & \fr{-3u^n_{i,j,k}
                      +4u^n_{i,j,k+1}- u^n_{i,j,k+2}}{2h}  
\end{eqnarray}
\caption[Forward FDAs of First Order Derivatives.]{
Forward FDAs of first order derivatives.
}
\label{table:1st_forward}
\end{table}

\begin{table}[htp]
\begin{eqnarray}
\fr{\pa u}{\pa x} & \rightarrow  & \fr{3u^n_{i,j,k}
                      -4u^n_{i-1,j,k}+u^n_{i-2,j,k}}{2h}  \\
\fr{\pa u}{\pa y} & \rightarrow  & \fr{3u^n_{i,j,k}
                      -4u^n_{i,j-1,k}+u^n_{i,j-2,k}}{2h}  \\
\fr{\pa u}{\pa z} & \rightarrow  & \fr{3u^n_{i,j,k}
                      -4u^n_{i,j,k-1}+u^n_{i,j,k-2}}{2h}  
\end{eqnarray}
\caption[Backward FDAs of First Order Derivatives.]{
Backward FDAs of First Order Derivatives.
}
\label{table:1st_backward}
\end{table}

\begin{table}[htp]
\begin{eqnarray}
\fr{\pa^2 u}{\pa x^2} & \rightarrow  & \fr{2u^n_{i,j,k}
         -5u^n_{i+1,j,k}+4u^n_{i+2,j,k}-u^n_{i+3,j,k}}{h^2}  \\
\fr{\pa^2 u}{\pa y^2} & \rightarrow  & \fr{2u^n_{i,j,k}
         -5u^n_{i,j+1,k}+4u^n_{i,j+2,k}-u^n_{i,j+3,k}}{h^2}  \\
\fr{\pa^2 u}{\pa z^2} & \rightarrow  & \fr{2u^n_{i,j,k}
         -5u^n_{i,j,k+1}+4u^n_{i,j,k+2}-u^n_{i,j,k+3}}{h^2}  
\end{eqnarray}
\caption[Forward FDAs of Second Order Derivatives.]{
Forward FDAs of Second Order Derivatives.
}
\label{table:2nd_forward}
\end{table}

\begin{table}[htp]
\begin{eqnarray}
\fr{\pa^2 u}{\pa x^2} & \rightarrow  & \fr{2u^n_{i,j,k}
         -5u^n_{i-1,j,k}+4u^n_{i-2,j,k}-u^n_{i-3,j,k}}{h^2}  \\
\fr{\pa^2 u}{\pa y^2} & \rightarrow  & \fr{2u^n_{i,j,k}
         -5u^n_{i,j-1,k}+4u^n_{i,j-2,k}-u^n_{i,j-3,k}}{h^2}  \\
\fr{\pa^2 u}{\pa z^2} & \rightarrow  & \fr{2u^n_{i,j,k}
         -5u^n_{i,j,k-1}+4u^n_{i,j,k-2}-u^n_{i,j,k-3}}{h^2}  
\end{eqnarray}
\caption[Backward FDAs of Second Order Derivatives.]{
Backward FDAs of Second Order Derivatives.
}
\label{table:2nd_backward}
\end{table}

\begin{table}[htp]
\begin{eqnarray}
\fr{\pa^2 u}{\pa x \pa y} & \rightarrow  & 
\fr{
    9u^n_{i,j,k} -12u^n_{i,j+1,k} +3u^n_{i,j+2,k}
   -12u^n_{i+1,j,k} +16u^n_{i+1,j+1,k}- 4u^n_{i+1,j+2,k}
   }{4h^2}
\nonumber \\
& &
\qquad \qquad \qquad \qquad \qquad \qquad \qquad 
+\fr{
3u^n_{i+2,j,k} - 4u^n_{i+2,j+1,k}+  u^n_{i+2,j+2,k}
    }{4h^2}  \\
\fr{\pa^2 u}{\pa y \pa z} & \rightarrow  & 
\fr{
    9u^n_{i,j,k} -12u^n_{i,j,k+1} +3u^n_{i,j,k+2}
   -12u^n_{i,j+1,k} +16u^n_{i,j+1,k+1}- 4u^n_{i,j+1,k+2}
   }{4h^2}
\nonumber \\
& &
\qquad \qquad \qquad \qquad \qquad \qquad \qquad 
+\fr{
3u^n_{i,j+2,k} - 4u^n_{i,j+2,k+1}+  u^n_{i,j+2,k+2}
    }{4h^2}  \\
\fr{\pa^2 u}{\pa x \pa z} & \rightarrow  & 
\fr{
    9u^n_{i,j,k} -12u^n_{i,j,k+1} +3u^n_{i,j,k+2}
   -12u^n_{i+1,j,k} +16u^n_{i+1,j,k+1}- 4u^n_{i+1,j,k+2}
   }{4h^2}
\nonumber \\
& &
\qquad \qquad \qquad \qquad \qquad \qquad \qquad 
+\fr{
3u^n_{i+2,j,k} - 4u^n_{i+2,j,k+1}+  u^n_{i+2,j,k+2}
    }{4h^2}  
\end{eqnarray}
\caption[Forward-Forward FDAs of Mixed Second Order 
Derivatives.]{
Forward-Forward FDAs of Mixed Second Order Derivatives.
}
\label{table:2nd_mixed_fw_fw}
\end{table}

\begin{table}[htp]
\begin{eqnarray}
\fr{\pa^2 u}{\pa x \pa y} & \rightarrow  & 
\fr{
   -3u^n_{i,j-2,k} +12u^n_{i,j-1,k} -9u^n_{i,j,k}
   +4u^n_{i+1,j-2,k} -16u^n_{i+1,j-1,k} 
   }{4h^2}
\nonumber \\
& &
\qquad \qquad \qquad \qquad \qquad 
+\fr{
12u^n_{i+1,j,k}
   -u^n_{i+2,j-2,k} + 4u^n_{i+2,j-1,k}- 3u^n_{i+2,j,k}
   }{4h^2}  \\
\fr{\pa^2 u}{\pa y \pa z} & \rightarrow  & 
\fr{
   -3u^n_{i,j,k-2} +12u^n_{i,j,k-1} -9u^n_{i,j,k}
   +4u^n_{i,j+1,k-2} -16u^n_{i,j+1,k-1} 
   }{4h^2}
\nonumber \\
& &
\qquad \qquad \qquad \qquad \qquad 
+\fr{
12u^n_{i,j+1,k}
   -u^n_{i,j+2,k-2} + 4u^n_{i,j+2,k-1}- 3u^n_{i,j+2,k}
   }{4h^2}  \\
\fr{\pa^2 u}{\pa x \pa z} & \rightarrow  & 
\fr{
   -3u^n_{i,j,k-2} +12u^n_{i,j,k-1} -9u^n_{i,j,k}
   +4u^n_{i+1,j,k-2} -16u^n_{i+1,j,k-1} 
   }{4h^2}
\nonumber \\
& &
\qquad \qquad \qquad \qquad \qquad 
+\fr{
12u^n_{i+1,j,k}
   -u^n_{i+2,j,k-2} + 4u^n_{i+2,j,k-1}- 3u^n_{i+2,j,k}
   }{4h^2}  
\end{eqnarray}
\caption[Forward-Backward FDAs of Mixed Second Order 
Derivatives.]{
Forward-Backward FDAs of Mixed Second Order Derivatives.
}
\label{table:2nd_mixed_fw_bw}
\end{table}

\begin{table}[htp]
\begin{eqnarray}
\fr{\pa^2 u}{\pa x \pa y} & \rightarrow  & 
\fr{
   -3u^n_{i-2,j,k} + 4u^n_{i-2,j+1,k} - u^n_{i-2,j+2,k}
   +12u^n_{i-1,j,k} -16u^n_{i-1,j+1,k} 
   }{4h^2}
\nonumber \\
& &
\qquad \qquad \qquad \qquad \qquad 
+\fr{
4u^n_{i-1,j+2,k}
   -9u^n_{i,j,k} +12u^n_{i,j+1,k}- 3u^n_{i,j+2,k}
   }{4h^2}  \\
\fr{\pa^2 u}{\pa y \pa z} & \rightarrow  & 
\fr{
   -3u^n_{i,j-2,k} + 4u^n_{i,j-2,k+1} - u^n_{i,j-2,k+2}
   +12u^n_{i,j-1,k} -16u^n_{i,j-1,k+1} 
   }{4h^2}
\nonumber \\
& &
\qquad \qquad \qquad \qquad \qquad 
+\fr{
4u^n_{i,j-1,k+2}
   -9u^n_{i,j,k} +12u^n_{i,j,k+1}- 3u^n_{i,j,k+2}
   }{4h^2}  \\
\fr{\pa^2 u}{\pa x \pa z} & \rightarrow  & 
\fr{
   -3u^n_{i-2,j,k} + 4u^n_{i-2,j,k+1} - u^n_{i-2,j,k+2}
   +12u^n_{i-1,j,k} -16u^n_{i-1,j,k+1} 
   }{4h^2}
\nonumber \\
& &
\qquad \qquad \qquad \qquad \qquad 
+\fr{
4u^n_{i-1,j,k+2}
   -9u^n_{i,j,k} +12u^n_{i,j,k+1}- 3u^n_{i,j,k+2}
   }{4h^2} 
\end{eqnarray}
\caption[Backward-Forward FDAs of Mixed Second Order 
Derivatives.]{
Backward-Forward FDAs of Mixed Second Order Derivatives.
}
\label{table:2nd_mixed_bw_fw}
\end{table}

\begin{table}[htp]
\begin{eqnarray}
\fr{\pa^2 u}{\pa x \pa y} & \rightarrow  & 
\fr{
     u^n_{i-2,j-2,k} - 4u^n_{i-2,j-1,k} +3u^n_{i-2,j,k}
   - 4u^n_{i-1,j-2,k} +16u^n_{i-1,j-1,k} -12u^n_{i-1,j,k}
   }{4h^2}
\nonumber \\
& &
\qquad \qquad \qquad \qquad \qquad \qquad \qquad \qquad 
+\fr{
   3u^n_{i,j-2,k} -12u^n_{i,j-1,k}+ 9u^n_{i,j,k}
   }{4h^2}  \\
\fr{\pa^2 u}{\pa y \pa z} & \rightarrow  & 
\fr{
     u^n_{i,j-2,k-2} - 4u^n_{i,j-2,k-1} +3u^n_{i,j-2,k}
   - 4u^n_{i,j-1,k-2} +16u^n_{i,j-1,k-1} -12u^n_{i,j-1,k}
   }{4h^2}
\nonumber \\
& &
\qquad \qquad \qquad \qquad \qquad \qquad \qquad \qquad 
+\fr{
   3u^n_{i,j,k-2} -12u^n_{i,j,k-1}+ 9u^n_{i,j,k}
   }{4h^2}  \\
\fr{\pa^2 u}{\pa x \pa z} & \rightarrow  & 
\fr{
     u^n_{i-2,j,k-2} - 4u^n_{i-2,j,k-1} +3u^n_{i-2,j,k}
   - 4u^n_{i-1,j,k-2} +16u^n_{i-1,j,k-1} -12u^n_{i-1,j,k}
   }{4h^2}
\nonumber \\
& &
\qquad \qquad \qquad \qquad \qquad \qquad \qquad \qquad 
+\fr{
   3u^n_{i,j,k-2} -12u^n_{i,j,k-1}+ 9u^n_{i,j,k}
   }{4h^2}  
\end{eqnarray}
\caption[Backward-Backward FDAs of Mixed Second Order 
Derivatives.]{
Backward-Backward FDAs of Mixed Second Order Derivatives.
}
\label{table:2nd_mixed_bw_bw}
\end{table}

\newpage
\section{FDAs for Operators of the Form $\pa/\pa (r^p)$} \label{sec:ddrp}

In this section we briefly discuss a technique that is useful for 
deriving finite difference approximations that yield discrete solutions
with good regularity behaviour in the vicinity of coordinate 
singularities.  We note that the method was introduced to the 
numerical relativity community by Evans in his PhD 
dissertation~\cite{evans:phd}.

The technique is best illustrated by way of example.  We thus consider 
the radial part of the Laplacian operator in spherical polar coordinates,
$(r,\theta,\phi)$.  Assuming our unknown function, $u$, depends only
on $r$, we then have
\beq
\nabla^2 u = \nabla^2 u(r) = 
	\fr{1}{r^2} \fr{\pa}{\pa r} \lb r^2 \fr{\pa u}{\pa r} \rb \, .
\label{eq:sph-lap}
\eeq
In order that the solution be regular (smooth) at the origin,
$r=0$, $u(r)$ must satisfy
\beq \label{eq:psi_regular}
\lim_{r\rightarrow 0}u(r)=u_0+u_2r^2 +O(r^4) \, ,
\eeq
where $u_0$ and $u_2$ are constants.

A key observation is that if we insert the above expansion 
into the right hand side of~(\ref{eq:sph-lap}) we find
\bea
\fr{1}{r^2} \fr{\pa}{\pa r} \lb r^2 \fr{\pa u}{\pa r} \rb &=&
\fr{1}{r^2} \fr{\pa}{\pa r} \lhb r^2 \fr{\pa }{\pa r}\lb u_0+u_2r^2+O(r^4)\rb\rhb\nonumber\\ &=&
\fr{1}{r^2} \fr{\pa}{\pa r} \lb 2 u_2 r^3 +O(r^5)\rb = 6u_2 + O(r^2) \, .
\label{eq:6u2}
\eea

We now discrete the right hand side of~(\ref{eq:sph-lap})
by first introducing a uniform radial grid, 
$r_j \equiv (j-1)h$, $j=1, 2, \ldots n_r$, where $h$ is the mesh spacing,
and then writing 
\beq
\fr{1}{r^2} \fr{\pa}{\pa r} \lb r^2 \fr{\pa u}{\pa r} \rb \simeq
\fr{1}{r^2_j} \fr{\lhb 
r_{j+\ha}^2 \lb u_{j+1}- u_{j} \rb -
r_{j-\ha}^2 \lb u_{j}- u_{j-1} \rb 
\rhb}{\lb r_{j+1/2}-r_{j-1/2} \rb^2} \, .
\label{eq:fda1}
\eeq
Here $r_{j+1/2}\equiv r_j+h/2$ and $r_{j-1/2}\equiv r_j-h/2$.  The 
approximation~(\ref{eq:fda1}) is second order and seems natural 
enough, especially if one requires that the expression involve only 
the values $u_{j-1}$, $u_j$ and $u_{j-1}$. 
However, we now consider evaluating~(\ref{eq:fda1}) for $u(r)$
given by truncating~(\ref{eq:psi_regular}) at $O(r^2)$, i.e. for 
$u(r)\equiv u_0 + u_2 r^2$, where, again, $u_0$ and $u_2$ are 
constants. Then we find after some elementary algebra that 
\beq
\fr{1}{r^2_j} \fr{\lhb 
r_{j+\ha}^2 \lb u_{j+1}- u_{j} \rb -
r_{j-\ha}^2 \lb u_{j}- u_{j-1} \rb 
\rhb}{\lb r_{j+1/2}-r_{j-1/2} \rb^2} = 
6u_2 + \ha u_2 \fr{h^2}{r_j^2} \, . 
\label{eq:fda1-leading}
\eeq
This is {\em not} the correct leading order behaviour given 
by~(\ref{eq:6u2}).  In particular, at the second grid point, $r_2=h$,
this evaluates to $(6\frac{1}{2}) u_2$, instead of $6 u_2$, so that 
irrespective of how small we make $h$, we will never capture the 
correct leading order behaviour of the differential operator in
the vicinity of $r=0$.  Especially in dynamical evolutions, this 
can lead a lack of solution smoothness near the origin, and even
to numerical instability.

However, with a simple change of variables, and an application
of the chain rule, we can rewrite the radial Laplacian operator as 
follows 
\beq
	\frac{1}{r^2}\frac{\pa}{\pa r}\lb r^2 \frac{\pa u}{\pa r} \rb 
	= 3\frac{\pa}{\pa(r^3)}\lb r^2 \frac{\pa u}{\pa r}\rb
\eeq
which can then be discretized as
\beq
3\fr{\pa}{\pa (r^3)} \lb r^2 \fr{\pa u}{\pa r} \rb \simeq 
\fr{3}{\lb r^3_{j+\ha} - r^3_{j-\ha}\rb} \lhb 
r^2_{j+\ha} \fr{\lb u_{j+1}- u_{j} \rb}{\lb r_{j}-r_{j-1} \rb}-
r^2_{j-\ha} \fr{\lb u_{j}- u_{j-1} \rb}{\lb r_{j}-r_{j-1} \rb}
\rhb \, .
\eeq
When this finite difference approximation
is evaluated for $u(r)\equiv u_0 + u_2 r^2$
we find 
\beq
\fr{3}{\lb r^3_{j+\ha} - r^3_{j-\ha}\rb} \lhb 
r^2_{j+\ha} \fr{\lb u_{j+1}- u_{j} \rb}{\lb r_{j}-r_{j-1} \rb}-
r^2_{j-\ha} \fr{\lb u_{j}- u_{j-1} \rb}{\lb r_{j}-r_{j-1} \rb}
\rhb = 6u_2 \, .
\eeq
Thus, the correct leading order behaviour as $r\to0$ {\em is} 
mirrored in the discretization through this change-of-variables trick. 

The generalization of this strategy is straightforward.  When
confronted with a differential expression of the general form 
\beq
	\frac{\pa f(r)}{\pa r} \, ,
\label{eq:dfdr}
\eeq
where $f(r)$ satisfies the regularity condition
\beq
	\lim_{r\to0} f(r) = c_p r^p + c_{p+2} r^{p+2} + O(r^{p+4}) \, ,
\eeq
for some integer $p>0$, we rewrite~(\ref{eq:dfdr}) as  
\beq
	p \,r^{p-1}\frac{\pa f}{\pa(r^p)} 
\eeq
prior to finite differencing.

 \resetcounters

\def\maple{{\rm Maple}}
\def\dfdandoff{{\tt dfdandoff}}
\def\resndoff{{\tt resndoff}}
\def\fdaeval{{\tt fdaeval}}
\chapter{PDEFDAOFF: A MAPLE Package for FDAS}\label{ap:pdefda}

This appendix documents the use of the package PDEFDAOFF (an acronym for
Partial Differential Equation to Finite Difference Approximation using 
OFF-centred and centred approximation schemes). The package is a set 
of \maple\footnote{\maple\ is a registered trademark of 
Waterloo Maple Inc.} 
procedures we wrote to assist in the process of 
discretizing the PDEs governing our model using finite difference 
methods.  The package contains two key \maple\ procedures, \dfdandoff\ and 
\resndoff\ which are discussed with illustrative examples 
in Secs.~\ref{sec:a3-1} and \ref{sec:a3-2} below.
We note that the examples were 
extracted directly from the demonstration worksheet that is included in the 
distribution~\cite{mundim:pdefdaoff_site}.

\section{FDAs of $n$-dimensional Differential Operators}
\label{sec:a3-1}

\dfdandoff~\footnote{\dfdandoff\ is an acronym for {\tt d}ifferential
{\tt f}inite {\tt d}ifference {\tt a}pproximation for 
{\tt n}-{\tt d}imensional functions including {\tt off}-centred 
schemes.}
is a \maple\
procedure that computes a finite difference 
approximation to a differential operator,
as applied to function of an arbitrary number of coordinates
$(x^1,x^2,\ldots,x^n)$.  Given a user-specified order of accuracy, $p$,
the routine returns a finite difference approximation which 
is $O(h^p)$ accurate in all coordinate directions.  
This procedure can calculate both centred and off-centred 
finite difference formula. 

The routine has a header given by 
\begin{verbatim}
   dfdandoff := proc(u::function, x::list(name), ud::name, j::list(name),
                     h::list(name), q::list(integer), p::integer,
                     off::list(integer))
\end{verbatim}
where the procedure arguments are defined as follows:
\begin{enumerate}
\item {\tt u}: A generic \maple\ function of $n$ variables.
\item {\tt x}: Length-$n$ list of independent variables (coordinates),
               in alphabetical order.
\item {\tt ud}: \maple\ name for the finite difference unknown.
\item {\tt j}: Length-$n$ list of names of indices corresponding to 
					each coordinate.
\item {\tt h}: Length-$n$ list of coordinate spacings in each direction.
\item {\tt q}: Length-$n$ list specifying the order of the differential
               operator with respect to each coordinate direction.
\item {\tt p}: Approximation order for finite difference scheme
\item {\tt off}: Length-$n$ list of offsets (one of $-1$, $0$ or $1$)
         which defines the off-centring of the approximation in 
         each coordinate direction.
\end{enumerate}

\subsection{1-dimensional Examples}

We first define a \maple\ alias for $u(x)$ which suppresses the functional
dependence on output, and allows us to omit the explicit dependence on 
input:
\begin{verbatim}
   > alias(u=u(x));
\end{verbatim}
\beq
\nonumber
u
\eeq

Our initial examples deal with first derivatives and illustrate the 
importance of proper specification of the ``off-centring'' argument,
{\tt off}.  Thus, if one naively attempts to find a first order 
centred scheme for $du/dx$, by using {\tt [0]} as the value 
of {\tt off}, the procedure executes as follows:
\begin{verbatim}
   > dfdandoff(u, [x], ud, [i], [hx], [1], 1, [0]);
     Error, (in dfdandoff) invalid input: rhs received 1, which is not valid 
     for its 1st argument, expr 
\end{verbatim}
We do not consider this a bug since there is {\em no} $O(h)$ centred 
scheme for $d/dx$.  Indeed, the 
simplest FDAs of $d/dx$ are the first order backward and forward 
approximations.  We can generate the backward formula using 
{\tt off := [-1]}:
\begin{verbatim}
   > dfdandoff(u, [x], ud, [i], [hx], [1], 1, [-1]);
\end{verbatim}
\beq
\nonumber
{\frac {{\pa}}{{\pa}x}}u_{{i}}=-{\frac {{\it ud}_{{i-1}}}{{\it hx}}}+{\frac {{\it ud}_{{i}}}{{\it hx}}}
\eeq

To verify the correctness of the 
finite difference operators that \dfdandoff\ computes---which 
means that we ensure that that approximations {\em are} $O(h^p)$ 
accurate---we use another \maple\ procedure, \fdaeval, due to 
Choptuik and which is included in {\tt PDEFDAOFF}.
\fdaeval\ has the header
\begin{verbatim}
   fdaeval := proc(c::array(integer), sigma::algebraic)
\end{verbatim}
with arguments defined as follows:
\begin{enumerate}
\item {\tt c}: A coefficient array such that ${\tt c[j]}$ is the 
   relative weight of the unknown, $u_{i+j}$, in the finite difference
   approximation centred at $x_i$.
\item {\tt sigma}: An overall scale factor to be applied to the 
	formula.
\end{enumerate}
The input to \fdaeval\ thus defines an expression of the 
form 
\beq
\label{eq:fdaeval}
	\sigma \sum_{j=j_{\rm min}}^{j_{\rm max}} c_j u_{i+j} \equiv
	\sigma \sum_{j=j_{\rm min}}^{j_{\rm max}} c_j u(x_i + j h)
\eeq
where $h$ is the (constant) mesh spacing that is hard-coded in the 
procedure. \fdaeval\ then replaces all of 
the $u_{i+j}\equiv u(x_i + j h)$  with Taylor series expansions 
(in $h$) about the value $u_i\equiv u(x_i)$.  Provided 
that~(\ref{eq:fdaeval}) {\em does} define a consistent approximation
to some differential operator applied at $x=x_i$, \fdaeval\ will
return that operator (acting on $u$, which again is hard coded into 
the routine), as well as the leading order truncation error terms.

Thus, we can check the above invocation of 
\dfdandoff\ as follows:
\begin{verbatim}
   > abwh := array(-1..0, [-1,1]);
\end{verbatim}
\beq
\nonumber
\hbox{{\it abwh}}:=\hbox{{\it ARRAY}} \left( -1..0, [0=1,-1=-1 ] \right)
\eeq
\begin{verbatim}
   > fdaeval(abwh, 1/h);
\end{verbatim}
\beq
\nonumber
\mbox {D} \left( u \right)  \left( x \right) -\ha \,  \mbox {D}^{ \left( 2 \right) } 
  \left( u \right)  \left( x \right) h+ \fr{1}{6}\,  \mbox {D}^{ \left( 3 \right) }  \left(
u \right)  \left( x \right) {h}^{2}
\eeq
This shows that, as expected, our backward formula 
{\em is} a valid first order finite difference approximation of 
$du/dx$.

Continuing, we can generate and check the $O(h^2)$ centred, forward 
and backward approximations of $du/dx$ as follows:
\begin{verbatim}
   > # Centred approximation 
   > dfdandoff(u, [x], ud, [i], [hx], [1], 2, [0]);
\end{verbatim}
\beq
\nonumber
{\frac {{\pa}}{{\pa}x}}u_{{i}}=-\ha\,{\frac {{\it ud}_{{i-1}}}{{\it hx}}}+\ha\,{\frac {{\it ud}_{{i+1}}}{{\it hx}}}
\eeq
\begin{verbatim}
   > ach2 := array(-1..1, [-1,0,1]):
   > fdaeval(ach2, 1/(2*h));
\end{verbatim}
\beq
\nonumber
\mbox {D} \left( u \right)  \left( x \right) +\fr{1}{6}\, \mbox {D}^{ \left( 3 \right) } \left( u \right)  \left(x \right) {h}^{2}
\eeq
\begin{verbatim}
   > # Backward approximation
   > dfdandoff(u, [x], ud,[i], [hx], [1], 2, [-1]); 
\end{verbatim}
\beq
\nonumber
{\frac{\pa}{\pa x}}u_{{i}}=\ha\,{\frac {{\it ud}_{{i-2}}}{{\it hx}}}-2\,{\frac {{\it ud}_{{i-1}}}{{\it hx}}}+\fr{3}{2}\,{\frac {{\it ud}_{{i}}}{{\it hx}}}
\eeq
\begin{verbatim}
   > abwh2 := array(-2..0, [1,-4,3]):
   > fdaeval(abwh2, 1/(2*h));
\end{verbatim}
\beq
\nonumber
\mbox {D} \left( u \right)  \left( x \right) -\fr{1}{3}\, \mbox {D}^{ \left( 3 \right) } \left( u \right)  \left( x \right) {h}^{2}+\fr{1}{4}\, \mbox {D}^{ \left( 4 \right) } \left( u \right)  \left( x \right) {h}^{3}
\eeq
\begin{verbatim}
   > # Forward approximation
   > dfdandoff(u, [x], ud, [i], [hx], [1], 2, [1]);
\end{verbatim}
\beq
\nonumber
{\frac {\pa}{\pa x}}u_{{i}}=-\fr{3}{2}\,{\frac {{\it ud}_{{i}}}{{\it hx}}}+2\,{\frac {{\it ud}_{{i+1}}}{{\it hx}}}-\ha\,{\frac {{\it ud}_{{i+2}}}{{\it hx}}}
\eeq
\begin{verbatim}
   > afwh2 := array(0..2, [-3,4,-1]):
   > fdaeval(afwh2, 1/(2*h));
\end{verbatim}
\beq
\nonumber
\mbox {D} \left( u \right)  \left( x \right) -\fr{1}{3}\, \mbox {D}^{ \left( 3 \right) } \left( u \right)  \left( x \right) {h}^{2}-\fr{1}{4}\, \mbox {D}^{ \left( 4 \right) } \left( u \right)  \left( x \right) {h}^{3}
\eeq

We conclude this section with examples illustrating the computation
of $O(h^4)$ accurate approximations for the second derivative,
$d^2u/dx^2$. 
\begin{verbatim}
   > # Centred approximation
   > dfdandoff(u, [x], ud, [i], [hx], [2], 4, [0]);
\end{verbatim}
\beq
\nonumber
{\frac {\pa^{2}}{\pa{x}^{2}}}u_{{i}}=-\fr{1}{12}\,{\frac {{\it ud}_{{i-2}}}{{{\it hx}}^{2}}}+\fr{4}{3}\,{\frac {{\it ud}_{{i-1}}}{{{\it hx}}^{2}}}-\fr{5}{2}\,{\frac {{\it ud}_{{i}}}{{{\it hx}}^{2}}}\\
\mbox{}+\fr{4}{3}\,{\frac {{\it ud}_{{i+1}}}{{{\it hx}}^{2}}}-\fr{1}{12}\,{\frac {{\it ud}_{{i+2}}}{{{\it hx}}^{2}}}
\eeq
\begin{verbatim}
   > a2ch4 := array(-2..2, [-1,16,-30,16,-1]):
   > fdaeval(a2ch4, 1/(12*h**2));
\end{verbatim}
\beq
\nonumber
 \mbox {D}^{ \left( 2 \right) } \left( u \right)  \left( x \right) -{\frac {1}{90}}\, \mbox {D}^{ \left( 6 \right) } \left( u \right)  \left( x \right) {h}^{4}
\eeq
\begin{verbatim}
   > # Forward approximation 
   > dfdandoff(u, [x], ud, [i], [hx], [2], 4, [2]);
\end{verbatim}
\beq
\nonumber
{\frac {\pa^{2}}{\pa{x}^{2}}}u_{{i}}={\frac {15}{4}}\,{\frac {{\it ud}_{{i}}}{{{\it hx}}^{2}}}-{\frac {77}{6}}\,{\frac {{\it ud}_{{i+1}}}{{{\it hx}}^{2}}}\\
\mbox{}+{\frac {107}{6}}\,{\frac {{\it ud}_{{i+2}}}{{{\it hx}}^{2}}}-13\,{\frac {{\it ud}_{{i+3}}}{{{\it hx}}^{2}}}+{\frac {61}{12}}\,{\frac {{\it ud}_{{i+4}}}{{{\it hx}}^{2}}}\\
\mbox{}-\fr{5}{6}\,{\frac {{\it ud}_{{i+5}}}{{{\it hx}}^{2}}}
\eeq
\begin{verbatim}
   > a2fw2h4 := array(0..5, [45,-154,214,-156,61,-10]):
   > fdaeval(a2fw2h4, 1/(12*h**2));
\end{verbatim}
\beq
\nonumber
\mbox {D}^{ \left( 2 \right) } \left( u \right)  \left( x \right) -{\frac {137}{180}}\, \mbox {D}^{ \left( 6 \right) } \left( u \right)  \left(x \right) {h}^{4}\\
\mbox{}-{\frac {19}{12}}\, \mbox {D}^{ \left( 7 \right) } \left( u \right)  \left(x \right) {h}^{5}
\eeq

\subsection{3-dimensional Example}

The following example illustrates the use of \dfdandoff\ for functions 
with dependence on three independent variables.
\begin{verbatim}
   > alias(w=w(x,y,z));
\end{verbatim}
\beq
\nonumber
w 
\eeq
\begin{verbatim}
   > # Second order forward approximation of d/dz
   > dfdandoff(w, [x,y,z], wd, [i,j,k], [hx,hy,hz], [0,0,1], 2, [0,0,1]);
\end{verbatim}
\beq
\nonumber
{\frac {\pa}{\pa z}} w_{{i,j,k}} =-\fr{3}{2}\,{
\frac {{\it wd}_{{i,j,k}}}{{\it hz}}}+2\,{\frac {{\it wd}_{{i,j,k+1}}}{{\it hz}}}\\
\mbox{}-\ha\,{\frac {{\it wd}_{{i,j,k+2}}}{{\it hz}}}
\eeq

\section{Residual Evaluators for $n$-dimensional PDEs}
\label{sec:a3-2}

\resndoff~\footnote{\resndoff\ is an acronym for {\tt res}idual
evaluator for {\tt n}-{\tt d}imensional PDEs 
including {\tt off}-centred finite difference schemes.}
is a \maple\ procedure that automatically generates residuals evaluators 
associated with the finite-difference discretization of a 
general partial differential equation in $n$-dimensions.
This routine was initially written with elliptic PDEs in mind, and
in conjunction with the multigrid method, where 
the evaluation of residuals plays an important role when a relaxation
technique is used as a smoother.  However \resndoff\ can also be 
employed in the context of time-dependent PDEs.
In particular, it is very useful for generating independent residual 
evaluators for {\em all} types of PDEs, including time-dependent ones.
As discussed in Chap.~\ref{numerical},
independent residual evaluation is a very powerful tool for verifying 
the implementation of finite difference codes, especially when 
one can be confident that the independent residual evaluators are 
themselves error-free.  Use of a routine such as \resndoff\ 
helps ensure that this is the case.

\resndoff\ has the following header:
\begin{verbatim}
   resndoff := proc(eqn::equation, u::list(function), x::list(name),
                  ud::list(name), j::list(name), h::list(name), p::integer,
                  off::list(integer))
\end{verbatim}
with procedure arguments defined as follows:
\begin{enumerate}
\item {\tt eqn}: Partial differential equation for which independent 
                 residual is to be computed.
\item {\tt u}: List of \maple\ names of the functions appearing in the PDE.
\item {\tt x}: List of independent variables (coordinates).
\item {\tt ud}: List of \maple\ names for the discrete representations 
   of the functions (i.e.~the grid function names).
\item {\tt j}: Length-$n$ list of names of indices corresponding to 
					each coordinate.
\item {\tt h}: Length-$n$ list of coordinate spacings in each direction.
\item {\tt p}: Approximation order for finite difference scheme.
\item {\tt off}: Length-$n$ list of offsets which defines the off-centring
               of the approximation in each coordinate direction.
\end{enumerate}

Its use is illustrated here for the case of a nonlinear Poisson equation
in 3 dimensions, where compactified Cartesian coordinates are adopted:
\begin{verbatim}
   > # Define aliases for the functional dependence of u and f.
   > alias(u=u(chi,eta,zeta), f=f(chi,eta,zeta));
\end{verbatim}
\beq
\nonumber
u,\,f
\eeq
\begin{verbatim}
   > # Define the PDE.
   > POI3DCP := (1-chi**2)*diff((1-chi**2)*diff(u,chi),chi)+
           (1-eta**2)*diff((1-eta**2)*diff(u,eta),eta)+
           (1-zeta**2)*diff((1-zeta**2)*diff(u,zeta),zeta)+sigma*u**2=f;
\end{verbatim}
\begin{multline}
\nonumber
\left( 1-{\chi}^{2} \right)  \left( -2\,\chi\,{\frac {d}{d\chi}}u+ \left( 1-{\chi}^{2} \right) {\frac {d^{2}}{d{\chi}^{2}}}u \right) + \left( 1-{\eta}^{2} \right)  \left( -2\,\eta\,{\frac {d}{d\eta}}u+ \left( 1-{\eta}^{2} \right) {\frac {d^{2}}{d{\eta}^{2}}}u \right) \\
\mbox{}+ \left( 1-{\zeta}^{2} \right)  \left( -2\,\zeta\,{\frac {d}{d\zeta}}u+ \left( 1-{\zeta}^{2} \right) {\frac {d^{2}}{d{\zeta}^{2}}}u \right) +\sigma\,{u}^{2}=f
\end{multline}
\begin{verbatim}
   > # Generate the residual
   > resndoff(POI3DCP, [u,f], [chi,eta,zeta], [ud,fd], [i,j,k], [hx,hy,hz],
              2, [0,0,0]);
\end{verbatim}
\begin{multline}
\nonumber
\left( 1-{\chi_{{i}}}^{2} \right)  \left[ -2\,\chi_{{i}} \left( -\fr{1}{2}\,{\frac {{\it ud}_{{i-1,j,k}}}{{\it hx}}}+\fr{1}{2}\,{\frac {{\it ud}_{{i+1,j,k}}}{{\it hx}}} \right) + \left( 1-{\chi_{{i}}}^{2} \right)  \left( {\frac {{\it ud}_{{i-1,j,k}}}{{{\it hx}}^{2}}}-2\,{\frac {{\it ud}_{{i,j,k}}}{{{\it hx}}^{2}}}+{\frac {{\it ud}_{{i+1,j,k}}}{{{\it hx}}^{2}}} \right) \right]\\
\mbox{} + \left( 1-{\eta_{{j}}}^{2} \right)  \left[ -2\,\eta_{{j}} \left( -\fr{1}{2}\,{\frac {{\it ud}_{{i,j-1,k}}}{{\it hy}}}+\fr{1}{2}\,{\frac {{\it ud}_{{i,j+1,k}}}{{\it hy}}} \right) + \left( 1-{\eta_{{j}}}^{2} \right)  \left( {\frac {{\it ud}_{{i,j-1,k}}}{{{\it hy}}^{2}}}-2\,{\frac {{\it ud}_{{i,j,k}}}{{{\it hy}}^{2}}}+{\frac {{\it ud}_{{i,j+1,k}}}{{{\it hy}}^{2}}} \right)\right] \\
\mbox{} + \left( 1-{\zeta_{{k}}}^{2} \right)  \left[ -2\,\zeta_{{k}} \left( -\fr{1}{2}\,{\frac {{\it ud}_{{i,j,k-1}}}{{\it hz}}}+\fr{1}{2}\,{\frac {{\it ud}_{{i,j,k+1}}}{{\it hz}}} \right) + \left( 1-{\zeta_{{k}}}^{2} \right)  \left( {\frac {{\it ud}_{{i,j,k-1}}}{{{\it hz}}^{2}}}-2\,{\frac {{\it ud}_{{i,j,k}}}{{{\it hz}}^{2}}}+{\frac {{\it ud}_{{i,j,k+1}}}{{{\it hz}}^{2}}} \right) \right]\\
\mbox{}+\sigma\,{{\it ud}_{{i,j,k}}}^{2}-{\it fd}_{{i,j,k}}=0
\end{multline}

 \resetcounters

\chapter{Review of Relaxation Methods} \label{ap:ngs}

\def\u{{\bf u}}
\def\ur{{\bf {\tilde u}}}
\def\L{{\bf L}}
\def\f{{\bf f}}
\def\c{{\bf c}}
\def\r{{\bf r}}
\def\rr{{\bf {\tilde r}}}
\def\T{{\bf T}}
\def\i#1{{}^{(#1)}}
\def\ii#1{{}^{[#1]}}

In this appendix we provide a quick review of relaxation techniques
as applied to large sparse systems of algebraic equations (both linear
and nonlinear).  Relaxation methods are used in two distinct capacities
in the 3D code that we describe in the main body of the thesis:
\begin{enumerate}
\item In the case of the hyperbolic PDEs, point-wise Newton-Gauss-Seidel
	relaxation is used to {\em solve} the time-implicit algebraic 
	equations that result from our use of an $O(h^2)$ Crank-Nicholson
	scheme.
\item For the elliptic PDEs, point-wise Newton-Gauss-Seidel is used 
	as a {\em smoother} in the context of the multigrid solution 
   of the discrete equations, which again arise from an $O(h^2)$ 
   approximation of the PDEs.
\end{enumerate}

The reader who is interested in full details concerning relaxation
methods and related iterative techniques for the solution of large 
systems of equations should consult the classic reference 
by Varga~\cite{Varga:MIA}.
\section{The Linear Case: Jacobi and Gauss-Seidel Relaxation}
We write a general linear system of equations in the form
\beq \label{eq:luf_GS}
\L\u = \f \, ,
\eeq
where $\L$ is a $N \times N$ matrix, and $\u$ and $\f$ are 
$N$-component column vectors.  In the typical case of interest, $\u$ 
will enumerate all of the discrete unknowns, $u(t^n,x_i,y_j,z_k)$,
associated with the finite difference approximation of some continuum 
function, $u(t,x,y,z)$, at some specific discrete time, $t=t^n$.  
The size of the system will then be given by $N=n_x\times n_y \times n_z$,
where $n_x$, $n_y$ and $n_z$ are the number of grid points in 
the $x$, $y$ and $z$ directions, respectively.   Due to the locality 
of finite difference operators, the matrix $\L$ will generally
be very sparse: that is, unless $n_x$, $n_y$ and $n_z$ are very 
small, most of its entries will be 0, and the fraction
of non-zeros will in fact tend to 0 as the mesh spacing 
$h$ approaches 0.   As is well known,
even if one adopts a specially crafted ordering of the unknowns 
(see~\cite{george_liu} for example), the solution 
of~(\ref{eq:luf_GS}) using standard methods of 
numerical linear algebra---that is, using 
some form of Gaussian elimination such as LU 
decomposition~\cite{burden}---is 
extremely inefficient.  This is already the case for discretizations
of problems in 2 spatial dimensions, and is even more so for 3D 
calculations.   For example, consider the situation where the 3D Poisson
equation in $(x,y,z)$ coordinates has been discretized on a uniform mesh with 
$n=n_x=n_y=n_z$ grid points per edge of the computational domain.
Then the size of the system is $N=n^3$.  We assume that we have used the
standard $O(h^2)$ 7-point approximation for the discrete Laplacian, 
which couples a particular unknown to its nearest neighbours in each 
of the three coordinate directions, and that we have numbered the 
unknowns in so called lexicographic order 
$(((u_{ijk}, \, i =1,\ldots n), \, j = 1,\ldots n), \, k = 1,\ldots n)$.
Then the bandwidth of $\L$ is $O(n^2)=O(N^{2/3})$, and the cost 
for solution via Gaussian elimination will be $O(N^{7/3})$ in computational
time and up to $O(N^{5/3})$ in memory.  In practice
this precludes direct solution of~(\ref{eq:luf_GS}) for any but the smallest
3D problems.

The class of {\em iterative} techniques known as relaxation methods 
were developed decades ago largely in response to the need to 
efficiently solve the large systems of algebraic equations arising from 
the finite-difference discretization of PDEs, particularly those 
of elliptic type.  Relaxation methods directly exploit the sparsity 
structure of these systems, and have the further advantage of having
extremely economical memory requirements, since the system of equations
is not explicitly stored.

We thus start with the general notion of an iterative, or {\em fixed point}
technique for the solution of~(\ref{eq:luf_GS}).  We note that the 
system can be rewritten as 
\beq \label{eq:uTu}
\u = \T \u + \c \, ,
\eeq
where $\T$ is the $N\times N$ {\em update matrix} and 
$\c$ is a constant vector that depends on the specific iterative 
method used.  The idea is to choose  $\T$ and $\c$ such that 
starting from some initial estimate, $\u\i 0$, of $\u$, the iteration 
\beq \label{eq:uTun}
\u\i{k+1} = \T \u\i k + \c \, ,
\eeq
has a fixed point $\u$ which satisfies~(\ref{eq:luf_GS}).  That is,
we want
\beq
\lim_{k\to\infty} \u\i k = \u \, .
\eeq
Whether or not the iteration~(\ref{eq:uTun}) converges is determined
by the eigenvalues of the update matrix $\T$.
More precisely, the necessary and sufficient condition for 
convergence is that the {\em spectral radius}, $\rho(\T)$, of 
$\T$ satisfy $\rho(\T)<1$.  We recall that the spectral radius 
may be defined as 
\beq
\rho(T) = \max_{1\leq i\leq N} |\la_i| ,
\eeq
where $\la_i$ are the eigenvalues of $\T$. 
For strictly diagonally dominant matrices $\T$, which satisfy
\beq
\label{eq:dd}
|T_{ii}| > \sum_{j=1 \atop j\neq i}^N |T_{ij}| ,
\eeq
two iterative schemes become particularly interesting since they are known 
to converge for any initial estimate $\u\i0$. This observation is 
pertinent to the case of equations resulting from FDA approximation since 
such systems often satisfy some form of diagonal dominance, which although
frequently weaker than~(\ref{eq:dd}), is still enough to ensure convergence
of the methods we will outline.

Although the formal analysis of relaxation methods (again, see~\cite{Varga:MIA})
typically proceeds via a matrix approach which involves a decomposition,
or splitting, of ${\bf L}$, for our purposes it is more intuitive 
to describe the techniques as they are typically {\em implemented}
as computer code.  We thus write~(\ref{eq:luf_GS}) in the form
\beq
\label{eq:luf-comp}
\sum_{j=1}^N L_{ij} u_j = f_i, \quad i = 1,2,\dots,N \, ,
\eeq
which amounts to explicitly writing out the action of each row
of $\L$ on $\u$, and equating it to the corresponding element of $\f$.
We then view any of the $N$ subsystems of equations defined 
by~(\ref{eq:luf-comp}) as an equation for the {\em single} unknown,
$u_i$.  Solving for $u_i$ we get 
\beq
u_i = \fr{1}{L_{ii}} \lhb -\sum_{j=1 \atop j\neq i}^N L_{ij} u_j + f_i  
    \rhb \, ,
\eeq
and the sequence of approximations starting from the initial guess $\u\i0$
can be generated component by component using
\beq
\label{eq:jac}
u^{(k)}_i = \fr{1}{L_{ii}} \lhb -\sum_{j=1 \atop j\neq i}^N L_{ij} u^{(k-1)}_j + f_i  \rhb
, \quad i = 1,2,\dots,N .
\eeq
The method defined by~(\ref{eq:jac}) is known as 
\emph{Jacobi iteration}.  We note that it requires that we 
store values for both the current approximation, $\u\i k$, as 
well as the previous estimate, $\u\i{k-1}$. 
Crucially, however, for 
systems derived from FDAs there is no need to explicitly 
store the elements of $\L$.  Rather, the computation of the right
hand side of~(\ref{eq:jac}) simply amounts to the evaluation of 
the finite difference formula at a given grid point, so that the 
components $L_{ij}$ are effectively ``hard coded'' into the 
computer program.  Thus, the memory requirements for the 
technique---and for all relaxation methods---are the optimal $O(N)$.
We also note at this point that the complete process of iterating 
the solution estimate from $\u\i k$ to $\u\i{k+1}$ is frequently
called a {\em relaxation sweep}, in appeal to the notion that 
one ``sweeps'' through the finite difference grid, updating one 
discrete unknown at a time.

The other relaxation technique we wish to consider can be viewed 
as a simple modification of the Jacobi iteration that aims
to speed convergence by using the most recently computed components
of the solution estimate when they are available.  So, assuming 
that we are updating the unknowns $u_i\i k$ in the order 
$i=1,2,\ldots,\,N$, then for $j<i$ we will have already computed
$u_j\i k$, while for $j>i$ we will need to use values $u_j\i{k-1}$.
We thus modify~(\ref{eq:jac}) as follows:
\beq \label{eq:GS}
u^{(k)}_i = \fr{1}{L_{ii}} \lhb - \sum_{j=1}^{i-1} L_{ij} u^{(k)}_j 
- \sum_{j=i+1}^N L_{ij} u^{(k-1)}_j + f_i  \rhb
, \quad i = 1,2,\dots,N \, .
\eeq
This defines the {\em Gauss-Seidel} iteration 
for the solution of~(\ref{eq:luf_GS}).  In practice, Gauss-Seidel 
relaxation {\em does} tend to require fewer sweeps than Jacobi 
to achieve convergence to a specified tolerance.~\footnote{It should
be noted, however, that both Gauss-Seidel and Jacobi generally 
have the same asymptotic convergence {\em rate}~\cite{Varga:MIA}.}
Additionally, Gauss-Seidel iteration only needs storage for a single
instance of the unknown vector, which is half of that required by
the Jacobi method.

We should also emphasize at this juncture that the 
methods given by~(\ref{eq:jac}) and (\ref{eq:GS}) can be further 
qualified as defining {\em point-wise} Jacobi and {\em point-wise}
Gauss-Seidel relaxations, respectively.  This distinguishes the 
techniques from extensions of the methods in which groups of 
unknowns---$u(x_i,y_j,z_k), \, i =1,\ldots,\,N$, for fixed $j$ and $k$,
for example---are updated simultaneously.  These generalizations 
are known as {\em line} relaxation methods, and although we 
have mentioned them briefly in Sec.~\ref{sec:space_compact},
we did not use them in any of the work described in this thesis.

In preparation for our discussion of nonlinear systems in the 
next section it is useful to describe the point-wise Gauss-Seidel
iteration~(\ref{eq:GS}) in terms of the {\em residual vector}
associated with the approximate solution.  First, given the 
approximation $\u\i k$ that is defined after the completion of 
the $k$-th relaxation sweep, we define the residual vector $\r\i k$
by
\beq
	\r\i k \equiv \L \u\i k - \f \, ,
\eeq
so that, assuming that the Gauss-Seidel iteration converges, 
we have 
\beq
	\lim_{k\to\infty} \r\i k = {\bf 0}\, ,
\eeq
where $\bf 0$ is the $N$-component 0--vector.  We again emphasize 
that we can only compute ${\bf r}^{k}$ {\em after} the $k$-th pass
through the solution unknowns has been completed.  We therefore 
further introduce the concept of a {\em running residual}, 
generically denoted $\rr$, which has components that are 
continually changing as 
individual unknowns are updated.  We thus define the (running)
unknown vector $\ur_i\i k$, which at any iteration $k$, has the following
``instantaneous'' values when we have just updated unknown $i-1$
and are about to update unknown $i$:
\beq
\label{eq:ur}
\ur^{(k)}_i = 
\left[ u^{(k)}_1,u^{(k)}_2,\dots,u^{(k)}_{i-1},u^{(k-1)}_i,
\dots,u^{(k-1)}_N \right]^T \, .
\eeq
With this definition, the running residual, $\rr_i\i k$,
is given by
\beq
\label{eq:def-rr}
	\rr_i\i k \equiv \L \ur_i\i k - \f \, ,
\eeq
where the subscript $i$ reminds us that $\rr_i\i k$ has components 
that are constantly changing as we sweep through the grid, updating
unknowns one by one.
Using (\ref{eq:GS}) and (\ref{eq:def-rr}), we can explicitly write the 
components, $[\rr_i\i k]_m$, of the running residual vector as 
\beq 
\label{eq:run_res}
 [\rr_i\i k]_m  = \sum_{j=1}^{i-1}L_{mj} u^{(k)}_j +
                \sum_{j=i+1}^{N}L_{mj} u^{(k-1)}_j+
                                L_{mi} u^{(k-1)}_i- f_m .
\eeq
For the particular case of the $i$-th residual this last expression
gives us 
\beq
[\rr_i\i k]_i = \sum_{j=1}^{i-1}L_{ij} u^{(k)}_j +
               \sum_{j=i+1}^{N}L_{ij} u^{(k-1)}_j+
                               L_{ii} u^{(k-1)}_i- f_i \, ,
\eeq
which upon insertion into~(\ref{eq:GS}) 
yields a more compact form of the point-wise Gauss-Seidel 
iteration:
\beq \label{eq:GS1}
u^{(k)}_i = u^{(k-1)}_i - \frac{[\rr_i\i k]_i}{L_{ii}} \, .
\eeq
This expression will be generalized to the case of nonlinear relaxation
in the next section.

\section{The Nonlinear Case: Newton-Gauss-Seidel Relaxation}
In situations such as ours, where the governing PDEs are nonlinear, 
finite difference approximation will naturally lead to large sparse,
systems of {\em nonlinear} algebraic equations.
Newton's method~\cite{burden} then provides a route to extend the 
techniques described above so that they can be used to solve 
nonlinear systems.

We begin then by reviewing Newton's method for the solution of 
single nonlinear equation in a single unknown, $u$.  We write 
the equation in the canonical form 
\beq
\label{eq:fu}
	f(u) = 0 \, ,
\eeq
and note that the technique is itself iterative. Thus 
we must start with some initial estimate, $u\ii 0$, of the solution,
and then generate iterates, $u\ii n$, such that, assuming that the 
method converges (which generally is dependent on the quality 
of the initial guess), we have $\lim_{n\to\infty}u\ii n=u$.
The Newton iteration for~(\ref{eq:fu}) is then given by 
\beq
\label{eq:newt}
	u\ii{n+1}=u\ii{n} - \frac{f\lb u\ii n \rb}{f'\lb u\ii n \rb} \, ,
\eeq
where $f'(u)\equiv df(u)/du$.
Defining the residual, $r\ii n$, associated with the iterate,
$u\ii n$, as 
\beq
	r\ii n \equiv f\lb u\ii n \rb \, ,
\eeq
we can rewrite~(\ref{eq:newt}) as 
\beq
\label{eq:newtr}
	u\ii{n+1}=u\ii{n} - \frac{r\ii n}{f'\lb u\ii n \rb} \, .
\eeq

We now assume that the system of $N$ equations that result 
from the finite difference discretization of our PDEs has been 
written in the canonical form 
\beq \label{eq:NL}
F_i[\u] = 0 \, ,\qquad i = 1,2,\dots,N\,,
\eeq
where each of the $F_i$, $i = 1,\ldots,N$ is a nonlinear function 
of the unknowns, $\u=[u_1,u_2,\ldots,u_N]^T$ (but only a few of
them in general, again due to the locality of finite difference 
operators).

The application of relaxation to nonlinear equations again involves 
relaxation sweeps through the unknowns, so focusing on the extension
of the Gauss-Seidel method, and in analogy to what we did in 
the previous section, we define the components,
$[\rr_i\i k]_m$, of the running residual vector by 
\beq
[\rr_i\i k]_m \equiv F_m[\ur_i\i k] \, .
\eeq
Here, the running solution vector, $\ur_i\i k$, is defined 
by~(\ref{eq:ur}) as previously, and the subscript $i$ again emphasizes 
that the elements of the running vectors are constantly changing as we
sweep through the grid, updating each unknown in turn.

With this definition, and using~(\ref{eq:newtr}),
the (one-step) point-wise Newton-Gauss-Seidel relaxation method is 
defined by 
\beq
	u_i\i k = u_i\i{k-1} - \frac{[\rr_i\i k]_i}{J_{ii}} \, ,
\eeq
where the ``diagonal Jacobian element'', $J_{ii}$ is given by 
\beq
	J_{ii} = \left.\frac{\pa F_i\left[\u\right]}{\pa u_i}\right\vert_{u_i=u_i\i {k-1}}
\eeq

The additional nomenclature ``one step'' indicates that we do not use
Newton's method to fully {\em solve} any of the individual nonlinear equations
as we visit the unknowns; instead, we apply only {\em one} iteration
of the Newton technique before moving on to the next unknown.  This 
strategy is largely motivated by the observation that a complete 
solution of any individual equation will generally represent some 
amount of wasted computational work, since as soon as one of the 
neighbouring unknowns is modified by the subsequent relaxation process,
the equation will no longer be satisfied. 

\end{document}